\documentclass[usenatbib,usegraphicx,apjfonts,floatfix]{emulateapj}
\usepackage{morefloats,epsf,amsmath,apjfonts,afterpage,pgf}

\newcommand{\kms}{\ensuremath{\,\mbox{km}\,\mbox{s}^{-1}}}

\newcommand{\HI}{H\,{\sc i}}
\newcommand{\degree}{\ensuremath{^{\circ}}}

\newcommand{\MLstar}{\ensuremath{\Upsilon_{\star}}}
\newcommand{\Ups}{\ensuremath{\Upsilon_{\star}^{3.6}}}
\newcommand{\tsm}{\ensuremath{3.6\ \mu{\rm m}}}

\shorttitle{THINGS rotation curves}
\shortauthors{de Blok et al.}

\begin{document}
\title{High-Resolution Rotation Curves and Galaxy Mass Models from THINGS}
\author{W.J.G.~de~Blok\altaffilmark{1}, F. Walter\altaffilmark{2}, E. Brinks\altaffilmark{3},
C. Trachternach\altaffilmark{4}, S-H. Oh\altaffilmark{5}, R.C. Kennicutt, Jr.\altaffilmark{6}}

\altaffiltext{1}{Dept.\ of Astronomy, University of Cape Town,
  Rondebosch 7700, South Africa}
\email{edeblok@ast.uct.ac.za}

\altaffiltext{2}{Max Planck Institute f\"ur Astronomy, K\"onigstuhl
  17, 69117 Heidelberg, Germany} \email{walter@mpia.de}

\altaffiltext{3}{Centre for Astrophysics Research, Science \&
  Technology Research Institute, University of Hertfordshire, Hatfield
  AL10 9AB, United Kingdom} \email{e.brinks@herts.ac.uk}

\altaffiltext{4}{Astronomisches Institut, Ruhr-Universit\" at Bochum,
  Universit\" atstrasse 150, 44780 Bochum, Germany}
\email{trachter@astro.rub.de}

\altaffiltext{5}{Research School of Astronomy \& Astrophysics, Mount Stromlo Observatory, Cotter Road, Weston Creek, ACT 2611, Australia}

\email{seheon@mso.anu.edu.au} 

\altaffiltext{6}{Institute of Astronomy, University of Cambridge, Madingley Road, Cambridge CB3 0HA, United Kingdom}
\email{robk@ast.cam.ac.uk}

\begin{abstract}
  We present rotation curves of 19 galaxies from THINGS, The \HI\
  Nearby Galaxy Survey.  The high spatial and velocity resolution of
  THINGS make these the highest quality \HI\ rotation curves available
  to date for a large sample of nearby galaxies, spanning a wide range
  of \HI\ masses and luminosities.  The high quality of the data
  allows us to derive the geometrical and dynamical parameters using
  H\,{\sc i} data alone.  We do not find any declining rotation curves
  unambiguously associated with a cut-off in the mass distribution out
  to the last measured point.  The rotation curves are combined with
  3.6 $\mu$m data from SINGS (\emph{Spitzer} Infrared Nearby Galaxies
  Survey) to construct mass models. Our best-fit, dynamical disk
  masses, derived from the rotation curves, are in good agreement with
  photometric disk masses derived from the 3.6 $\mu$m images in
  combination with stellar population synthesis arguments and two
  different assumptions for the stellar Initial Mass Function (IMF).
  We test the Cold Dark Matter-motivated cusp model, and the
  observationally motivated central density core model and find that
  (independent of IMF) for massive, disk-dominated galaxies, all halo
  models fit apparently equally well; for low-mass galaxies, however,
  a core-dominated halo is clearly preferred over a cuspy halo.  The
  empirically derived densities of the dark matter halos of the
  late-type galaxies in our sample are half of what is predicted by
  CDM simulations, again independent of the assumed IMF.
\end{abstract}


\section{Introduction}

Early observations of the rotation curves of spiral galaxies clearly
indicated the presence of  ``missing'' matter in these objects (e.g.,
\citealt{bosma_phd,rubinford,bosma81}; see also
\citealt{rotcurreview}).  This ``missing'' or ``dark'' matter, as it
is now more commonly called, has since become one of the pillars of
modern cosmology. An elaborate framework, explaining and describing
the properties of the Universe has been built, involving a
cosmological constant $\Lambda$, and a form of collisionless Cold Dark
Matter (CDM). The $\Lambda$CDM paradigm provides a comprehensive
description of the Universe at large (as shown by, e.g., the WMAP
results, see \citealt{spergel2006}), but problems on smaller (galaxy)
scales remain.

A well-known problem is that galaxy-scale structures formed in
cosmological $\Lambda$CDM simulations are too compact. One symptom of
this is the well-known missing satellite problem (simulations produce
an overabundance of dwarf galaxy halos, cf.~\citealt{moore99}).  This
seems to have been alleviated somewhat by the recent discovery of many
faint dwarfs in our Local Group -- some of which seem to be dominated by
dark matter -- but the discrepancy is still significant
\citep{simongeha}.

The so-called cusp-core discrepancy (the predicted dark matter density
profiles in galaxy halos are too steep; cf.~\citealt{dB2001}) is
another manifestation of the (too) compact structures in CDM
simulations.  Galaxy rotation curves have in the last decade taken
center stage in the debate about this problem.  Going into some more
detail, results from cosmological $\Lambda$CDM simulations suggest
that the density profiles of dark matter halos should be nearly
universal, independent of the mass of the halo. This density profile
has a characteristic steep mass-density slope in the inner parts which
can be approximated with a power-law $\rho \sim r^{\alpha}$ with
$\alpha \la -1$; this is the so-called ``cusp''
\citep[e.g.,][]{NFW96,NFW97,klypin2001,hayashi,diemand2005}.

Extensive observational determinations of this inner mass-density
distribution seem, however, to indicate that the mass-density profiles
of dark matter halos can be better described using an approximately
constant-density inner ``core'' ($\rho \sim r^{\alpha}$ with $\alpha
\simeq 0$). This core has a typical size of order a kpc
\citep[e.g.,][]{moore94,deblok1996,dBMcG1997,dB2001,dBB2002,mar2002,weldrake03,gentile2005,rachel,gentile2007}.
The discrepancy between these two outcomes has been much debated, as the
presence of cusps is a fundamental feature of current cosmological
simulations, confirmed many times by large increases in resolution and computing power
that have been achieved over the years.

A large part of this debate has focused on possible
systematic effects in the observations. Resolution of the data, 
accuracy of the positions of centers of galaxies, as well as presence
of non-circular motions have featured prominently in these discussions
\citep[e.g.,][]{swaters_phd,frankvdb,SMT,simon2003,rhee2004}.
Nevertheless, multiple, repeated, independent long-slit measurements
of rotation curves as well as analyses using two-dimensional optical
velocity fields consistently yield results suggesting the presence of
a dark matter core in the inner parts of disk galaxies
\citep[e.g.,][]{mar2002,
  edb_sydney,dBB2002,rachel,gentile2007,spano}\footnote{Note that the paper
  by \citet{swaters03}, which is often quoted as an observational
  counter-example, states that their results ``cannot rule out steep
  slopes'', but also that ``halos with constant density cores provide somewhat
  better fits''.}.

\HI\ synthesis observations of disk galaxies would, in principle, be
eminently suited to address this problem. The \HI\ disks of galaxies
have a higher filling factor than the often patchy H$\alpha$
distributions used for optical rotation curves. Measuring accurate
velocities and modeling them across the entire, two-dimensional,
filled \HI\ disks is therefore one of the strengths of radio
synthesis observations. This means that possible effects of
non-circular motions, which may not always be recognized in one-dimensional
slit observations, can be analyzed within the context of the
surrounding rotating disk \citep{oh2007,clemens2007}.  Furthermore, a  determination
of the position of the dynamical center can be made, enabling a direct and independent
comparison with the position of the photometric center \citep{clemens2007}. In short, the rotation curve,
as well as the uncertainties that could affect it, can be constrained
by the same observation.

The one major disadvantage which hitherto hampered making
the fullest possible use of \HI\ data in these analyses, has been
their modest angular resolution, especially when compared to optical
observations: to directly address cosmologically relevant questions
regarding the distribution of dark matter requires high spatial
resolution.  For example, unambiguously distinguishing between a core
and a cusp, requires physical resolutions better than $\sim 1$ kpc
\citep{edb_sydney}.  Also, to determine the importance of non-circular
motions within the disks of galaxies one needs to resolve the size
scales associated with features that cause these motions, such as
bars, spiral arms and oval distortions.

The majority of published \HI\ synthesis observations have at best beam sizes
$\ga 15''$, meaning they can reach the desired sub-kpc resolution only
out to distances of $\sim 4$ Mpc.  This, of course, severely limits
the number of galaxies available for study.  THINGS (The \HI\ 
Nearby Galaxy Survey) has addressed this by observing in \HI\ a sample
of 34 nearby galaxies at the desired sub-kpc resolution. To achieve
this, it used the NRAO\footnote{The National Radio Astronomy
  Observatory is a facility of the National Science Foundation
  operated under cooperative agreement by Associated Universities,
  Inc.}  Very Large Array (VLA) in its B, C and D configurations.  The
total observing time (including data retrieved from the archive) was
$\sim 500$ hours.  The maximum angular resolution that can be achieved
with these data is $\sim 6''$ for robust weighting and $\sim 12''$ for
natural weighting. The velocity resolution is 5.2 km s$^{-1}$ or
better. With the sample galaxies at distances between 2 and 15 Mpc,
the typical linear resolution varies between $\sim 100$ and $\sim 500$
pc.  For further discussion on the sample, aims and scientific goals
of THINGS, as well as the technical details of the observations, we
refer to \citet{THINGS1}.

The current paper focuses on the rotation curves of THINGS galaxies.
We use a sub-sample of the THINGS galaxies (discussed in Sec.~2) and
present their velocity fields, tilted-ring models and the rotation
curves. Additionally, we combine our \HI\ data with 3.6 $\mu$m data
from SINGS, the \emph{Spitzer} Infrared Nearby Galaxies Survey
\citep{sings} and derive well-constrained stellar mass-to-light ratios
for the stellar disks.  We use these to investigate the mass
distributions in the sample galaxies using core- and cusp-dominated
dark matter halo models.  A number of galaxies presented here have
been observed in the past, in some cases as part of the pioneering
work on \HI\ rotation curves in the 1970s and 1980s\footnote{See,
  e.g., Fig.~1 in \citet{bosma81} for a proto-THINGS depiction of the
  velocity fields of 22 nearby spiral galaxies. Half of these are also
  part of THINGS.}, but this is the first time that such a large
sample of galaxies is analyzed and modeled in a uniform way at sub-kpc
resolution.

In Sect.\ 2 we summarize the selection of our sample. Section 3
describes the creation of the velocity fields as well as the
derivation of the rotation curves. Section 4 presents the rotation
curves and velocity fields of the individual galaxies in more
detail. In Sect.\ 5 the mass models are presented, and in Sect.\ 6
these models are discussed in more detail for individual galaxies. The
results are discussed in Sect.\ 7 and summarized in
Sect.\ 8. Technical aspects of the analysis are presented in
Appendix~A.

\section{The Sample}

Our goal in this paper is to derive rotation curves of THINGS galaxies
which depend on a minimum amount of external information. That is, we
aim to derive the geometrical and kinematical parameters of the
galaxies (inclination, position angle, rotation velocity, etc.) as
much as possible from the THINGS \HI\ data alone.  This limits our
analysis to rotationally dominated galaxies with favorable
orientations. Although using additional information (such as derived
from optical data) would, in principle, increase the number of THINGS
galaxies for which a rotation curve can be determined,
the unquantifiable systematic effects associated with this introduce
their own difficulties, as discussed below.

There are practical upper and lower limits to the inclinations
of galaxies for which one can reliably derive  rotation curves
using standard methods.  At very high inclinations ($i \ga
80^{\circ}$) the line of sight crosses a large range of projected
velocities and it becomes difficult to quantify the typical rotation
velocity (although not impossible, see e.g., \citealt{kregel}). Such
high-inclination galaxies are, however, not present in the THINGS
sample.

At low inclinations, the practical limit lies at $\sim 40^{\circ}$
\citep{begeman87,begeman89}.  At lower inclinations the velocity
dispersion of the gas starts to become an important contaminant. As
rotation takes place in a two-dimensional disk, the line-of-sight
(projected) component of this rotation velocity decreases in amplitude
with decreasing inclination, whereas the contribution due to random
motions, which take place in three dimensions, remains the same. This
leads to unacceptably large uncertainties in the deprojected
rotational velocity.

Independent of inclination, we also require our sample galaxies to be
dominated by rotation. This is not the case for a small number of
THINGS galaxies.  These are M81dwA \citep{sargent} and M81dwB, as well
as NGC 1569 (dynamics disturbed by a starburst; \citealt{stil}), NGC
3077 (heavily interacting with M81; \citealt{walter02}) and NGC 4449
(tidally disturbed; \citealt{hunter4449}). We do not consider these
galaxies further.

To quantify our sample selection further, we consider indicative
\HI\ inclinations derived from ellipse fits to the outer \HI\ disks,
in addition to optical inclinations from LEDA \citep{leda}. These
values are listed in Table~\ref{table1}. In determining the
\HI\ inclinations, using the maps presented in \citet{THINGS1}, care
was taken that these fits were not affected by the tails, clouds, and
other features often seen in the outer parts of the \HI\ disks. The
\HI\ inclinations listed here are deemed to be as reliable as their
optical counterparts, and in some cases even preferable (cf.\ the
values for DDO 154 and the associated images in \citealt{THINGS1}).
We stress that determining precise values for these inclinations is
not the issue here, as we merely use them to clarify our sample
selection. They are not used for any subsequent dynamical analyses.

\begin{deluxetable*}{lcclcc}
\tablewidth{0pt}
\tablecaption{Indicative global inclinations of the THINGS galaxies}
\tablehead{
\colhead{Name} & \colhead{$i_{\rm HI}\ (^\circ)$} &\colhead{$i_{\rm LEDA}\ (^\circ)$}& \colhead{Name} & \colhead{$i_{\rm HI}\ (^\circ)$}&\colhead{$i_{\rm LEDA}\ (^\circ)$}}
\startdata
NGC 628        &  15   &21    & NGC 4449       &\nodata&54\\
{\bf NGC 925}  &  50   &58    & {\bf NGC 4736} &  44   &29  \\
NGC 1569      & \nodata &55   & {\bf NGC 4826} &  64   &58  \\
{\bf NGC 2366}& 65      &71   & {\bf NGC 5055} &  51   &53  \\
{\bf NGC 2403}& 55      &61   & NGC 5194       &  30   &32  \\
{\bf NGC 2841}& 69      &61   & NGC 5236       &  31   &21  \\
{\bf NGC 2903}& 66      &64   & NGC 5457       &  30   &12  \\
{\bf NGC 2976}& 54      &58   & {\bf NGC 6946} &  35   &17  \\
{\bf NGC 3031}& 59      &58   & {\bf NGC 7331} &  77   &67  \\
NGC 3077      & \nodata &35   & {\bf NGC 7793} &  43   &56  \\
NGC 3184      & 29      &12   & {\bf IC 2574} & 51      &67 \\
{\bf NGC 3198}& 72      &74   & DDO 53        & 33      &46 \\
NGC 3351      &  39   &53     & {\bf DDO 154} & 70      &29 \\
{\bf NGC 3521} &  69   &58    & Ho I          & 27      &17 \\
{\bf NGC 3621} &  62   &66    & Ho II         & 31      &45 \\
{\bf NGC 3627} &  61   &63    & M81 dwA       & 27 &\nodata \\
NGC 4214       &  38   &39    & M81 dwB       & 28      &63 \\
\enddata
\tablecomments{Galaxies that have rotation curves presented in this
  paper are printed in bold. For some galaxies a representative
  inclination could not be determined. All inclinations listed here
  are indicative and are not used in the dynamical analysis presented in
  this paper.}
\label{table1}
\end{deluxetable*}

A first analysis of the velocity fields presented in \citet{THINGS1}
showed that THINGS galaxies with indicative \HI\ inclinations higher
than 40$^{\circ}$ are all suitable for rotation curve analysis. These
galaxies are all discussed in the current paper. Galaxies with
\HI\ inclinations less than 30$^{\circ}$ are unsuitable and will not
be discussed any further here.  Of the galaxies with \HI\ inclinations
between $30^{\circ}$ and $40^{\circ}$, only NGC 6946 proved suitable.
The reason why for this galaxy a rotation curve can be derived,
whereas this was not possible for, e.g., NGC 4214 (which has a
comparable indicative \HI\ inclination), is the large ratio of galaxy
size to beam size.  The resulting larger number of resolution elements
in the NGC 6946 velocity field help constrain the kinematics better
than for the other galaxies in that (\HI) inclination range.

A look at Table~\ref{table1} shows that there are a few cases where
the \HI\ and optical inclinations differ substantially.  These can,
however, be readily explained.  DDO 154 ($i_{\rm HI}> 40\degree$,
$i_{\rm LEDA} < 40\degree$) has a well-defined and regular \HI\ disk,
but a low surface brightness ill-defined optical component. The
\HI\ inclination is thus far more reliable.  M81 dwB ($i_{\rm HI}<
40\degree$, $i_{\rm LEDA} > 40\degree$) has a high-surface brightness
central bar-like structure which affects the optically derived
inclination.  NGC 3351 ($i_{\rm HI}< 40\degree$, $i_{\rm LEDA} >
40\degree$) has a fairly face-on and regular \HI\ disk, but optically
is dominated by a very strong central bar.  NGC 4736 ($i_{\rm HI}>
40\degree$, $i_{\rm LEDA} < 40\degree$) is the only galaxy where the
outer \HI\ disk seems to have a genuinely higher inclination than the
central optical component\footnote{This is consistent with the results
  we will derive later in this paper. NGC 4736 shows a well defined
  inclination trend, varying from $i \sim 30^{\circ}$ in the inner
  parts to $i \sim 50^{\circ}$ in the outer parts.}.

One might argue that our choice to only consider galaxies for which
the geometrical parameters can be kinematically determined is
needlessly strict, and that rotation curves could in principle be
derived for the lower inclination galaxies by using additional
information, such as the axis ratios of the optical or \HI\ disks.
However, in these cases, the best one can do is assume a
\emph{constant} inclination.  This will then ignore intrinsic
inclination changes (due to, e.g., the presence of warps), which can
strongly affect the derived rotation velocities: an estimate which is
only a few degrees off at these low inclinations readily leads to
spurious rotation velocities. To illustrate this, at an inclination of
$ \sim 20^{\circ}$, a $5^{\circ}$ uncertainty, equivalent to a 3
percent uncertainty in the axial ratio, leads to a 25 percent
uncertainty in the deprojected rotation velocity.  Furthermore, as
discussed earlier, at low inclinations random or streaming motions can
be as important as the \emph{projected} rotation velocity and these will
therefore severely affect the resulting rotation curve.

In the following, we will limit ourselves to the subsample of galaxies
indicated in Table~\ref{table1}, i.e., all high-inclination THINGS
galaxies ($i \ga 40^{\circ}$) that are dominated by rotation, and, in
addition, NGC 6946, leading to a total sample size of 19 galaxies.  In
our analysis we will use the natural-weighted THINGS data sets.  The
robust weighting does give a higher angular resolution, but at a
decreased column density sensitivity. We found that the small loss in
resolution due to the use of the natural-weighted data is more than
compensated for by the increased column density sensitivity and,
hence, increased usable area of the velocity fields.

We did apply neither residual-scaling corrections, nor primary beam
corrections (as described in \citealt{THINGS1}). The reason for not
using these corrections is that for the construction of our velocity
fields (described in the next section), we need a constant noise level
throughout each galaxy's data cube. Primary beam corrections introduce
a noise and flux scaling dependent on distance to the pointing
center. Residual-scaling, in essence, artificially decreases the noise
in the data cubes in order to produce correct flux densities.  All
velocities mentioned in this paper are heliocentric using the optical
velocity definition.

\section{Velocity Fields and Rotation Curves}

\subsection{Velocity field types\label{sect:velfi}}

\HI\ rotation curves are most often derived from velocity fields. A
velocity field aims to give a compact and accurate ``short-hand''
description of the dynamics of a galaxy by assigning a ``typical''
velocity to every spatial position. That is, for every position one
uses the velocity that most accurately represents the
circular motion of the bulk of the quiescent component of the gas as
it moves around the center of the galaxy.  In an ideal case
(quiescent, purely rotating disk, infinite resolution, intermediate
inclination) this velocity corresponds to the velocity associated with
the peak emission in the velocity profile.

In practice, however, the observed velocity profiles are affected by
several systematic effects. These can be divided into effects inherent
to the observations themselves, and those due to physical processes
within the galaxy.

The first category, instrumental and observational effects, is
especially important in lower resolution observations. Here the effect
of the relatively large beam (compared to the size of the galaxy)
causes asymmetric velocity profiles, with the long tail always
pointing towards the systemic velocity. This is known as ``beam
smearing''.  Generally, one attempts to correct for this by choosing a
typical velocity close to the maximum velocity (i.e.\ furthest away
from the systemic velocity) found in the velocity profile
\citep[e.g.][]{sancisi_allen}.  When left uncorrected, these
resolution-related effects (including finite velocity resolution)
always \emph{lower} the apparent velocity.

The second category is due to physical processes within the galaxy
which cause non-circular motions. These will show up as deviations
both towards and away from the systemic velocity. For our galaxies,
this is discussed extensively in accompanying papers by \citet{oh2007} and
\citet{clemens2007}.

To minimize the impact these effects can have, different methods to
construct velocity fields have been used in the past (discussed below).  The THINGS observations have high enough
spatial and velocity resolutions that the resolution effects are no
longer a problem (cf.\ Sect.~\ref{beamsmear}); they also give us a
detailed enough view of the small-scale movements of the \HI\ within
the galaxies to enable us to quantify the non-circular motions.
Nevertheless, each method determines the ``typical'' 
velocity in a slightly different way, and it is worth exploring these
to quantify their impact on the final results.

In principle, our data are of high enough quality that (barring large
non-circular motions) the velocity of the peak emission in a profile
is representative of the total rotational motion. The most direct way
to create a velocity field is therefore to simply determine the
velocity at which this peak occurs.  \citet{usero2007} derive these
velocity fields in their study of the distribution of brightness
temperatures within galaxy disks.  For our purposes, these velocity
fields are less suited, due to the decreased signal-to-noise in the
outer parts of the galaxies and their smaller usable areas.  As
mentioned, in the following we briefly discuss alternative methods
most commonly used to construct velocity fields, list their pros and
cons, and present our favored method.

\begin{figure*}[t]
  \epsfxsize=0.6\hsize \hfil\epsfbox[20 180 320
  720]{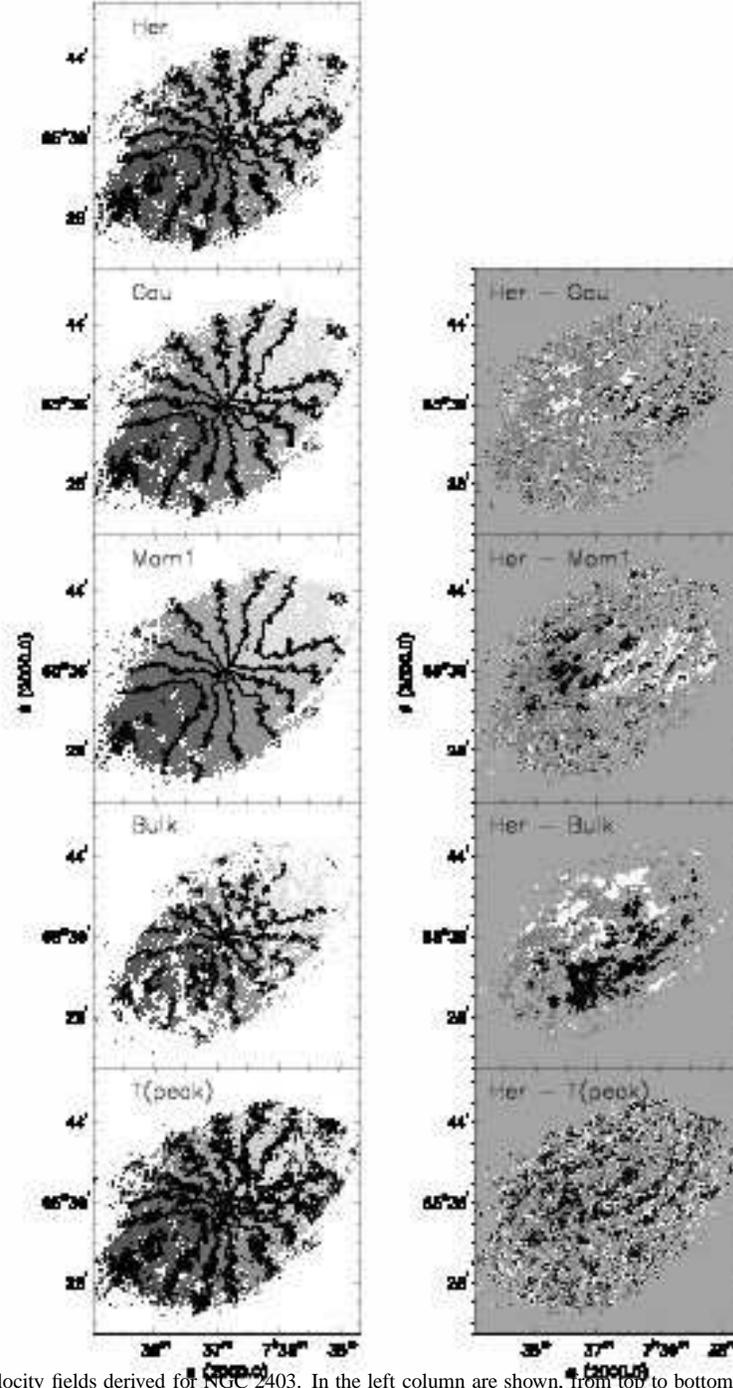} \hfil \figcaption{Different types of
    velocity fields derived for NGC 2403.  In the left column are
    shown, from top to bottom, the Hermite velocity [Her], the single
    Gaussian [Gau], the intensity-weighted first-moment (Mom1), the
    bulk [Bulk] and the peak amplitude [T(peak)]) velocity fields.
    The velocity contours run from $10$ \kms\ (top-right; light
    grayscales) to $250$ \kms\ (bottom-left; dark grayscales) in steps
    of 30 \kms. The minor axis contour is at $130$ \kms. The right
    column shows the difference velocity fields with respect to the
    Hermite velocity field, as indicated in the figure.  Grayscales
    run from $-15$ (black) to $+15$ (white) \kms, in steps of 5
    \kms. The white contour shows the $-10$ \kms\ level, the black
    contour the $+10$ \kms\ level.
 \label{fig:diffvelfi1}}
 \end{figure*}

\begin{figure*}[t]
  \epsfxsize=0.6\hsize \hfil\epsfbox[20 180 320 720]{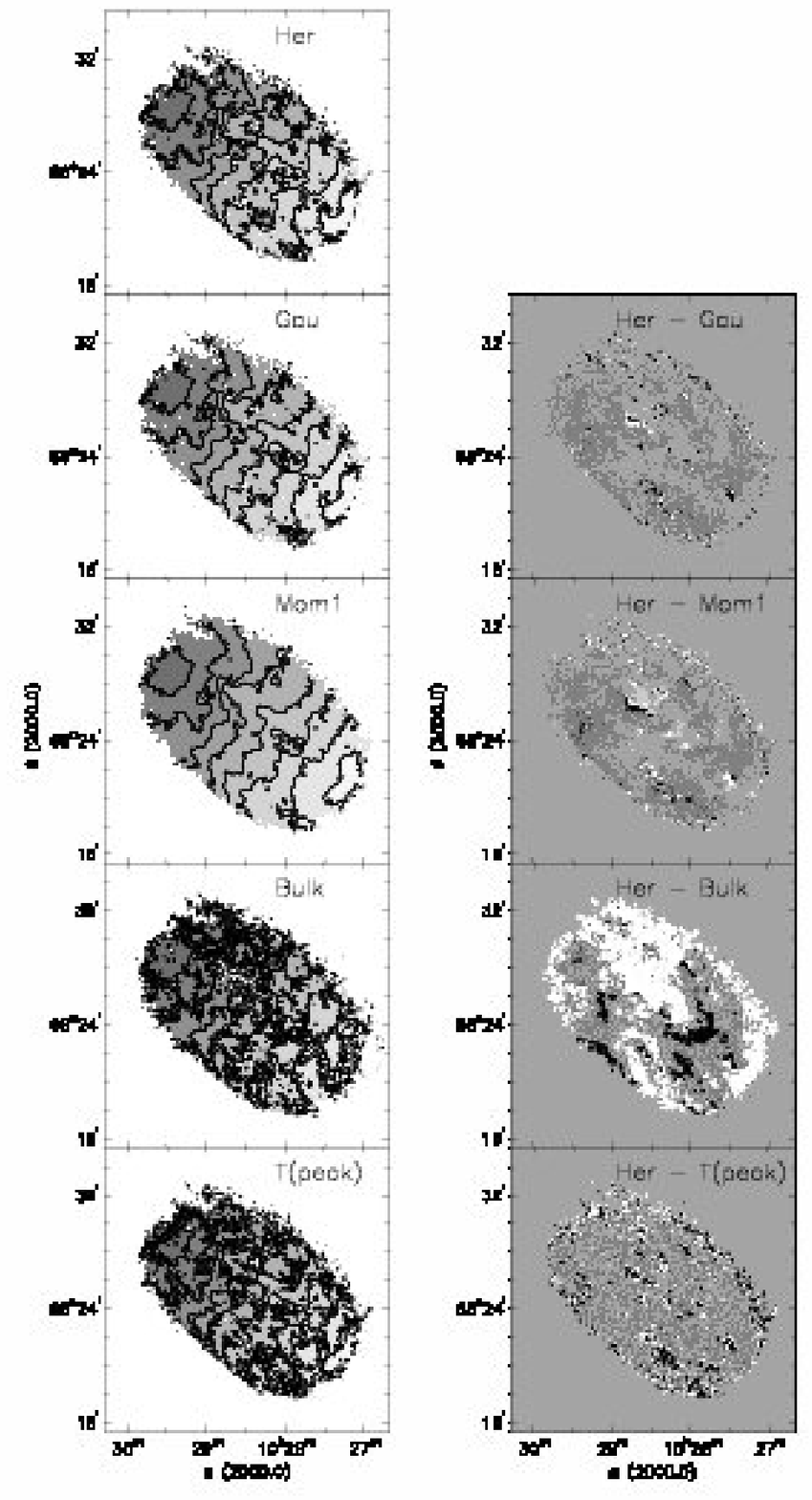} \hfil 
  \figcaption{Different types of velocity fields derived for IC2574.
    In the left column are shown, from top to bottom, the Hermite
    velocity (Her), the single Gaussian (Gau), the intensity-weighted
    first-moment (Mom1), the bulk (Bulk) and the peak amplitude
    [T(peak)]) velocity fields.  The velocity contours run from $-5$
    \kms\ (bottom-right; light grayscales) to $+100$ \kms\ (top-left; dark grayscales) in steps of 15
    \kms. The right column shows the difference velocity fields with
    respect to the Hermite velocity field, as indicated in the figure.
    Gray-scales run from $-15$ (black) to $+15$ (white) \kms, in steps
    of 5 \kms. The white contour shows the $-10$ \kms\ level, the
    black contour the $+10$ \kms\ level.
 \label{fig:diffvelfi2}}
 \end{figure*}

\begin{figure*}[t]
   \epsfxsize=0.95\hsize \hfil\epsfbox[18 427 592 718]{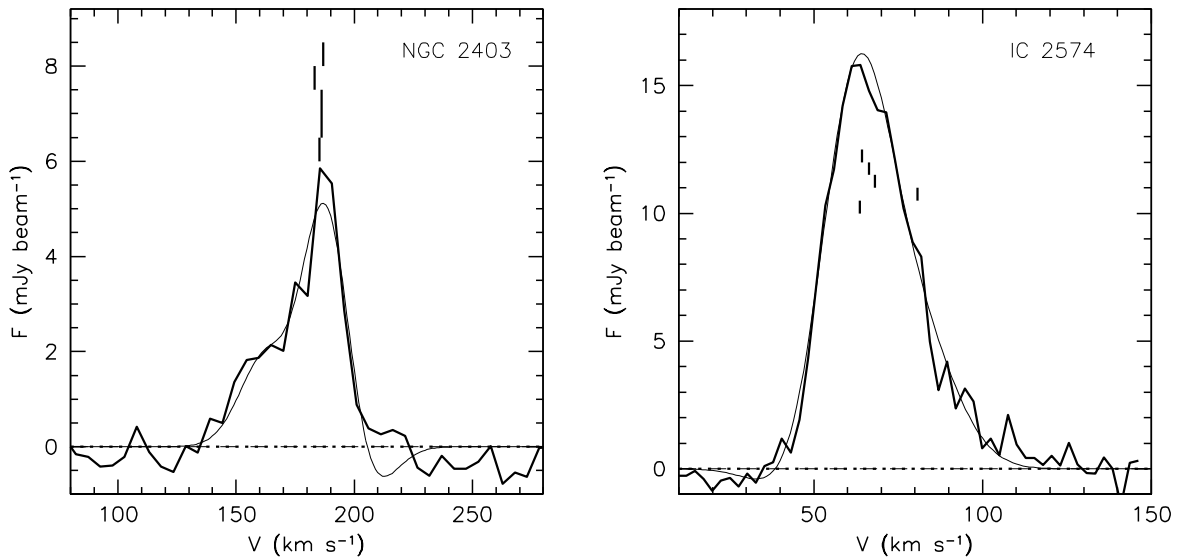}\hfil
\figcaption{\emph{Left:} Observed profile of the \HI\ emission in NGC 2403 at
     position $(\alpha, \delta)(2000.0) = 07^h37^m36.7^s,
     65^{\circ}36'46.3''$ (thick line). We adopt the notation given in
     Fig.~\ref{fig:diffvelfi1}. The five vertical lines indicate the
     derived typical velocities using, from top to bottom, the Her
     ($V=186.9$ \kms), the Gau ($V=183.2$ \kms), the Mom1 ($V=186.2$
     \kms), the Bulk ($V=186.1$ \kms) and the T(peak) velocity fields
     ($V=185.4$ \kms).  The curve shows the Hermite fit. \emph{Right:} Observed
     profile of the \HI\ emission in IC 2574 at position $(\alpha,
     \delta)(2000.0) = 10^h28^m51.4^s, 68^{\circ}28'13.0''$ (thick
     line). The five vertical lines indicate the derived typical
     velocities using, from top to bottom, the Her ($V= 64.2$ \kms),
     the Gau ($V=66.3.2$ \kms), the Mom1 ($V=68.1.2$ \kms), the Bulk
     ($V=80.8$ \kms) and the T(peak) velocity field ($V=63.8$
     \kms). 
 \label{fig:profiles1}} 
\end{figure*}

\begin{figure*}[t]
   \epsfxsize=0.95\hsize \hfil\epsfbox{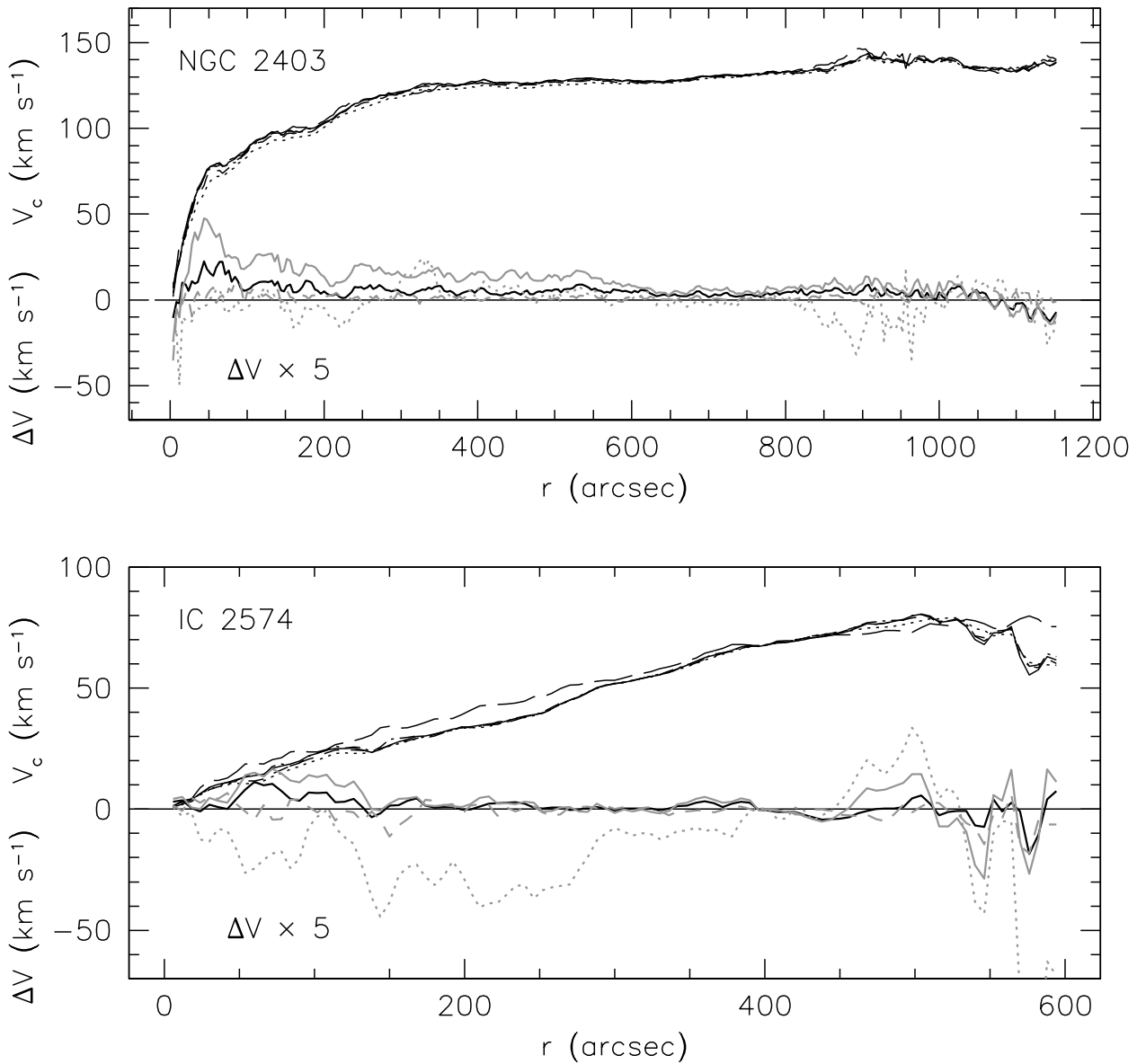}\hfil
   \figcaption{Rotation curves for NGC 2403 derived using the five
     different velocity fields. Dotted curve: Mom1; dashed curve: Gau;
     full curve: Her; long-dashed curve: Bulk; dot-short dash:
     T(peak).  Also shown are the differences with respect to the
     Hermite curve. Thick black curve: Her--Gau; thick gray full
     curve: Her--Mom1; thick dotted gray curve: Her--Bulk; thick
     dashed gray curve: Her--T(peak).  The differences have been
     multiplied by a factor of 5 for clarity. \emph{Bottom:} Rotation
     curves and differences for IC2574. Line types as described above.
 \label{fig:profiles2}} 
\end{figure*}

A practical comparison of the outcomes of the various methods is shown
in Figs.~\ref{fig:diffvelfi1}-\ref{fig:profiles2} and discussed in
more detail below as well as in Sect.\ 3.2 and 3.3.

{\bf First-moment or intensity-weighted mean velocity:} One of the
most well-known and widely-used methods involves taking the first
moment or intensity-weighted mean along the velocity axis in the data
cube. The procedure is straightforward, but has the disadvantage that,
if the velocity distributions are not symmetric, the ``typical''
velocities are biased towards the longest tail of the velocity
profile.  Also, to avoid noise at outlying velocity values affecting
the mean, one needs to carefully identify the regions with significant
emission in the data cube and use only those areas when constructing
the velocity field.  This method is the one most commonly implemented
in astronomical software.

{\bf Peak velocity fields: } Using the velocity where the \HI\ profile
flux peaks has the great advantage that it is totally independent from
assumptions on profile shapes. A disadvantage is that as the profile
peak value decreases the velocity values become more and more affected
by noise. The peak velocity field is therefore best used in regions of
high signal-to-noise.

{\bf Gaussian profiles: } An alternative is to fit single Gaussian
functions to the velocity profiles. The velocity of the peak of the
Gaussian profile is then taken to be the velocity field value. This
method is less sensitive to noise and (moderate) asymmetries in the
profiles than the first-moment method, but can still suffer from
biased velocities if the asymmetries are strong or if multiple
components are present.  The single Gaussian fit method is best used
in cases where the instrumental resolution is comparable to the
typical width of the profiles. Many of the ``classical'' \HI\ rotation
curve studies from the 1970s and 1980s fall in this category, and used
this method to derive rotation curves.

{\bf Multiple Gaussian profiles: } A logical extension is to fit
multiple Gaussian functions to the velocity profiles.  In many cases,
however, the narrow width of the \HI\ profiles combined with limited
velocity resolution yields only a relatively small number of
independent profile data points with sufficiently high
signal-to-noise.  \emph{Simultaneous} fits with two or more Gaussian
functions can in these case become sensitive to the choice of initial
values, or have more free parameters than data points. A careful
fine-tuning is in any case needed to provide optimum fits. An
additional uncertainty is how to decide which of the components
represents the true rotation velocity, and which one characterizes
non-circular, random motions (induced by, e.g., star formation or
spiral arms).  As stated, these uncertainties apply to
\emph{simultaneous} fitting of multiple Gaussian functions.  An
alternative method is to optimally fit a single Gaussian function to
the profile, subtract it, and subsequently make a second Gaussian fit
to the residual profile.  A robust application of this method is
presented in \citet{oh2007}.  This method is remarkably effective in
separating random motions from the average rotation velocity (or
`bulk' velocity, to use the terminology from \citealt{oh2007}).

{\bf Hermite $h_3$ polynomials: } A robust way to minimize the number
of components is to include in the fitting function a prescription of
the typical asymmetry encountered.  A convenient function to use for
this purpose is a \emph{Gauss-Hermite polynomial} that includes an
$h_3$ (skewness) term (see e.g.,\ \citealt{vdmarel93}). The Hermite
method has already been applied successfully to derive rotation curves
of early-type galaxies in \cite{noordermeer2006}, and we will use it
as our primary method to derive velocity fields.  We do not include an
additional $h_4$ term in the fit, which would provide a measure of the
kurtosis of the profile (i.e., whether the profile is ``fatter'' or
more ``pinched'' than a Gaussian function). Attempts to include an
$h_4$ term show that, for the purposes of determining a velocity
field, this leads to diminishing returns: the fits become less stable
and need increasingly fine-tuned initial estimates.  For high
signal-to-noise profiles, the velocities derived using the hermit
profile will be close to those as derived from the peak-velocity
fields.  One could therefore argue that the latter are therefore
preferable, as they are model-independent. However, in low
signal-to-noise regions the Hermite velocity field results are more
stable as more data points (from neighboring channels) are used to
minimize the effects of the noise. We therefore
prefer to use the Hermite velocity fields.

We illustrate the various methods for two representative galaxies.
The first is NGC 2403 (Sect.~3.2, Fig.~\ref{fig:diffvelfi1}), which is dominated
by rotation, and where most of the asymmetric profiles are associated
with streaming motions along the spiral arms. The second example is
the dwarf galaxy IC 2574 (Sect.~3.3, Fig.~\ref{fig:diffvelfi2}) where a large
fraction of the profiles is affected by asymmetries and random
motions.

\begin{figure*}[t]
  \epsfxsize=0.95\hsize \hfil\epsfbox{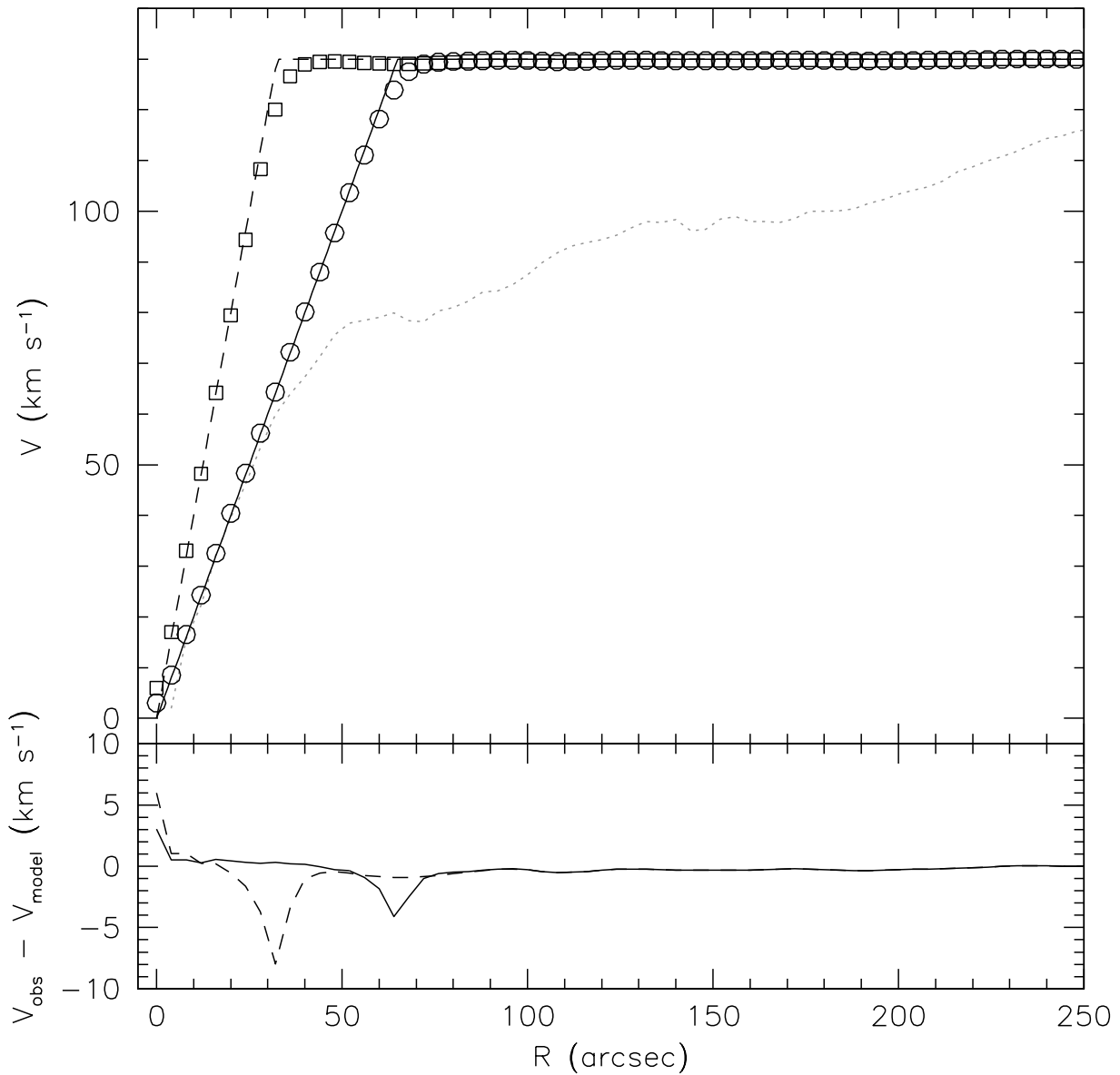}\hfil
  \figcaption{\emph{Top panel:} Comparison of input model rotation
    curves with ``observed'' curves as derived from model data cubes
    smoothed to the THINGS resolution. The full curve represents the
    input model rotation curve of a model that rises to its maximum
    rotation velocity within 1 kpc.  The open circles represent the
    corresponding rotation curve as observed at the THINGS resolution.
    The dashed curve and open squares are the corresponding curves for
    a model rising to its maximum value within 0.5 kpc.  \emph{Bottom
      panel:} Difference between ``observed'' rotation velocity and
    input model velocity. The full line represents the 1 kpc model;
    the dashed curve the 0.5 kpc model.
 \label{fig:beamsmear}} 
\end{figure*}

\begin{figure*}[t]
  \epsfxsize=0.9\hsize \epsfbox{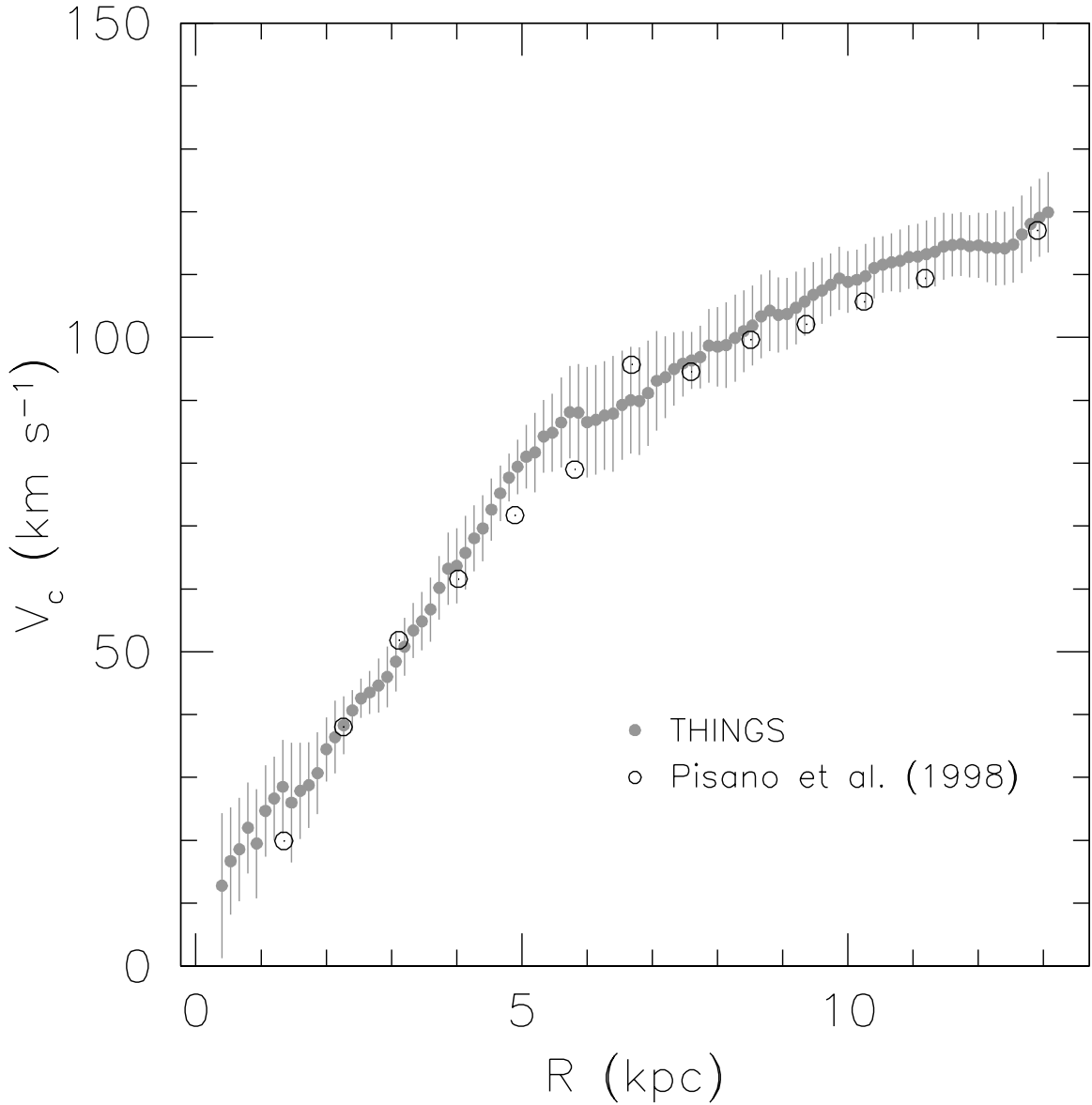}
  \figcaption{Comparison of the NGC 925 THINGS rotation curve with the
    curve derived in \citet{DJ}.  Symbols as indicated in the
    figure. The uncertainties given in \citet{DJ} are smaller than the
    symbol sizes. However, there uncertainties are the formal
    least-square fit errors and severely underestimate the true
    uncertainties in the measured rotation velocity. Cf.~discussion in Sect.~3.5.
\label{fig:n925_comp}}
\end{figure*}

 \begin{figure*}[t]
 \epsfxsize=\hsize \epsfbox{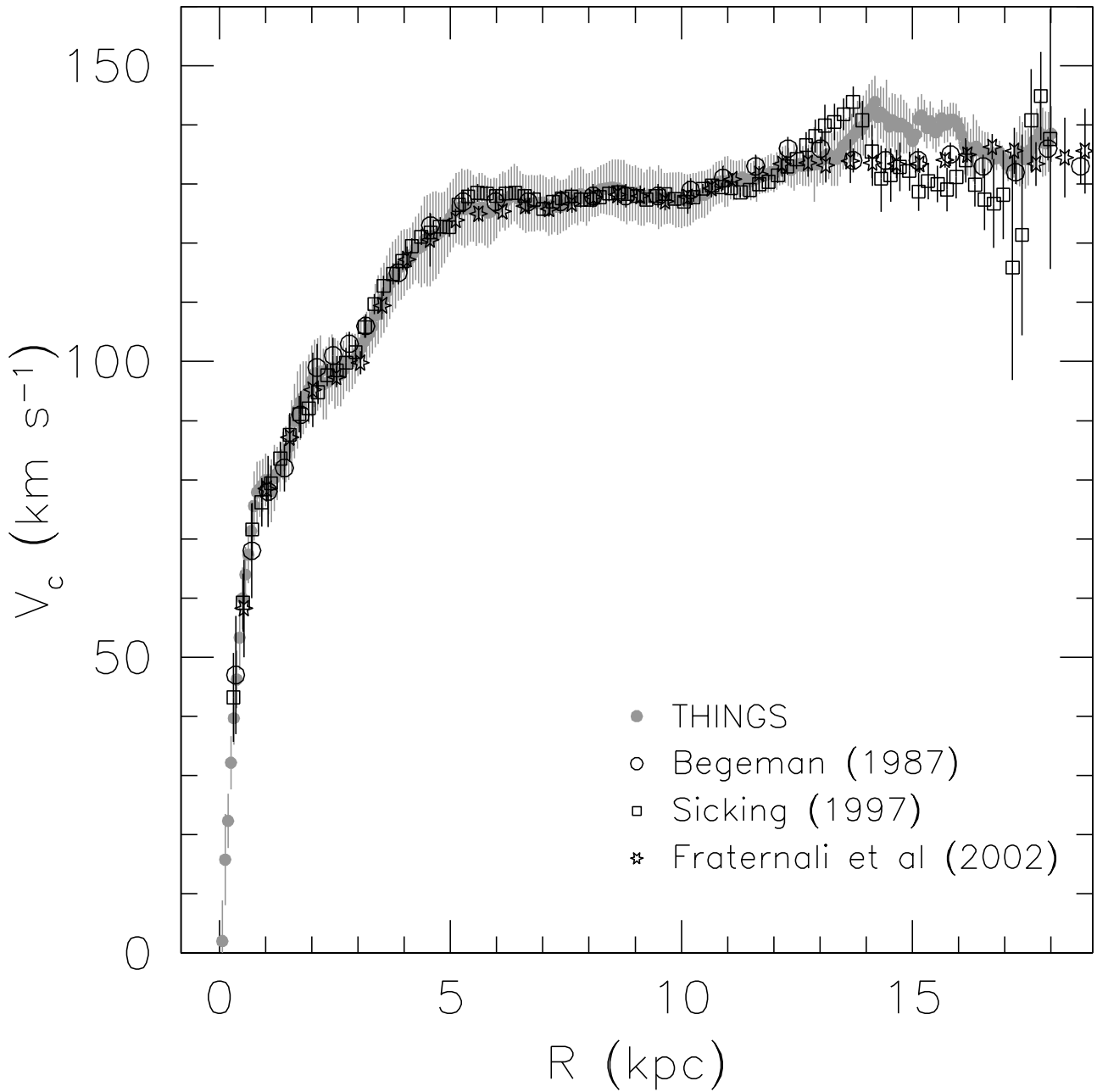} \figcaption{Comparison of the
 NGC 2403 THINGS rotation curve with recent determinations from the literature.
 Symbols and references are indicated in the figure.
 \label{fig:n2403_comp}}
 \end{figure*}

 \begin{figure*}[t]
   \epsfxsize=\hsize \epsfbox{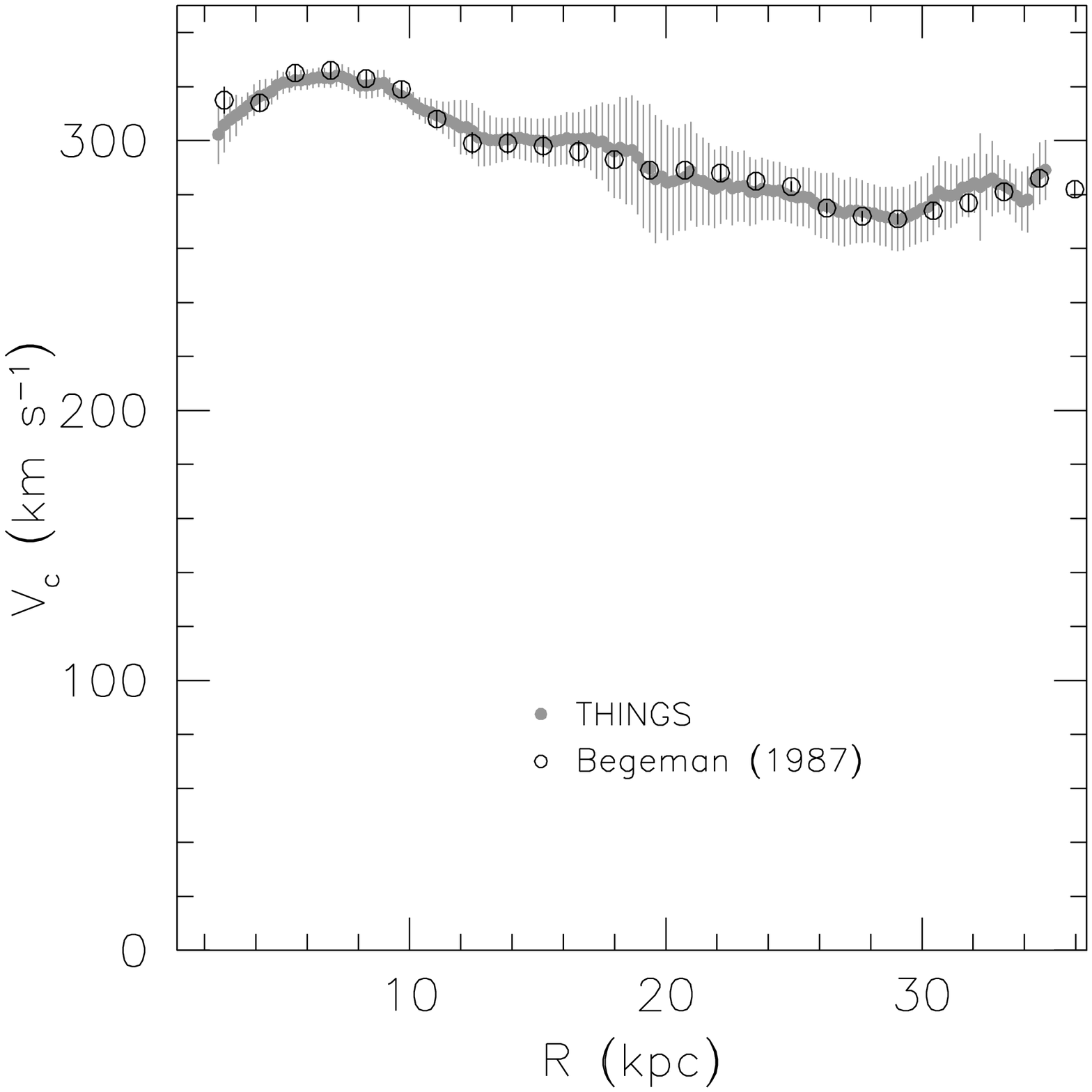}
   \figcaption{Comparison of the NGC 2841 THINGS rotation curve with
     the analysis by \citet{begeman87}.  Symbols and references are
     indicated in the figure. Note the different nature of the
     uncertainties: the \citet{begeman87} result uses the formal
     errors in the least-squares fit, whereas our definition takes
     into account the differences between the two sides of the galaxy,
     and the dispersion in velocity values found along the rings.
 \label{fig:n2841_comp}}
 \end{figure*}

 \begin{figure*}[t]
 \epsfxsize=\hsize \epsfbox{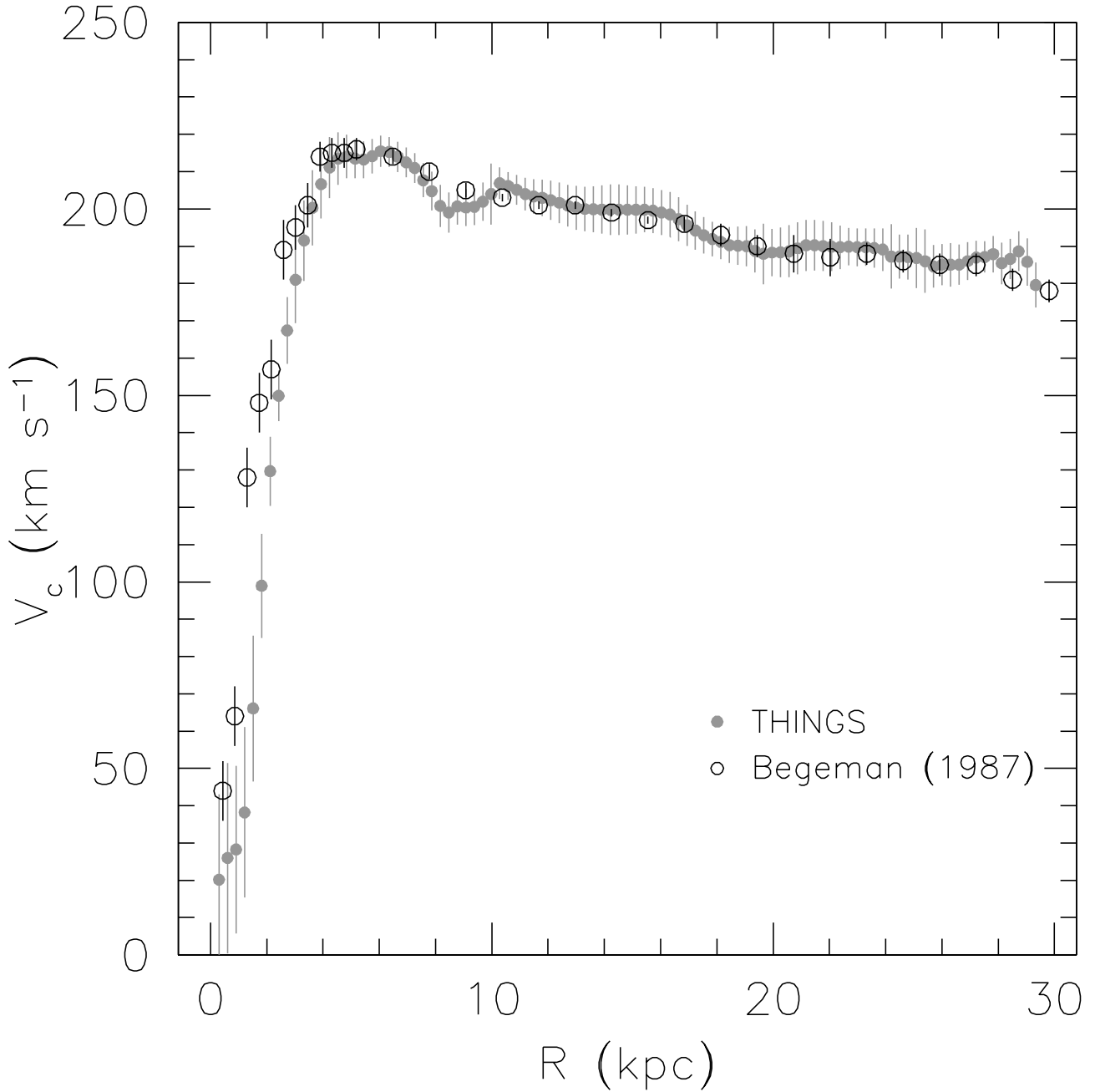} \figcaption{Comparison
   of the NGC 2903 THINGS rotation curve with the analysis by
   \citet{begeman87}.  Symbols and references are indicated in the
   figure. See Fig.~\ref{fig:n2841_comp} for a description of the
   uncertainties.
 \label{fig:n2903_comp}}
 \end{figure*}

 \begin{figure*}[t]
 \epsfxsize=\hsize \epsfbox{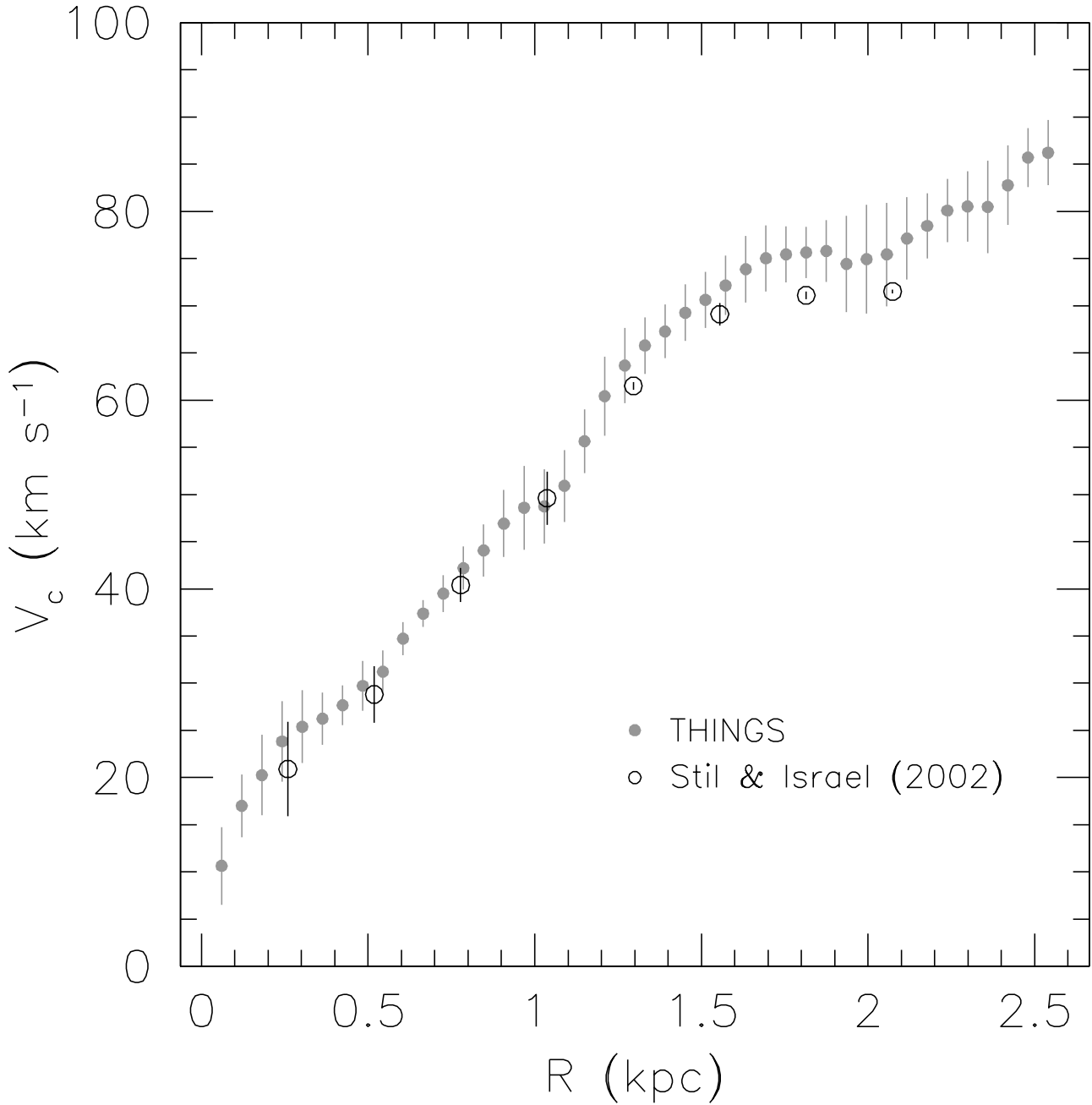} \figcaption{Comparison
   of the NGC 2976 THINGS rotation curve with the analysis by
   \citet{stil}. Symbols and references are indicated in the
   figure. Note the different nature of the uncertainties: 
   \citet{stil}  use the difference between approaching and
   receding side as a measure for the uncertainty, whereas our
   definition additionally takes into account the dispersion in
   velocity values found along the rings.
 \label{fig:n2976_comp}}
 \end{figure*}

 \begin{figure*}[t]
 \epsfxsize=\hsize \epsfbox{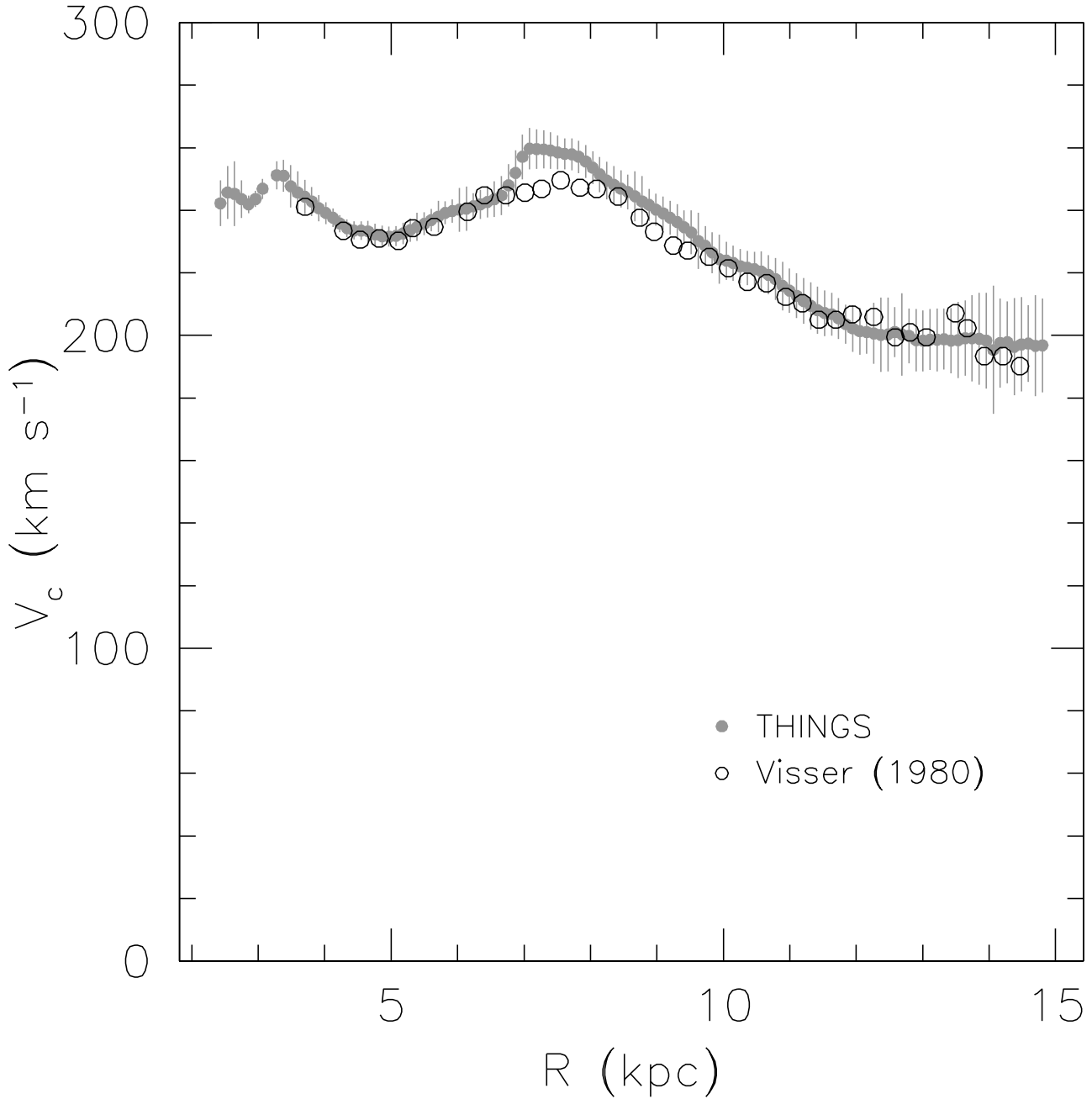} \figcaption{Comparison
   of the NGC 3031 THINGS rotation curve with the analysis by
   \citet{visser80}. Symbols and references are indicated in the
   figure. Note that no uncertainties are given in \citet{visser80}.
 \label{fig:n3031_comp}}
 \end{figure*}

 \begin{figure*}[t]
   \epsfxsize=0.9\hsize \centerline{\epsfbox{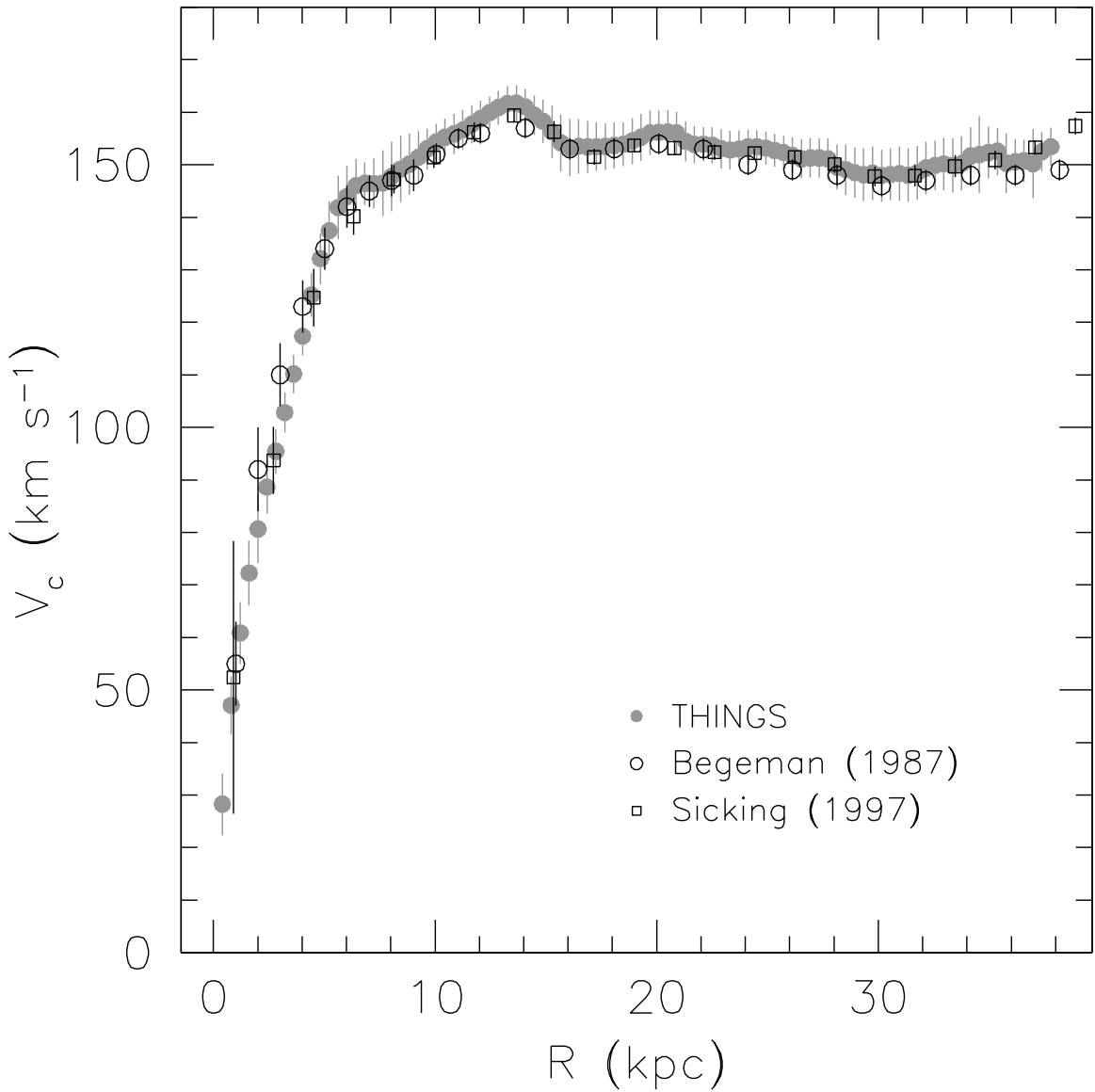}}
   \figcaption{\emph{Left:} Comparison of the NGC 3198 THINGS rotation
     curve with recent determinations from the literature.  All curves
     have been corrected to a distance of 13.80 Mpc.  Symbols and
     references are indicated in the figure.
 \label{fig:n3198_comp}}
 \end{figure*}

\begin{figure*}[t]
\epsfxsize=\hsize \epsfbox{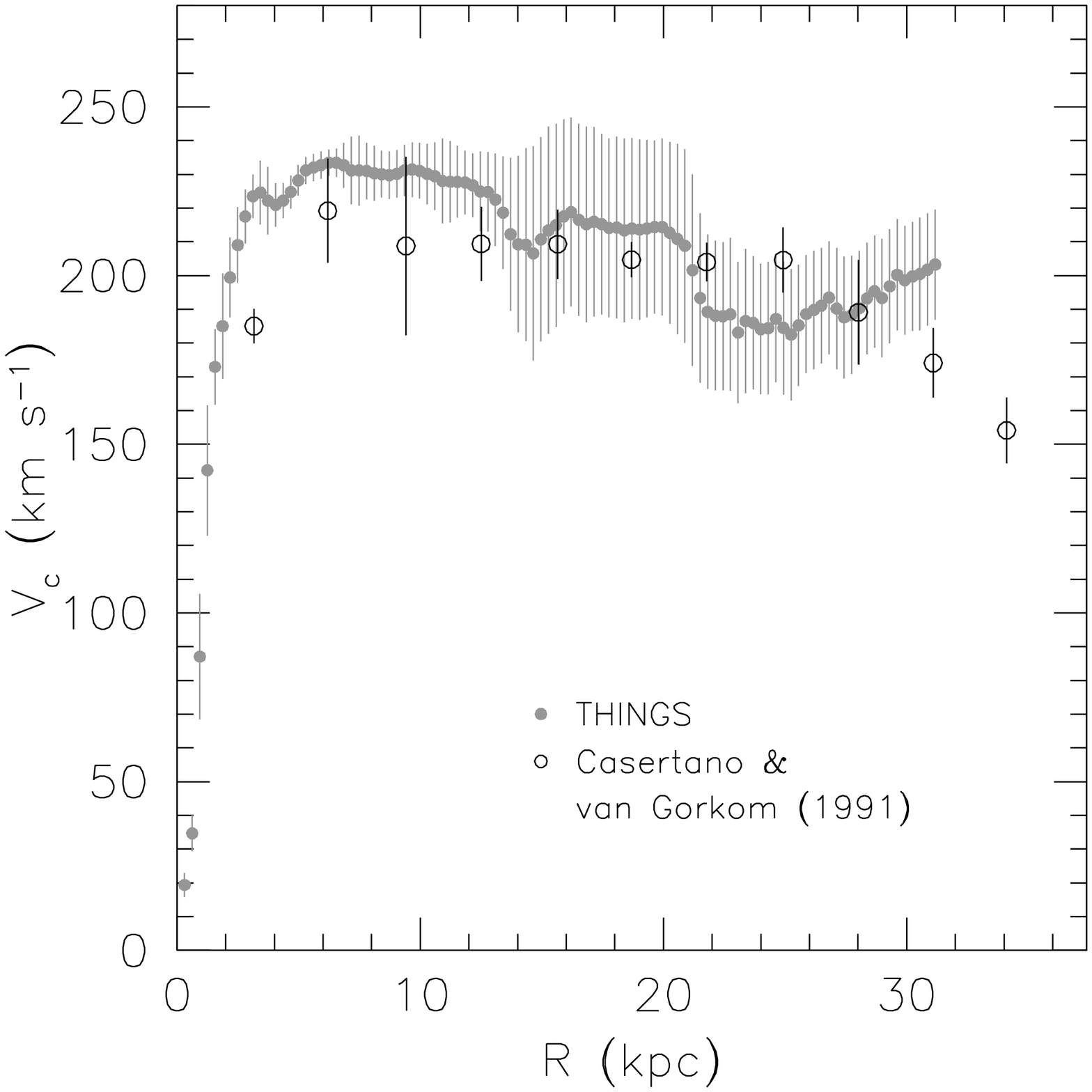} \figcaption{Comparison of the
NGC 3521 THINGS rotation curve with the analysis by \citet{cas_vg91}.
Symbols and references are indicated in the figure. 
\label{fig:n3521_comp}}
\end{figure*}

\begin{figure*}[t]
  \epsfxsize=\hsize \epsfbox{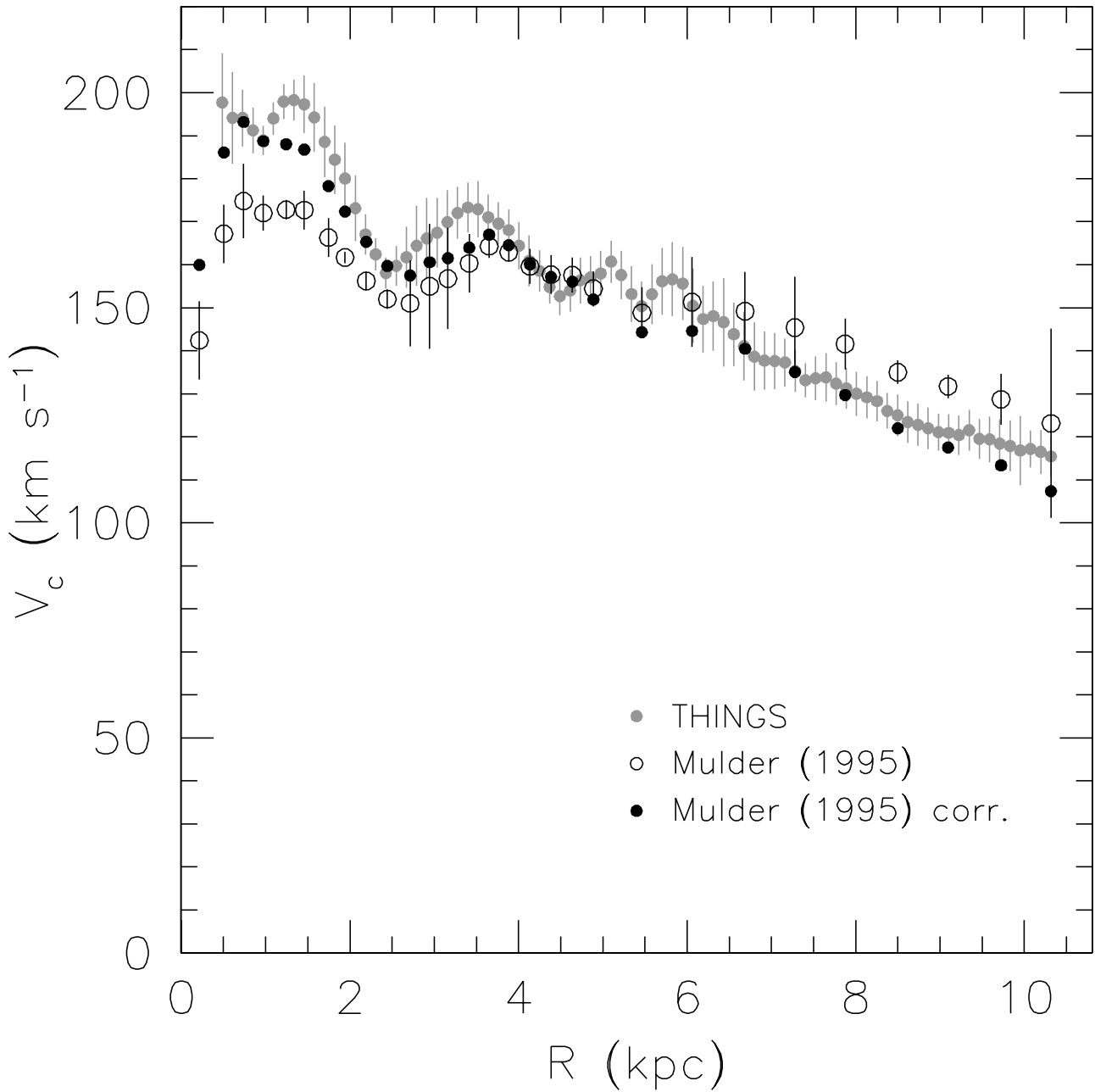}
  \figcaption{Comparison of the NGC 4736 THINGS rotation curve (grey
    points) with the analysis by \citet{mulder}. The open circles
    show the rotation curve as published in \citet{mulder} assuming a
    constant inclination of $40^{\circ}$. The filled black circles
    show the same data but corrected using our inclinations.
\label{fig:n4736_comp}}
\end{figure*}

\begin{figure*}[t]
\epsfxsize=0.9\hsize \epsfbox{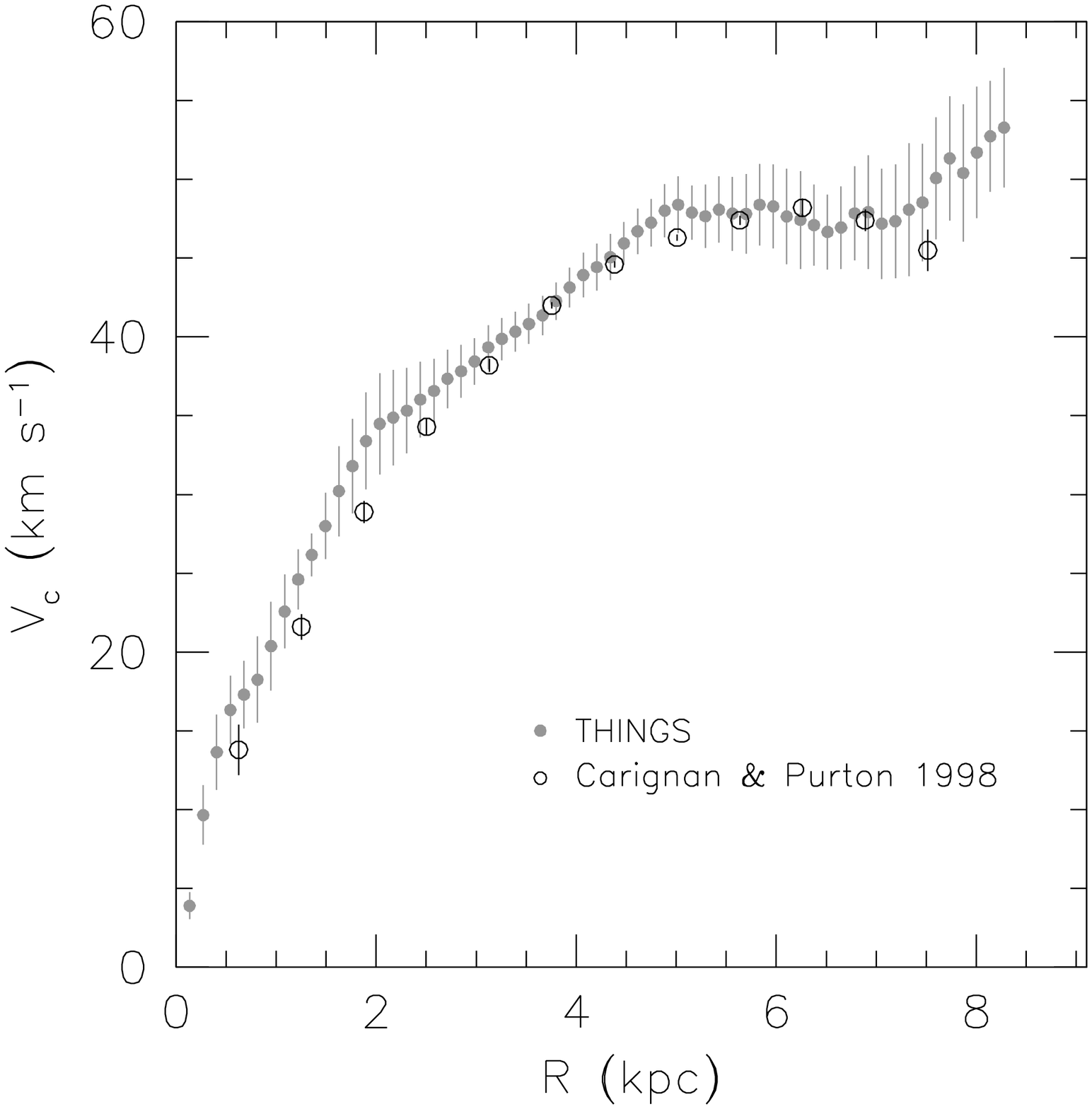}
\figcaption{Comparison of the DDO 154 THINGS rotation curve with the
  curve derived in \citet{car_purton}.  Symbols and references are
  indicated in the figure.
\label{fig:ddo154_comp}}
\end{figure*}

\begin{figure*}[t]
  \epsfxsize=\hsize \epsfbox{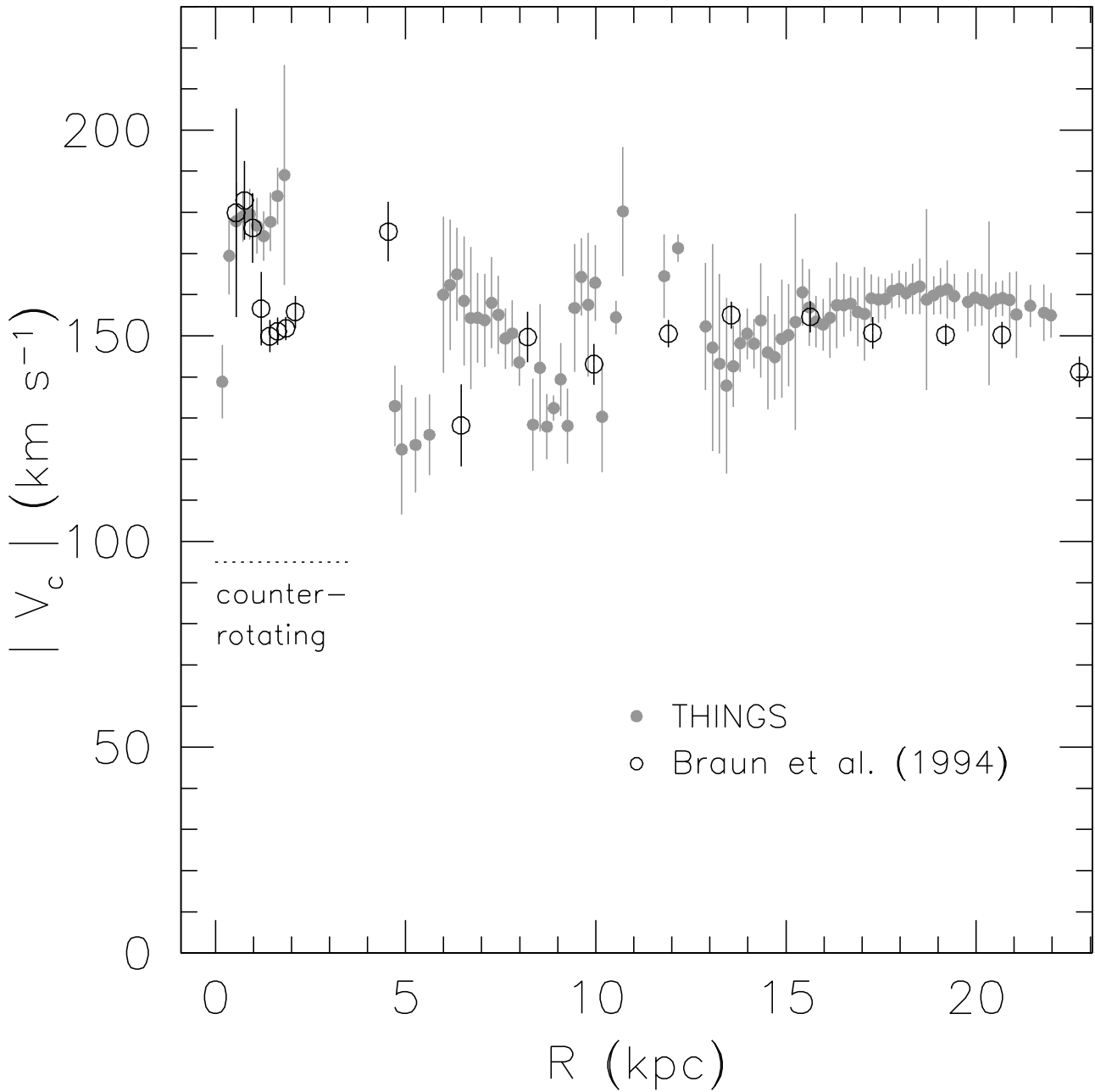} \figcaption{Comparison
    of the NGC 4826 THINGS rotation curve with the analysis by
    \citet{braun94}. Symbols as indicated in the figure. Plotted are
    the \emph{absolute} rotation velocities, the inner disk ($R<4$
    kpc; indicated by the horizontal dotted line) rotates in the
    opposite direction compared to the outer disk.
\label{fig:n4826_comp}}
\end{figure*}

\begin{figure*}[t]
\epsfxsize=\hsize \epsfbox{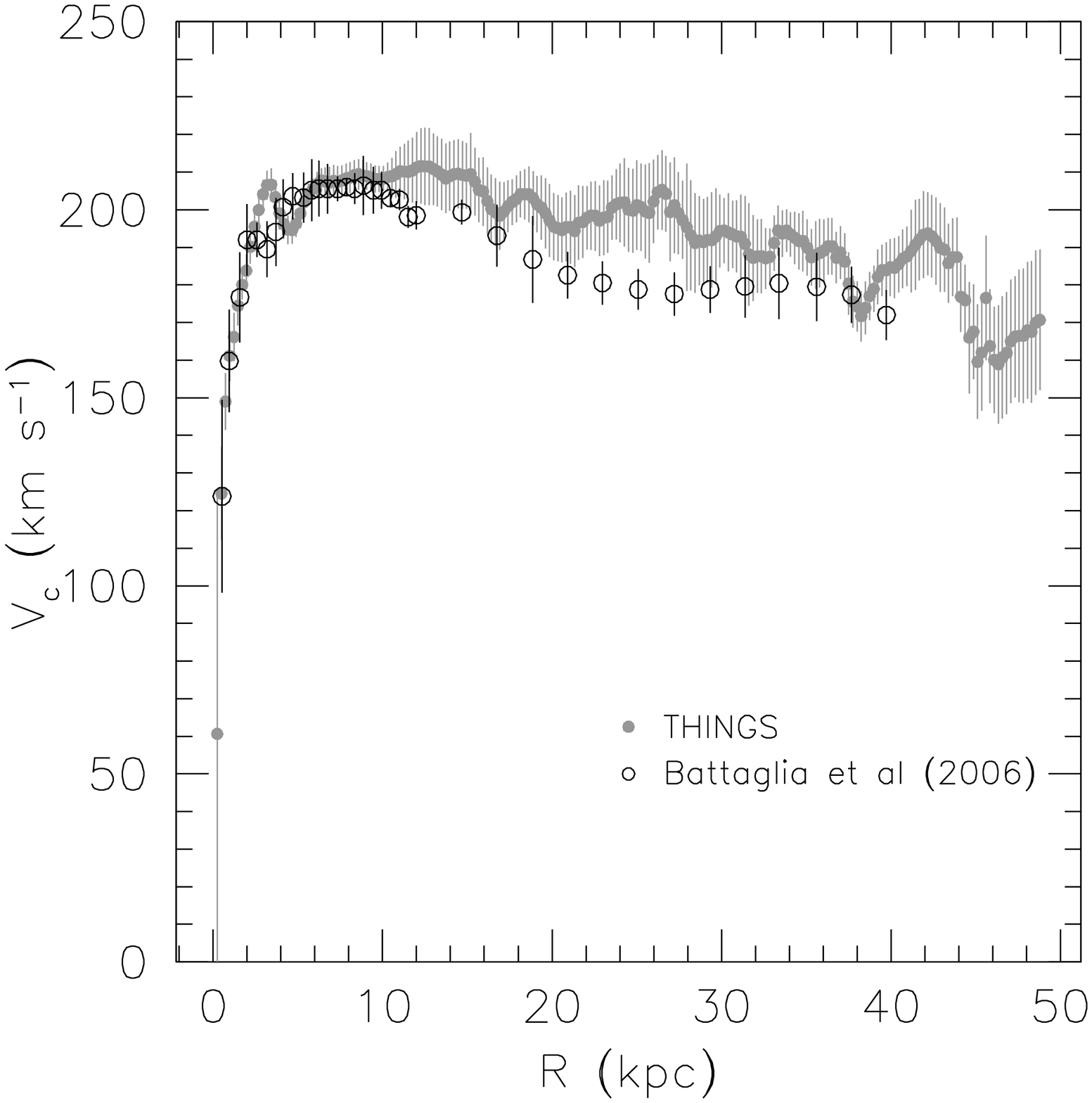} \figcaption{Comparison of the
NGC 5055 THINGS rotation curve with the analysis by \citet{battaglia}.
Symbols and references are indicated in the figure. 
\label{fig:n5055_comp}}
\end{figure*}

\begin{figure*}[t]
  \epsfxsize=\hsize \epsfbox{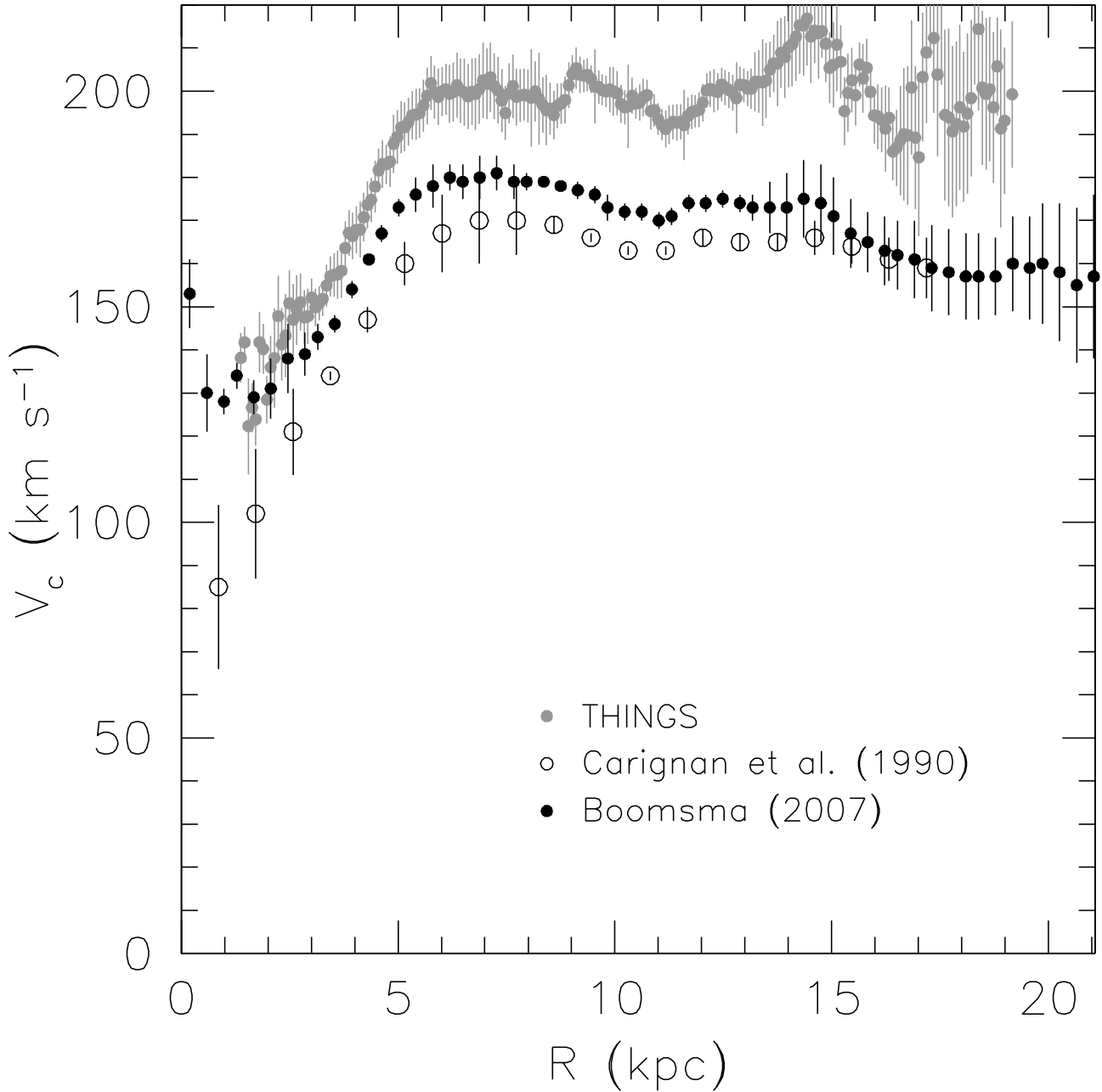}
  \figcaption{Comparison of the NGC 6946 THINGS rotation curve with
    the analyses by \citet{carignan6946} and \citet{boomsma}.  Symbols and references are indicated in the
    figure.
\label{fig:n6946_comp}}
\end{figure*}

\begin{figure*}[t]
  \epsfxsize=0.9\hsize \epsfbox{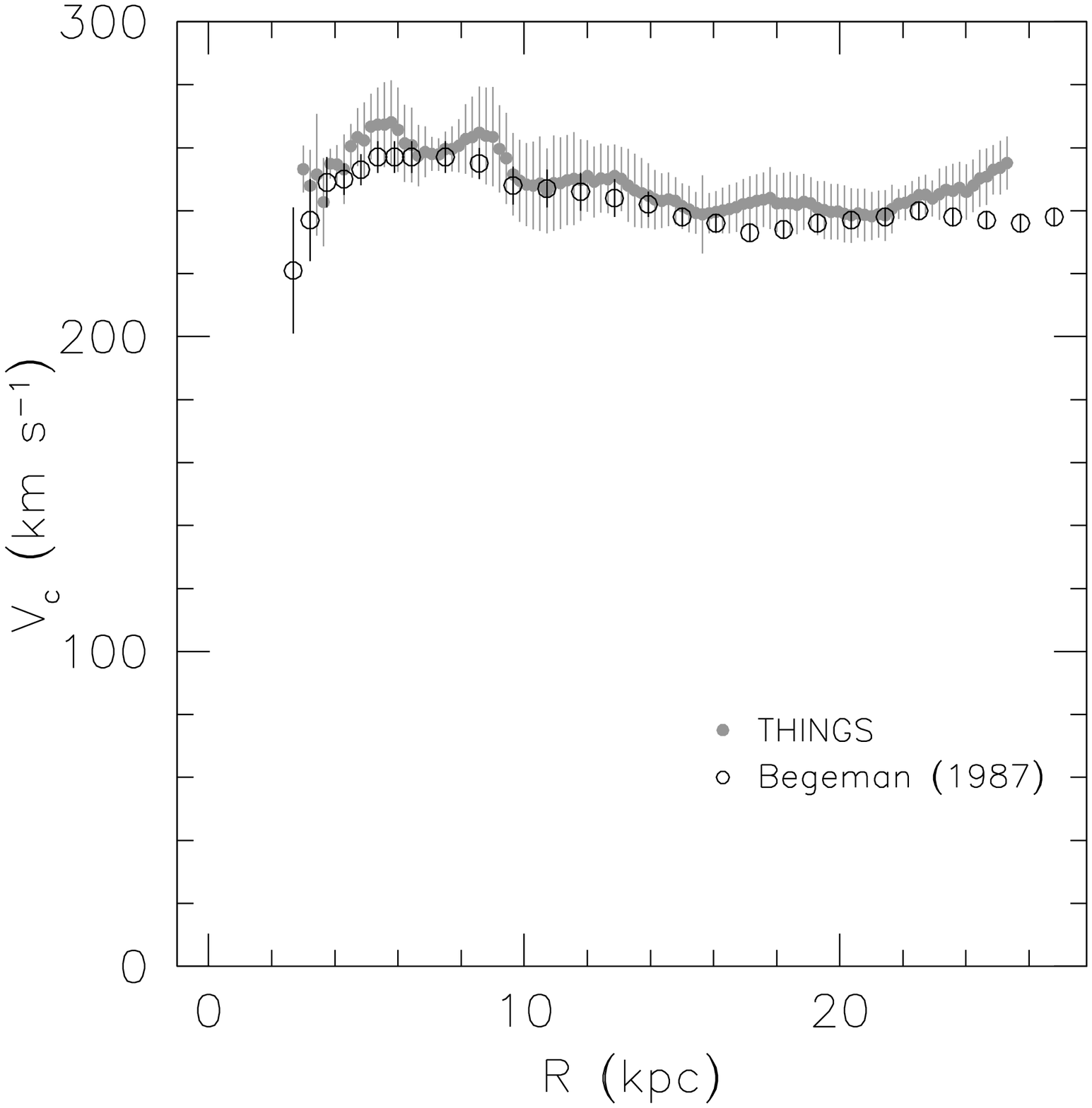}
  \figcaption{Comparison of the NGC 7331 THINGS rotation curve with
    result by \citet{begeman87}.  Symbols and references as indicated
    in the figure.  Note the different nature of the uncertainties:
    the \citet{begeman87} result uses the formal errors in the
    least-squares fit, whereas our definition takes into account the
    differences between the two sides of the galaxy, and the
    dispersion in velocity values found along the rings.
\label{fig:n7331_comp}}
\end{figure*}

\subsection{NGC 2403: a galaxy dominated by regular rotation}

Figure~\ref{fig:diffvelfi1} shows the velocity fields for NGC 2403,
derived using the procedures described above, together with difference
velocity fields with respect to the Hermite velocity field. The
absolute differences between the Hermite velocity field and the
Gaussian and first-moment velocity fields are larger than 10 km
s$^{-1}$ for only a small fraction of the area. For the peak velocity
field the difference with the Hermite field is consistent with pure
noise (mostly caused by the lower signal-to-noise of the peak velocity
field).

Figure~\ref{fig:profiles1} shows an example of an asymmetric profile
in NGC 2403, together with the values for the typical velocity, again
derived using the methods described above.  Shown in
Fig.~\ref{fig:profiles2} are the rotation curves derived from the
respective velocity fields, using otherwise identical assumptions and
methods (cf.\ Sect.~3.5). Any difference is thus entirely due to the
manner in which the velocity field was derived.  The curve based on
the Hermite velocity field systematically shows the highest rotation
velocities and agrees best with the peak velocity field curve.  The
curves based on the Gaussian and first-moment fields are progressively
more affected: the latter curve systematically underestimates the
rotation velocity by $\sim 4$ km s$^{-1}$ over the inner half of the
disk.  The bulk velocity field curve is also consistent with the
Hermite field rotation curve, but with a larger scatter.

\subsection{IC 2574: a galaxy with substantial random motions}

A second example is IC 2574, where the velocity field is clearly
affected by non-circular, random motions (cf.\ \citealt{walter99}).
These are clearly visible as ``kinks'' in the iso-velocity contours in
all of the Hermite, Gaussian, first-moment and peak velocity fields
shown in Fig.~\ref{fig:diffvelfi2}. It is clear that before a rotation
curve can be derived some correction for the non-circular motions must
first be made. This process is described in detail for IC 2574 in
\citet{oh2007}. The effect of this correction is illustrated in
Fig.~\ref{fig:profiles1}. The resulting ``bulk'' velocity field,
showing the rotational part of the kinematics, is also shown in
Fig.~\ref{fig:diffvelfi2}. Though noisier than the other velocity
fields, the kinks have largely disappeared, and the velocity field is
consistent with simple solid-body rotation.  The resulting bulk
rotation curve also differs significantly from those derived from the
uncorrected velocity field as shown in the bottom panel of
Fig.~\ref{fig:profiles2}.

Further analysis shows that in our sample only NGC 2366 and IC 2574
are significantly affected by these random motions and they are
therefore the only two galaxies where the bulk velocity field differs
significantly from the Hermite velocity field \citep{oh2007}. For
these two galaxies we will in the rest of this paper use the latter.
For the other galaxies the differences are negligible and 
we will use the Hermite velocity fields.

\subsection{Resolution Effects\label{beamsmear}}

The THINGS data are extremely well resolved. For example, the ratio of
galaxy size to beam size for NGC 2403 is $\sim$ 350:1. Nevertheless,
some galaxies, including NGC 2403, show steep increases in velocity in
their inner parts, and despite the high resolution, these small radii
could potentially still be affected by beam smearing.  We quantify
this effect by constructing two model data cubes using the Groningen
Image Processing SYstem (GIPSY; \citealt{gipsy}) task {\sc
  galmod}. This task takes an input rotation curve, radial \HI\
distribution, inclination and position angle, as well as \HI\ scale
height and velocity dispersion and creates a model data cube at
arbitrarily high resolution by distributing a large number of \HI\
``clouds'' through this data cube using the input parameters as
probability distributions. This cube can then be ``observed'', e.g.,
by smoothing it to a lower resolution, and be used to construct
velocity fields, derive rotation curves, etc.

To make the model galaxies as realistic as possible, we adopt the \HI\
distribution and average inclination and position angle of NGC
2403. We assume a velocity dispersion of 8 \kms, and a vertical
(Gaussian) scale height of 100 pc. None of these assumptions
critically affect the results. For the input rotation curves we adopt
two extreme versions of the observed NGC 2403 rotation curve. For the
first model we assume a steep linear rise to 130 \kms\ (the maximum
rotation velocity of NGC 2403) within 1 kpc (65$''$), and a flat 130
\kms\ rotation curve outside 1 kpc. The second model is identical,
except that it rises to its maximum rotation velocity within 0.5 kpc
(33$''$) (compare these with the observed rotation curve shown in
Fig.~\ref{fig:n2403_comp}, and indicated in Fig.~\ref{fig:beamsmear}).

The resulting model cubes were smoothed to a resolution of $8''$ (the
natural-weighted beam size for the NGC 2403 observations;
cf.~\citealt{THINGS1}) and a channel spacing of 5.2 \kms\ was adopted.
The resulting ``observed'' cubes were then used to construct Hermite
velocity fields and rotation curves were derived
(cf.~Sect.~\ref{derivevelfie} and \ref{deriverotcur}).

We compare the derived curves with the respective input models in
Fig.~\ref{fig:beamsmear}. Though we did calculate the cube for the
entire radial extent of our pseudo-NGC 2403, we only show the inner
portions of the rotation curves in Fig.~\ref{fig:beamsmear} to focus
on the rising part of these curves. It is clear that beam smearing is
not a serious problem, with most of the curves differing by less than
1 \kms\ from the input model. The only significant difference occurs
around $R = 65''$ for the 1 kpc model and around $R=33''$ for the 0.5
kpc model, but this is entirely due to the unrealistically sharp break
in our input curves.  The innermost point of both curves is also
somewhat affected, but only at the level of $\sim 3$ \kms\ for the 1
kpc model and at $\sim 6$ \kms\ for the 0.5 kpc model.  In summary,
this demonstrates that resolution effects such as beam smearing have
no significant effect on the THINGS data.

\subsection{Deriving the velocity fields\label{derivevelfie}}

As mentioned before, we use data sets without primary-beam and
residual-scaling corrections as we need the original noise properties
of the data to perform the profile fitting.  For each galaxy, Hermite
$h_3$ polynomials were fitted to all velocity profiles with the GIPSY
task {\sc xgaufit}.

To ensure high-quality velocity fields, we only retain profiles with
fitted intensity maxima higher than $3 \sigma_{\rm chan}$ (where
$\sigma_{\rm chan}$ is the average noise in the line-free velocity
channels in the relevant cube) and where the equivalent Gaussian
dispersion of the fit is larger than the channel separation (see
\citealt{THINGS1} for the noise values and channel separations).  To
objectively remove a small number of noise pixels admitted by these
filter criteria, we use the integrated \HI\ column density map as an
additional mask and only retain the fits at positions where the total
flux value in the integrated \HI\ map is higher than $3\sigma_N$,
where $\sigma_N$ is the noise in the integrated \HI\ map. For
independent channels (as is the case with the THINGS cubes) $\sigma_N$
is defined as $\sigma_N = \sqrt{N} \sigma_{\rm chan}$, where $N$ is
the number of channels with signal contributing to each
pixel\footnote{Note that in creating an integrated \HI\ map one
  \emph{adds} emission across a number of channels, rather than
  calculating the \emph{average}. Provided the channels are
  independent, i.e., the noise is not correlated between channels, the
  noise thus \emph{increases} as the square root of the number of
  channels added.}.  These various steps result in objectively defined
Hermite velocity fields that form the basis of our analysis.

\subsection{Deriving the rotation curves\label{deriverotcur}}

The high spatial and velocity resolutions of the THINGS observations,
as well as the favorable inclinations of the sample galaxies, make
these objects ideal targets for a tilted-ring rotation curve analysis.
In a standard tilted-ring analysis
a galaxy is described using a set of concentric rings, each with their
own inclination $i$, position angle PA, systemic velocity $V_{\rm sys}$,
center position $(x_0,y_0)$ and rotation velocity $V_C$.  Assuming
that the gas moves in purely circular orbits, one can then describe
the line-of-sight velocity for any position $(x,y)$ on a ring with
radius $R$ as
\begin{equation}
V(x,y)=V_{\rm sys}+ V_C(R)\sin(i)\cos(\theta).
\end{equation} 
In this equation, $\theta$ is the position angle with respect to the
receding major axis measured in the plane of the galaxy. This quantity is
related to the position angle PA in the plane of the sky by
\begin{subequations}
\begin{eqnarray}
\cos(\theta) = \frac{-(x-x_0)\sin({\rm PA})+(y-y_0)\cos({\rm PA})}{R}\\
\sin(\theta) = \frac{-(x-x_0)\cos({\rm PA})-(y-y_0)\sin({\rm PA})}{R\cos(i)}.
\end{eqnarray}
\end{subequations}
PA is defined as the angle measured counter-clockwise between the
north direction on the sky and the major axis of the receding half of
the galaxy. For each ring the parameters are varied using a
least-squares algorithm until an optimum fit with the velocity field
is achieved.

We use the GIPSY task {\sc rotcur} to make the tilted-ring fits and
derive the rotation curves. As positions closer to the major axis
carry more rotational information than positions near the minor axis,
we use a $|\cos(\theta)|$ weighting for all fits.  In defining the
radii and widths of the rings we sample the rotation curve at a rate
of two points per synthesized beam width.

In this section we give a general description of the procedure we
used.  Detailed, and more technical, descriptions of the derivation of
individual rotation curves  are given in the Appendix.
For each galaxy we firstly fix the position of the dynamical center to
that determined in the analysis of \citet{clemens2007} who used the
same data.  For completeness, the resulting coordinates of the
dynamical centers are given in Table~\ref{bigtable}.  After fixing the
center position, an additional {\sc rotcur} run is used to determine
$V_{\rm sys}$ (with PA and $i$ still left as free parameters). In
general the systemic velocity has very little uncertainty associated
with it.  The adopted values are given in Table~\ref{bigtable}.  After
fixing the central position and systemic velocity, {\sc rotcur} is run
once more with PA and $i$ as free parameters.  The trends of $i$ and
(especially) PA with radius are generally well-defined, varying
smoothly with radius and exhibiting little scatter.  Of the two
fitting parameters, the PA is very stable, and any small perturbations
in its value result in only second-order changes in the rotation
curve.  In some cases, small-scale fluctuations in the inclination are
seen. These are mostly caused by effects such as streaming motions
along spiral arms which the fitting program tries to ``compensate''
for by changing $i$.

To prevent these small spurious changes and to retrieve the underlying
``bulk'' rotation, we describe the PA and $i$ distributions by
slightly smoothed radial distributions that ignore these small-scale
``wiggles''. A simple box-car smoothing with a kernel width of 3 or 5
data points was used in most cases.  This only affects the
point-to-point scatter of the $i$ and PA values for the few galaxies
where these wiggles are relevant, and does not affect the resolution
of their radial distributions. With PA and $i$ fixed using these
radial distributions, we derive the final rotation curve. We also
construct model velocity fields based on the tilted-ring models, as
well as corresponding residual velocity fields containing the
non-circular component.

The inclination wiggles described above, as caused by streaming
motions, affect only a small part of the area of the velocity field
for a small number of galaxies and do not affect the shape of the
rotation curves in any critical way. Our treatment of the $i$ and PA
distributions assigns the dominant non-circular motions to the
residual velocity field. An explicit comparison between the {\sc
  rotcur} residual velocity fields and the results from a full
rigorous treatment of the non-circular motions is made in
\citet{clemens2007} and shows that the above procedure is justified.
Care was also taken to double-check that the solutions produced using
our PA and $i$ distributions were self-consistent (i.e., we checked
whether running {\sc rotcur} again with only PA and $i$ free, and
$V_C$ fixed, yielded PA and $i$ distributions consistent with our
input values).

The method just described assumes galaxies are azimuthally symmetric.
However, asymmetries in the disk, or lopsidedness, can introduce
intrinsic differences.  To quantify the uncertainties introduced by
this, we derive separate fits to the rotation curves of the
approaching and receding sides of the galaxies.

A final issue is the definition of the final uncertainties in the rotation
curves. Unfortunately, there is no consensus in the literature on how
to quantify these.  Sometimes the formal $1\sigma$ $\chi^2$-fit
uncertainty in $V_C$ is used.  However, by their nature, these uncertainties
are much smaller than the dispersion of individual velocity values
found along a tilted ring.  This dispersion is, therefore, also
commonly used as an alternative definition, as it is more
representative of the physical uncertainties than the formal fit
error. For the NGC 3198 rotation curve (discussed in Sect.~4.8), for
example, the average $1\sigma\ \chi^2$ fit error is $\sim 17$ times
smaller than the average dispersion along the rings.  Using only
formal $\chi^2$-based error bars thus severely underestimates the true
``physical'' uncertainty.

An additional source of uncertainty is formed by possible differences
in rotation velocity between the approaching and receding sides.
\citet{swaters_phd} assumes that the differences between the rotation
curves derived for the entire disk and those for either the
approaching or receding side represent a $2\sigma$ difference, and
defines a pseudo-$1\sigma$ uncertainty due to asymmetries, as one
fourth of the difference between the approaching and receding side
velocities. Though this is only an assumption, we follow this
convention and define the uncertainties in the rotation curves as the
quadratic addition of the dispersions found along the rings and the
pseudo-$1\sigma$ uncertainties due to asymmetries between approaching
and receding sides. Note that this is a very conservative definition:
the difference in velocity between approaching and receding sides will
already be partly reflected in the dispersion found along a ring.
The uncertainties in the rotation velocities presented here thus give
a realistic, and possibly even conservative, picture of the error
budget of our rotation curve determinations.

\section{Individual galaxies}

Detailed descriptions of the data as well as the derivation of the
tilted ring models and the resulting rotation curves of individual
galaxies are given in the Appendix.  In this section we compare
these results with previous determinations of the various H{\sc i}
rotation curves from the literature. (For NGC 3621 and NGC 3627 we
could find no previous, published determinations of their
\HI\ rotation curves. These are therefore  not discussed in this
Section and we refer to the Appendix for more information.) We do not
compare with H$\alpha$ or CO rotation curves.  These are determined
using different methods, and originate from different phases of the
ISM; systematic effects due to these differences are not
straight-forward to quantify and beyond the scope of the current
paper.

For reference, a summary of the resulting tilted-ring parameters is
given in Table \ref{bigtable}. These include the position of the
dynamical center, the rotation curve sampling interval, systemic
velocity and radially averaged values of PA and $i$. For the central
positions we use the results presented in \citet{clemens2007} as
derived using the same data. We adopt luminosities and distances as
given in \citet{THINGS1}, and where necessary literature measurements
have been corrected to the same distance scale.

\subsection{NGC 925\label{sec:indiv_gal_begin}}

NGC 925 is classified as a late-type barred spiral. It was previously
observed in \HI\ by \citet{DJ}. Figure \ref{fig:n925_comp} compares
the rotation curves.  The main difference with our analysis, namely
the position of the dynamical center, is discussed in
\citet{clemens2007}. This difference also explains the slightly higher
value for $V_{\rm sys}$ of $551.5$ \kms\ found by \citet{DJ}, compared
to our 546.3 \kms. Other than this, we derive broadly similar trends
of $i$ and PA with radius. The curves themselves agree within the
uncertainties, showing that for solid-body (inner) rotation curves,
such as this one, the shape does not critically depend on the position
of the center. The uncertainties quoted in \citet{DJ} are the formal
fit errors. As discussed before, these are unrealistically small,
representing the error in the mean velocity along a ring, rather than
the scatter. In this case these formal errors are smaller than the
sizes of the symbols in Fig.~\ref{fig:n925_comp}.  The uncertainties
we have adopted are more representative of the true physical
uncertainties.

\subsection{NGC 2366}

The data and analysis of NGC 2366 are given in \citet{oh2007}. We
refer to their paper for a complete discussion.

\subsection{NGC 2403}

NGC 2403 is a late-type Sc spiral and member of the M81 group.  Its
\HI\ rotation curve has been derived in many previous studies. The
first measurement using a synthesis telescope was presented in
\citet{shostak1973}.  NGC 2403 has since been re-observed and
reanalyzed at ever increasing sensitivity and resolution by amongst
others \citet{bosma_phd}, \citet{begeman87}, \citet{sicking_phd} and
\citet{fraternali}. In Fig.~\ref{fig:n2403_comp} we compare our curve
with some of these analyses, all corrected to the same distance.
There is generally good agreement between the various
determinations. Beyond $\sim 13$ kpc the curves differ somewhat, but
most of the velocity information at these radii originates from close
to the minor axis of the velocity field, and thus depends on details
such as the weighting used in the tilted ring fit, the flux limit used
to mask the velocity field, etc.

Other parameters also compare favorably: the average $i$ and PA values
found by \citet{begeman87} were $60.2^{\circ}$ and $122.5^{\circ}$,
respectively.  \citet{sicking_phd} found $60.9^{\circ}$ and
$123.9^{\circ}$, whereas \citet{fraternali} derived $62.9^{\circ}$ and
$124.5^{\circ}$. Our average values (defined as the average unweighted
values of the model distributions) agree well, at $62.9^{\circ}$ and
$123.7^{\circ}$, respectively.  Similarly, the average absolute
difference between our $V_{\rm sys}$ value and the three literature
values is only $0.5$ \kms.

The ``bump'' at $R\simeq 15$ kpc in the THINGS curve is real, in the
sense that it cannot be explained by an incorrectly chosen $i$ or
PA. Most of the information in this part of the curve originates from
the two arms that are seen to stretch away in the outermost northern
and southern parts of the \HI\ map. Motions along these arms could
explain the feature, as suggested by the kinks in the velocity
contours at those radii.  The velocity fields of NGC 2403 are also
discussed in Sec.~3.2.

\subsection{NGC 2841}

NGC 2841 is an early-type (Sb) spiral and was previously observed at
lower resolution in \HI\ by \citet{begeman87}.  We compare the curves
in Fig.\ \ref{fig:n2841_comp} and find good correspondence overall.
\citet{begeman87} finds a systemic velocity of $631.1 \pm 1.4$ \kms, a
difference of only 2.6 \kms\ with our value of $633.7 \pm 1.8$
\kms. The average PA and $i$ values found by \citet{begeman87} are
$154.5^{\circ}$ and $73.1^{\circ}$, respectively, and differ only by a
few degrees from our values (cf.\ Table \ref{bigtable}).

\subsection{NGC 2903}

NGC 2903 was previously observed in \HI\ by \citet{begeman87}.
Fig.\ \ref{fig:n2903_comp} compares the two rotation curves.  The
\citet{begeman87} value of $V_{\rm sys} = 557.3 \pm 1.3$ \kms\ agrees
well with our value of $555.6 \pm 1.3$ \kms.  The outer declining
parts of the rotation curves also agree well with each other.  There
are however significant differences in the inner, rising parts of the
rotation curve, where the curve by \citet{begeman87} rises much more
steeply.  His data did not allow a full tilted-ring model in the inner
part of the galaxy, and the sudden changes in PA and $i$ are therefore
also not detected in his analysis. Rather, the innermost data points
of the \citet{begeman87} curve were derived from a major axis
position-velocity diagram.  The higher spatial and velocity resolution
of our data and the different methods used to extract the velocities
both contribute to this large difference.

\subsection{NGC 2976}

NGC 2976 is classified as an Sc galaxy and was previously observed in
\HI\ by \citet{stil}.  They chose the center of mass of the \HI\ disk as
their dynamical center, and there is an offset of $\sim 10''$ with
respect to our choice. Their choice for $V_{\rm sys}$ is also slightly
different from ours: $4\pm 2$ \kms, versus our $1.1 \pm 1.3$ \kms.
\citet{stil} find $i=65^{\circ}$ and PA $= 326^{\circ}$, versus our
values of $64.5^{\circ}$ and $334.5^{\circ}$, respectively.  Figure
\ref{fig:n2976_comp} compares the two curves.  The \citet{stil} curve
suggests a flattening towards the outer radii, whereas our curve keeps
rising until the last measured point. The curves do, however, agree
within their respective uncertainties.

\subsection{NGC 3031}

Better known as M81, this galaxy is the proto-typical grand-design
spiral and together with M82 and NGC3077 forms an interacting system,
\citep[e.g.,][]{vdhulst1979,yun1994}.  A previous determination of the
\HI\ rotation curve was by \citet{visser78,visser80}.  We compare our
results in Fig.\ \ref{fig:n3031_comp} after correction to the same
distance scale. We find excellent agreement overall between his and
our rotation curves where they overlap in radius. The differences we
find are all attributable to different local values for PA and $i$: we
allow $i$ and PA to change with radius, whereas \citet{visser80} uses
a constant inclination of $59^{\circ}$ and a PA value of
329$^{\circ}$. These constant values agree, however, well with our
average values of $59^{\circ}$ and 330.2$^{\circ}$, respectively.

\subsection{NGC 3198}

NGC 3198 is known for its proto-typical flat rotation curve (see
e.g.\ \citealt{tjeerd85}).  We compare our rotation curve with those
derived by \citet{begeman87,begeman89} and \citet{sicking_phd} in
Fig.\ \ref{fig:n3198_comp}.  The systemic velocities agree within
their respective $1\sigma$ uncertainties: our result is $660.7 \pm
2.6$ \kms, \citet{begeman89} finds $660.4 \pm 0.8$ \kms, whereas
\citet{sicking_phd} finds $659.4 \pm 2.6$ \kms.  The respective
average PA and $i$ values only show differences $\la 1^{\circ}$.

Turning now to the rotation curves themselves and starting with the
outer parts, we find that our curve agrees well with the
\citet{sicking_phd} result, but is a few \kms\ higher than the
\citet{begeman89} data.  Noting that the uncertainties in neither the
Sicking nor the Begeman analysis do take into account the difference
between approaching and receding sides, the results are still
consistent with each other.

In the inner parts, the \citet{begeman89} data show consistently
higher rotation velocities.  These were, however, derived from
major-axis position velocity diagrams, rather than from a tilted-ring
model. Due to the much lower resolution they were also explicitly
corrected for beam smearing. The systematic offset could therefore
imply an overcorrection of the \citet{begeman89} data, or a systematic
deviation due to the use of major axis profile
velocities. \citet{begeman87} mentions that the beam smearing
corrections are only significant within $1.5'$ ($\sim 6$ kpc),
precisely the radius within which the curves show disagreement.  The
inner parts of our curve are fully consistent with the
\citet{sicking_phd} rotation curve.

\subsection{IC 2574}

The data and analysis of IC 2574 are described by \citet{oh2007} and
we refer to their paper for a more complete discussion. The velocity
fields of IC 2574 have also been discussed in Sec.~3.3.

\subsection{NGC 3521}

NGC 3521 was previously observed in \HI\ by \citet{cas_vg91}. Its
rotation curve was presented as one of the first cases where a
genuinely declining \HI\ rotation curve was found in a spiral galaxy.
We compare our curve with the \citet{cas_vg91} curve in
Fig.~\ref{fig:n3521_comp}.  The agreement is not perfect, the
literature curve underestimates the rotation velocity in the inner
parts, whereas our curve does not show the steep drop in the outer
parts.  Unfortunately, \citet{cas_vg91} do not list the values they
assume for $i$ and PA, which precludes a more detailed comparison.
The high inclinations, and resulting insensitivity of the rotation
velocities to the exact inclination values, make it unlikely that the
difference is a pure inclination issue.  For example, a change in $i$
of 5 degrees, starting from an initial value of $i = 75^{\circ}$, only
amounts to a change of some 2 percent or $\sim 4$ \kms, not the $\sim
30$ \kms\ that is needed to bring the \citet{cas_vg91} in agreement
with ours. In summary, even though we find rotation velocities in the
outer parts that are lower than in the inner parts, we do not find
evidence for the steep drop, nor for the negative gradient in 
outer velocities as presented in \citet{cas_vg91}.

\subsection{NGC 4736}

NGC 4736 was previously observed in \HI\ by \citet{mulder} who also
derived a rotation curve. Fig.\ \ref{fig:n4736_comp} shows a
comparison between the respective rotation curves.  Due to the lower
resolution and sensitivity of his data, \citet{mulder} did not
detect a trend of inclination with radius, and therefore assumed a
constant value of $40^{\circ}$. In Fig.\ \ref{fig:n4736_comp} we
therefore also show his rotation curve corrected using our inclination
values. Although the large-scale trends are similar to those found for
our curve, the THINGS curve in general shows more pronounced
small-scale features.

\subsection{DDO 154}

DDO 154 is a gas-rich dwarf galaxy whose rotation curve has been
extensively studied \citep[see][]{car_freeman, car_purton}. It was
also was one of the first galaxies whose rotation curve was used to
illustrate the conflict with Cold Dark Matter predictions
\citep{moore94}. DDO 154 shows clear evidence for a warp in the outer
parts. There is no evidence for large non-circular motions in the
velocity field \citep{clemens2007}.
This is evidenced by the residual
velocity field where the absolute value of the residuals nowhere
exceeds $\sim 6$ \kms.

Previous determinations of the rotation curve of DDO 154 \citep[see
e.g.,][]{car_freeman, car_purton}, show an apparent drop in the
rotation velocity in the outermost parts, which was interpreted as
evidence that at these radii one has almost reached the edge of the
dark matter halo.  In Fig.\ \ref{fig:ddo154_comp} we compare our
rotation curve with one of these determinations. Contrary to the
\citet{car_purton} and \citet{car_freeman} results, we do not find
strong evidence for a decline and our rotation curve in the outer
parts is consistent with being flat.

Given the differences found in the tilted ring parameters of the
approaching and receding sides separately, the error bars belonging to
the \citet{car_purton} rotation curve most likely underestimate the
true uncertainties. Also note that our rotation curve in general shows
higher rotation velocities than the \citet{car_purton} one. This is
not merely an inclination effect, as our values for $i$ are in general
higher by a few degrees than the ones presented in
\citet{car_purton}. Correcting the curves to the same inclinations
would therefore only increase the difference.  Note that a model with
a decreasing inclination in the outer parts, as perhaps suggested by
our analysis (cf.\ the Appendix), would lead to an \emph{increasing}
rotation velocity in the outer parts. The only way in which our
determination of the rotation curve could be made consistent with a
\emph{declining} rotation curve is to increase the inclination, which
is ruled out by our data.

\subsection{NGC 4826}

NGC 4826, or the ``Evil-Eye'' galaxy is an early-type spiral with as
its most remarkable kinematical feature the presence of two
counter-rotating gas disks \citep{braun94}: the inner high-column
density \HI\ disk is associated with the bright stellar disk and
rotates in the same direction. The outer, much lower column-density
\HI\ disk rotates in the direction opposite to that of the stars.  The
transition between the two disks occurs around $\sim 100''$
radius. \citet{braun94} speculate that the outer disk might find its
origin in a merger of a gas-poor spiral with a gas-rich and star-poor
dwarf with a retrograde spin (with the mass ratio of the dwarf and the
spiral $\la 0.1$; \citealt{rix95}).

In Fig.\ \ref{fig:n4826_comp} we compare our rotation curve with that from
\citet{braun94}. Even though to first order there is reasonable
agreement, the scatter in the velocities is much larger than in any of
the other rotation curves presented here.  Braun et al.\ remark that
between $100'' \la R \la 300''$ ($4\ {\rm kpc} \la R \la 12\ {\rm
  kpc}$) the kinematical parameters are difficult to determine due to
what they call a ``severe distortion'' in the velocity field, showing
that the transition region between the two counter-rotating disks
deviates from circular motions. Similar conclusions were reached for
the ionized gas by \citet{rix95}.

\subsection{NGC 5055}

NGC 5055 is an Sbc galaxy with an extended and warped tenuous outer
\HI\ disk.  Its \HI\ distribution and kinematics were recently
analyzed by \citet{battaglia}.  For $V_{\rm sys}$, \citet{battaglia}
find a value of $497.6 \pm 4.8$ \kms, which is in close agreement with
our value of $496.8 \pm 0.7$ \kms. Their distributions of $i$ and PA
are also in agreement, and they note a similar difference between the
models for approaching and receding sides.

We compare the two rotation curves in Fig.\ \ref{fig:n5055_comp}. The
first inner maximum at $R\sim 2$ kpc is more pronounced in our data
set. This is likely due to a combination of increased resolution and
the use of Hermite polynomials as opposed to the Gaussian functions that
\citet{battaglia} used. This is also the most likely explanation for
the higher rotation velocities we find in the outer parts.

\subsection{NGC 6946}

NGC 6946 is a late-type spiral which, in terms of \HI\ studies, is
better known for its population of \HI\ structures such as \HI\ holes
and high-velocity gas \citep[e.g.,][]{kamphuis,boomsma}. Its global
dynamics have been relatively little studied. This can be partly
attributed to its fairly low inclination which is at the limit of what
is feasible using tilted-ring studies.

The rotation curve of NGC 6946 was determined before by
\citet{carignan6946} and more recently by \citet{boomsma}.  The
comparison with our curve is made in Fig.\ \ref{fig:n6946_comp}. Both
\citet{carignan6946} and \citet{boomsma} used a constant value of $i =
38^{\circ}$ in their analyses and these different inclinations, compared to our average value of 32.6$^\circ$,  explain
the offset between the curves.  Apart from this, the only
prominent difference is the large rotation velocity \citet{boomsma}
finds for the innermost point of the rotation curve. In our data the
corresponding area of the velocity field is below our $3\sigma$ \HI\
column density level, and we can therefore not comment on the presence
of this feature in our data, though there is a hint of a turn-up in
the major axis position-velocity diagram (cf.~Fig.~\ref{fig:n6946}).

\subsection{NGC 7331}

NGC 7331 is an early-type Sb spiral with prominent spiral arms.  It
was previously observed in \HI\ by \citet{begeman87}.  Figure
\ref{fig:n7331_comp} compares both curves. It is clear that the two
curves broadly agree, but there are many small differences.  Most of
them can be attributed to the very different spatial and velocity
resolutions and different choices for the inclination distribution
(see discussion in the Appendix).  Note that the inner 8 points of the
\citet{begeman87} curve are based on position-velocity diagrams and
not on a tilted-ring fit to a velocity field. The differences in the
outermost parts are due to the different inclinations used there.

\subsection{NGC 7793\label{sect:7793}\label{sec:indiv_gal_end}}

\begin{figure*}[t]
\epsfxsize=0.9\hsize \epsfbox{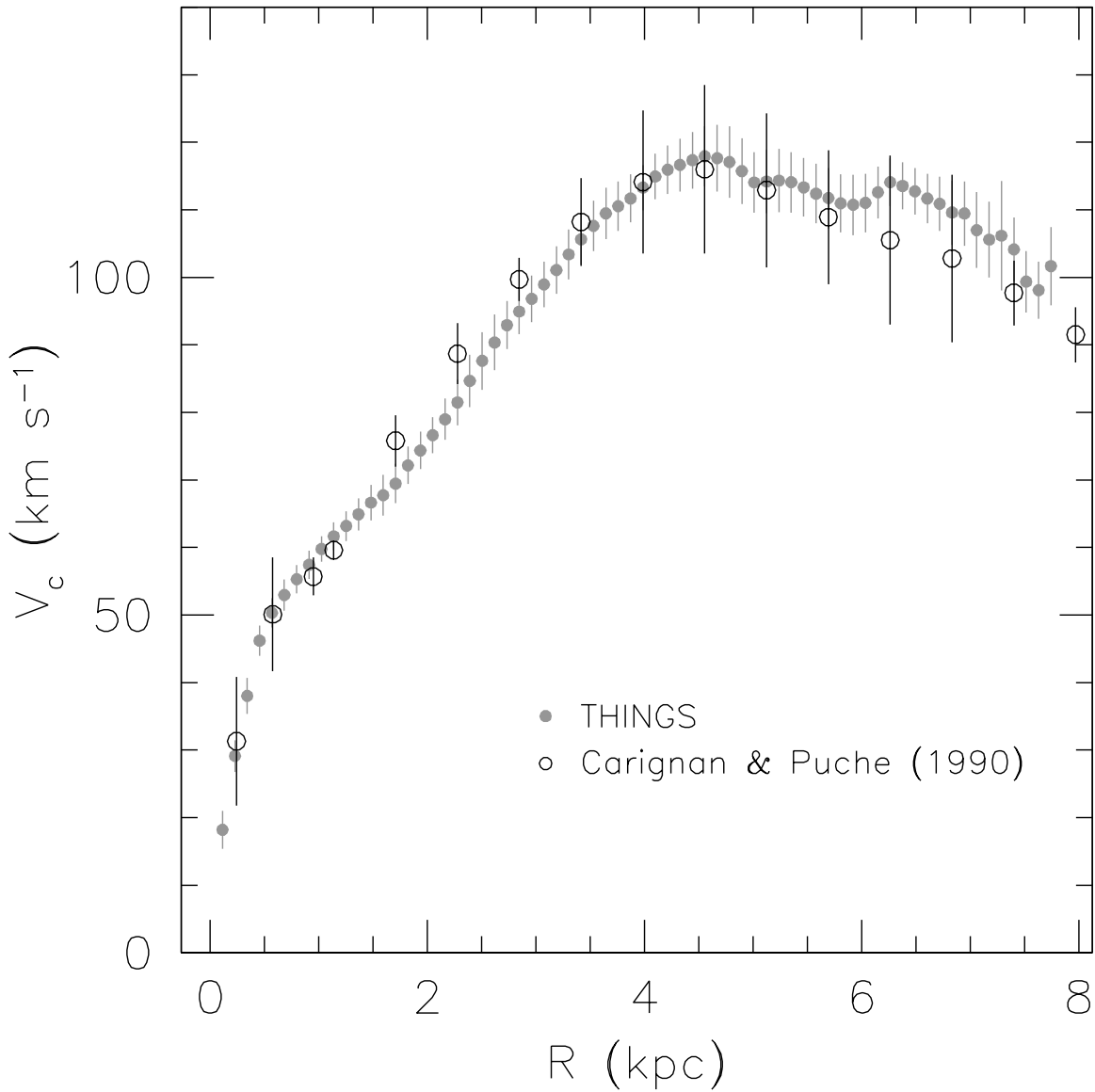} 
\figcaption{Comparison
  of the NGC 7793 THINGS rotation curve with the curve derived in
  \citet{carignan90}.  Symbols and references are indicated in the
  figure.
\label{fig:n7793_comp}}
\end{figure*}

NGC 7793 is a late-type Sd spiral galaxy in the Sculptor group and was
observed before in \HI\ by \citet{carignan90}.  In
Fig.\ \ref{fig:n7793_comp} we compare our rotation curve with their
results.  The dynamical center and systemic velocity, as well as PA,
as found by \citet{carignan90} agree with ours within the
uncertainties.  The behavior of the inclination is subtly different:
whereas the \citet{carignan90} analysis indicates a rising inclination
in the outer parts, we find a (better defined) gradual decrease there.
The resulting rotation curves on the whole agree reasonably well with
each other, but differ in small but important respects. The most
important one is the more gradual decline of our rotation curve in the
outer parts. Even though \citet{carignan90} note that this is ``one of
the very few cases with a \emph{truly declining} rotation curve''
(their emphasis), we argue that any drop in rotation velocity is much
less extreme, especially as there is some evidence that the inclination
might decrease more strongly in the outer parts than we have assumed
(cf.\ the distribution of inclination with radius of the approaching
and receding sides in Fig.\ \ref{fig:n7793}). Adopting this steeper
drop would raise the rotation velocities we find in the outer parts by
$\sim 25$ \kms. We stress that these inclination values are uncertain,
and determined by only a small number of pixels, but they seem to
suggest that a declining rotation curve cannot be unambiguously
established.

\subsection{Summary}

In summary, we find in general good agreement with previous
determinations of the rotation curves of the galaxies in our
sample. The most notable differences are:
\begin{itemize}

\item We find some differences with previous determinations of the
  inner rotation curves of NGC 2903 and NGC 3198. Both galaxies
  contain a small central bar. The presence of this feature combined
  with the lower resolution of the literature data can contribute to
  this difference.

\item The outer rotation curve of NGC 5055 is higher than previously
  determined. Most likely our use of a Hermite velocity field
  contributes to this difference.

\item We do not find steep declines in velocity in the outer rotation
  curves of NGC 3521, NGC 7793, DDO 154 and NGC 2366 (see
  \citealt{oh2007} for the latter curve). Where declines are observed,
  they are more gentle, and (within the uncertainties in rotation
  velocity and inclination) consistent with flat rotation curves.
\end{itemize}

\begin{deluxetable*}{llrrrrrr}
\tablewidth{0pt}
\tablecaption{Tilted-ring model parameters of the THINGS galaxies}
\tablehead{
\colhead{Name} & \colhead{$\alpha(2000.0)$} & \colhead{$\delta(2000.0)$}  & \colhead{$D$} & \colhead{$\Delta R$} & \colhead{$V_{\rm sys}$} & \colhead{$\langle i \rangle$} & \colhead{$\langle {\rm PA} \rangle$} \\
\colhead{} & \colhead{(\ $^h\ ^m\ ^s$)} & \colhead{$(\ ^{\circ}\ '\ '')$}  & \colhead{(Mpc)} & \colhead{$('')$} & \colhead{(\kms)} & \colhead{$(^{\circ})$} & \colhead{$(^{\circ})$}\\ 
\colhead{(1)}&\colhead{(2)}&\colhead{(3)}&\colhead{(4)}&\colhead{(5)}&\colhead{(6)}&\colhead{(7)}&\colhead{(8)}
 }
\startdata
NGC 925  & 02 27 16.5 & +33 34 43.5   & 9.2 & 3.0 & 546.3 & 66.0 & 286.6  \\
NGC 2366 & 07 28 53.4 & +69 12 51.1  & 3.4  & 6.0 & 104.0 & 63.8 & 39.8 \\
NGC 2403 & 07 36 51.1 & +65 36 02.9  & 3.2  & 4.0 & 132.8 & 62.9 & 123.7 \\
NGC 2841 & 09 22 02.6 & +50 58 35.4  & 14.1 & 5.0 & 633.7 & 73.7  & 152.6  \\
NGC 2903 & 09 32 10.1 & +21 30 04.3   & 8.9 & 7.0 & 555.6 & 65.2  & 204.3 \\
NGC 2976 & 09 47 15.3 & +67 55 00.0  & 3.6 &3.5 &  1.1   & 64.5  & 334.5 \\
NGC 3031 & 09 55 33.1 & +69 03 54.7   & 3.6 &6.0 & --39.8 & 59.0  & 330.2  \\ 
NGC 3198 & 10 19 55.0 & +45 32 58.9   & 13.8  & 6.0 &  660.7 & 71.5  & 215.0\\
IC 2574  & 10 28 27.7 & +68 24 59.4   & 4.0 & 6.0 & 53.1  & 53.4  & 55.7  \\
NGC 3521 & 11 05 48.6 & --00 02 09.2  & 10.7 & 6.0 & 803.5 & 72.7 & 339.8 \\ 
NGC 3621 & 11 18 16.5 & --32 48 50.9 &  6.6 & 6.5& 728.5 & 64.7 & 345.4  \\ 
NGC 3627 & 11 20 15.0 & +12 59 29.6   &  9.3 & 5.0 & 708.2 & 61.8 & 173.0 \\ 
NGC 4736 & 12 50 53.0 & +41 07 13.2   &  4.7 & 5.0 & 306.7 & 41.4  & 296.1 \\ 
DDO 154  & 12 54 05.7 & +27 09 09.9   & 4.3 & 6.5 & 375.9 & 66.0  & 229.7  \\
NGC 4826 & 12 56 43.6 & +21 41 00.3  &  7.5  & 5.0 & 407.4 & 65.2  & 120.9  \\ 
NGC 5055 & 13 15 49.2 & +42 01 45.3  &  10.1 & 5.0 & 496.8 & 59.0  & 101.8  \\ 
NGC 6946 & 20 34 52.2 & +60 09 14.4   & 5.9 & 3.0 & 43.7  & 32.6  & 242.7\\
NGC 7331 & 22 37 04.1 & +34 24 56.5  & 14.7 & 3.0 & 818.3 & 75.8 & 167.7  \\
NGC 7793 & 23 57 49.7 & --32 35 27.9  & 3.9 &6.0  & 226.2 & 49.6 & 290.1 \\
\enddata
\tablecomments{(1) Name of galaxy; (2) Right Ascension (J2000.0); (3)
  Declination (J2000.0), center positions from \citet{clemens2007};
  (4) Distance as given in \citet{THINGS1}; (5) Sampling increment
  used to derive the rotation curve; (6) Adopted systemic velocity;
  (7) Average value of the inclination; (8) Average value of the position
  angle of the receding side, measured from north to east, and in the plane of the sky.}
\label{bigtable}
\end{deluxetable*}

\section{Mass models}

We now use the THINGS rotation curves, in combination with information on the
distribution of gas and stars, to construct mass models of our sample
galaxies. These models are then used to quantify the distribution of
dark matter within the galaxies.
We calculate our best model estimate for the observed velocity as 
\begin{equation}
V_{\rm obs}^2 = V_{\rm gas}^2 + \Upsilon_{\star}V_{\star}^2 + V_{\rm halo}^2.
\label{eq:mm}
\end{equation}

The mass models thus need as input the observed rotation curve,
$V_{\rm obs}$, the rotation curve of the gas component, $V_{\rm gas}$,
and the rotation curve of the stellar component, $V_{\star}$. The
rotation curve from the halo, $V_{\rm halo}$, can then be calculated
from these known input curves. An additional free parameter is the
stellar mass-to-light ratio, \MLstar, which is introduced as we
generally can only measure the distribution of the \emph{light} of the
stellar population, rather than the required \emph{mass}.  This
parameter introduces one of the largest uncertainties in the procedure
and will be discussed in detail in Sect.~5.2.

While the \emph{concept} of mass modeling is well described by
Eq.~\ref{eq:mm}, there are some intricacies that warrant a brief
discussion.  The rotation velocities under consideration are in all
cases the velocities of respective test particles moving in circular
orbits in the plane of the galaxy. The rotation velocities of the gas,
$V_{\rm gas}$, and the stars, $V_{\star}$, are defined as the
velocities that each of the components would induce on a test particle
in the plane of the galaxy, if these disks were sitting in isolation,
without any external influences. These velocities in the plane are
calculated from the observed baryonic mass density distributions,
taking into account the (assumed) vertical density distribution.  For
the gas disk one generally assumes a thin disk, while for the stellar
disk a thick, sech$^2$ distribution is chosen.  These procedures and
choices of parameters are described in more detail on Sect.~5.1 for
the rotation curve of the gas and in Sect.~5.2 for that of the
stars. The latter section also describes in detail possible choices
for \MLstar.

The halo velocities, $V_{\rm halo}$, in Eq.~\ref{eq:mm} (or rather the
velocities induced on a test particle by the halo potential) are also
measured in the plane of the galaxy and assume a spherical halo.
Sect.~5.3 briefly discusses the dark matter models used, and in
Sect.~5.4 the procedure of constructing the models is presented.
Specific discussions dealing with individual galaxies are given in
Sect.~6.

Initial analysis of the rotation curves of NGC 3627 and NGC 4826
showed their dynamics to be severely affected by non-circular motions
\citep{clemens2007} and  we will therefore not consider these two
galaxies further.

\subsection{Neutral Gas Distribution}

To derive the rotation curve associated with the neutral gas
component, we use the primary beam and residual-scaled integrated
H{\sc i} maps from \citet{THINGS1}. We derive neutral hydrogen surface
density profiles using our previously derived tilted ring geometrical
parameters. The surface densities are corrected by a factor 1.4 to
take into account the presence of helium and metals (associated with
the atomic hydrogen; the helium component by itself scales with a
factor of 1.33).  The {\sc rotmod} task in GIPSY is then used to
compute the corresponding rotation curve, under the assumption of an
infinitely thin gas disk.

As we are dealing with a mass distribution in the form of a
\emph{disk}, rather than a sphere, the rules-of-thumb that apply to
the gravitational force of a spherical mass distribution do not apply.
Most importantly, the convenient result that the mass distribution
\emph{outside} a particular radius does not affect the effective force
inside that radius is not universally valid for a disk.  Specifically,
in the case of a disk with a sufficiently prominent central depression
in the mass distribution, a test particle located in or near it can
feel a net \emph{outward} force. A literal interpretation in the
context of normal gravity would imply negative mass, an imaginary
rotation velocity and therefore a negative value of $V^2$ Here,
following the convention in rotation curve literature, we plot the
imaginary rotation velocities as negative velocities, keeping in
mind that it really is $V^2$ that is
negative.  We stress that this convention for representing the
imaginary velocities is merely adopted for convenience and does not
imply counter rotation, nor the presence of repulsive matter or other
exotic phenomena: it simply is a short-hand to enable illustrating the
effective force on a test particle caused by a non-spherical mass
distribution with a central minimum. The mass model of NGC 3031 in
Sect.~6.6 is a good example of this phenomenon.

\subsection{Stellar Distribution}

One of the main uncertainties in deriving mass models is the
conversion from observed stellar luminosity to stellar mass (through
the stellar mass-to-light ratio $\Upsilon_{\star}$).  In the optical
bands, $\Upsilon_{\star}$ is dominated by the young stellar
population, recent star formation events and the effects of dust and/or
metallicity.  Values of $\Upsilon_{\star}$ derived in, e.g., the
$B$-band thus have inherently a large uncertainty which can be up to a
factor of a few.  These limitations are the reason why extreme
mass-models involving so-called ``maximum'' or ``minimum disks''
\citep{maxdisk86} are often used. The maximum disk assumption
maximizes the contribution of the baryonic matter by assuming the
maximum value of \MLstar\ allowed by the total rotation curve. Minimum
disk is the opposite and minimizes the baryonic contribution by
setting $\MLstar = 0$. Sometimes the contribution of the gas is also
ignored (and all rotation is therefore assumed to be induced by the
dark matter halo). These two extreme \MLstar\ assumptions can thus be
regarded as ``minimum halo'' and ``maximum halo'' models,
respectively, minimizing or maximizing the contribution of the
dark matter. Though not necessarily physically plausible, they do
provide hard upper and lower limits on the concentration and
distribution of the dark matter.

Fortunately, the uncertainties in \MLstar\ decrease dramatically
towards the near-infrared (from a factor $\sim 10$ in the $B$-band to
a factor $\sim 3$ in $K$; cf.\ \citealt{bell_dejong}).  At near-IR
wavelengths one mostly probes the old stellar population containing
the bulk of the stellar mass, as described in \citet{bell_dejong} and
\citet{verheijen_phd}. The range in plausible values for
$\Upsilon_{\star}$ is in the near-IR only a shallow function of the
color of the disk, and observations at these wavelengths are therefore
useful in constraining the mass of the disk. \citet{verheijen_phd}
finds that maximum disk fits to a sample of HSB galaxies in the UMa
cluster require only a small range of $\Upsilon_{\star}^{K'} = 0.7 \pm
0.2$. Similarly, \citet{bell_dejong} find, from population synthesis
modeling, that disks of galaxies exhibit a typical range of
$\Upsilon_{\star}^K = 0.5 \pm 0.2$ measured over the full range of
observed galaxy colors.

To characterize the stellar distributions in our sample galaxies, we
use the 3.6 $\mu$m images from SINGS \citep{sings}. (For NGC 2366, NGC
2903 and DDO 154, which are not part of the SINGS sample, we have
retrieved data from the \emph{Spitzer} archive.) These provide a good
proxy for the emission of the (old) stellar disk (cf.\ \citealt{pahre2004}).
Note that some of the 3.6 $\mu$m emission may originate in AGB stars,
PAHs and hot young stars, but as we will only be using the radial
distributions of the emission, their contribution (if present) will be
strongly diminished by the azimuthal averaging.

After the images were cleaned of foreground stars, the tilted ring
parameters were used to derive the radial luminosity profiles. We used
the standard photometric calibration provided with the SINGS images to
construct the surface brightness profiles.  In most cases the 3.6
$\mu$m surface brightness distributions can be well described by a
single exponential disk. In a small number of galaxies, an additional
central component was seen, containing only a small fraction of the
total luminosity of the galaxy.  It is a matter of debate whether it
is appropriate to call this central component a ``bulge''. The surface
brightness distributions measured are not as steep as the canonical
$R^{1/4}$ bulge surface distribution, which may reflect the THINGS
sample selection criteria, with its emphasis on more late-type
galaxies.  Beyond this, the radial range over which this inner profile
dominates the total emission is not large enough to determine
accurately enough the value of $n$ in an $R^{1/n}$ parameterization of
the surface brightness profile of that central component. This is,
however, not a major problem due to its rapid drop in surface
brightness and we found that most often an exponential surface
brightness distribution provides an adequate and convenient fit. We
experimented with a number of different functional power-law forms,
but found that this did not affect the shape of the (compact) rotation
curve of this component to any relevant degree, and does not impact on
our final mass models.  Where necessary, we therefore decomposed the
surface brightness profile in inner and outer exponential
components. In the rest of this analysis we refer to the outer
component as the ``outer disk'', or simply ``disk''. The inner
component we interchangeably call ``inner disk'' or ``bulge''.  This
simply is a short-hand to indicate the components, and,
especially for the inner component, should not be taken as a statement
on its implied evolution or other physical properties. Central surface
brightness values and scale lengths, where mentioned in the text, are
not corrected for inclination. Needless to say that this correction
\emph{has} been applied prior to calculation of the rotation curve.

As far as we are aware, no exhaustive investigation has yet been made
into the relation between 3.6 $\mu$m emission and $\Upsilon_{\star}$.
We therefore use an empirical approach based on results derived using
the 2MASS $J$, $H$ and $K$ bands to derive $\Upsilon_{\star}$ in the
3.6 $\mu$m band (hereafter \Ups) from the \emph{Spitzer} observations.
All our sample galaxies are included in the 2MASS Large Galaxy Atlas
\citep{2mass}, and we use the $J-K$ colors given there, together with
the relations between color and $\Upsilon_{\star}$ given in
\citet{bell_dejong}, as well as the relation between $\Upsilon_*^K$
and $\Upsilon_*^{3.6}$ from \citet{oh2007}, to derive the relation
between \Ups\ and $J-K$. We find
\begin{equation} 
\log(\Upsilon_{\star}^K) = 1.43\, (J-K) - 1.38
\end{equation}
and
\begin{equation}
\Ups = 0.92\, \Upsilon_{\star}^K - 0.05.
\end{equation}

One of the largest uncertainties in determining \MLstar\ is the
stellar Initial Mass Function (IMF). Some aspects of this, within the
context of THINGS, are discussed in \citet{adam}; see also
\citet{bell_dejong} and \citet{bell03} for a more general discussion.
Using stellar population synthesis models, \citet{bell_dejong} show
that a \citet{salpeter} IMF results in stellar disk masses that are
too heavy to be consistent with dynamical (maximum disk) constraints
from rotation curves. By scaling the stellar disk masses down by a
factor of 0.7, in order to make them consistent with the rotation
curves, they introduce a ``diet''-Salpeter IMF, equivalent to a
standard Salpeter IMF, but with a reduced number of stars below 0.35
$M_{\odot}$. This diet-Salpeter is thus the maximum IMF (i.e.,
yielding the highest disk masses for a given photometric constraint)
allowed by galaxy dynamics studies.  Studies of the stellar population
in the Milky Way suggest an IMF that produces lower disk masses
\citep{kroupa}. The diet-Salpeter disk masses are 0.15 dex (a factor
1.4) more massive than the Kroupa ones.  We refer to \citet{adam} and
\citet{bell03} for a discussion on other IMF indicators.

The disk masses as implied by Eq.~(4) assume the diet-Salpeter IMF,
and are therefore the maximum disk masses allowed for a given $J-K$
color.  To gauge the effects of a ``lighter'' IMF on our rotation curve
results we also derive disk masses assuming a Kroupa IMF by decreasing
the constant term in Eq.~(4) by 0.15 dex.

Disk galaxies typically show radial color gradients of a few tenths of
a magnitude, becoming bluer with radius. This is interpreted as a
combined stellar age and metallicity gradient, with the outer regions
younger and of lower metallicity; dust seems to play only a minor role
in this regard \citep{dejong96}.  In the optical, these gradients are
thought to imply radial $\Upsilon_{\star}$ differences of a factor of
$\sim 2$ between inner and outer parts of galaxies.  Even in the
$K$-band where many of the effects of recent star formation and dust
are minimized, changes in $\Upsilon_{\star}$ of $\sim 20$ to $\sim 30$
per cent are not uncommon.  We therefore associate the $(J-K)$ color
gradients observed in the disks of our sample galaxies with variations
in $\Upsilon_{\star}$.  \citet{oh2007} show that similar gradients in
$\Ups$ are expected in the 3.6 $\mu$m band\footnote{Note that, as
  mentioned before, a small amount of dust emission due to, e.g., PAHs
  may be present; this emission will, however, be localized and have a
  small surface covering factor. The azimuthal averaging used will
  also tend to minimize its contribution.}.  We refer to \citet{dejong96} for
an extensive discussion of color gradients in the optical and near-IR. Our
goal is to create mass models taking into account radial
$\Upsilon_{\star}$ variations within the stellar disk.  We determine
``\Ups-weighted'' mass-surface density profiles for our galaxies and
use these as input for {\sc rotmod} to compute the rotation curve of
the stellar disk. For these stellar disks we assume a vertical
sech$^2$ distribution and use $z_0=h/5$ where $h$ is the exponential
radial scale length \citep{piet1,piet2}\footnote{This canonical ratio
  of $h/z_0 = 5$ has also been confirmed by \citet{kregel_hz} who for
  a sample of 34 edge-on galaxies find a value of $h/z_0 = 4.8 \pm
  1.3$. This is consistent with $h/z_0=5$ which we will continue to
  use.}. The rotation curve derived in this way does not critically
depend on the value of $h/z_0$.

One component we have not explicitly addressed is the molecular gas.
However, as the molecular gas is (at least in non-dwarf) THINGS disk
galaxies distributed like the stars \citep{adam}, and with a surface
density only a few percent of that of the stars \citep{portas}, its
(small) contribution is reflected in a small increase in
\MLstar.

\begin{figure*}[t]
  \epsfxsize=0.9\hsize \epsfbox{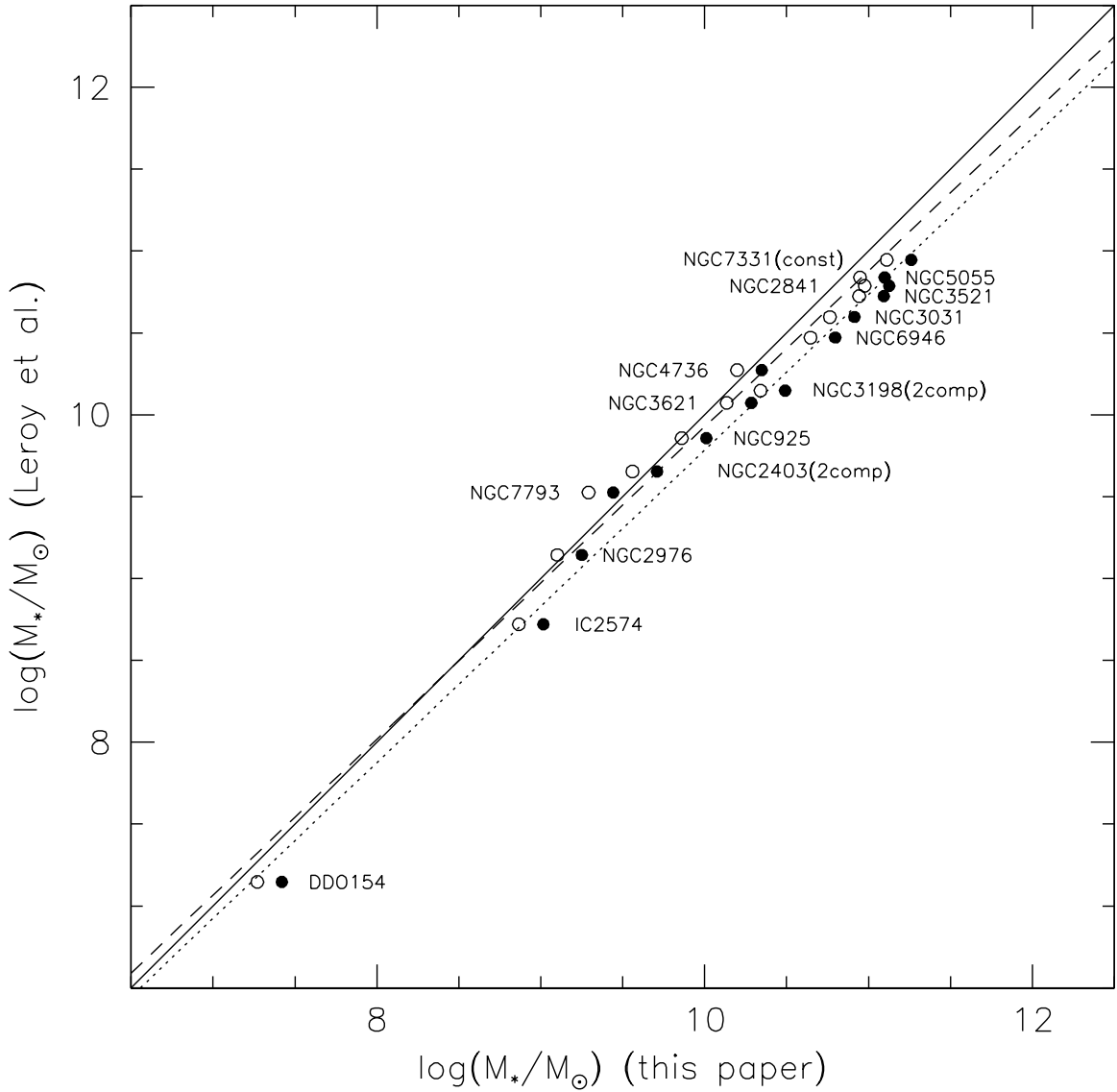} \figcaption{Comparison
    between the stellar disk masses derived using our method
    (horizontal axis) and the method described in \citet{adam}
    (vertical axis).  The filled circles indicate our diet-Salpeter
    IMF masses, the open circles the Kroupa IMF masses.  The drawn
    line is the line of equality, the dotted line a least-squares fit
    to the diet-Salpeter IMF masses with a slope of 0.94 and an offset of
    0.44. The scatter is 0.12 dex.  The dashed line shows the fit to
    the Kroupa IMF masses with a slope of 0.95, an offset of 0.39 and
    a scatter of 0.13 dex.
\label{fig:adam}}
\end{figure*}

In an accompanying paper, \citet{adam} present an alternative
empirical method to derive $\Upsilon_{\star}$ values from the 3.6
$\mu$m images.  They use the 3.6 $\mu$m and 2MASS $K$-band images
themselves to directly derive an empirical pixel-to-pixel relation
between the two bands, enabling them to predict $K$-band fluxes from
the IRAC images.  Combining these $K$-band predictions with a constant
$\Upsilon_{\star}^K = 0.5$ then yields the stellar masses.  We compare
with our results (discussed in Sect.~6) in Fig.~\ref{fig:adam}, for
both our assumptions for the IMF. The masses we derive tend to be
slightly larger, a least-squares fit yields a slope of 0.94 with a
scatter of 0.12 dex.  Given the empirical nature of both methods, and
the absence of a full set of rigorously computed population synthesis
models, this should be regarded as satisfactory agreement.

\subsection{Dark matter halos\label{sect:dmhalos}}

Here we describe the two well-known models for the dark matter
distribution that we will use in our analysis.  Firstly, the NFW halo
\citep{NFW96,NFW97}, derived from CDM simulations, and dominated by a
central density cusp. Secondly, we discuss the venerable,
observationally motivated pseudo-isothermal (ISO) halo, dominated by a
central constant-density core.

\subsubsection{NFW halo}

The NFW mass density distribution takes the form
\begin{equation}
\rho_{\rm NFW}(R) = \frac{\rho_i}{\left(R/R_s\right)
\left(1+ R/R_s\right)^2}
\end{equation}   
where $R_s$ is the characteristic radius of the halo and
$\rho_i$ is related to the density of the universe at the time of collapse of the dark matter halo.
This mass distribution gives rise to a halo rotation curve  
\begin{equation}
V(R) = V_{200} \left[\frac{\ln(1+cx)-cx/(1+cx)}
{x[\ln(1+c)-c/(1+c)]}\right]^{1/2},
\end{equation}  
where $x = R/R_{200}$.  It is characterized by a concentration
parameter $c = R_{200}/R_s$ and a velocity $V_{200}$. These are directly
related to $R_s$ and $\rho_i$ and used to
parameterize the rotation curve.  The radius $R_{200}$ is the radius
where the density contrast with respect to the critical density of the
universe exceeds 200, roughly the virial radius \citep{NFW96}. The
characteristic velocity $V_{200}$ is the velocity at this radius.  

We can describe the steepness of the inner slope of the mass density
profile with a power law $\rho \sim r^{\alpha}$. The NFW models then
imply a value of $\alpha=-1$, the so-called ``cusp''.  Note that many
different parameterizations of CDM halos exist in the literature
\citep[e.g.,][]{NFW96,NFW97,moore99b,navarro04,diemand2005}.  These
all agree on the presence of a density cusp, but the steepness of the
inner density profile depends slightly on the simulation used, with
many models producing slopes that are steeper than the original NFW
$\alpha=-1$ value.  From an observational point of view, however,
these models all give similar results, and we therefore use the (in
terms of inner slope, conservative) NFW parameterization.

The NFW $c$ and $V_{200}$ parameters are not independent and are set
by the cosmology. \citet{stacy07} presents the relation between $c$
and $V_{200}$ for two recent and representative $\Lambda$CDM
models. The first one is the ``vanilla CDM'' model from
\citet{tegmark}; the second one is based on  \emph{WMAP}
results \citep{spergel2006}. Both models predict a nearly flat
distribution of $c$ as a function of $V_{200}$.  Simulations suggest a
scatter in $\log(c)$ of $\sim 0.18$ \citep{bullock01}. The main
difference between the two models is a different normalization,
resulting in lower concentration values for the \citet{spergel2006}
model. If we consider the union of both models and take their
respective $1\sigma$ scatters into account, we expect to find values
$c \simeq 8_{-3}^{+5}$ for $V_{200} = 50$ km s$^{-1}$, to values $c
\simeq 6_{-2}^{+4}$ for $V_{200} = 200$ km s$^{-1}$. As we are
considering the union of two models with different normalizations,
this represents a generous allowable range of $c$ values, comparable
to a $2\sigma$ scatter for any of the two models
separately. Anticipating our results, we note that the
$c$ parameter indicates the amount of collapse a halo has
undergone. Larger values of $c$ indicate a larger collapse factor.  A
value $c=1$ indicates no collapse; a value $c<1$ makes no sense within
the CDM context, and we will refer to these as unphysical values in subsequent modeling.

\subsubsection{ISO halo}

The spherical pseudo-isothermal (ISO) halo has a density profile
\begin{equation}
\rho_{\rm ISO}(R) = \rho_0 \Bigl[ 1 + \Bigl(
{{R}\over{R_C}} \Bigr)^2 \Bigr]^{-1},
\end{equation}
where $\rho_0$ is
the central density of the halo, and $R_C$ the core radius of the
halo.  The corresponding dark matter rotation curve is given by
\begin{equation} V(R) = \sqrt{ 4\pi G\rho_0 R_C^2 \Bigl[ 1 -
{{R_C}\over{R}}\arctan \Bigl( {{R}\over{R_C}} \Bigr) \Bigr] }.
\end{equation}
The asymptotic velocity of the halo, $V_{\infty}$, is given by
\begin{equation}
  V_{\infty} = \sqrt{ 4 \pi G \rho_0 R_C^2 }.
\end{equation}
As the inner mass-density distribution is dominated by an (almost) constant-density core,
it can be approximated by a power-law description $\rho \sim r^\alpha$ with $\alpha = 0$.

\subsection{Constructing the mass models}

We use the observed rotation curve, the derived gas rotation curve and
the rotation curve(s) of the stellar component(s) as input for the
mass modeling with the GIPSY task {\sc rotmas}.  This task subtracts
the (squared) rotation curves of the baryonic components (after
appropriate scaling with $\Ups$) from the (squared) observed rotation
curve, and fits the desired dark matter halo model to the
residuals. We use inverse-squared weighting of the rotation curve
uncertainties in making the fits.  For each combination of galaxy and
dark matter halo model we derive two sets of fits.  Firstly, we fix
the \Ups\ values of the stellar components to the values predicted by
our empirical color-\Ups-relations for both the diet-Salpeter and the
Kroupa IMFs and determine the best-fitting NFW and ISO halo
parameters. We will refer to these fits as the ``fixed \Ups'' models.
Secondly, we derive a set of fits where \Ups\ is left as a free
parameter. That is, we ignore any prior knowledge derived from
population synthesis models, and let the fitting program choose its
best value. These will be referred to as the ``free \Ups'' models.

\begin{figure*}[t]
\epsfxsize=0.95\hsize \epsfbox{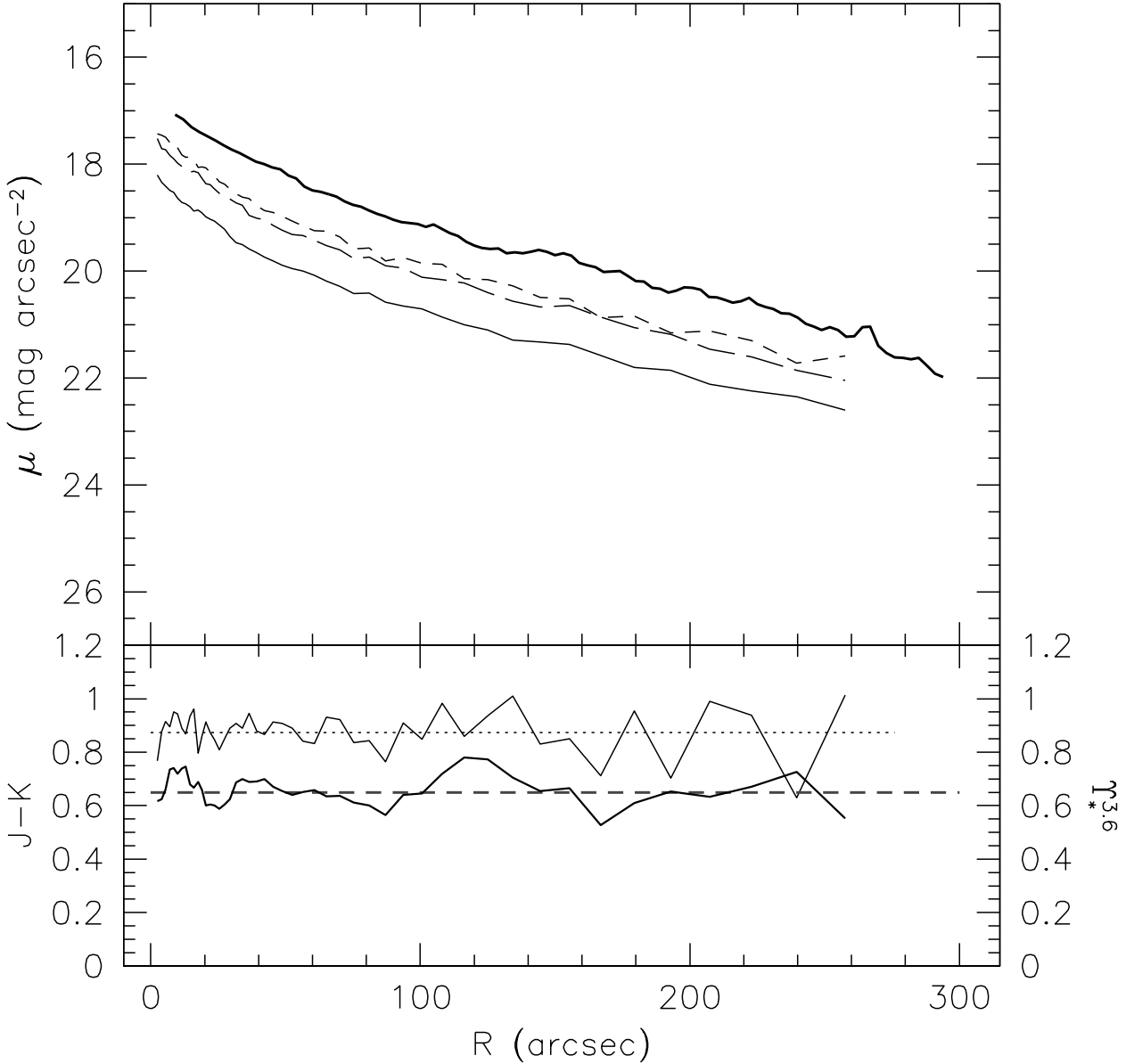} \figcaption{Surface
  brightness and \Ups\ profiles for NGC 925. Top panel: The thin full
  line, long dashed and short dashed curve show the 2MASS $J$, $H$ and
  $K$ surface brightness profiles, respectively. The thick full curve
  shows the Spitzer IRAC 3.6 $\mu$m surface brightness
  distribution. None are corrected for inclination. Bottom panel: The thin
  full curve shows the observed 2MASS $(J-K)$ distribution, with the
  thin dotted line indicating the average value.  The thick full line
  shows the distribution of \Ups, derived from $(J-K)$ as described in
  the text. The thick dashed curve shows our assumed model for
  the \Ups\ distribution of the disk.
\label{fig:n925_prof}}
\end{figure*}

\begin{figure*}[t]
  \epsfxsize=0.95\hsize \epsfbox{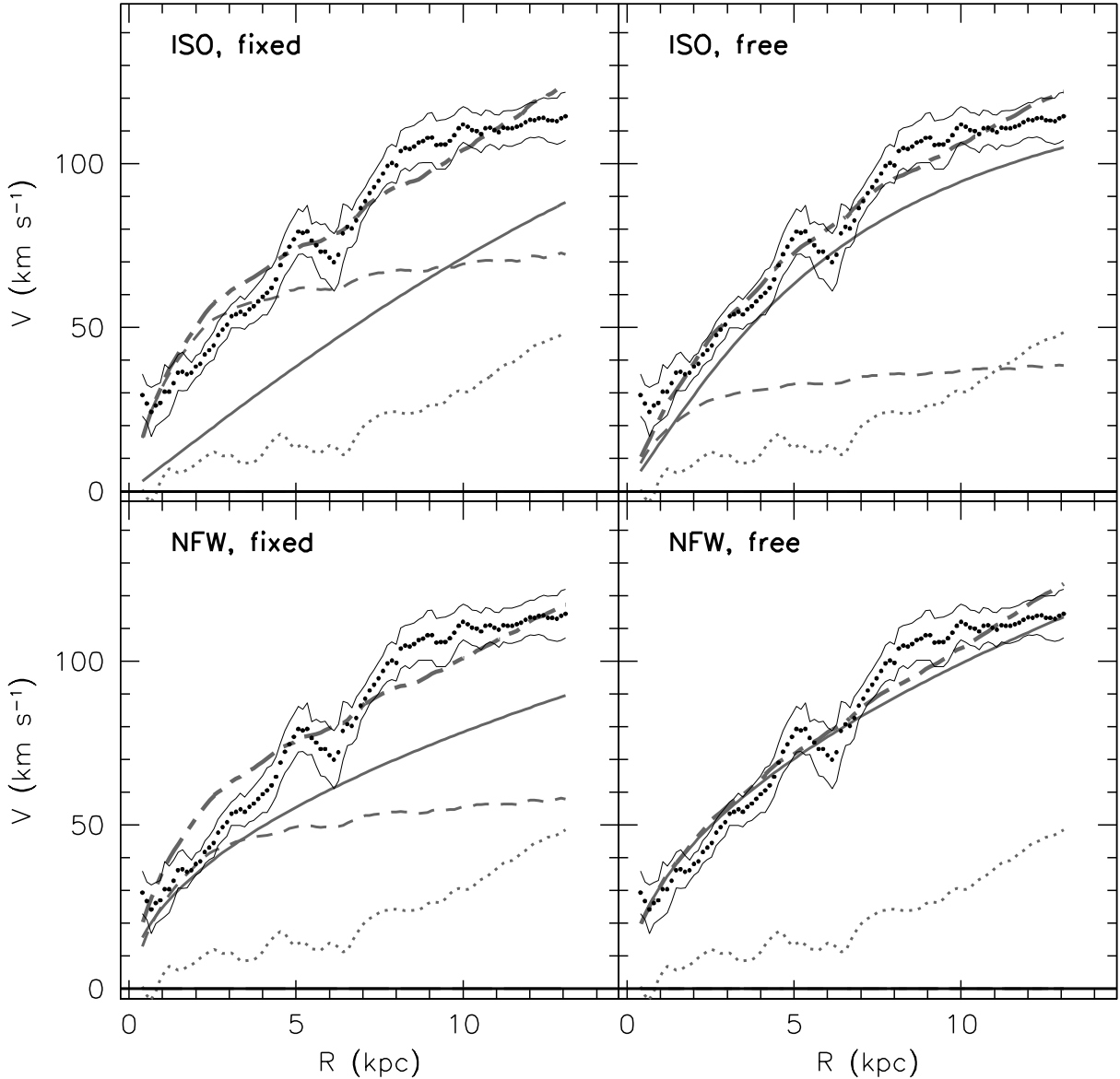} \figcaption{ISO
    and NFW rotation curve fits for NGC 925. The top row shows fits
    using the ISO halo. The bottom row those assuming the NFW
    model. The left-hand column shows the fits assuming fixed
    (predicted) values of \Ups. The right-hand column shows the fits
    produced with a free \Ups. In all panels the dots represent the
    observed curve and the thin full lines the uncertainties. The grey
    dotted curve shows the rotation curve of the gas; the thin dashed
    grey curve that of the stellar disk. The thick, grey full curve
    shows the resulting rotation curve of the halo. The thick
    long-short dashed curve the resulting best-fit model rotation
    curve.
\label{fig:n925_curve}}
\end{figure*}


\begin{figure*}[t]
  \epsfxsize=0.95\hsize \epsfbox{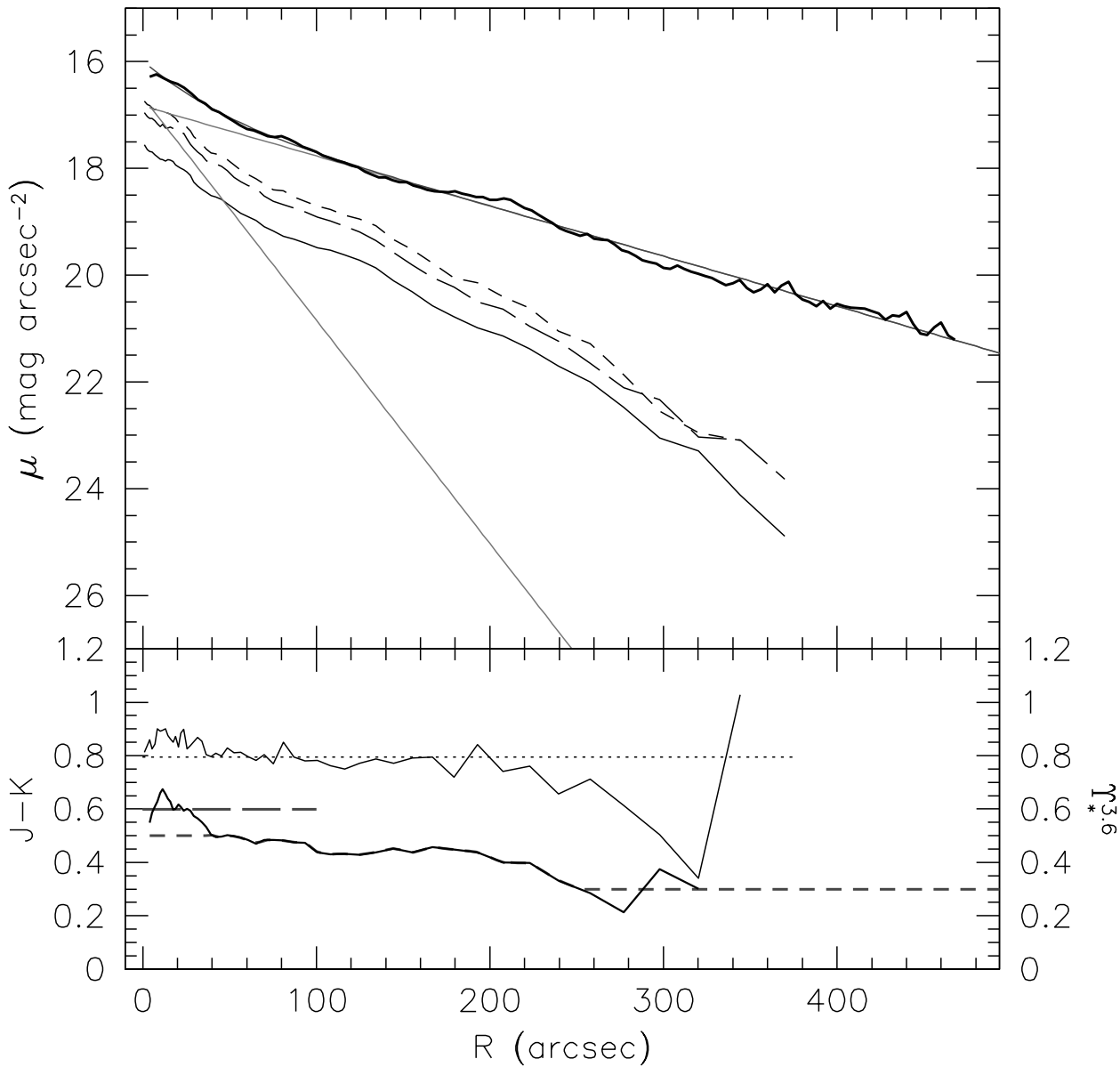} \figcaption{
    Surface brightness and \Ups\ profiles for NGC 2403. Lines and
    symbols are as in Fig.~\ref{fig:n925_prof}. In addition, the thin
    gray lines in the top panel show the decomposition of the profile
    in multiple exponential disk models.  The long-dashed grey line in
    the bottom panel shows the assumed value of \Ups\ for the inner
    disk.
\label{fig:n2403_prof}}
\end{figure*}

\begin{figure*}[t]
  \epsfxsize=0.95\hsize \epsfbox{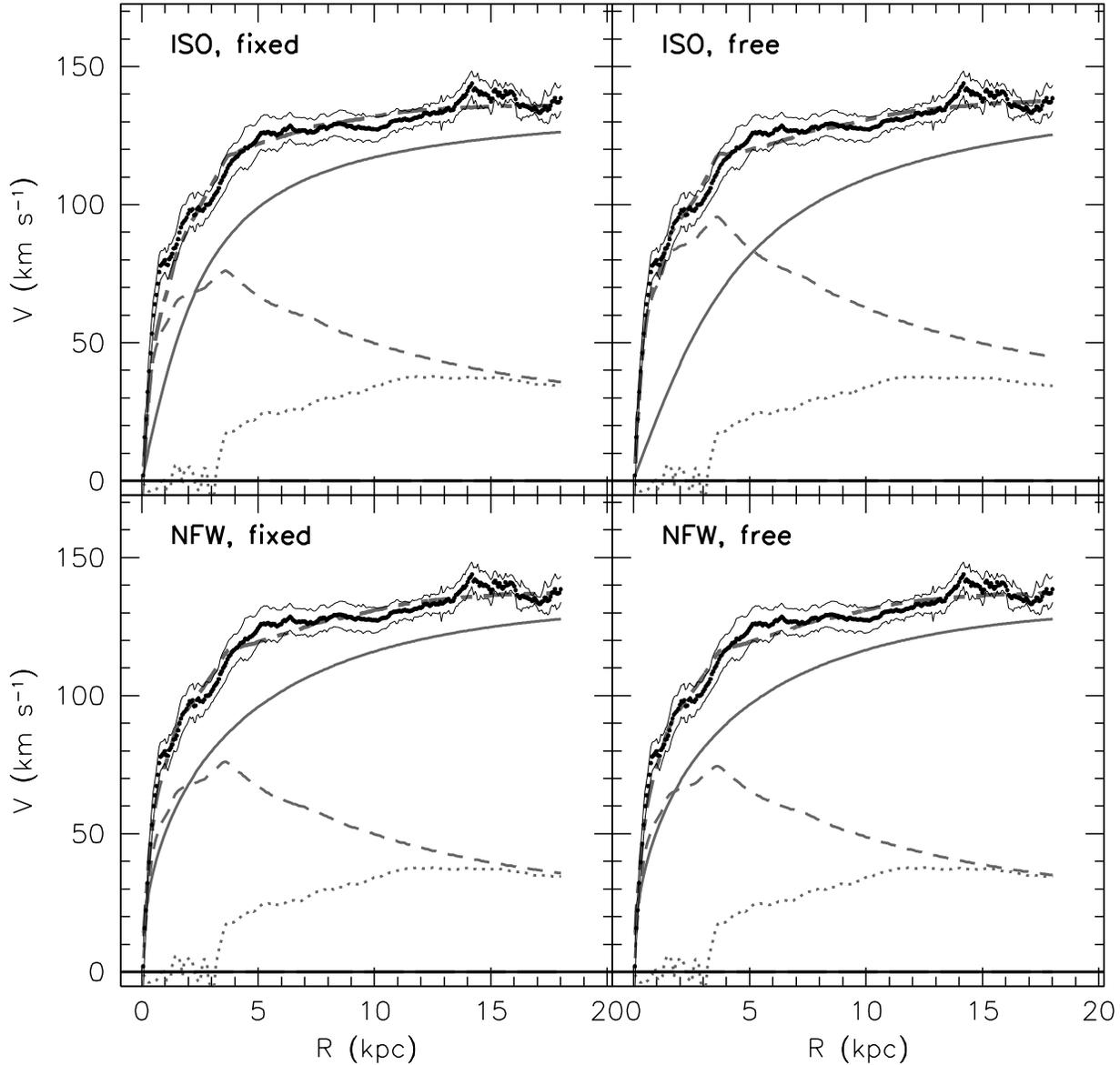}
  \figcaption{ ISO and NFW rotation curve fits for the single-disk
    model of NGC 2403. Symbols and lines as in
    Fig.~\ref{fig:n925_curve}.
\label{fig:n2403_1curve}}
\end{figure*}

\begin{figure*}[t]
  \epsfxsize=0.95\hsize \epsfbox{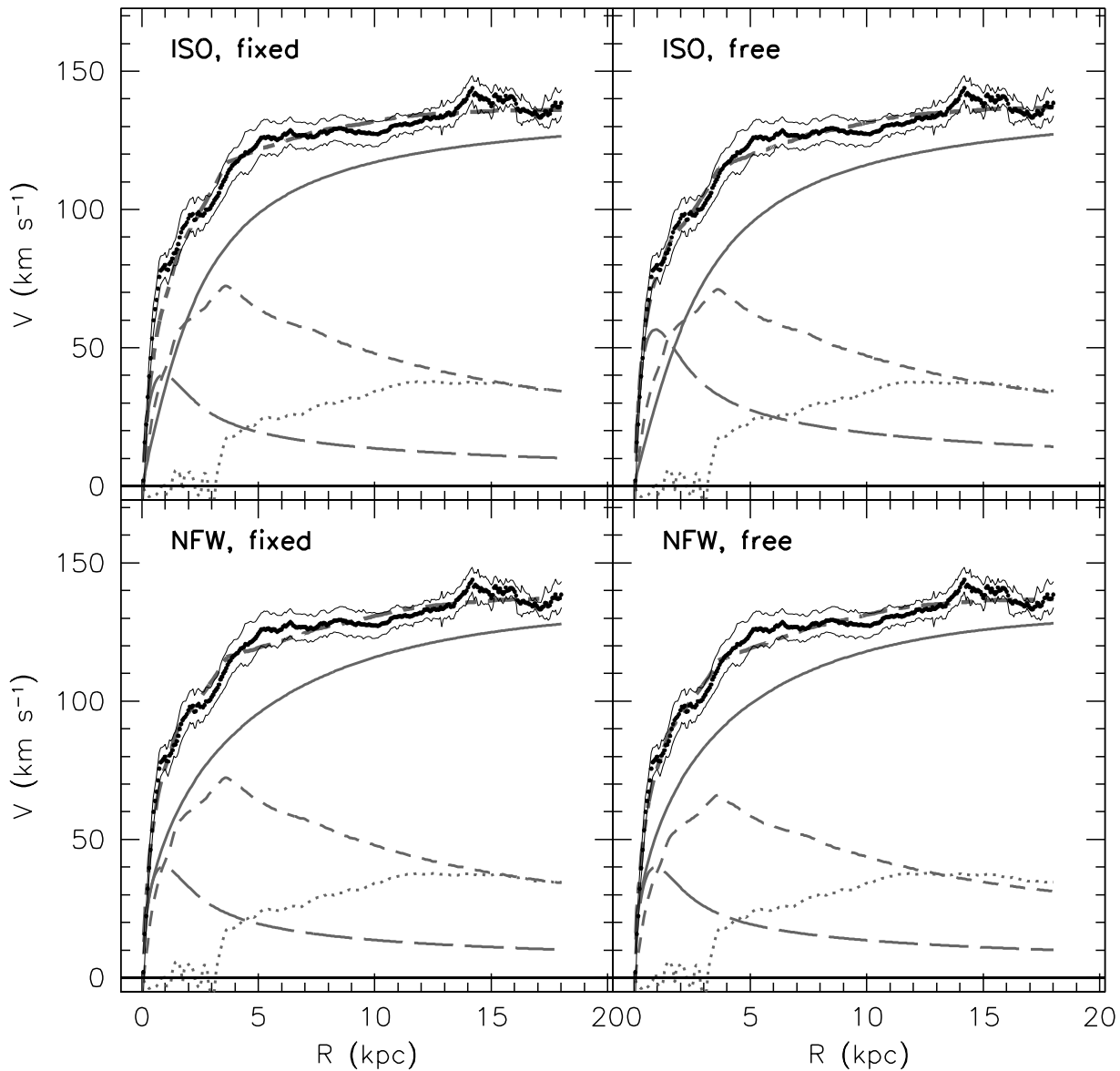} \figcaption{ ISO
    and NFW rotation curve fits for the multiple-disk model of NGC
    2403.  Symbols and lines as in Fig.~\ref{fig:n925_curve}. In
    addition, the long-dashed grey curve represents the rotation curve
    of the inner disk.
\label{fig:n2403_2curve}}
\end{figure*}

\begin{figure*}[t]
\epsfxsize=0.95\hsize \epsfbox{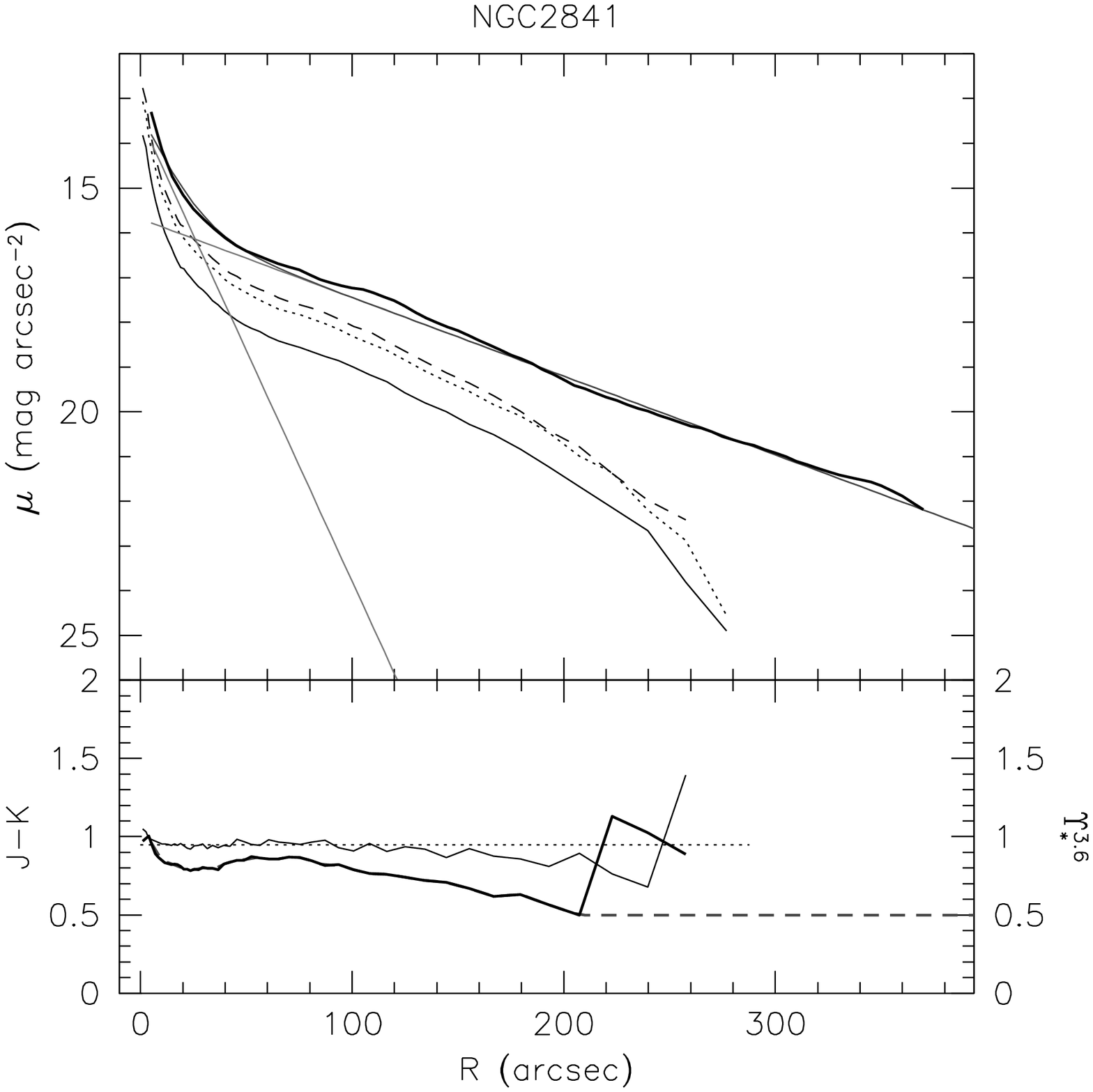} \figcaption{ Surface
  brightness and \Ups\ profiles for NGC 2841. Lines and symbols as in
Fig.~\ref{fig:n2403_prof}.
\label{fig:n2841_prof}}
\end{figure*}

\begin{figure*}[t]
\epsfxsize=0.95\hsize \epsfbox{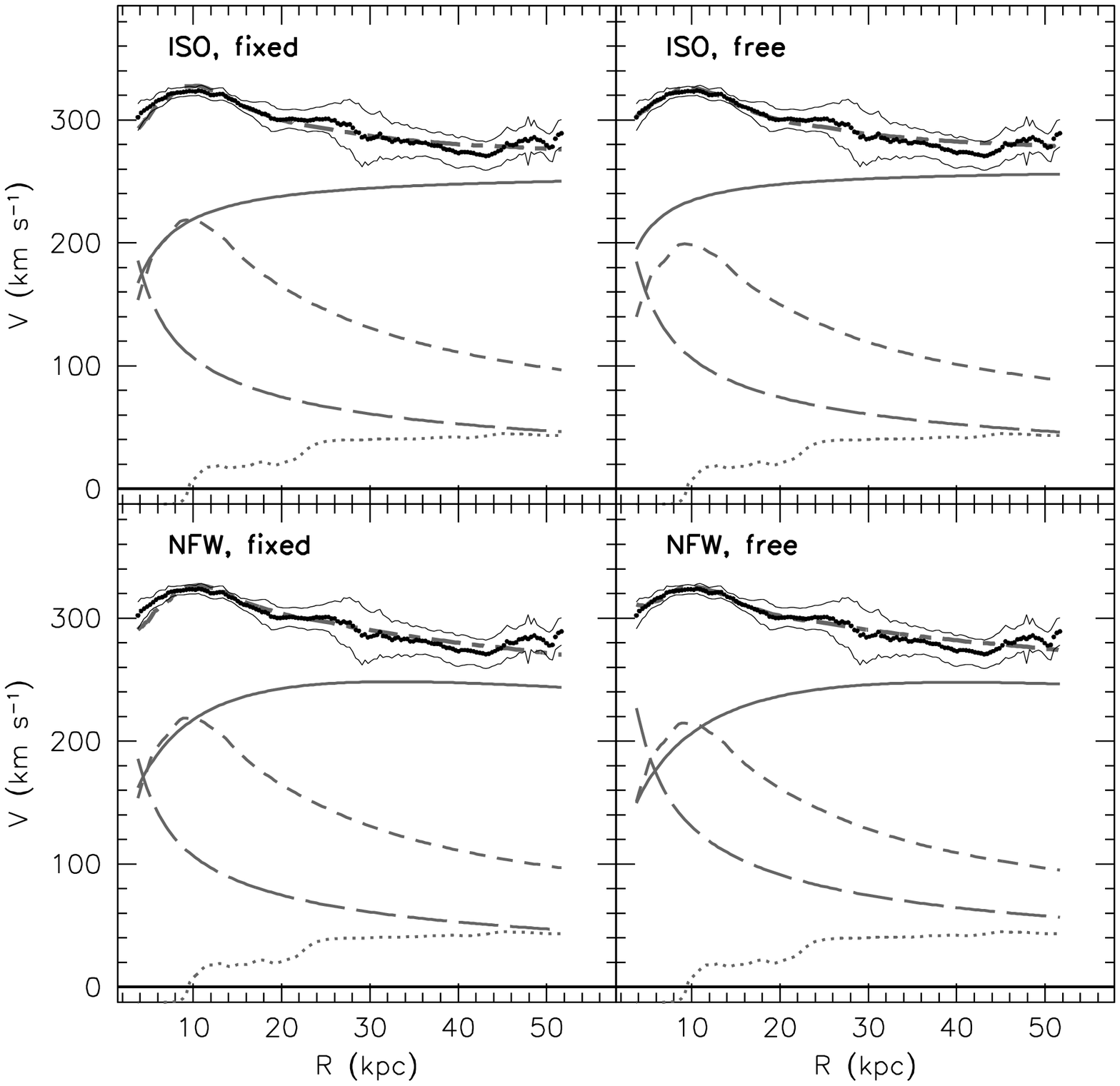} \figcaption{ ISO and
  NFW rotation curve fits for NGC 2841.  Lines and symbols as in
  Fig.~\ref{fig:n2403_2curve}.
\label{fig:n2841_curve}}
\end{figure*}

\begin{figure*}[t]
\epsfxsize=0.95\hsize \epsfbox{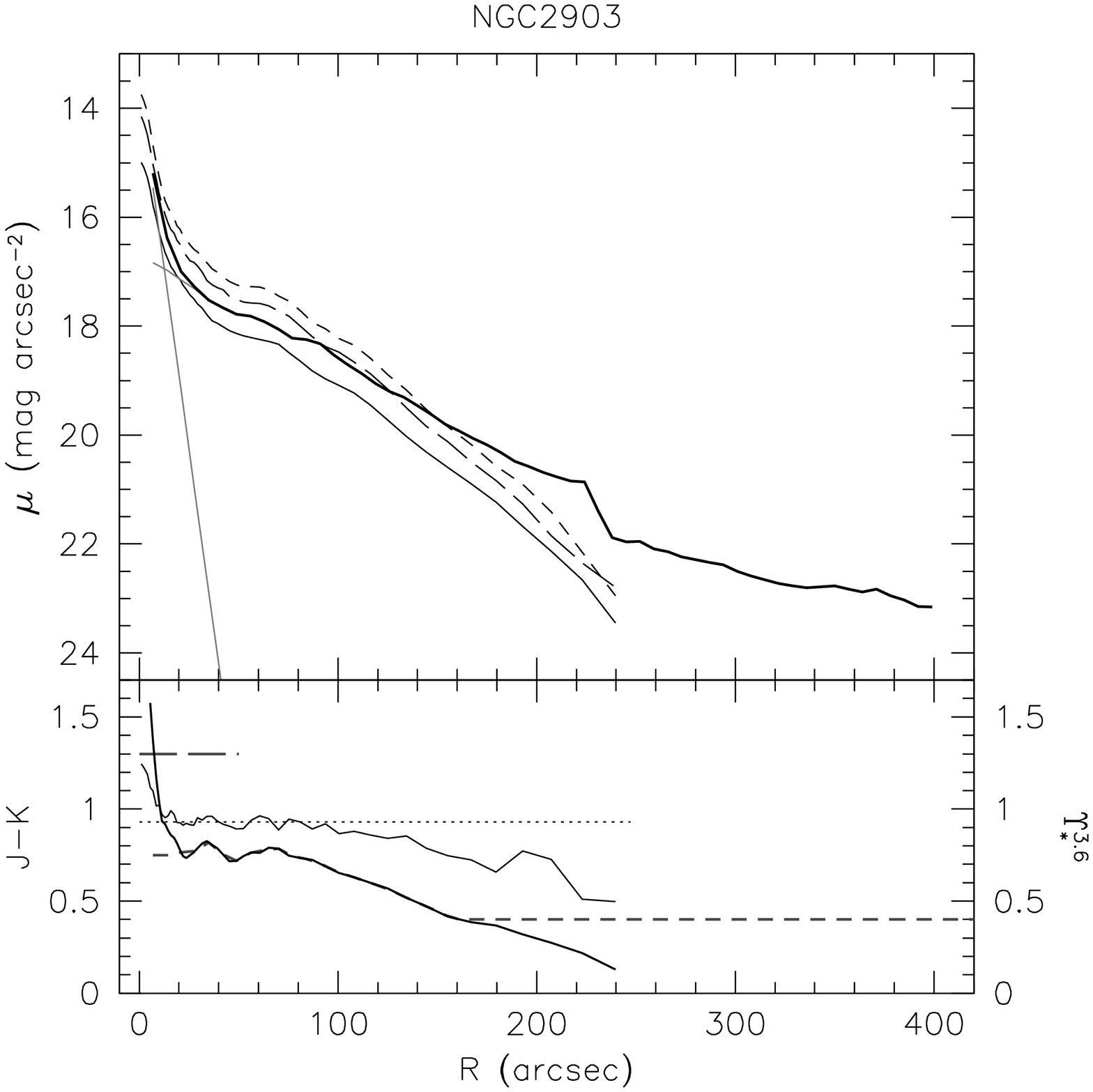} \figcaption{ Surface
  brightness and \Ups\ profiles for NGC 2903. Lines and symbols as in
Fig.~\ref{fig:n2403_prof}.
\label{fig:n2903_prof}}
\end{figure*}

\begin{figure*}[t]
  \epsfxsize=0.95\hsize \epsfbox{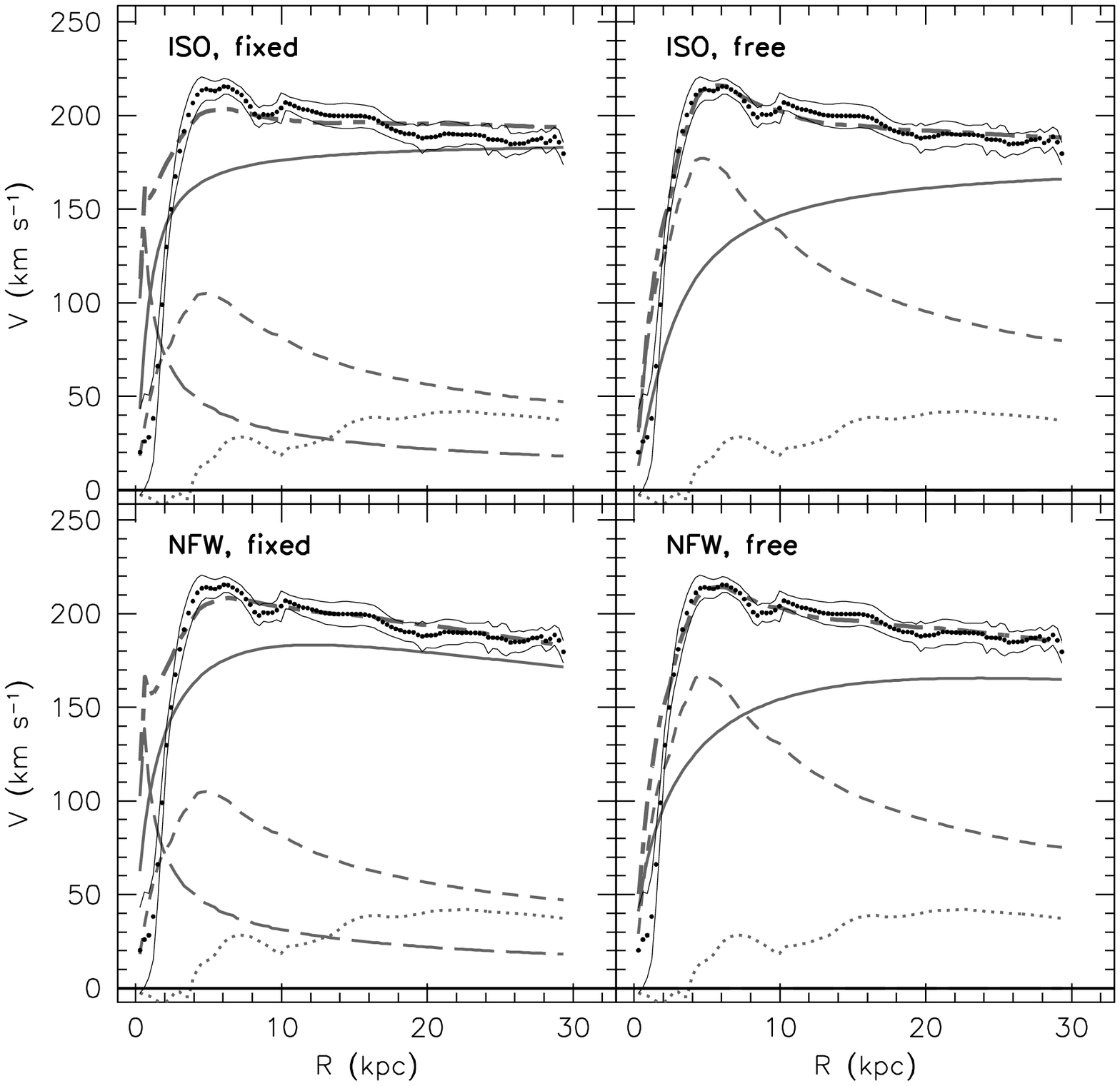} \figcaption{ ISO
    and NFW rotation curve fits for NGC 2903 for the entire radial
    range.  Lines and symbols as in Fig.~\ref{fig:n2403_2curve}.
\label{fig:n2903_allcurve}}
\end{figure*}

\section{Results for individual galaxies}

Below we describe the mass models for each galaxy along with a
motivation for specific choices made for individual galaxies.  The
results for IC 2574 and NGC 2366 are described in \citet{oh2007}.  The
\Ups\ values and halo model parameters (including those for IC 2574
and NGC 2366) are given in Table~\ref{table2} for the fixed \Ups\
models with a diet-Salpeter IMF, in Table~\ref{table2kroupa} for a fixed
\Ups\ with a Kroupa IMF, in Table~\ref{table3a} for the free \Ups\ ISO
models, and in Table~\ref{table3} for the free \Ups\ NFW models.  The
free \Ups\ models do not depend on the IMF.  The 3.6 $\mu$m surface
brightness and \Ups\ profiles, as well as the fixed and free \Ups\
models assuming NFW and ISO models are shown in
Figs.~\ref{fig:n925_prof}--\ref{fig:n7793_2curve}.  The relations of
\Ups\ with $(J-K)$ shown in these Figures are valid for a
diet-Salpeter IMF. Values for \Ups\ assuming a Kroupa IMF can be
obtained by shifting the relations down by 0.15 dex.  In the interests
of space, we only show the diet-Salpeter mass models, as they are
generally similar to the ones derived assuming a Kroupa
IMF. Noteworthy exceptions are discussed below in the sections for
individual galaxies. Readers more interested in the outcomes for the
sample as a whole can at this point skip to Sect.~7.
In the following discussion all \Ups\ values assume a diet-Salpeter IMF, unless
noted otherwise.

\subsection{NGC 925}

The surface brightness profiles of NGC 925 are shown in
Fig.~\ref{fig:n925_prof}. The 2MASS $J$, $H$ and $K$ profiles can be
traced out to $\sim 250''$; the 3.6 $\mu$m profile can be traced over
the entire extent of the \HI\ disk.  NGC 925 shows no evidence for a
bright central component.  There is also no evidence for a strong
color gradient, and we assume a constant $\Ups = 0.65$ for the disk.

The rotation curves are shown in Fig.~\ref{fig:n925_curve}. The
predicted \Ups\ value make NGC 925 slightly exceed maximum disk. The
ISO model can to some degree accommodate this, but the NFW model less
so. This does not change when adopting the Kroupa \Ups\ values. Fits
with \Ups\ free prefer lower values for \Ups. For the NFW case a
negative \Ups\ is preferred; we have assumed a fixed $\Ups=0$ in this
case.  For both fixed and free $\Ups$ the NFW model yields parameters
that are inconsistent with the CDM expectations (cf.\ Sect.~5.3.1),
resulting in $V_{200}$ values many times larger than the maximum
rotation of the galaxy and $c$ values approaching zero. This is a
general symptom of an NFW halo trying to fit a linearly rising
rotation curve \citep{mcgaugh98}. The ISO models in general fit the
data better than the NFW models.

\subsection{NGC 2403}

Fig.\ \ref{fig:n2403_prof} shows the surface brightness profiles
derived from the \tsm\ data, as well as the 2MASS $J$, $H$ and $K$
profiles. The \tsm\ profile can be reliably traced to $\sim 430''$
galactocentric radius, the 2MASS profiles out to $\sim 380''$.  The
$(J-K)$ color shows a clear trend, from $\sim 0.9$ in the inner parts
to $\sim 0.6$ at the outermost measured radii.  This translates into a
trend of $\Ups \simeq 0.6$ in the center to an outer $\Ups \simeq
0.3$. Beyond the radius where $(J-K)$ (and therefore \Ups) could be
reliably determined we assume a constant value $\Ups = 0.3$.
The surface brightness profile shows evidence for a second component
in the inner parts. Based on its rather modest appearance it is not
clear though whether this constitutes a dynamically separate
component. We therefore consider both cases. For the single-component
case we simply multiplied the surface brightness profile with the
derived \Ups\ values. For the double-component case we decomposed the
profile in two exponential profiles. The outer component is
described by a central surface brightness $\mu_0^{3.6} = 16.9$ mag
arcsec$^{-2}$ and a scale length $h = 1.81$ kpc.  The inner component
has $\mu_0^{3.6} = 16.7$ mag arcsec$^{-2}$ and $h= 0.41$ kpc.  We
used the fitted model as a description for the inner disk. For
the outer disk we used the observed surface brightness distribution,
with the inner disk model subtracted. We extended the observed outer
disk profile with the corresponding fit at large radii where the disk
was no longer directly observable.

The $(J-K)$ color shows a slight increase in the inner parts. For the
inner disk we therefore assumed a constant value of $\Ups = 0.6$,
whereas for the outer, main disk we assumed the single-disk color
distribution, the only difference being that we adopted a constant
$\Ups = 0.5$ for the innermost ($R<40''$) parts.

Figures~\ref{fig:n2403_1curve} and \ref{fig:n2403_2curve} present the
fits for the single and multiple disk models, respectively. The NFW
models (fixed and free) are of the same quality as the free ISO model.
The fixed \Ups\ ISO models are of somewhat lower quality. NGC 2403 in
combination with the ISO model prefers higher \Ups\ values than
predicted from the infra-red color.

\subsection{NGC 2841}

The surface brightness profiles of NGC 2841 are shown in
Fig.~\ref{fig:n2841_prof}. The 2MASS $J$, $H$ and $K$ profiles can be
traced out to $\sim 290''$, the IRAC 3.6 $\mu$m profile out to $\sim
375''$.  All surface brightness profiles show an extended outer
exponential disk, as well as a more compact central component. We
decompose the profile using a double exponential disk. For the outer
disk we find $\mu_0 = 15.7$ mag arcsec$^{-2}$ and $h = 4.20$ kpc. The
inner disk has parameters $\mu_0 = 13.5$ mag arcsec$^{-2}$ and
$h=0.72$ kpc.  In subsequent modeling we use the inner disk model for
the central component. For the outer component we use the total
observed surface brightness profile with the inner disk model
subtracted, extrapolated at large radii using the outer exponential
model parameters.  The surface brightness profile shows some evidence
for central emission in excess of the exponential model. However, the
central parts of the velocity field cannot be modeled due to lack of
\HI\ emission. 

NGC 2841 possesses a modest color gradient, which results in a \Ups\
trend from $\Ups \simeq 0.8$ in the inner parts to $\Ups \sim 0.5$ in
the outer parts.  For radii beyond which the 2MASS color could not be
reliably determined, we assume a constant $\Ups = 0.5$.  The color
profile shows a slight reddening in the innermost part, but this
occurs at radii where no velocity information is available.

The mass models are presented in Fig.~\ref{fig:n2841_curve}.  For the
ISO model, the disk mass obtained through a free fit is very close to
the mass derived from the population synthesis modeling assuming a
diet-Salpeter IMF.  The resulting halo has the highest
(well-determined) central density in the sample. The NFW model with
fixed \Ups\ is of slightly lower quality, its main problem is the
under-prediction of the velocity at the smallest radii. An increase in
\Ups\ of the inner component when \Ups\ is released compensates for
this, and yields a model of equal quality as the ISO model.  For the
Kroupa fixed \Ups\ the ISO model results in an extremely compact halo,
probably indicating that \Ups\ is underestimated.

\subsection{NGC 2903}

The surface brightness profiles of NGC 2903 are shown in
Fig.~\ref{fig:n2903_prof}. The 2MASS $J$, $H$ and $K$ profiles can be
traced out to $\sim 240''$; the IRAC 3.6 $\mu$m profile out to $\sim
400''$.  The break in the profile around $230''$ is due to the sudden
change in inclination at this radius.  The profile can be
decomposed into two components: the inner one is modeled using an
exponential disk with $\mu_0= 13.6$ mag arcsec$^{-2}$ and $h = 0.17$
kpc. For the outer disk we used the observed profile with the inner
disk profile subtracted. At radii larger than $400''$ we used an
exponential extrapolation of the main disk.

NGC 2903 shows a well-defined color gradient. For the inner disk we
assumed $\Ups = 1.3$ which is the average value of \Ups\ within
$20''$. For the outer disk we assumed $\Ups=0.75$ in the innermost
parts and $\Ups=0.4$ in the outermost parts. Between $20''$ and
$150''$ we adopted the observed \Ups\ profile.

The rotation curve models for NGC 2903 are shown in
Fig.~\ref{fig:n2903_allcurve}. It is immediately obvious that the
inner part of NGC 2903 is not well fitted by the models.  The models
with \Ups\ fixed severely over-predict the data, independent of IMF.
Models with \Ups\ as a free parameters can only be made to
approximately fit the data by setting $\Ups=0$ for the inner disk.
This large discrepancy is most likely due to the bar. A harmonic
decomposition of the velocity field shows significant non-circular
motions in the inner 1 kpc \citep{clemens2007}. Because of the low
quality of the fits we do not list them in Tables \ref{table2},
\ref{table3a} and \ref{table3}.  An additional source of uncertainty
could be the presence of a very dense inner molecular disk, with
significant amounts of associated star formation \citep{2903disk}.

\begin{figure*}[t]
  \epsfxsize=0.95\hsize \epsfbox{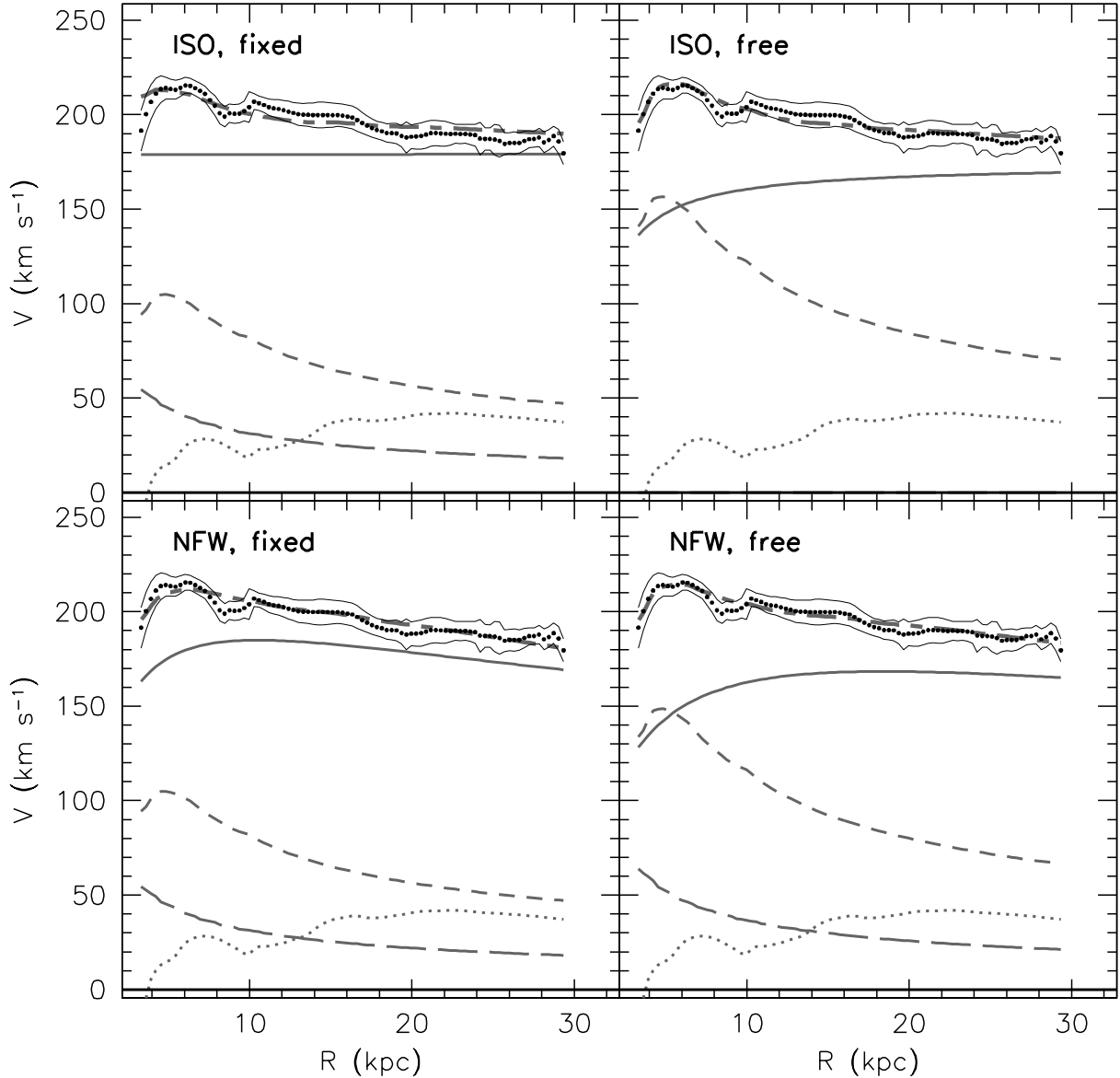} \figcaption{ ISO
    and NFW rotation curve fits for NGC 2903. Only radii $R>3.3$ kpc
    are considered here.  Lines and symbols as in
    Fig.~\ref{fig:n2403_2curve}. The horizontal line in the top-left
    panel represents the degenerate halo rotation curve; see text for
    a description.
\label{fig:n2903_curve}}
\end{figure*}

We performed an extra set of fits to those parts of the rotation curve
that are unaffected by the central non-circular motions 
(the flat part of the curve at $R>3.3$ kpc). These are shown in 
Fig.~\ref{fig:n2903_curve}.  We find that with \Ups\ as a free
parameter, the resulting disk is dynamically more significant than
predicted by population synthesis models and $(J-K)$ colors. The ISO
model prefers a dynamically insignificant central component, whereas
the NFW model prefers a central \Ups\ value close to the predicted
one. Here also, the main disk is found to be more massive than
predicted by the colors. The NFW model fits better than the ISO model
but due to the presence of the non-circular motions, it is not clear
whether this can be generalized to the entire rotation curve.
For fixed \Ups, the ISO model demands an extremely small core radius $R_C$ and
a very large central density $\rho_0$, in order to produce the flat
halo rotation curve shown in Fig.~\ref{fig:n2903_curve}.

\subsection{NGC 2976}

\begin{figure*}[t]
\epsfxsize=0.95\hsize \epsfbox{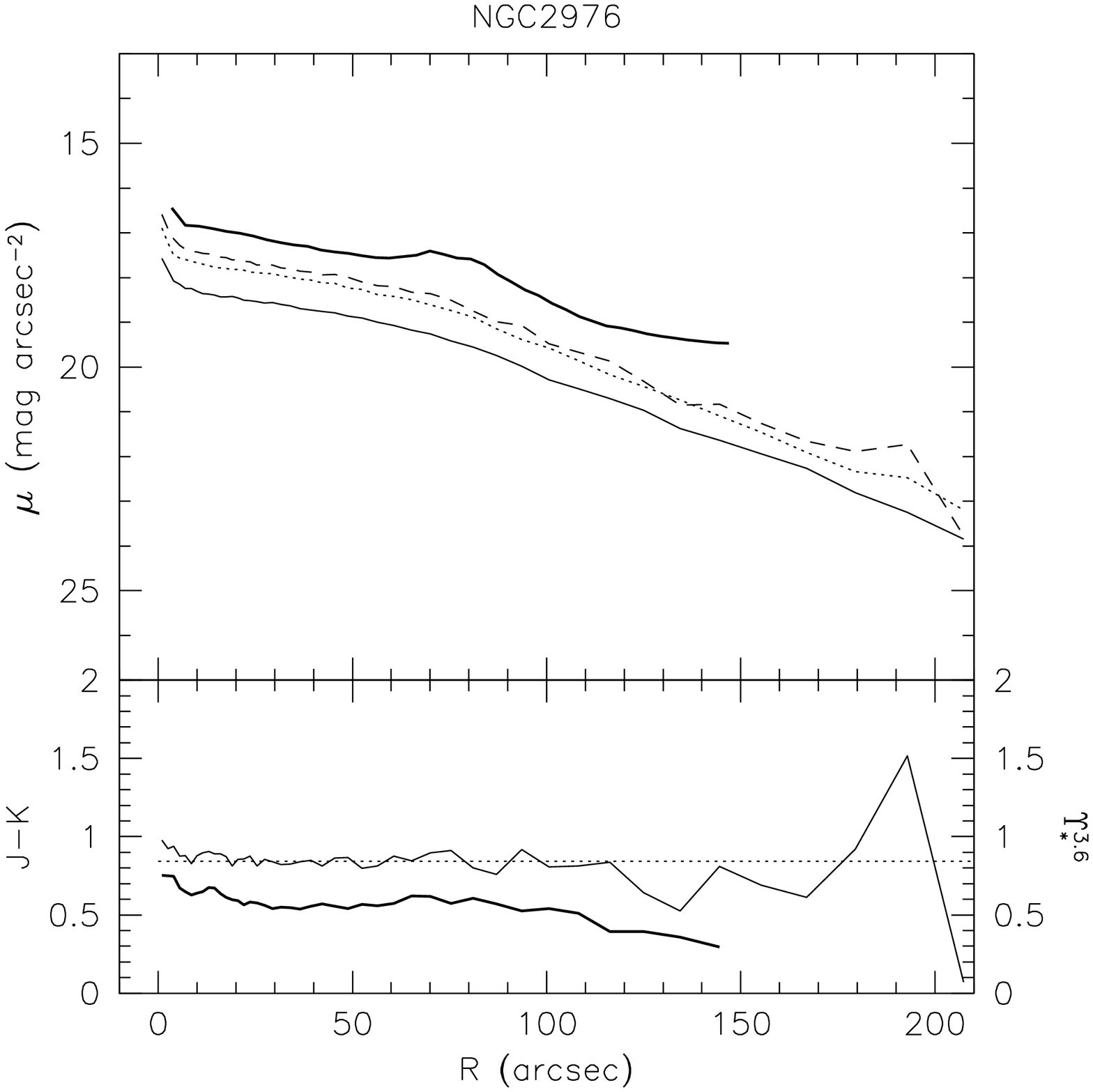} \figcaption{ Surface
  brightness and \Ups\ profiles for NGC 2976. Lines and symbols as in
Fig.~\ref{fig:n2403_prof}.
\label{fig:n2976_prof}}
\end{figure*}

The surface brightness profiles of NGC 2976 are shown in
Fig.~\ref{fig:n2976_prof}. The 2MASS $J$, $H$ and $K$ profiles can be
traced out to $\sim 210''$. The \HI\ disk and rotation curve extend
only out to $150''$, and no extrapolations using exponential disks
were necessary. The galaxy shows a well-defined color gradient,
leading to a \Ups\ variation from $\Ups \simeq 0.75$ in the inner
parts to $\Ups \simeq 0.3$ in the outer parts. We adopted the \Ups\
distribution as shown in Fig.~\ref{fig:n2976_prof} without any further
modifications.

\begin{figure*}[t]
\epsfxsize=0.95\hsize \epsfbox{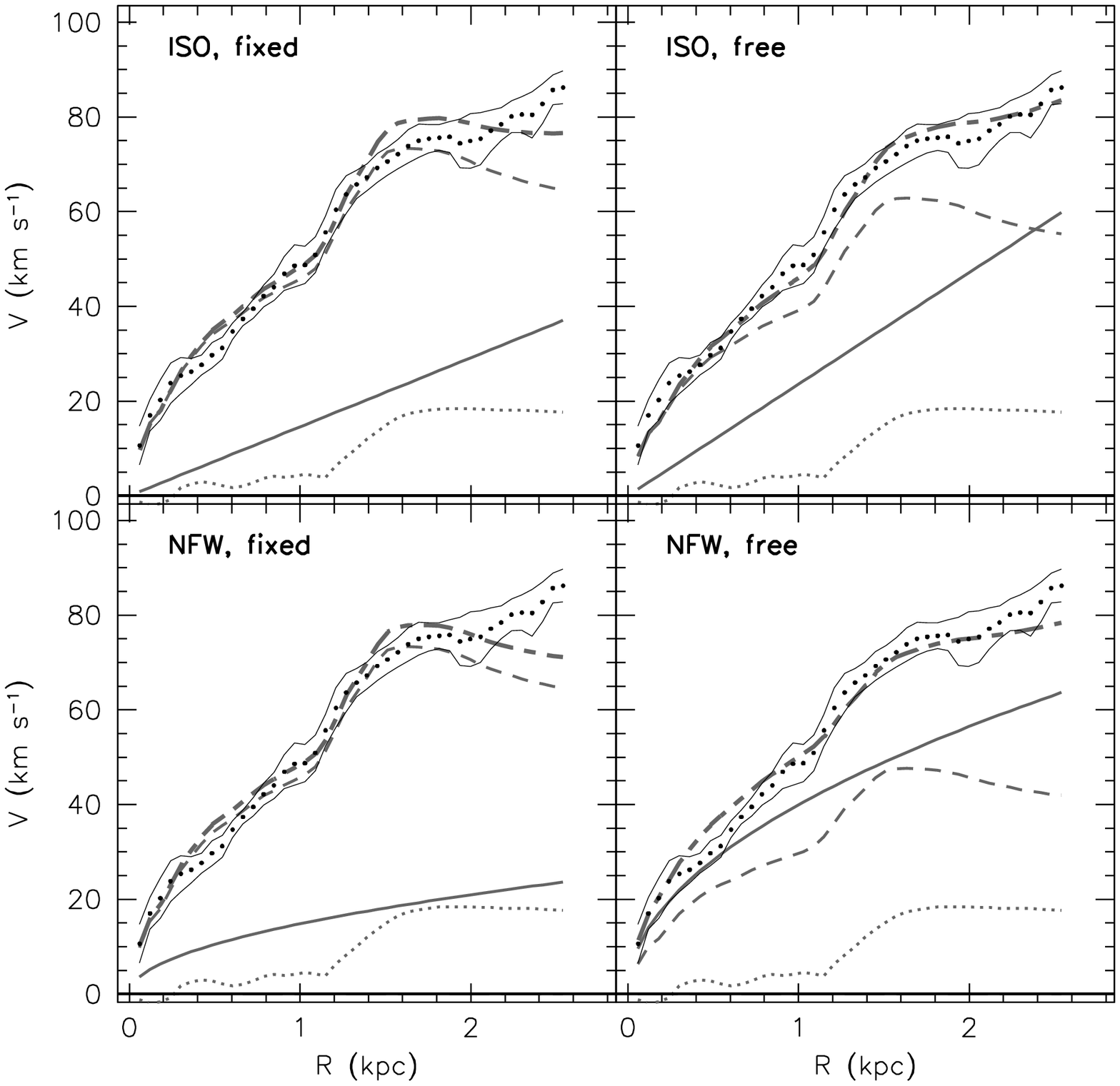} \figcaption{ISO and
  NFW rotation curve fits for NGC 2976. Lines and symbols as in
  Fig.~\ref{fig:n2403_1curve}.
\label{fig:n2976_curve}}
\end{figure*}

The rotation curve fits are shown in Fig.~\ref{fig:n2976_curve}.  The
models with the fixed diet-Salpeter \Ups\ value slightly over-predict
the observed rotation curve. The free-\Ups, ISO model also prefers a
model very close to maximum disk, with a value for \Ups\ close to the
fixed \Ups\ Kroupa value.  The large value of $R_C$, seems to indicate
NGC 2976 is located well within the core radius of its halo.  The
free-\Ups, NFW fit yields unphysical halo parameters trying to
approximate a solid-body curve.  It can only do this by choosing a
large value of $V_{200}$ and a small value of $c$.  As mentioned
before, $c$ represents a collapse factor, and it is clear that a value
$c<1$ makes no physical sense. Similarly, it is unlikely that a fairly
modest galaxy such as NGC 2976 is embedded in a halo with a typical
velocity $V_{200} > 500$ km s$^{-1}$.

\subsection{NGC 3031}

\begin{figure*}[t]
\epsfxsize=0.95\hsize \epsfbox{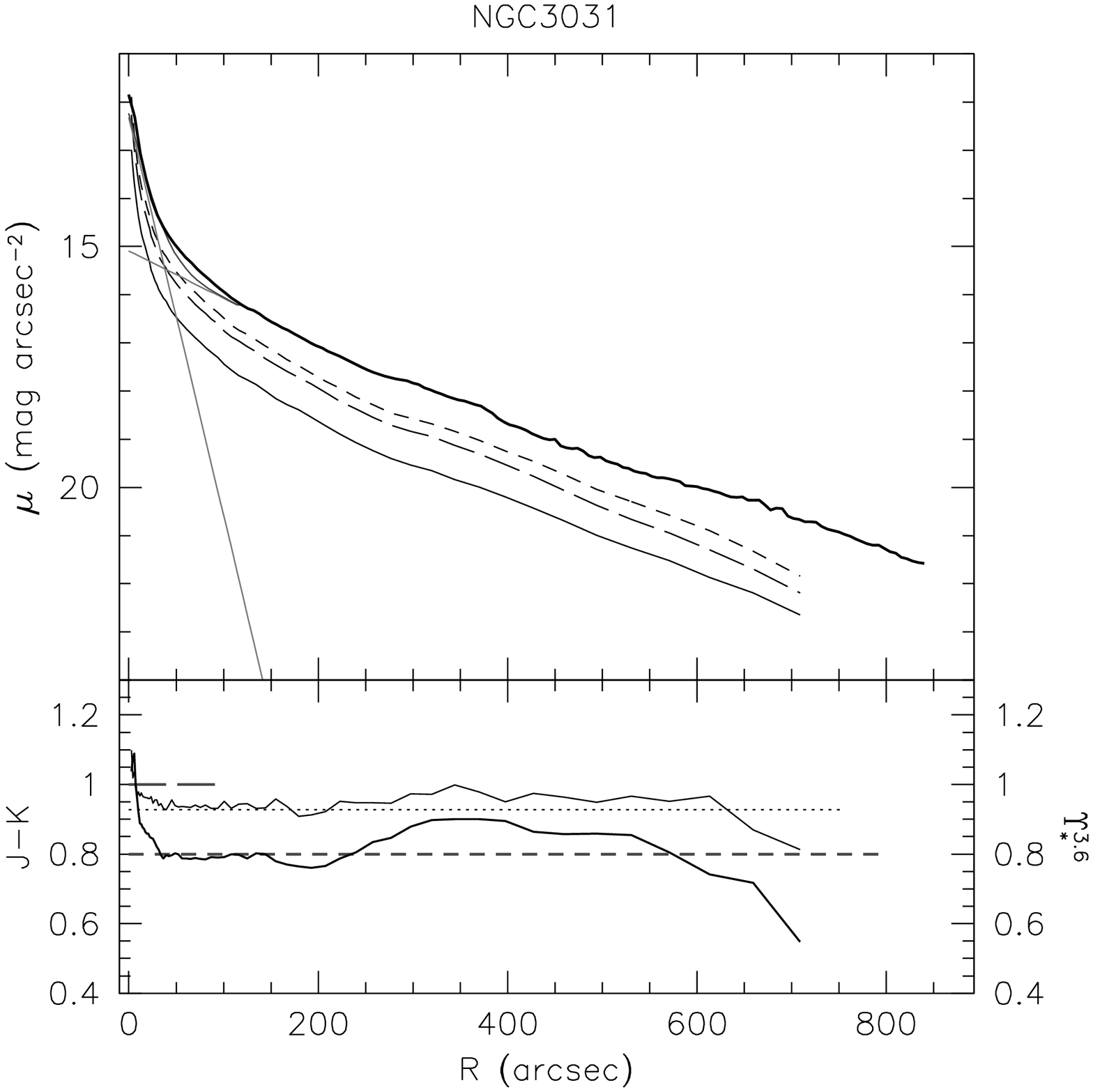} \figcaption{ Surface
  brightness and \Ups\ profiles for NGC 3031. Lines and symbols as in
Fig.~\ref{fig:n2403_prof}.
\label{fig:n3031_prof}}
\end{figure*}

The surface brightness profiles of NGC 3031 are shown in
Fig.~\ref{fig:n3031_prof}. The 2MASS $J$, $H$ and $K$ profiles can be
traced out to $\sim 750''$; the 3.6 $\mu$m profile out to $\sim
850''$.  The profiles show clear evidence for a central component.
The outer disk shows a gradual flattening of the surface brightness
profile towards larger radii. We modeled the central component using
an exponential disk with parameters $\mu_0=12.2$ mag arcsec$^{-2}$
and $h=0.25$ kpc.  To describe the outer disk we simply subtracted the
central exponential disk from the total light profile.  The slight
$(\sim 0.2$ mag) residuals seen around $R\sim 100''$ do not impact on
any subsequent modeling.  The $(J-K)$ color profile shows no
significant color gradient, except for a central redder component. We
assume a constant $\Ups=0.8$ for the disk and $\Ups=1.0$ for the
central component.

\begin{figure*}[t]
\epsfxsize=0.95\hsize \epsfbox{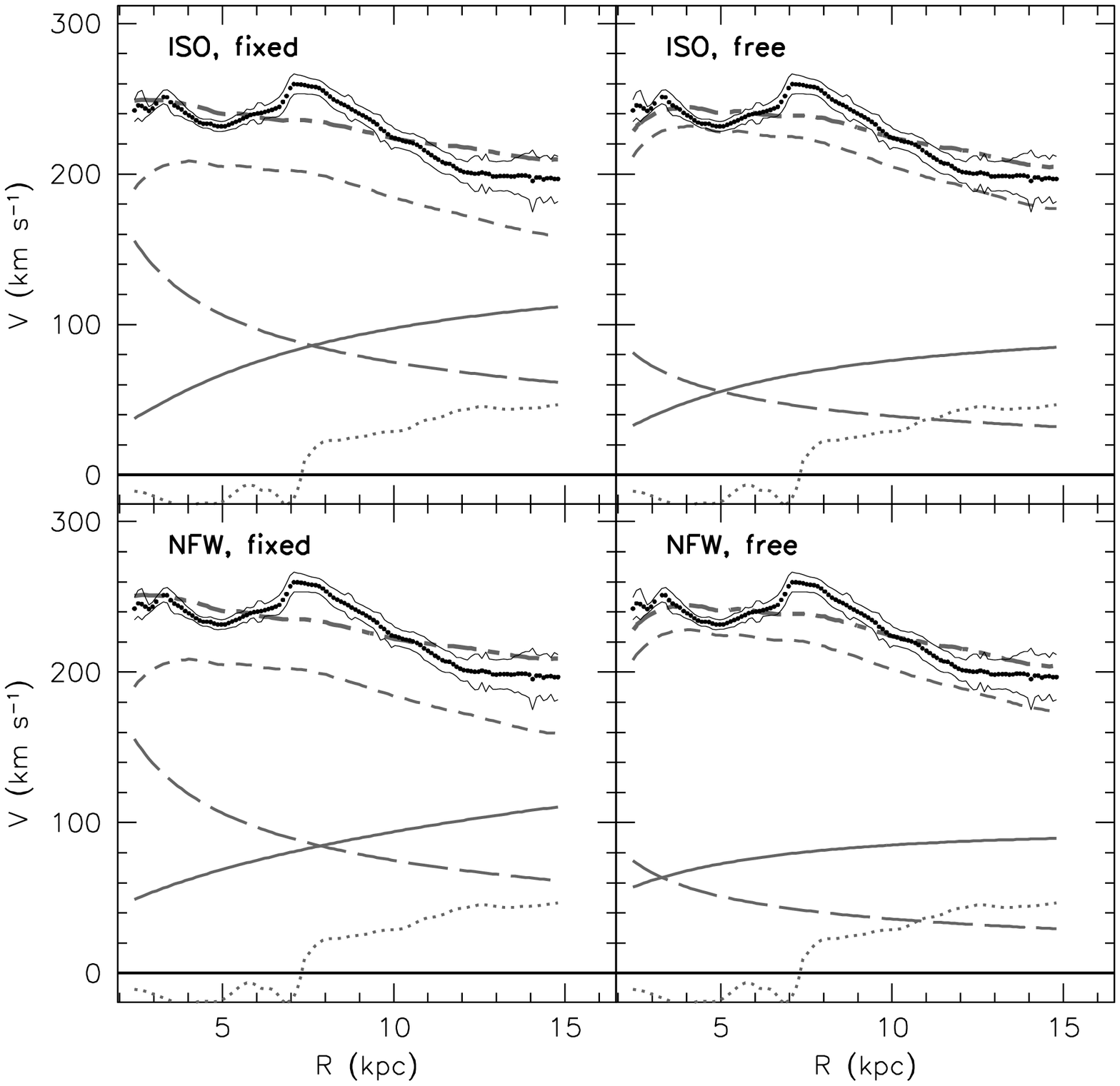} \figcaption{ ISO and
  NFW rotation curve fits for NGC 3031.  Lines and symbols as in Fig.~\ref{fig:n2403_2curve}.
\label{fig:n3031_curve}}
\end{figure*}

The rotation curve mass models are shown in
Fig.~\ref{fig:n3031_curve}. Already at first glance, it is clear that
the model fits deviate significantly from the observed rotation curve.
Although the general trends in the curve are described well, smaller
scale deviations clearly indicate the presence of significant
non-circular motions. Given the presence of the prominent,
grand-design spiral arms (causing the bump at $R \sim 7.5$ kpc), and
M81's location in an interacting system, this should not come as a
surprise.  The fit with fixed \Ups\ yields $\chi^2_r$ values that are
only slightly inferior to the free \Ups\ fits.

\subsection{NGC 3198}

\begin{figure*}[t]
\epsfxsize=0.95\hsize \epsfbox{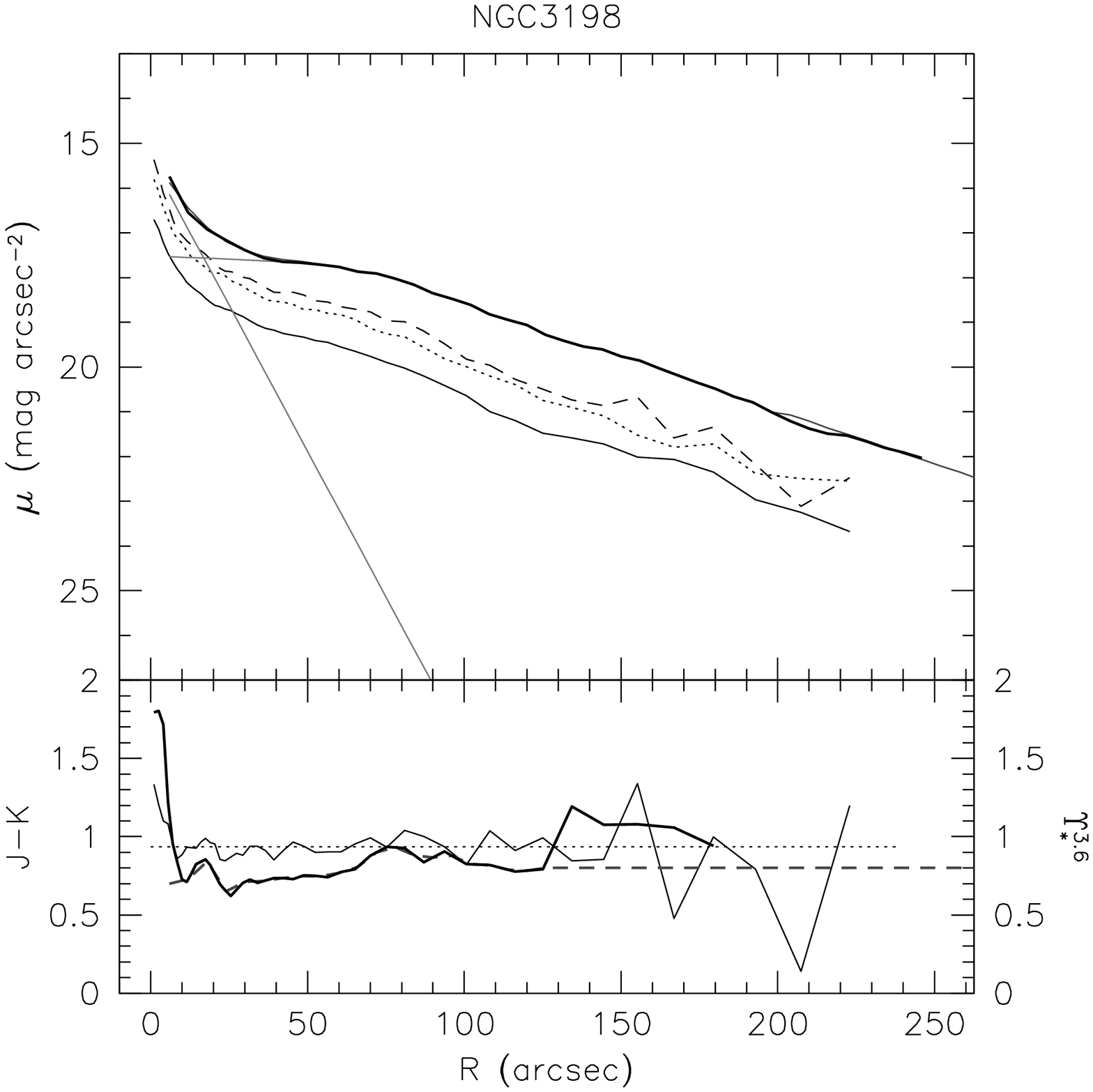} \figcaption{ Surface
  brightness and \Ups\ profiles for NGC 3198. Lines and symbols as in
Fig.~\ref{fig:n2403_prof}.
\label{fig:n3198_prof}}
\end{figure*}

The surface brightness profiles of NGC 3198 are shown in
Fig.~\ref{fig:n3198_prof}. The 2MASS $J$, $H$ and $K$ profiles can be
traced out to $\sim 240''$; the 3.6 $\mu$m profile out to $\sim
250''$.  The profiles show a central component, as well as a
``shoulder'' in the surface brightness around $R\sim 50''$. We
initially model the system using two components.  For the innermost
component we use an exponential disk with parameters $\mu_0 = 15.4$
mag arcsec$^{-2}$ and $h=0.56$ kpc. The profile of the main disk is
derived by subtracting the inner component model from the total
luminosity profile and is extended beyond $R\sim 240''$ by an
exponential disk model with $\mu_0 = 16.2$ mag arcsec$^{-2}$ and
$h=3.06$ kpc.
 
The most striking feature in the $(J-K)$ color distribution is the
sudden reddening in the innermost couple of arc-seconds. At face-value
this would imply a steep increase in \Ups\ in the innermost parts.
This might indicate an additional (third) compact, central
component. Alternatively, it could indicate the presence of a
centrally concentrated pronounced dust component.  Whatever the cause,
the steep increase in color only affects the very innermost point of
the rotation curve and we will not model it. Adding a third component
would increase the number of possible parameter combinations, but
without obvious benefit.  We therefore adopt the \Ups\ profile as
observed, but assume a constant $\Ups = 0.8$ for $R>125''$, as well as
a constant $\Ups = 0.7$ for $R<10''$, as indicated in
Fig.~\ref{fig:n3198_prof}.

\begin{figure*}[t]
\epsfxsize=0.95\hsize \epsfbox{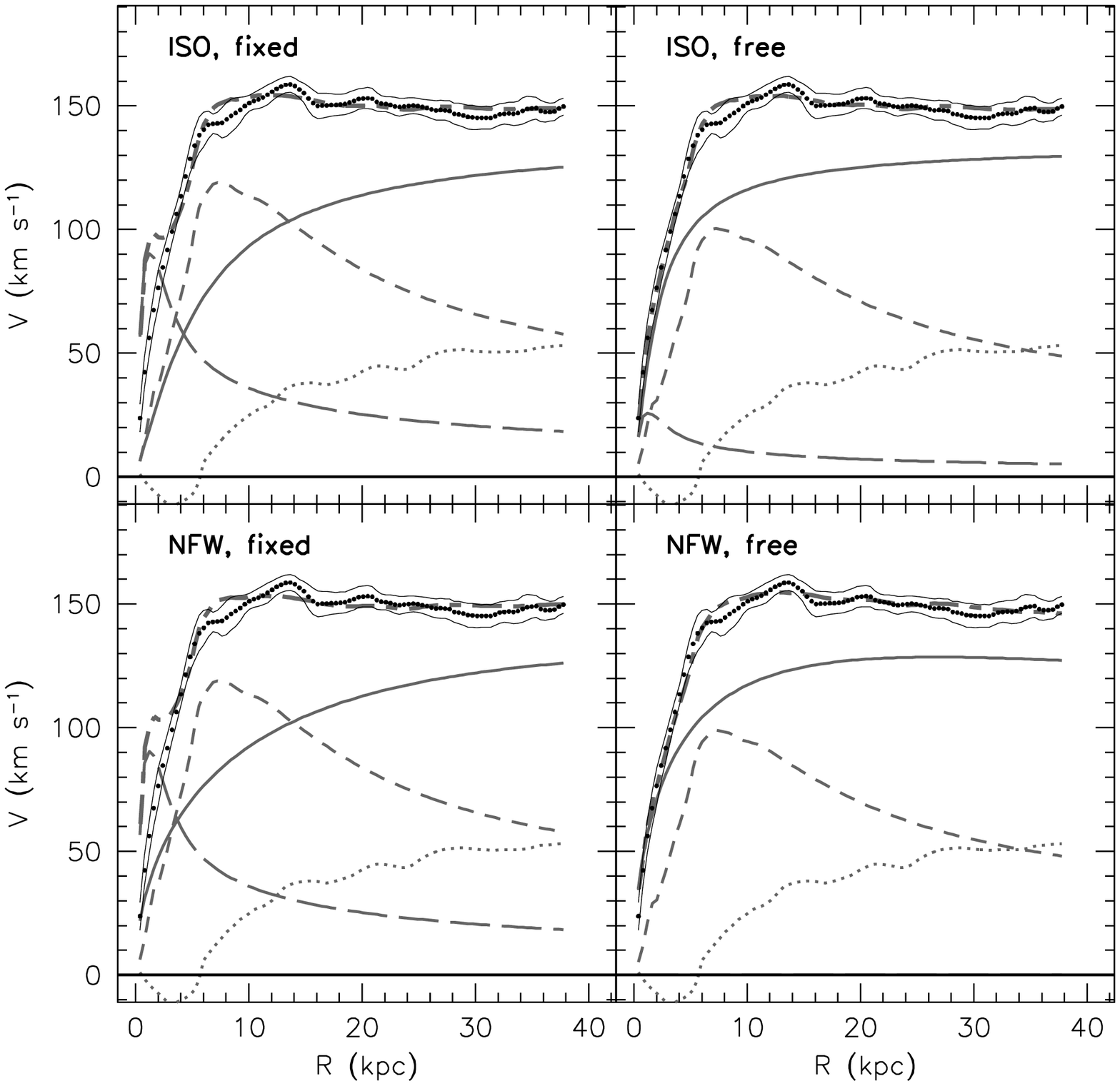}
\figcaption{ ISO and NFW rotation curve fits for the multiple-disk model of NGC 3198.  Lines and
  symbols as in Fig.~\ref{fig:n2403_2curve}.
\label{fig:n3198_2curve}}
\end{figure*}

\begin{figure*}[t]
\epsfxsize=0.95\hsize \epsfbox{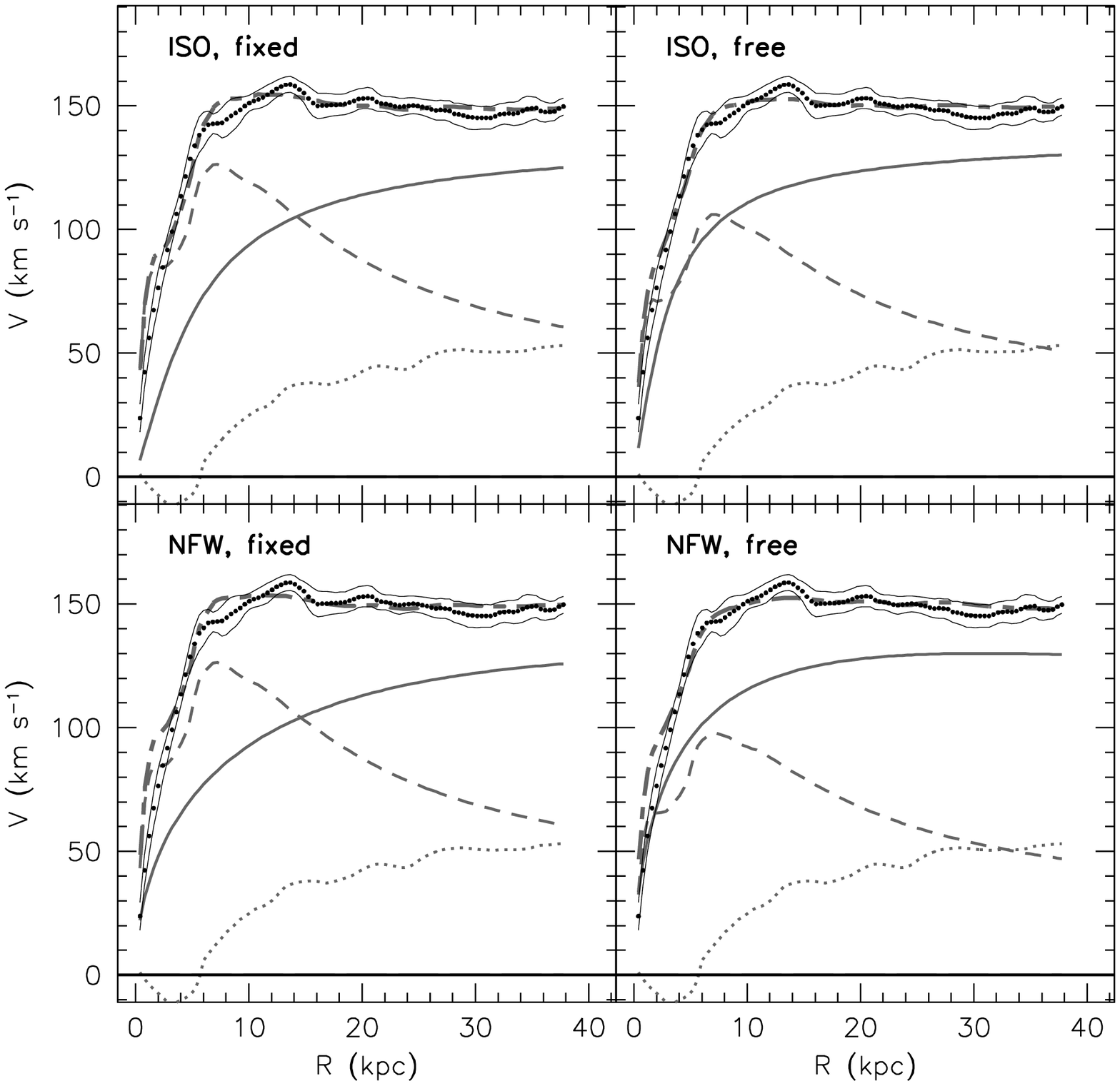}
\figcaption{ISO and NFW rotation curve fits for the single-disk model of NGC 3198. 
Lines and
  symbols as in Fig.~\ref{fig:n2403_1curve}.
\label{fig:n3198_1curve}}
\end{figure*}

The rotation curve fits are shown in Fig.~\ref{fig:n3198_2curve}. In
the case of fixed \Ups, the inner component clearly overestimates the
rotation velocity, indicating that it is likely to be dynamically less
important than suggested by the models (assuming the rotation curve is
not affected; see below).  With \Ups\ as a free parameter the outer
disk ends up being less massive than predicted for both models. The
central component is also fitted with a much smaller \Ups. For the ISO
model we find an inner component $\sim 10$ times lighter than
predicted, whereas the NFW model results in a negative, hence
unphysical, value for $\Ups$. We therefore fixed its value to $\Ups =
0$, with little impact on the results for the outer disk. We checked
whether modeling the inner component as a spherical bulge instead of
an exponential disk changed the results, but found this had a
negligible effect on the outcome.  Because of the apparent
insignificance of the inner component, we also investigated if a
single-disk model could explain the data.  The results are also given
in Fig.~\ref{fig:n3198_1curve}. It is clear that independent of the
choice for halo model, the inner radii cannot be well fitted.

The small value of \Ups\ for the inner disk makes the free \Ups\ fits
less satisfactory.  The low \Ups\ values in the inner parts seems to
contradict the sudden reddening in the inner $10''$, suggesting a
\emph{higher} \Ups\ value. A close inspection of the IRAC 3.6 $\mu$m image
indicates the presence of a short bar at a small angle with the minor
axis, which could mean that the rotation velocities are affected by
non-circular motions, thus giving incorrect values for \Ups.  However,
a harmonic decomposition of the velocity field shows only small
non-circular motions (of order a few \kms) in the inner parts of the
disk; see \citet{clemens2007}.

In an independent analysis, \citet{bottema} also note the presence of
a possible bar in a $K'$ image of NGC 3198.  They argue that having a
bar roughly parallel to the minor axis would \emph{increase} the
apparent rotation velocities in the inner region \citep{teuben}. If
this is indeed what is happening in NGC 3198, then the true rotation
velocity in the inner parts must be even lower than we now observe,
thus making the inner disk even less massive. Clearly there is scope
here for a more extensive study of the dynamics in the inner part of
NGC 3198.

\subsection{NGC 3521}

\begin{figure*}[t]
\epsfxsize=0.95\hsize \epsfbox{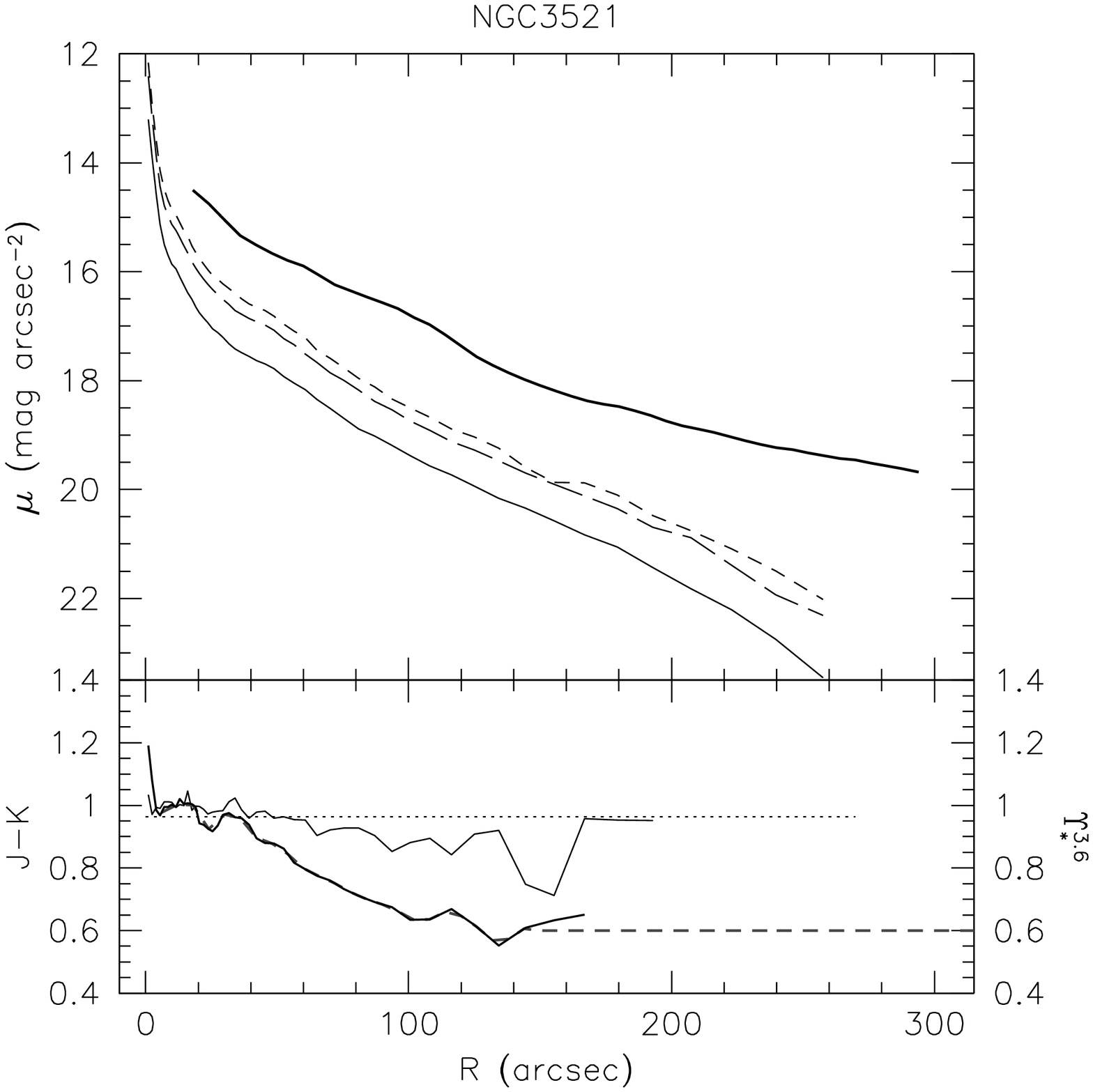} \figcaption{Surface
  brightness and \Ups\ profiles for NGC 3521. Lines and symbols as in
Fig.~\ref{fig:n2403_prof}.
\label{fig:n3521_prof}}
\end{figure*}

The surface brightness profiles of NGC 3521 are shown in
Fig.~\ref{fig:n3521_prof}. The 2MASS $J$, $H$ and $K$ profiles can be
traced out to $\sim 250''$; the 3.6 $\mu$m profile out to the edge of
the \HI\ disk. The photometric analysis is made  difficult by
the faint stellar halo surrounding the main disk.  It is significantly
extended along the minor axis, which, through the ellipse integration
used, results in a flattening of the surface brightness profile at
larger radii.  For this reason results from the stellar decomposition
are more uncertain than for other galaxies in the current sample.

\begin{figure*}[t]
\epsfxsize=0.95\hsize \epsfbox{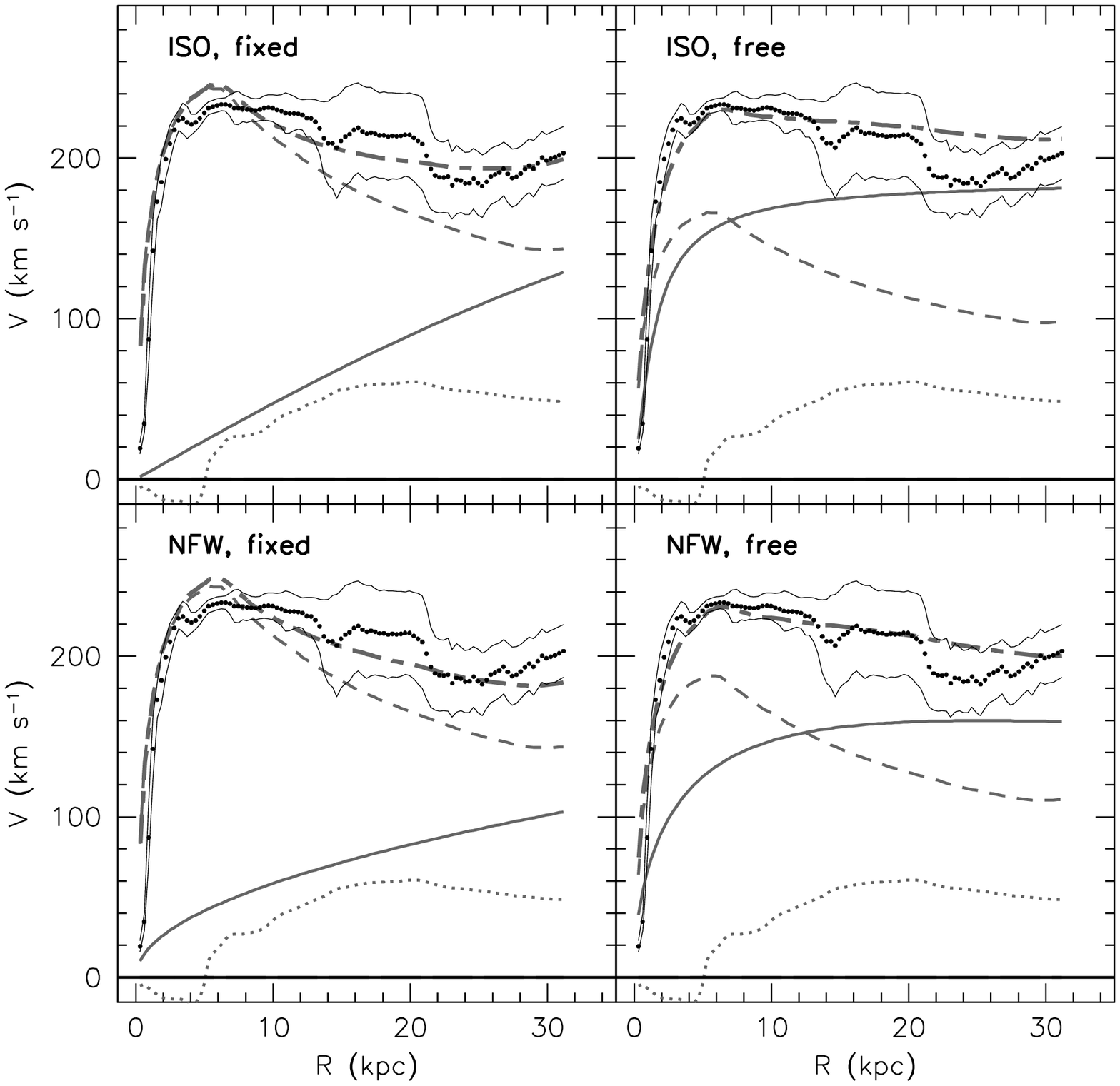} \figcaption{ISO and
  NFW rotation curve fits for NGC 3521.  Lines and symbols as in
  Fig.~\ref{fig:n2403_1curve}.
\label{fig:n3521_curve}}
\end{figure*}

We treat the stellar disk as a single component. A clear color
gradient is present.  The value for \Ups\ changes from $\sim 1$ in the
center to $\sim 0.6$ in the outer parts. Beyond the last reliable
2MASS radius we assume a constant $\Ups=0.6$.  The rotation curve
models are presented in Fig.~\ref{fig:n3521_curve}.  Neither model
fits perfectly, but both give reasonable \Ups\ ratios. Assuming the
fixed \Ups\ values, it follows that the dynamics of NGC 3521 is dominated
by its stars.

\subsection{NGC 3621}

\begin{figure*}[t]
\epsfxsize=0.95\hsize \epsfbox{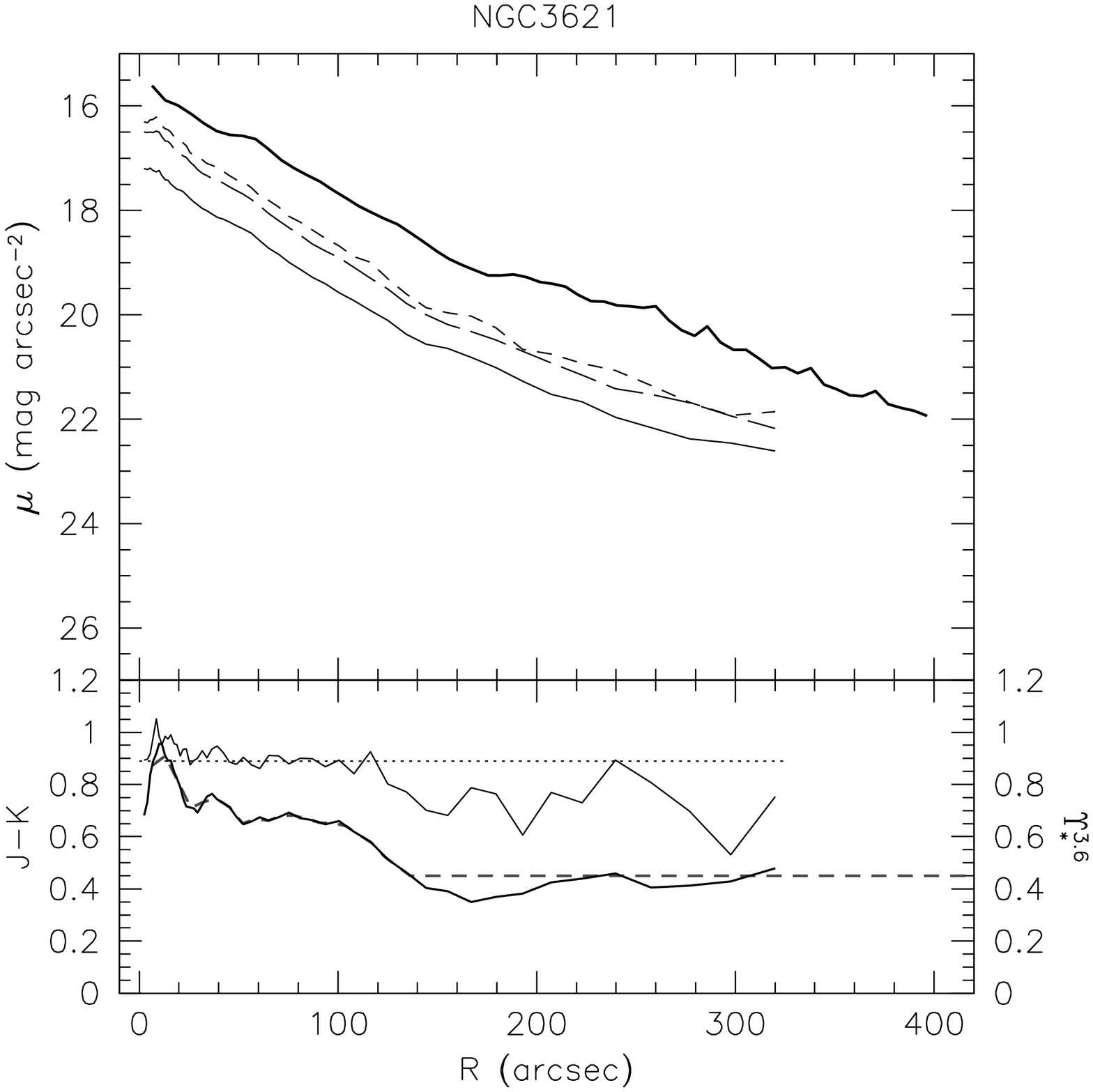} \figcaption{Surface
  brightness and \Ups\ profiles for NGC 3621. Lines and symbols as in
Fig.~\ref{fig:n2403_prof}.
\label{fig:n3621_prof}}
\end{figure*}

The surface brightness profiles of NGC 3621 are shown in
Fig.~\ref{fig:n3621_prof}. The 2MASS $J$, $H$ and $K$ profiles can be
traced out to $\sim 320''$; the 3.6 $\mu$m profile out to $\sim
400''$.  The latter profile shows slight evidence for a change in scale
length around $R\sim 200''$. However, neither the lack of a clear
break in the color gradient, nor the morphology, nor the (small)
deviation from a single exponential disk are enough justification for
a two disk model. We extended the disk to larger radii using an
exponential fit to radii $200''<R<400''$ and with parameters
$\mu_0 =16.6$ mag arcsec$^{-2}$ and $h=2.61$ kpc.  A small color
gradient is visible in the 2MASS data which results in a gradient in
\Ups\ from $\Ups \simeq 0.8$ in the inner parts to $\Ups \simeq 0.4$
in the outer parts. Due to the increased uncertainty in the 2MASS
color measurement in the outermost parts, we assumed $\Ups = 0.45$
for $R>150''$, as indicated in Fig.~\ref{fig:n3621_prof}.

\begin{figure*}[t]
\epsfxsize=0.95\hsize \epsfbox{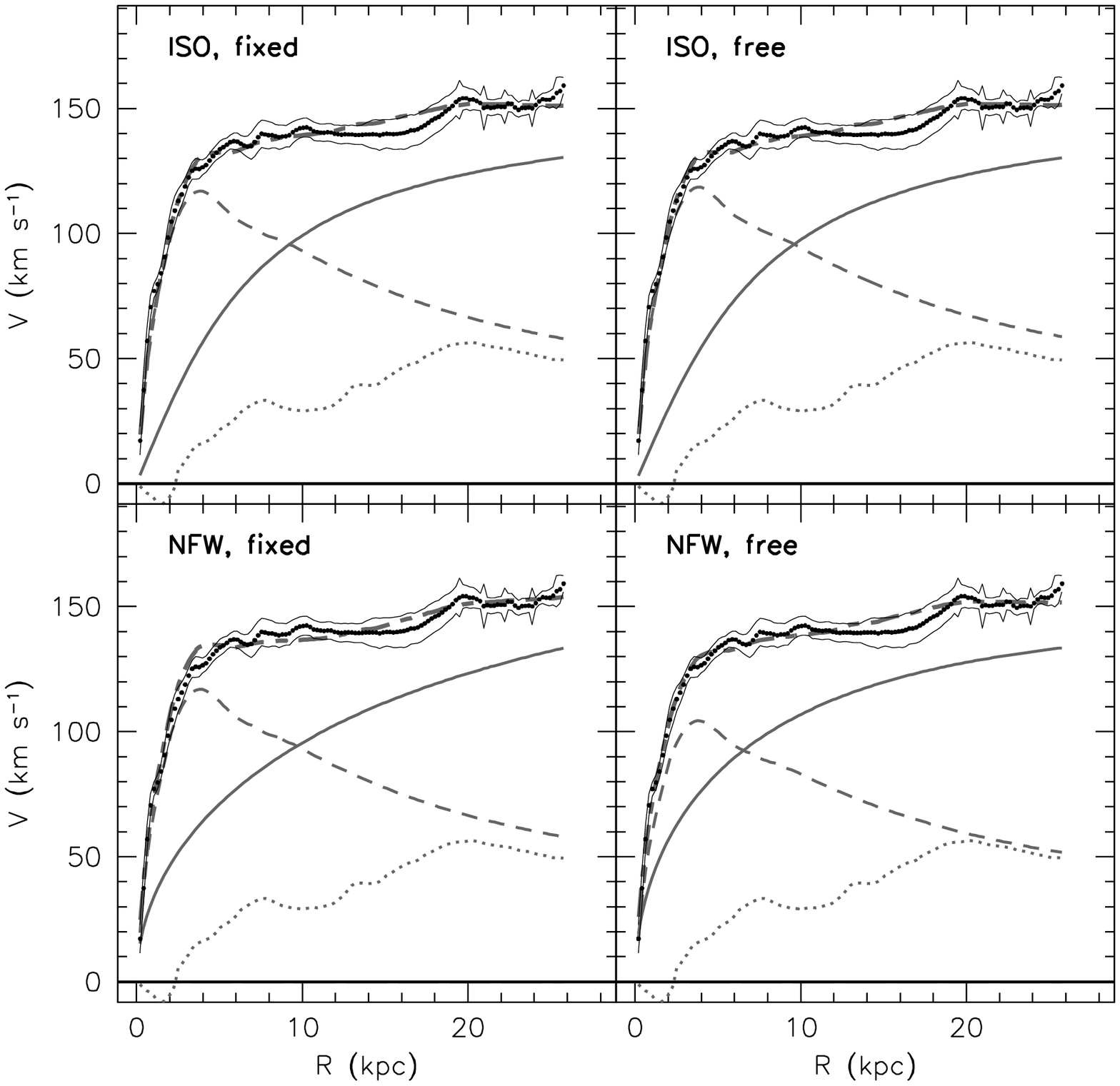} \figcaption{ISO and
  NFW rotation curve fits for NGC 3621. Lines and symbols as in
  Fig.~\ref{fig:n2403_1curve}.
\label{fig:n3621_curve}}
\end{figure*}

The rotation curve fits are presented in Fig.~\ref{fig:n3621_curve}.
It is interesting how well the best (free) fit and the fixed
diet-Salpeter fit for the ISO halo agree with each other. Similarly,
the free NFW fit yields a value of \Ups\ very close to the Kroupa
fixed \Ups\ value, a clear illustration of how the NFW model prefers
lighter disks in order to accommodate its steeper halo mass density
profile.  The free ISO and NFW fits are of identical quality, and on
the basis of these data it is difficult to single out a favorite model
for NGC 3621.

\subsection{NGC 4736}

\begin{figure*}[t]
\epsfxsize=0.95\hsize \epsfbox{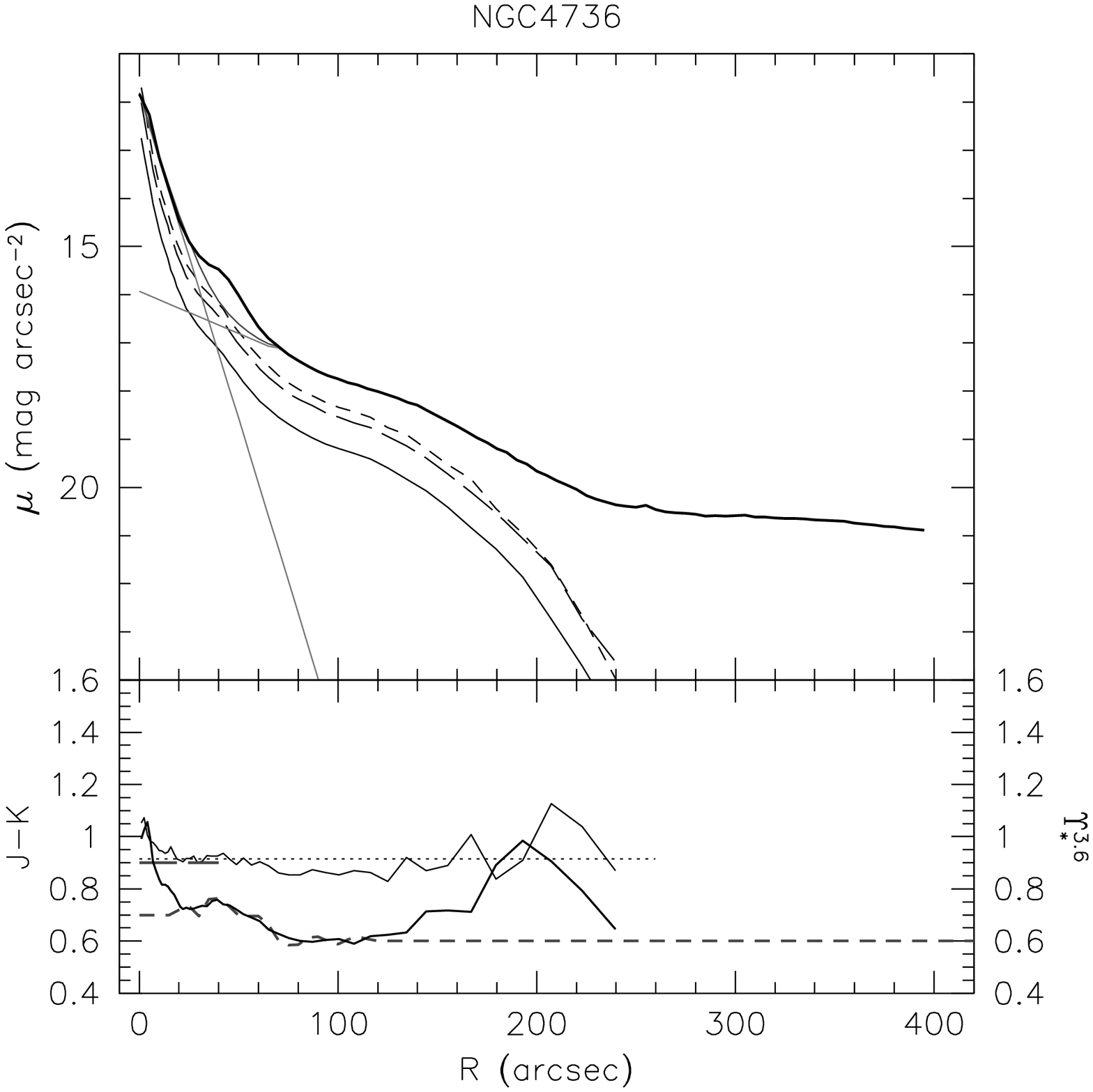} \figcaption{ Surface
  brightness and \Ups\ profiles for NGC 4736. Lines and symbols as in
Fig.~\ref{fig:n2403_prof}.
\label{fig:n4736_prof}}
\end{figure*}

The surface brightness profiles of NGC 4736 are shown in
Fig.~\ref{fig:n4736_prof}. The 2MASS $J$, $H$ and $K$ profiles can be
traced out to $\sim 250''$, whereas the 3.6 $\mu$m profile can be
traced out to the edge of the \HI\ disk at $\sim 400''$.

The stellar component of NGC 4736 has a complex structure.  A
prominent stellar ring can be seen in various wavelength at a radius
$R\sim 40''$. The ring, visible in the 3.6 $\mu$m image, is also
prominently visible in the 20-cm radio continuum.  The presence of
this fairly strong continuum suggests active star formation, and may
mean that at this particular radius the assumption that the 3.6 $\mu$m
emission is proportional to the stellar mass is not valid. For this
reason we will not take this feature into account in our subsequent
analysis of the surface brightness profile.  The IRAC 3.6 $\mu$m
profile flattens out at $250'' \la R \la 400''$, caused by the faint
stellar ring surrounding the main body of NGC 4736.  At larger radii
(beyond the edge of the \HI\ disk) the profile starts decreasing
again.  Also visible is a significant steepening in the inner parts.
We will treat this as evidence for a separate central component with
parameters $\mu_0 = 11.8$ mag arcsec$^{-2}$ and $h= 0.26$ kpc. To
model the outer disk we fitted an exponential disk to the radial range
$80''<R<240''$ and find parameters $\mu_0 = 15.9$ mag arcsec$^{-2}$
and $h=1.99$ kpc. We used the extrapolated fit for radii $R<80''$ and
the observed radial surface brightness for $R>80''$.

NGC 4736 exhibits a small color gradient in the outer parts with a
steeper inner reddening. We have adopted a constant value of $\Ups =
0.6$ for the outer disk beyond the last reliable 2MASS $(J-K)$ value,
and have additionally assumed a constant $\Ups=0.7$ for the innermost
$20''$. For the central disk we assume a constant $\Ups=0.9$ which is
the average value for $R>20''$.

\begin{figure*}[t]
\epsfxsize=0.95\hsize \epsfbox{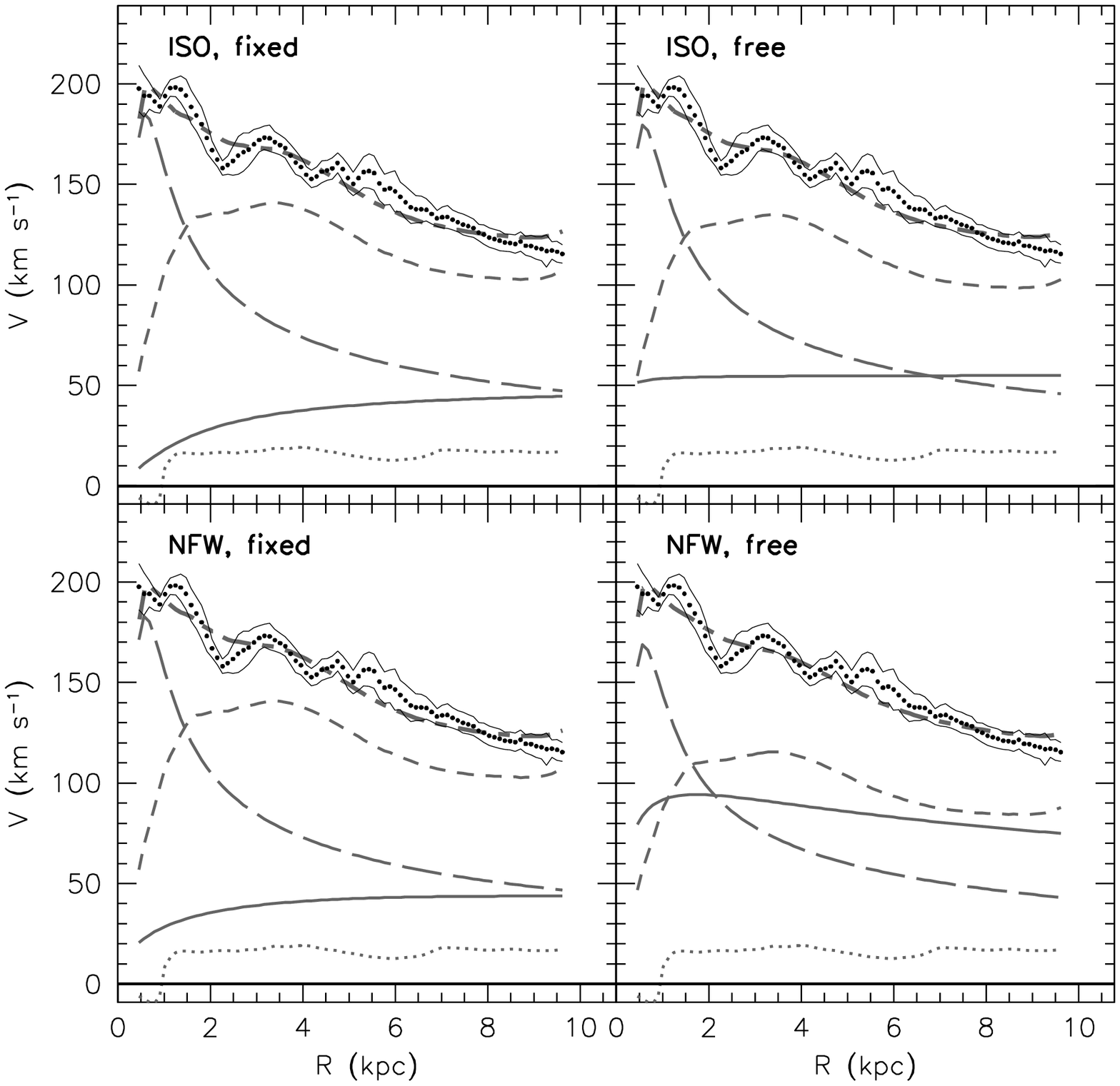} \figcaption{ISO and
  NFW rotation curve fits for NGC 4736.  Lines and symbols as in
  Fig.~\ref{fig:n2403_2curve}.
\label{fig:n4736_curve}}
\end{figure*}

\begin{figure*}[t]
\epsfxsize=0.95\hsize \epsfbox{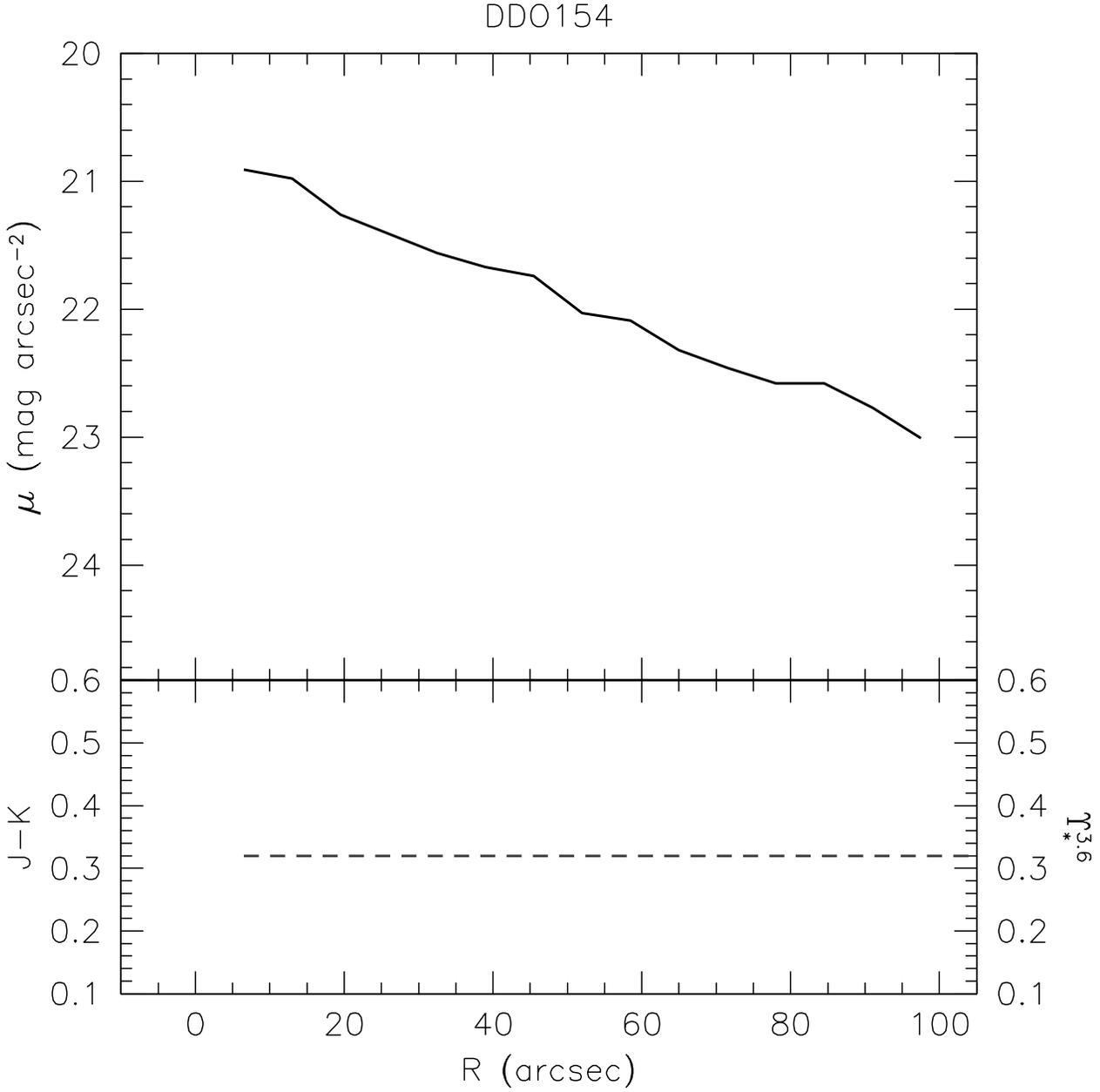} \figcaption{Surface
  brightness and \Ups\ profiles for DDO 154. Lines and symbols as in
Fig.~\ref{fig:n2403_prof}.
\label{fig:ddo154_prof}}
\end{figure*}

The rotation curves are shown in Fig.~\ref{fig:n4736_curve}. It is
immediately obvious that NGC 4736 must be close to maximum disk. The
predicted \Ups\ value for the inner disk yields a curve that
overestimates the rotation velocity (not shown here). We therefore
left \Ups\ of the inner disk as a free parameter after fixing the
\Ups\ value of the main disk. This resulted in an inner disk mass
about a factor of $\sim 2$ less than predicted (and is the model shown
in Fig.~\ref{fig:n4736_curve}).  Whilst the discrepancies are slighly
less for the Kroupa \Ups\ values, the resulting models still
overpredict the rotation velocity.  Because of these uncertainties and
because of the large non-circular motions in this galaxy
\citep{clemens2007}, it is virtually impossible to say anything
definite about the distribution of dark matter in this galaxy based
purely on rotation curve arguments.

\subsection{DDO 154}

DDO 154 is barely detected in the 2MASS images and we cannot therefore
use the same $(J-K)$ color method as with the other galaxies.  We
therefore use the $(B-V)$ and $(B-R)$ colors as proxies, as described
in \citet{oh2007}. For the color we use the values given in
\citet{car_beaulieu}: $(B-V) = 0.37 \pm 0.06$ and $(B-R) = 0.64 \pm
0.07$.  No significant color gradient is detected, and we therefore
assume a constant color as function of radius in our models.  Using
the equations from \citet{oh2007} we find values $\Ups = 0.34 \pm
0.04$ using the $(B-V)$ colors and $\Ups=0.31 \pm 0.02$ using the
$(B-R)$ colors. We adopt the weighted average $\Ups = 0.32$ as our
best estimate.  The IRAC 3.6 $\mu$m profile can be described by a
single exponential disk with $\mu_0 = 20.8$ mag arcsec$^{-2}$ and
$h=0.72$ kpc.

\begin{figure*}[t]
\epsfxsize=0.95\hsize \epsfbox{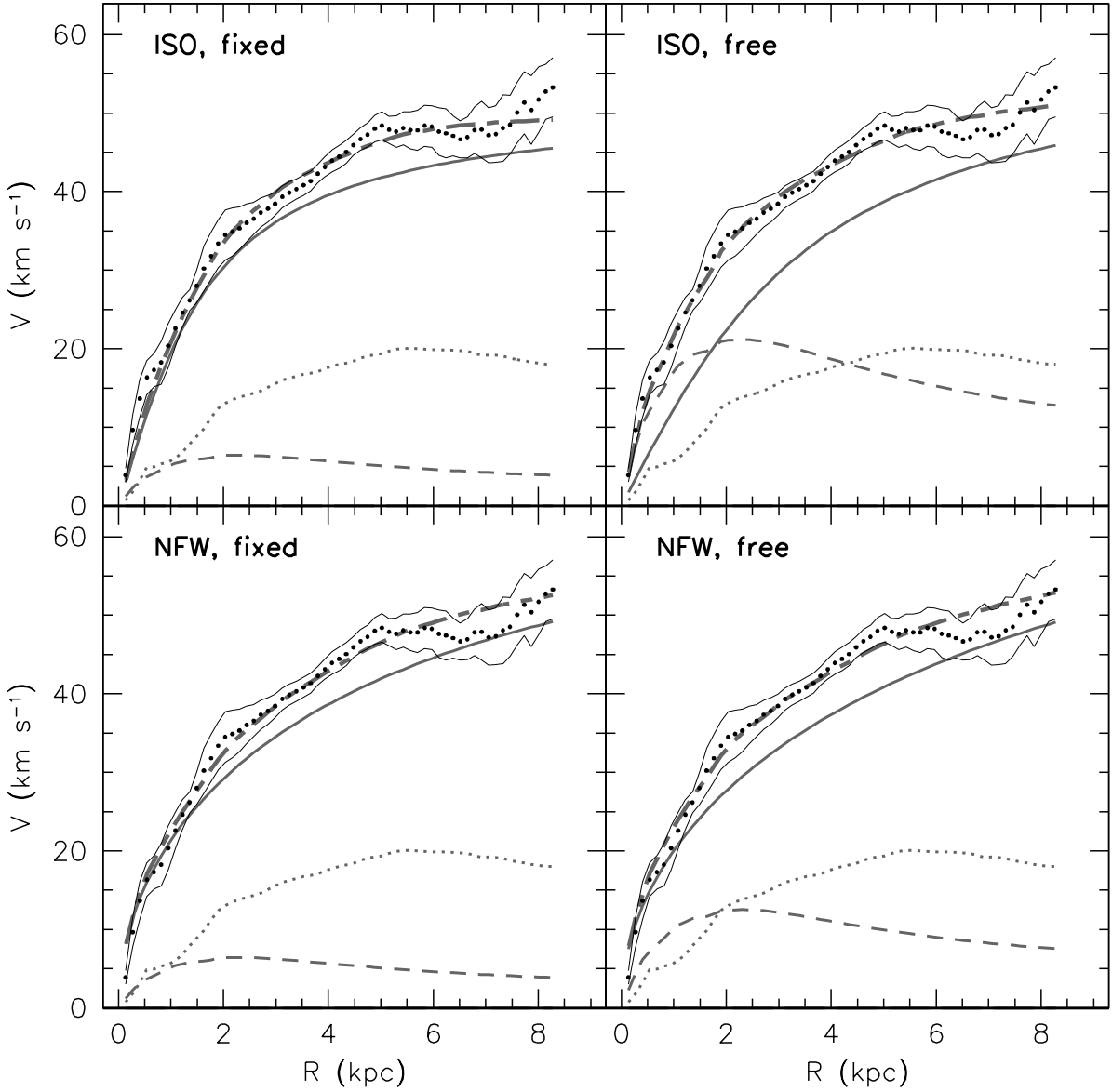} \figcaption{ISO and
  NFW rotation curve fits for DDO 154.  Lines and symbols as in
  Fig.~\ref{fig:n2403_1curve}.
\label{fig:ddo154_curve}}
\end{figure*}

The rotation curves are shown in Fig.~\ref{fig:ddo154_curve}. The ISO
model consistently fits better than the NFW model. The latter
over-predicts the velocity in the inner parts. Note the very large
values for \Ups\ when it is left as a free parameter. These will be
discussed in more detail in Sect.~7. Because of the dark matter
dominance in DDO 154 the derived halo parameters are virtually
independent from the assumed IMF.

\subsection{NGC 5055}

\begin{figure*}[t]
\epsfxsize=0.95\hsize \epsfbox{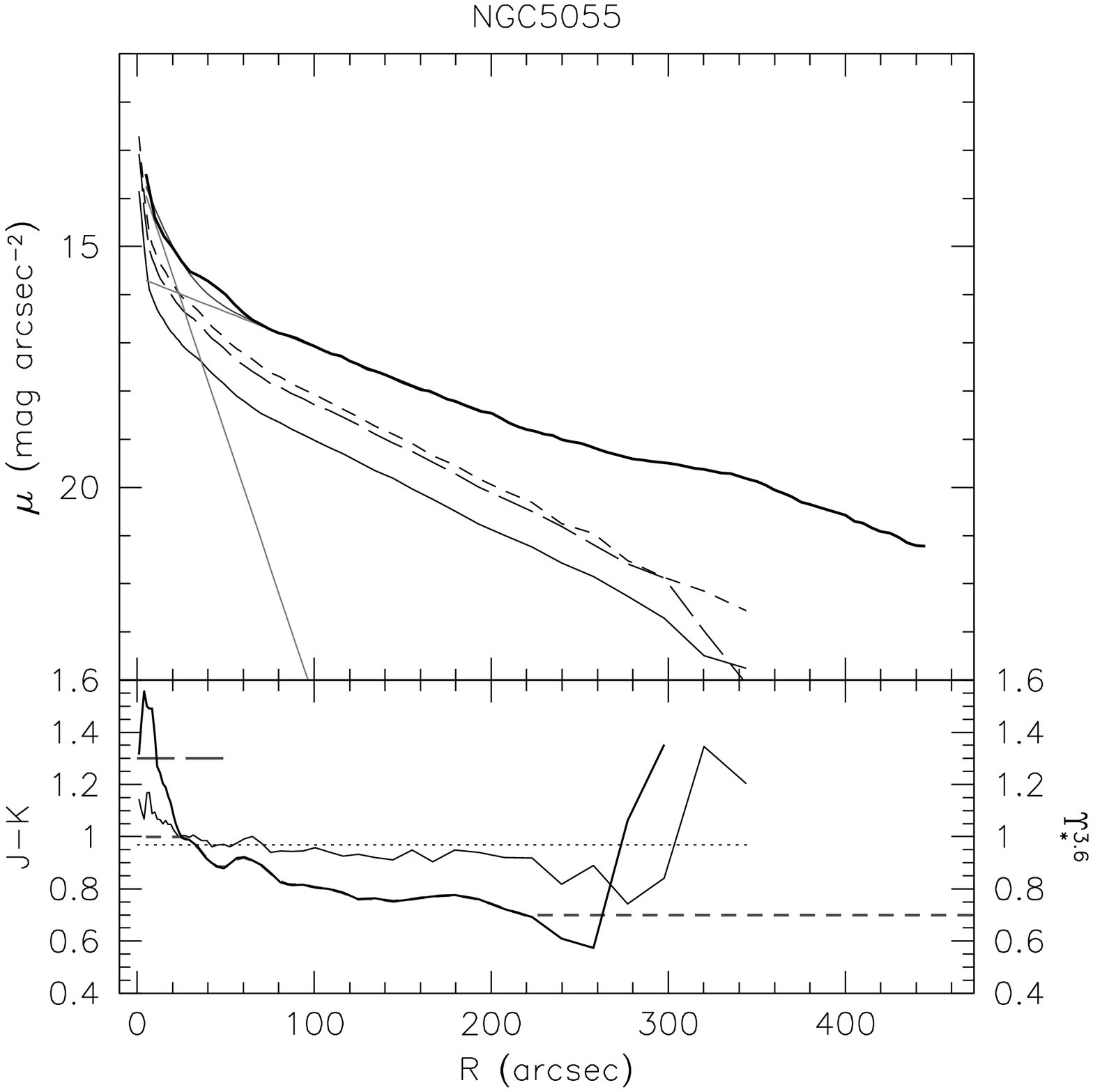} \figcaption{Surface
  brightness and \Ups\ profiles for NGC 5055. Lines and symbols as in
Fig.~\ref{fig:n2403_prof}.
\label{fig:n5055_prof}}
\end{figure*}

The surface brightness profiles of NGC 5055 are shown in
Fig.~\ref{fig:n5055_prof}. The 2MASS $J$, $H$ and $K$ profiles can be
traced out to $\sim 350''$; the 3.6 $\mu$m profile out to $450''$.
There is clear evidence for a compact central component which we model
as an exponential disk with parameters $\mu_0 = 13.4$ mag
arcsec$^{-2}$ and $h=0.35$ kpc.  For the outer disk, we simply used
the observed profile with the inner component subtracted, as indicated
in Fig.~\ref{fig:n5055_prof}. Small residuals are present, notably
around $R \sim 50''$, but these are of minor importance. An
exponential disk fit to radii $350'' < R < 450''$ was used to extend
the outer disk profile.  NGC 5055 has a significantly redder center,
with a well-defined color gradient at larger radii.  For the central
component we assume a constant $\Ups = 1.3$, whereas for the outer
disk we find values from an inner $\Ups \simeq 1$ to an outer $\Ups
\sim 0.5$.  Beyond the radii where the 2MASS colors could be reliably
determined we assume a constant value of $\Ups=0.5$. In the very inner
parts we assumed an extrapolated $\Ups = 1.0$ for the outer disk.

\begin{figure*}[t]
\epsfxsize=0.95\hsize \epsfbox{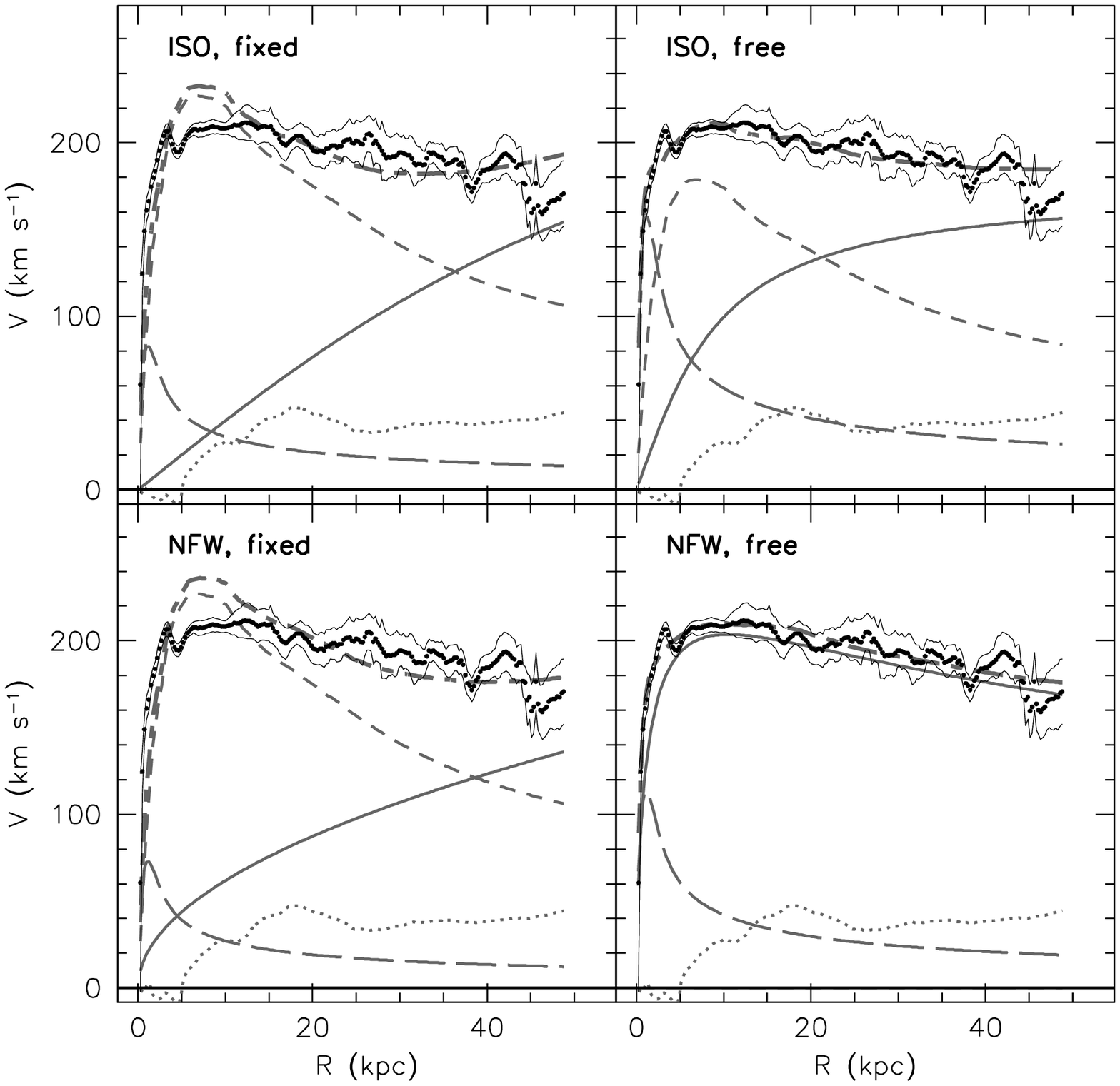} \figcaption{ISO and
  NFW rotation curve fits for NGC 5055.  Lines and symbols as in
  Fig.~\ref{fig:n2403_2curve}.
\label{fig:n5055_curve}}
\end{figure*}

The rotation curve fits are presented in Fig.~\ref{fig:n5055_curve}.
In the case of fixed $\Ups$ the inner disk over-predicts the curve by a
large factor (not shown here), and we kept it as free parameter,
yielding values of $\Ups \sim 0.2$ for the inner disk.  The NFW model
seems to have particular difficulties fitting this galaxy. With \Ups\
as a free parameter, the fit demands a negative value for the outer
disk. In the fits presented here we have therefore set $\Ups = 0$ for
this component.

\subsection{NGC 6946}

\begin{figure*}[t]
\epsfxsize=0.95\hsize \epsfbox{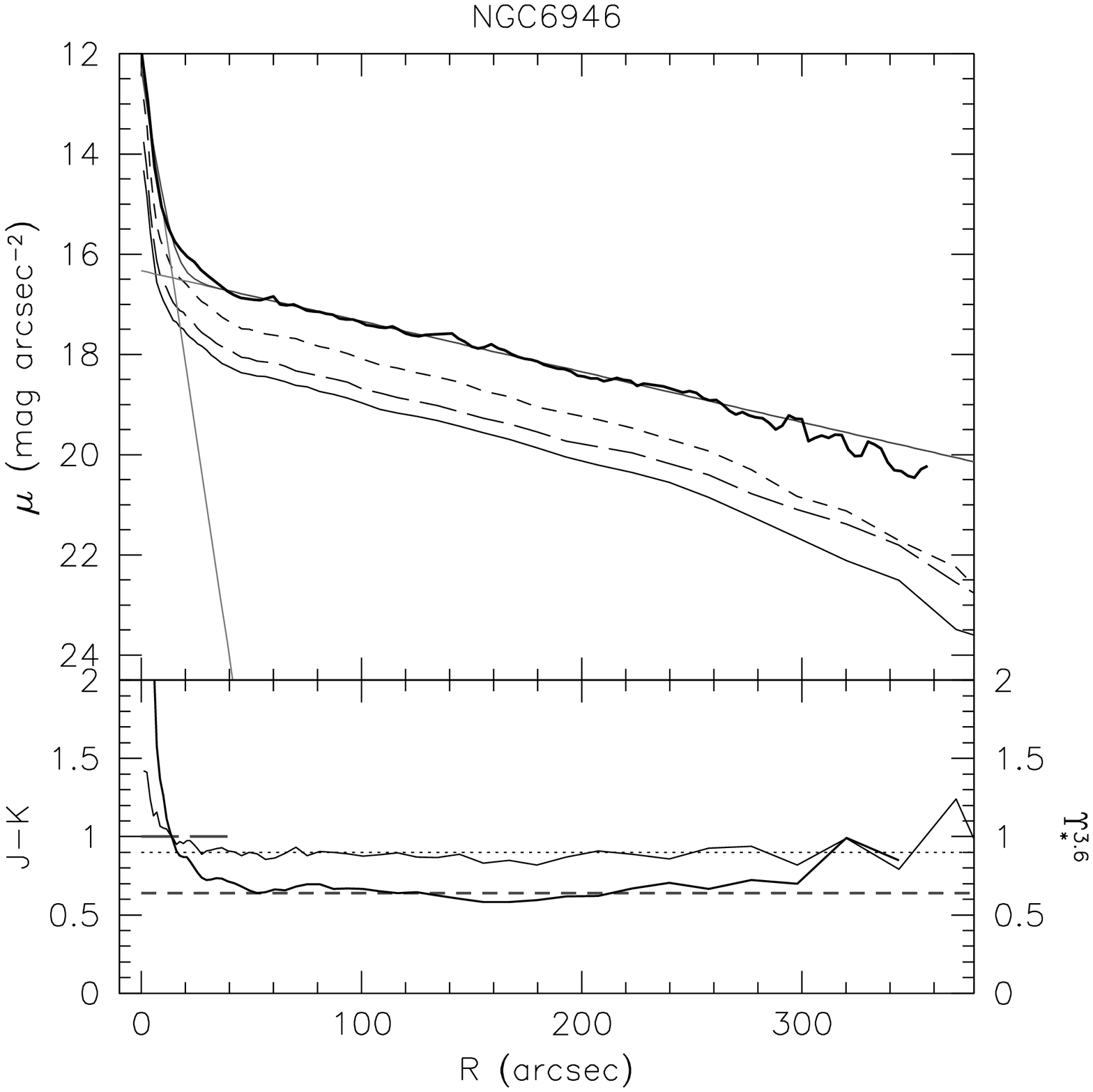} \figcaption{Surface
  brightness and \Ups\ profiles for NGC 6946. Lines and symbols as in
Fig.~\ref{fig:n2403_prof}.
\label{fig:n6946_prof}}
\end{figure*}

The surface brightness profiles of NGC 6946 are shown in
Fig.~\ref{fig:n6946_prof}. The 2MASS $J$, $H$ and $K$ profiles as well
as the 3.6 $\mu$m profile can all be traced out to $\sim 360''$.  The
profiles show clear evidence for a compact central component.  We
modeled this component as an exponential disk with parameters $\mu_0 =
12.8$ mag arcsec$^{-2}$ and $h=0.15$ kpc.  For the outer disk we used
the observed profile with the inner component subtracted. For larger
radii we extended the profile with an exponential fit with parameters
$\mu_0 = 16.3$ mag arcsec$^{-2}$ and $h=2.97$ kpc. There is evidence
for a slight excess of light in the ``shoulder'' of the profile at
$R\sim 50''$. This small excess does, however, have negligible effect
on the results of the rotation curve fit, and we therefore use a
two-component model for the stellar disk.

The outer disk of NGC 6946 shows no strong evidence for a $(J-K)$
color gradient and we use a constant $\Ups = 0.64$.  Within $R \sim
30''$, the change in color is, however, very pronounced, resulting in
very red central components [$(J-K)>1.4$ at the innermost
point]. Taking these colors at face-value would imply $\Ups >
2.0$. Such colors are only found for extreme star formation histories
(cf.\ \citealt{oh2007}), and it is therefore likely that other effects
contribute to this extreme color. For the purposes of the rotation
curve analysis we therefore assume $\Ups=1.0$ for the inner disk
(which is also the value one gets when averaging over the entire
radial extent of the stellar component).

\begin{figure*}[t]
\epsfxsize=0.95\hsize \epsfbox{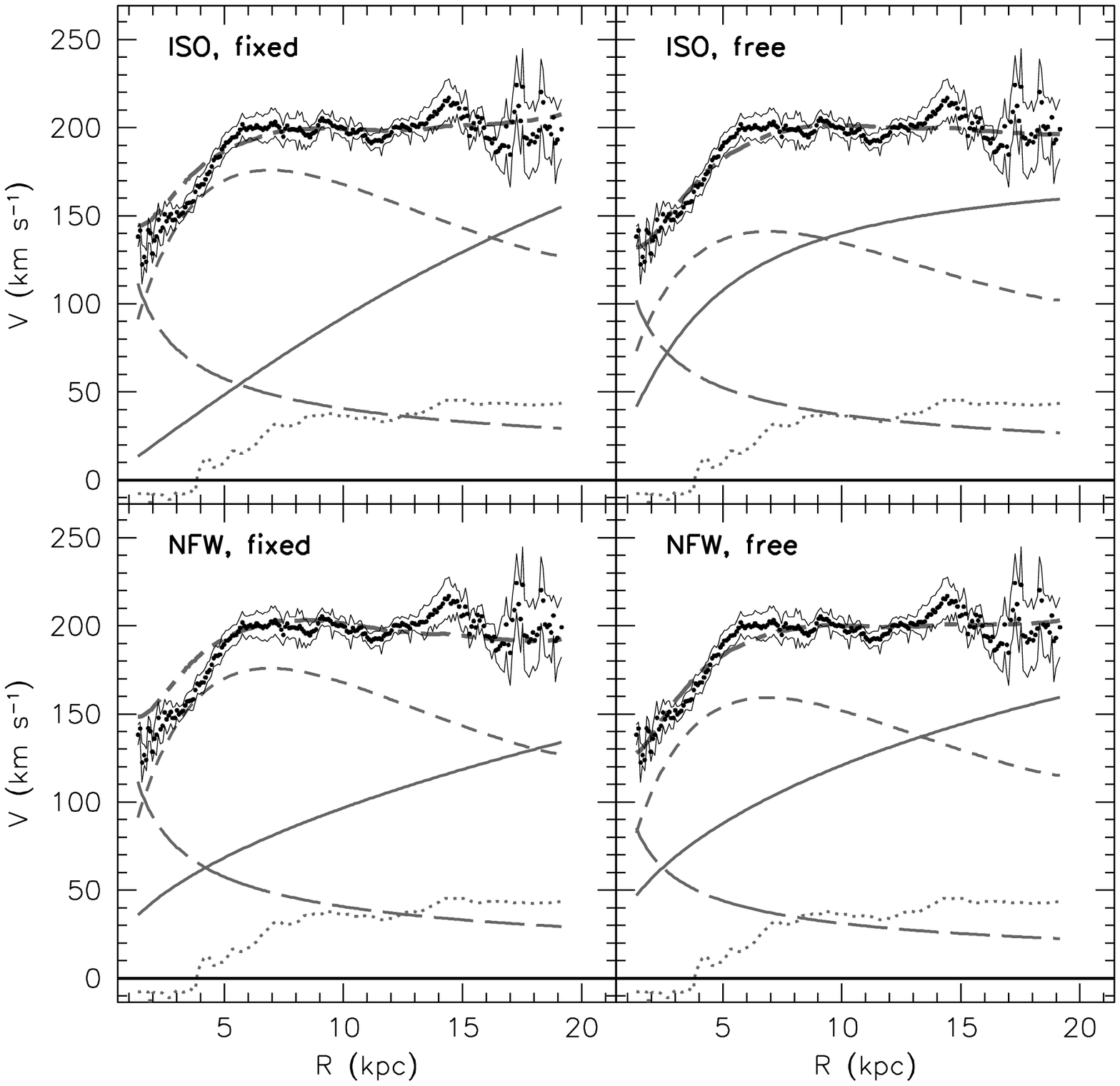} \figcaption{ISO and
  NFW rotation curve fits for NGC 6946.  Lines and symbols as in
  Fig.~\ref{fig:n2403_2curve}.
\label{fig:n6946_curve}}
\end{figure*}

The halo model fits are presented in Fig.~\ref{fig:n6946_curve}.  They
show that the inner disk component prefers a value for \Ups\ that is
lower than the $(J-K)$ color would suggest. The values derived for
the outer disk are reasonably close to the predicted values.  The NFW
and ISO models all produce very similar quality fits.

\subsection{NGC 7331}

\begin{figure*}[t]
\epsfxsize=0.95\hsize \epsfbox{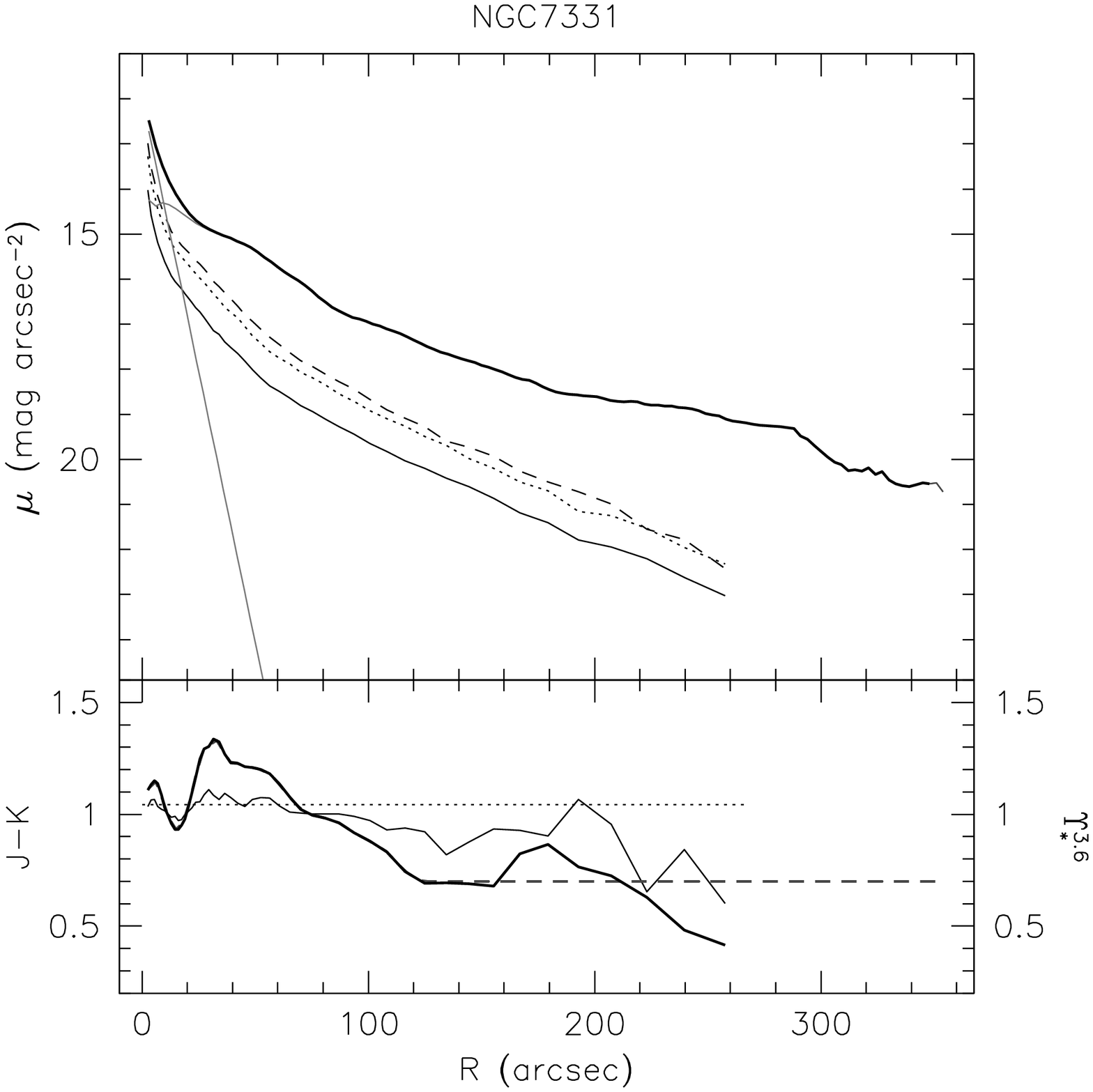} \figcaption{Surface
  brightness and \Ups\ profiles for NGC 7331. Lines and symbols as in
Fig.~\ref{fig:n2403_prof}.
\label{fig:n7331_prof}}
\end{figure*}

The surface brightness profiles of NGC 7331 are shown in
Fig.~\ref{fig:n7331_prof}. The 2MASS $J$, $H$ and $K$ profiles can be
traced out to $\sim 265''$; the 3.6 $\mu$m profile can be traced out
to $350''$, or the entire extent of the \HI\ disk.  The profiles show
clear evidence for a compact central component.  We modeled this
component as an exponential disk with parameters $\mu_0 = 12.0$ mag
arcsec$^{-2}$ and $h=0.32$ kpc.  For the outer disk we simply used the
observed profile with the inner component subtracted, as indicated in
Fig.~\ref{fig:n7331_prof}. NGC 7331 has a well-defined color gradient,
and shows some of the largest \Ups\ values, as well as one of the
steepest \Ups\ gradients in the entire sample. Beyond the radii where
the 2MASS colors could be reliably determined we assume a constant
value of $\Ups=0.7$.

\begin{figure*}[t]
\epsfxsize=0.95\hsize \epsfbox{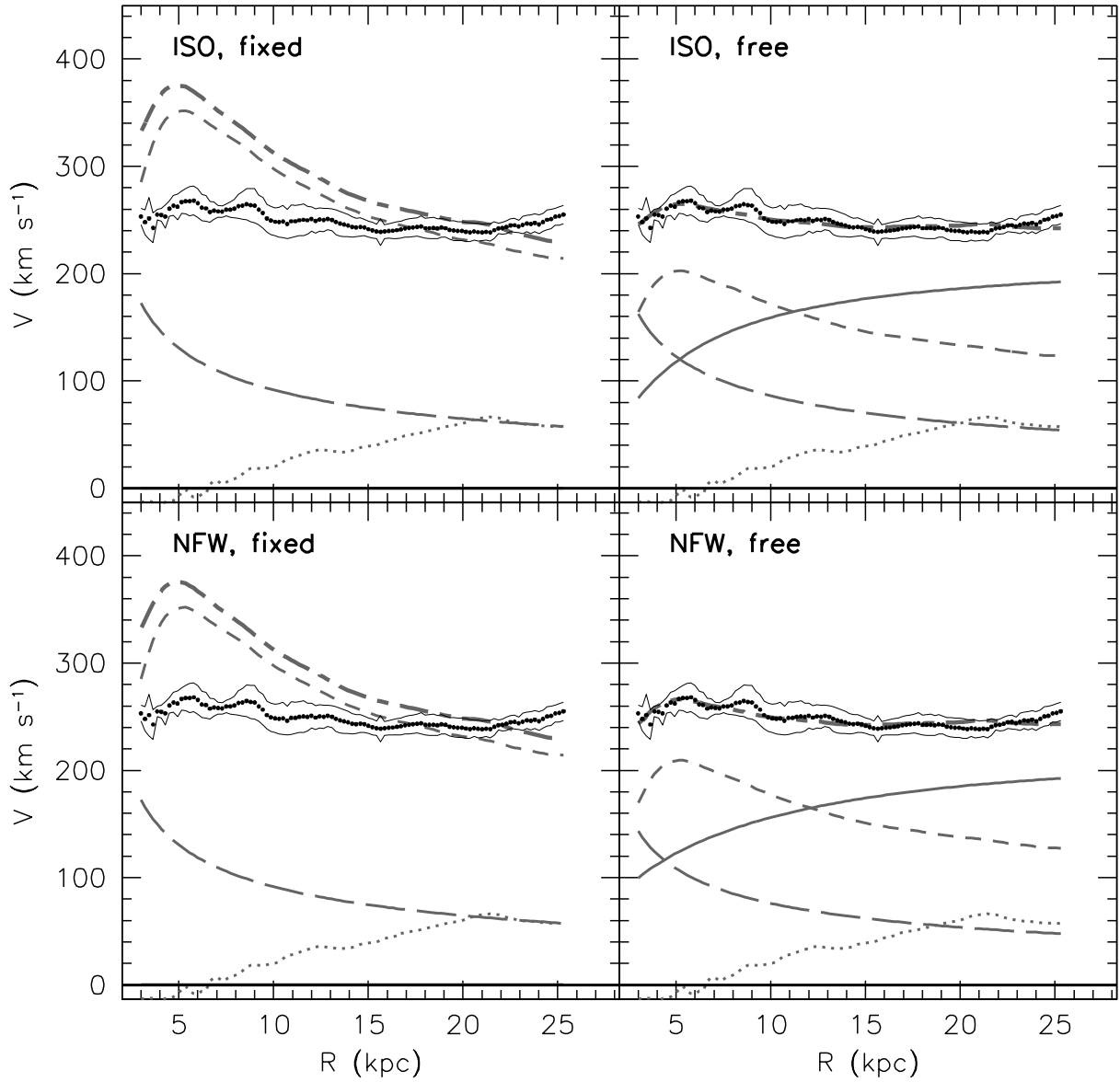} \figcaption{ISO and NFW
  rotation curve fits for the color-gradient model of NGC 7331. Lines and
  symbols as in Fig.~\ref{fig:n2403_2curve}.
\label{fig:n7331_1curve}}
\end{figure*}

These high \Ups\ values lead to incompatible mass models: using the
predicted \Ups\ values results in disks that are too massive.  This is
shown in Fig.~\ref{fig:n7331_1curve}, clearly suggesting that for this
galaxy the color-based values are not correct (for that reason we do
not list the model parameters in Tables \ref{table2}, \ref{table3a}
and \ref{table3}).  Leaving \Ups\ free in the fit, the data clearly
prefer lower values.  A possible explanation could be the presence of
a strong dust ring in the inner disk of NGC 7331
\citep{sings7331}. The radius at which this ring is found corresponds
with the radius where the highest \Ups\ values are found. This would
suggest that the very inner \Ups\ maximum and steep drop are
associated with the central component, whereas the subsequent steep
rise and gradual drop are associated with the dust ring, and therefore
do not reflect stellar population changes.

\begin{figure*}[t]
  \epsfxsize=0.95\hsize \epsfbox{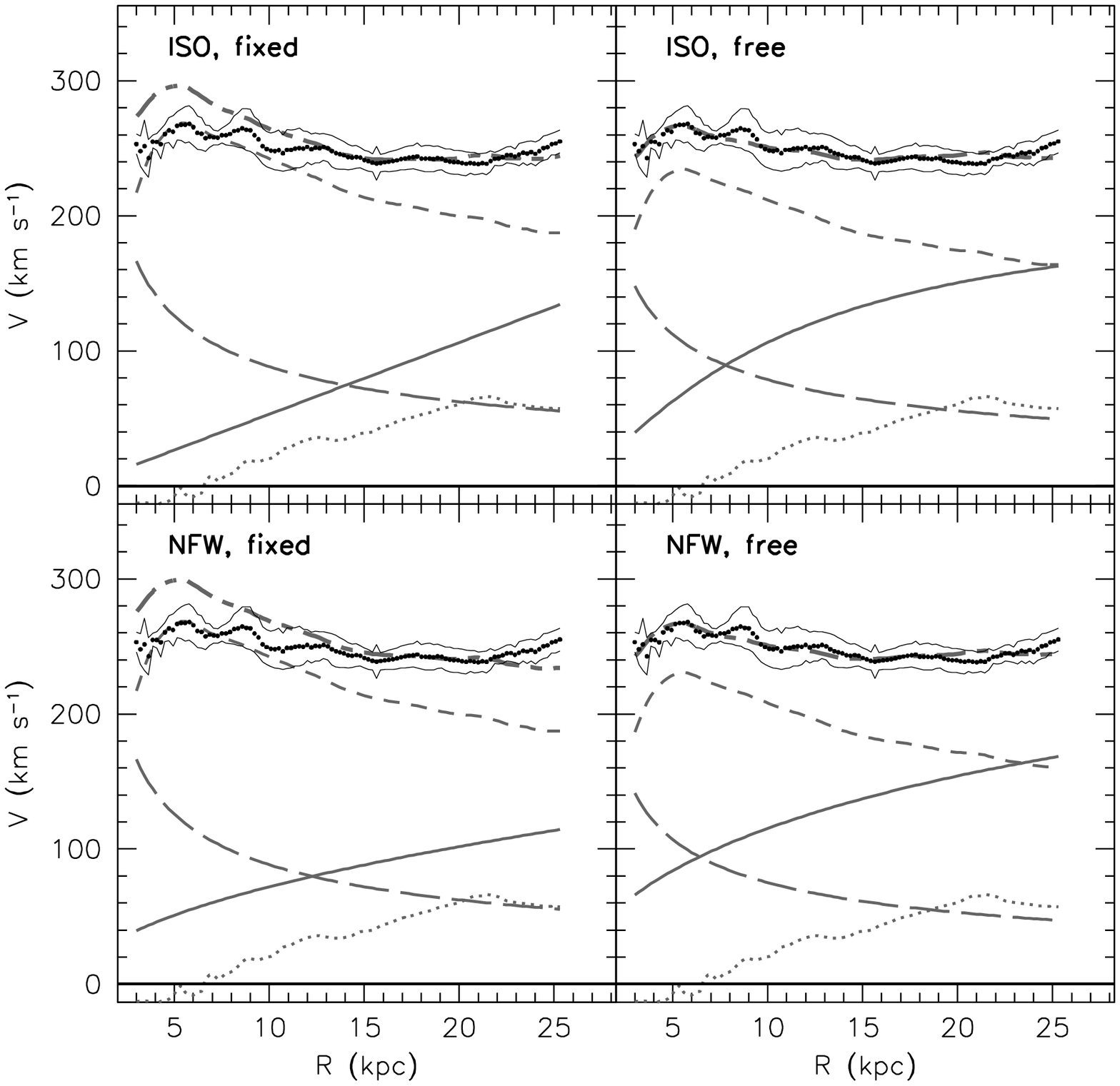}
  \figcaption{ISO and NFW rotation curve fits for the constant \Ups\
    model of NGC 7331. Lines and symbols as in
    Fig.~\ref{fig:n2403_2curve}.
\label{fig:n7331_2curve}}
\end{figure*}

We therefore also investigate a model where the two components each
have a radially constant \Ups. Using Fig.~\ref{fig:n7331_prof}, we
find that $\Ups=1.0$ and $\Ups=0.7$ are good estimates for the inner
and outer disks, respectively. These fits are presented in
Fig.~\ref{fig:n7331_2curve}.  We then find that both ISO and NFW
models result in better fits.  NGC 7331 is therefore the one galaxy in
our sample where the $(J-K)$-\Ups\ relation clearly fails.  Note that
for the fixed \Ups\ case, the Kroupa models yield much better fits
than the diet-Salpeter ones.

\subsection{NGC 7793}

\begin{figure*}[t]
\epsfxsize=0.95\hsize \epsfbox{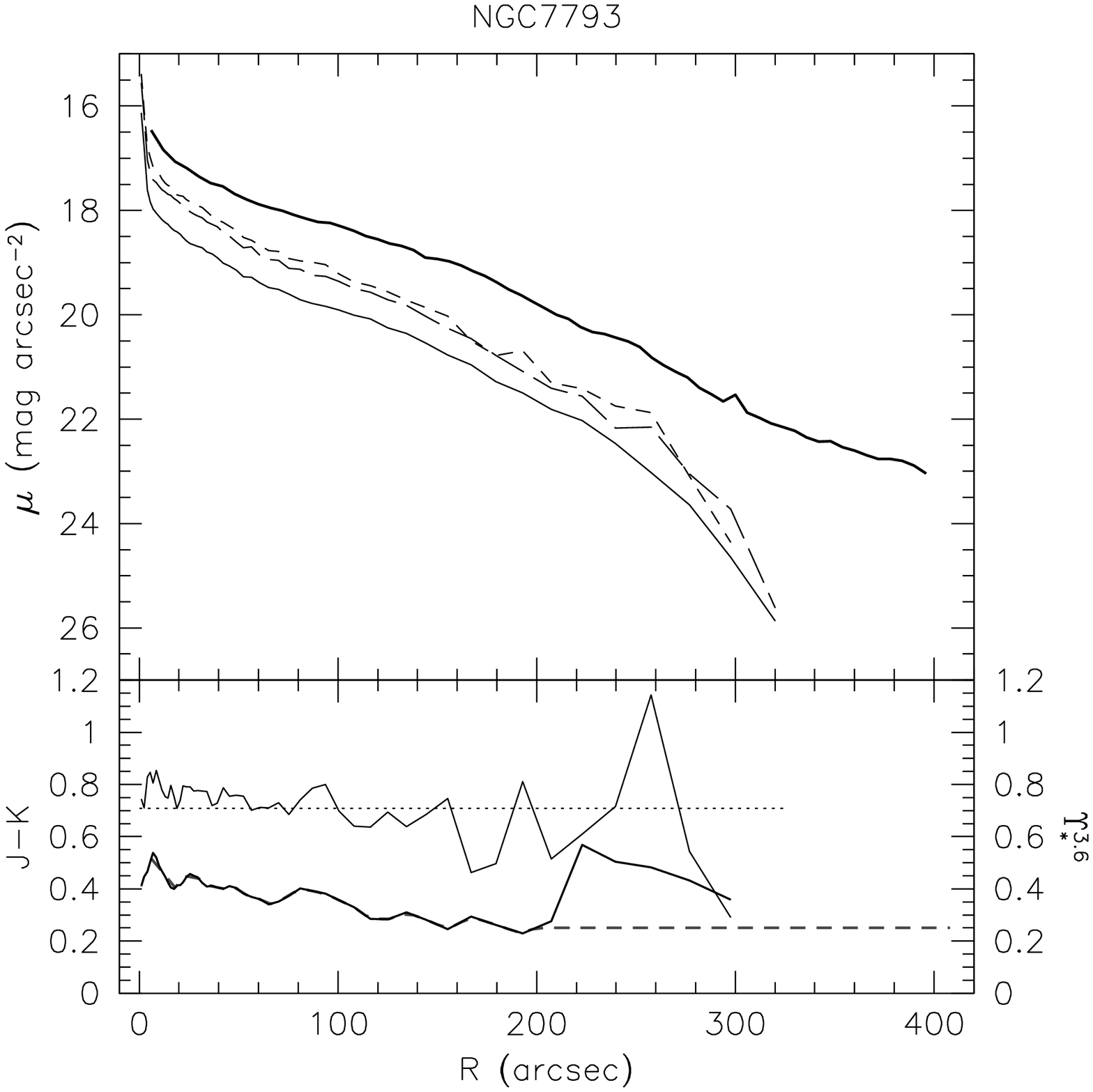} \figcaption{Surface
  brightness and \Ups\ profiles for NGC 7793. Lines and symbols as in
Fig.~\ref{fig:n2403_prof}.
\label{fig:n7793_prof}}
\end{figure*}

The surface brightness profiles of NGC 7793 are shown in
Fig.~\ref{fig:n7793_prof}. The 2MASS $J$, $H$ and $K$ profiles can be
traced out to $\sim 325''$; the 3.6 $\mu$m profile can be traced out
to $400''$ -- the entire extent of the \HI\ disk.  The 2MASS profiles
show a clear and sudden increase in surface brightness in the very
inner part. This is the signature of the nuclear star cluster in NGC
7793 \citep{boker02}. Its photometric and dynamical importance has
already become negligible at the radius of the innermost point of the
rotation curve.  The extent of the IRAC 3.6 $\mu$m profile equals that
of the \HI\ data, and no exponential extrapolations were necessary.
The $(J-K)$ profile shows a small but consistent color gradient,
translating in a gradient in \Ups\ from $\sim 0.5$ in the inner parts,
to $\sim 0.25$ at the outermost radius where the 2MASS colors were
deemed reliable. Beyond this radius we assume a constant $\Ups = 0.25$
as indicated in Fig.\ \ref{fig:n7793_prof}.

\begin{figure*}[t]
  \epsfxsize=0.95\hsize \epsfbox{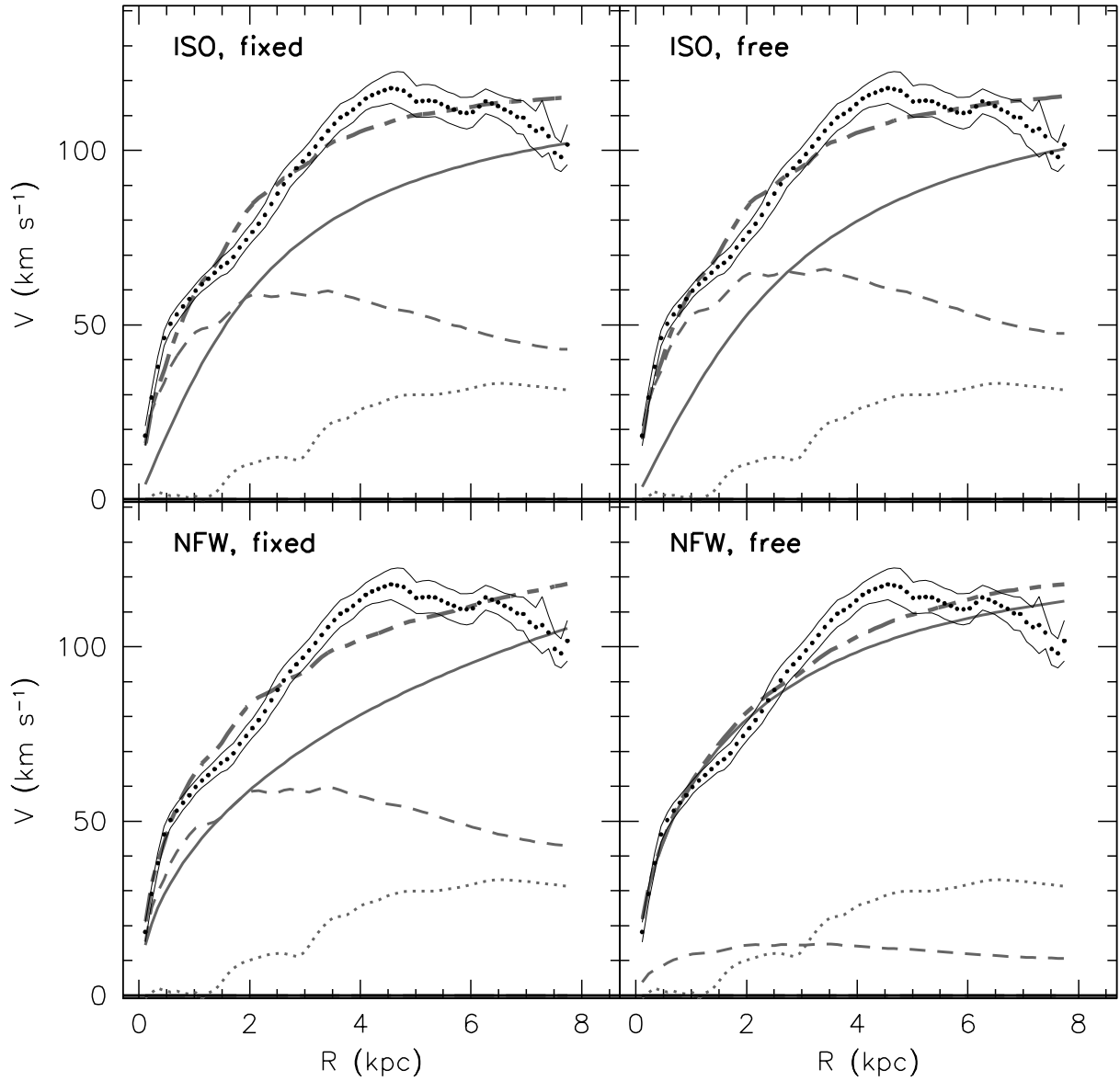} \figcaption{ISO
    and NFW rotation curve fits for NGC 7793 using the entire observed
    extent. Lines and symbols as in Fig.~\ref{fig:n2403_1curve}.
\label{fig:n7793_1curve}}
\end{figure*}

The rotation curve mass models are presented in
Fig.~\ref{fig:n7793_1curve}.  For the fixed \Ups\ model both ISO and
NFW models give roughly equal results: neither is fully able to
describe the observed rotation curve (independent of IMF).  The
``bump'' in the rotation curve within $R=2$ kpc, and its resemblance
with the shape of the stellar disk rotation curve suggests that the
inner part of NGC 7793 could be close to maximum disk. The ISO model
with \Ups\ as a free parameter does indeed prefer \Ups\ values that
are slightly higher than the fixed \Ups\ diet-Salpeter value. For the
NFW model, releasing \Ups\ has forced its value down to unrealistic
values in order to accommodate the more concentrated dark halo.

\begin{figure*}[t]
\epsfxsize=0.95\hsize \epsfbox{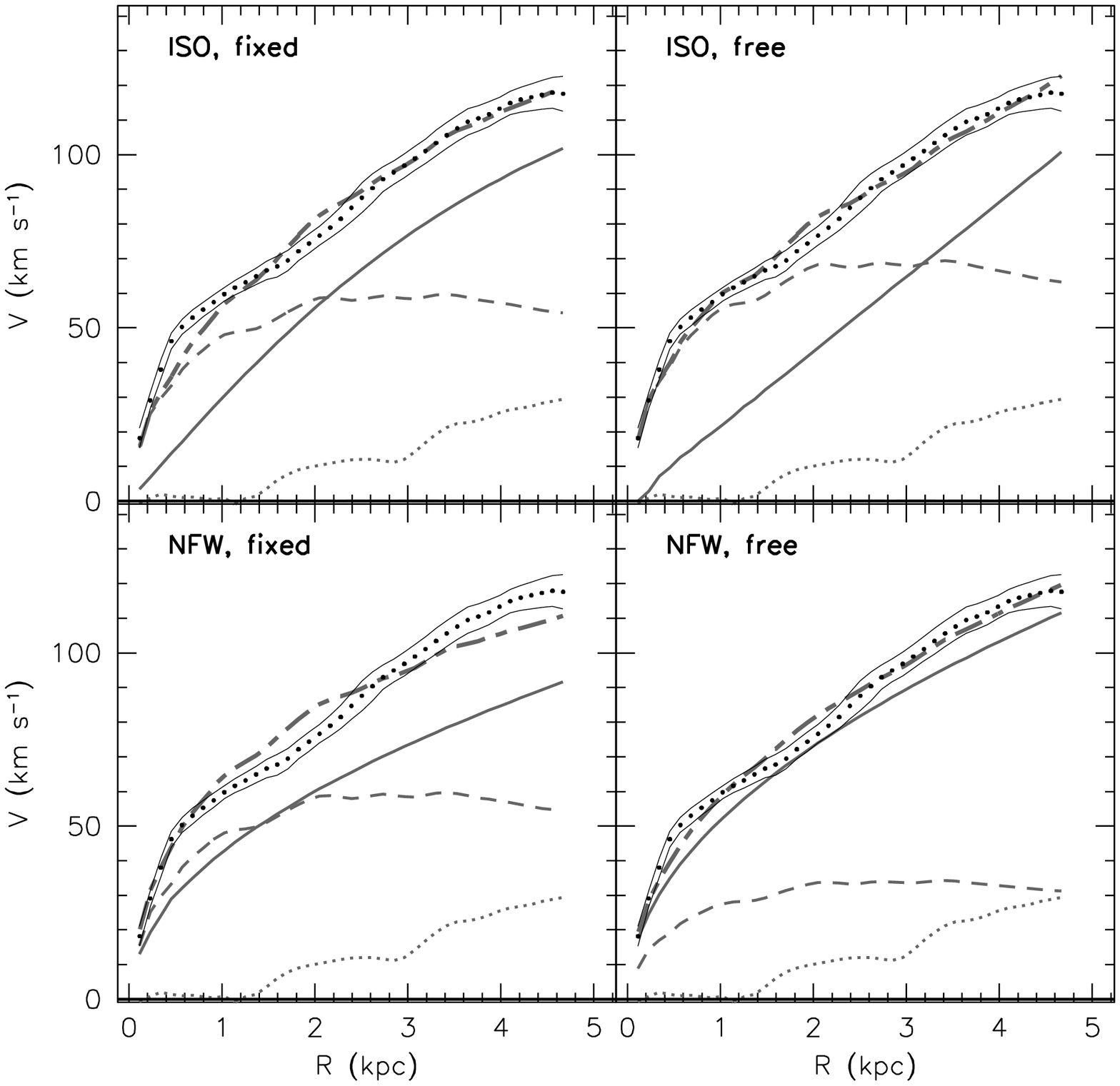} \figcaption{ISO and NFW
  rotation curve fits for NGC 7793 using only radii $R<4.7$ kpc.  Lines and
  symbols as in Fig.~\ref{fig:n2403_1curve}.
\label{fig:n7793_2curve}}
\end{figure*}

As remarked earlier (Sect.\ \ref{sect:7793}), the possible observed
decline in the outer rotation curve is uncertain.  We tested whether
this decline could cause the low quality of the fits by repeating the
analysis with radii $R>4.7$ kpc excluded. This analysis is shown in
Fig.~\ref{fig:n7793_2curve}.  For the fixed \Ups\ case the situation
does not change much. The biggest improvement is achieved for the ISO
model with a free \Ups. The preferred value is close to maximum disk,
with a significantly better overall fit.

\begin{deluxetable*}{lrcrcrrrrrrr}
\tablewidth{0pt}
\tablecaption{Mass Models with fixed \Ups and diet-Salpeter IMF}
\tabletypesize{\scriptsize}
\tablehead{
  & & & & & \multicolumn{3}{c}{ISO halo}&&\multicolumn{3}{c}{NFW halo}\\
\cline{6-8} \cline{10-12}\\
\colhead{Name} & \colhead{$\log M_{\star}^D$} & \colhead{$\langle \Upsilon_{\star,D}^{3.6} \rangle$} & \colhead{$\log M_{\star}^B$} & \colhead{$\langle \Upsilon_{\star,B}^{3.6} \rangle$} & \colhead{$R_C$} & \colhead{$\rho_0$} & \colhead{$\chi^2_r$} && \colhead{$c$} & \colhead{$V_{200}$} & \colhead{$\chi^2_r$}\\
\colhead{(1)} &\colhead{(2)} & \colhead{(3)} &\colhead{(4)} &\colhead{(5)} &\colhead{(6)} &\colhead{(7)} &\colhead{(8)} &&\colhead{(9)} &\colhead{(10)} &
\colhead{(11)}
}
\tablecolumns{12}
\startdata
NGC 925           &  10.01& 0.65 & \nodata & \nodata & 16.86 $\pm$ 7.47  &  3.4 $\pm$ \phn 0.5     & 2.15 && $<$0.1\ \phs\ \phn\phd\phn & $>$500\ \phs\ \phn\phn\phn\phd\phn & 2.81 \\
NGC 2366          &  8.41 & 0.33 & \nodata & \nodata & 1.36  $\pm$ 0.07  & 34.8 $\pm$ \phn 2.4     & 0.17 && $<$0.1\ \phs\ \phn\phd\phn & $>$500\ \phs\ \phn\phn\phn\phd\phn & 0.98 \\
NGC 2403 (1 comp) &  9.71 & 0.41 & \nodata & \nodata & 2.09  $\pm$ 0.05  & 81.2   $\pm$ \phn 3.6   & 0.88 && 9.9    $\pm$ 0.2    & 109.5  $\pm$ \phn\phn 1.0      & 0.55 \\
NGC 2403 (2 comp) &  9.67 & 0.39 & 8.63    & 0.60    & 2.14  $\pm$ 0.05  &  77.9  $\pm$ \phn 3.3   & 0.79 &&  9.8   $\pm$ 0.2    & 110.2  $\pm$ \phn\phn 1.0      & 0.56 \\
NGC 2841          &  11.04& 0.74 & 10.40   & 0.84    & 2.03  $\pm$ 0.05  & 298.7  $\pm$  14.9      & 0.27 && 16.1   $\pm$ 0.2    & 183.2  $\pm$ \phn\phn 1.2      & 0.42 \\
NGC 2903 (outer)  &  10.15& 0.61 & 9.33    & 1.30    &$<$0.01 $\pm$ 0.09 & $>$1000\ \phs\ \phn\phd\phn\phn & 0.63 && 30.9   $\pm$ 0.6    & 112.9  $\pm$ \phn\phn 0.6      & 0.36 \\
NGC 2976          &  9.25 & 0.55 & \nodata & \nodata &$>$1000\ \phs\ \phn\phd\phn\phn & 11.7   $\pm$ \phn 2.1   & 1.76 && $<$0.1\ \phs\ \phn\phd\phn & $>$500\ \phs\ \phn\phn\phn\phd\phn & 2.78 \\
NGC 3031          &  10.84& 0.80 & 10.11   & 1.00    & 5.25  $\pm$ 1.36  & 14.8   $\pm$ \phn 4.2   & 3.93 && 3.0    $\pm$ 2.9    &  190.9 $\pm$ 161.1    & 4.36 \\
NGC 3198 (1 comp) &  10.40& 0.80 & \nodata & \nodata & 3.22  $\pm$ 0.16  &  33.5  $\pm$ \phn 3.0   & 0.84 && 7.5    $\pm$ 0.4    & 112.4  $\pm$ \phn\phn 2.1      & 1.37 \\
NGC 3198 (2 comp) &  10.45& 0.80 & 9.46    & 0.73    & 4.97  $\pm$ 0.41  & 14.4   $\pm$ \phn 2.0   & 2.15 && 5.1    $\pm$ 0.5    & 122.7  $\pm$ \phn\phn 4.9      & 2.88 \\
IC 2574           &  9.02 & 0.44 & \nodata & \nodata & 7.23  $\pm$ 0.30  & 4.1    $\pm$ \phn 0.1   & 0.17 && $<$0.1\ \phs\ \phn\phd\phn & $>$500\ \phs\ \phn\phn\phn\phd\phn & 1.81 \\
NGC 3521          &  11.09& 0.73 & \nodata & \nodata & 39.4  $\pm$ 90.4  & 1.3    $\pm$ \phn 1.2   & 8.07 && $<$0.1\ \phs\ \phn\phd\phn& 403.2  $\pm$ 123.2    & 8.52 \\
NGC 3621          &  10.29& 0.59 & \nodata & \nodata &5.54   $\pm$ 0.16  & 14.4   $\pm$ \phn 0.6   & 0.62 && 3.7    $\pm$ 0.2    & 165.5  $\pm$ \phn\phn 5.9      & 0.81 \\
NGC 4736          &  10.27& 0.63 & 9.59\tablenotemark{a}& 0.33\tablenotemark{a}&1.44   $\pm$ 1.57  & 22.4   $\pm$ 41.5       & 1.52 && 11.4   $\pm$ 9.8    & 35.2   $\pm$ \phn\phn 0.3      & 1.51 \\
DDO 154           &  7.42 & 0.32 & \nodata & \nodata & 1.34  $\pm$ 0.06  & 27.6   $\pm$ \phn 1.6   & 0.44 &&4.4     $\pm$ 0.4    & 58.7   $\pm$ \phn\phn 4.3      & 0.82 \\
NGC 5055          &  11.09& 0.79 & 9.32\tablenotemark{b}& 0.11\tablenotemark{b}&45.63  $\pm$ 0.24  &0.9     $\pm$ \phn 0.2   & 8.13 && $<$0.1\ \phs\ \phn\phd\phn & 450.1   $\pm$ \phn 32.4      & 10.31\\
NGC 6946          &  10.77& 0.64 & 9.58    & 1.00    & 20.58 $\pm$ 3.77  & 5.4    $\pm$ \phn 0.4   & 1.45 && $<$0.1\ \phs\ \phn\phd\phn & $>$500\ \phs\ \phn\phn\phn\phd\phn & 2.59 \\
NGC 7331 (const)\tablenotemark{c}& 11.22& 0.70& 10.24   & 1.00    &$>$1000\ \phs\ \phn\phd\phn\phn            & 1.6   $\pm$  \phn 0.1   &2.93  && $<$0.1\ \phs\ \phn\phd\phn & $>$500\ \phs\ \phn\phn\phn\phd\phn& 4.08\\ 
NGC 7793          &  9.44 & 0.31 & \nodata & \nodata & 1.93  $\pm$ 0.15  & 77.2   $\pm$ \phn 7.7   & 2.97 && 5.8    $\pm$ 1.4    & 156.6  $\pm$ \phn 39.1     & 4.17 \\
NGC 7793 (rising) &  9.44 & 0.31 & \nodata & \nodata & 3.52  $\pm$ 0.51  & 50.9   $\pm$ \phn 4.8   & 2.26 && $<$0.1\ \phs\ \phn\phd\phn & $>$500\ \phs\ \phn\phn\phn\phd\phn & 3.67 
\enddata 
\tablecomments{(2): Logarithm of the predicted stellar mass
  of the disk ($M_{\odot}$). (3): Average stellar mass-to-light ratio
  in the 3.6 $\mu$m band $(M_{\odot}/L_{\odot})$. (4): Logarithm of
  the predicted stellar mass of the bulge ($M_{\odot}$). (5): Average
  stellar mass-to-light ratio of the bulge component in the 3.6 $\mu$m band
  $(M_{\odot}/L_{\odot})$. (6) Core radius $R_C$ and associated
  uncertainty (kpc). (7) Central density $\rho_0$ and associated
  uncertainty ($10^{-3} M_{\odot}$ pc$^{-3}$). (8) Reduced
  $\chi^2_r$. (9) NFW $c$ parameter and associated
  uncertainty. (10) NFW parameter $V_{200}$ and associated
  uncertainty (km s$^{-1}$).  (11) Reduced $\chi^2_r$.}  \tablenotetext{a}{$M_{\rm
    bulge}$ is a free parameter. Predicted model values are $\Ups =
  0.9$ and $\log M_{\star}^B = 10.02$.}  \tablenotetext{b}{$M_{\rm bulge}$ is a
  free parameter. Predicted model values are $\Ups = 1.3$ and $\log
  M_{\star}^B = 10.40$.} \tablenotetext{c}{The model with the color gradient
  included severely over-predicts the data and is not listed here. Its
  parameters are $\Ups =0.96$ and $\log M_{\star}^D = 11.36$ for the
  main disk and $\Ups =1.08$ and $\log M_{\star}^B = 10.28$ for the
  inner disk.}
\label{table2}

\end{deluxetable*}

\begin{deluxetable*}{lrcrcrrrrrrr}
\tablewidth{0pt}
\tablecaption{Mass Models with fixed \Ups and Kroupa IMF}
\tabletypesize{\scriptsize}
\tablehead{
  & & & & & \multicolumn{3}{c}{ISO halo}&&\multicolumn{3}{c}{NFW halo}\\
\cline{6-8} \cline{10-12}\\
\colhead{Name} & \colhead{$\log M_{\star}^D$} & \colhead{$\langle \Upsilon_{\star,D}^{3.6} \rangle$} & \colhead{$\log M_{\star}^B$} & \colhead{$\langle \Upsilon_{\star,B}^{3.6} \rangle$} & \colhead{$R_C$} & \colhead{$\rho_0$} & \colhead{$\chi^2_r$} && \colhead{$c$} & \colhead{$V_{200}$} & \colhead{$\chi^2_r$}\\
\colhead{(1)} &\colhead{(2)} & \colhead{(3)} &\colhead{(4)} &\colhead{(5)} &\colhead{(6)} &\colhead{(7)} &\colhead{(8)} &&\colhead{(9)} &\colhead{(10)} &
\colhead{(11)}
}
\tablecolumns{12}
\startdata
NGC 925           &  9.86& 0.47  & \nodata & \nodata & 9.67 $\pm$ 1.27   &  5.9 $\pm$ \phn\phn 0.5     & 1.14 && $<$0.1\ \phs\ \phn\phd\phn\phn & $>$500\ \phs\ \phn\phn\phn\phd\phn & 2.17 \\
NGC 2366          &  8.26 & 0.23 & \nodata & \nodata & 1.32  $\pm$ 0.07  & 37.3 $\pm$ \phn\phn 2.4     & 0.16 && $<$0.1\ \phs\ \phn\phd\phn\phn & $>$500\ \phs\ \phn\phn\phn\phd\phn & 1.01 \\
NGC 2403 (1 comp) &  9.56 & 0.29 & \nodata & \nodata & 1.49  $\pm$ 0.05  & 152.8  $\pm$ \phn\phn 7.5   & 1.04 && 12.4    $\pm$ \phn 0.2    & 101.7  $\pm$ \phn 0.7      & 0.57 \\
NGC 2403 (2 comp) &  9.52 & 0.26 & 8.48    & 0.43    & 1.52  $\pm$ 0.04  & 145.8  $\pm$ \phn\phn 6.9   & 0.97 && 12.3   $\pm$ \phn 0.2    & 102.2  $\pm$ \phn 0.7      & 0.57 \\
NGC 2841          &  10.88& 0.53 & 10.25   & 0.60    & 0.63  $\pm$ 0.04  & 3215.3 $\pm$  371.8      & 0.22 && 18.9   $\pm$ \phn 0.4    & 181.4  $\pm$ \phn1.0      & 0.23 \\
NGC 2903 (outer)  &  10.00& 0.43 & 9.18    & 0.92    &$<$0.01 $\pm$ 0.09 & $>$1000\ \phs\ \phn\phd\phn\phn\phn & 1.14 && 35.5  $\pm$ \phn 0.7    & 111.8  $\pm$ \phn 0.6      & 0.41 \\
NGC 2976          &  9.10 & 0.39 & \nodata & \nodata & 5.09 $\pm$ 2.54   & 35.5   $\pm$ \phn\phn 3.1   & 0.50 && $<$0.1\ \phs\ \phn\phd\phn & $>$500\ \phs\ \phn\phn\phn\phd\phn & 1.90 \\
NGC 3031          &  10.69& 0.57 & 9.96   & 0.71    & 0.78  $\pm$ 0.19  & 754.2   $\pm$ 323.6   & 3.88 && 26.4    $\pm$ \phn 2.5    &  94.6 $\pm$ \phn 3.9    & 3.61 \\
NGC 3198 (1 comp) &  10.25& 0.57 & \nodata & \nodata & 2.72 $\pm$ 0.13  &  46.9  $\pm$ \phn\phn 4.0   & 0.80 && 8.7    $\pm$ \phn 0.4    & 109.7  $\pm$ \phn 1.7      & 1.30 \\
NGC 3198 (2 comp) &  10.30& 0.57 & 9.31    & 0.52    & 2.82  $\pm$ 0.19  & 44.0   $\pm$ \phn\phn 5.1   & 1.41 && 8.5    $\pm$ \phn 0.5    & 110.4  $\pm$ \phn 2.2      & 2.06 \\
IC 2574           &  8.87 & 0.31 & \nodata & \nodata & 6.18  $\pm$ 0.21  & 5.0    $\pm$ \phn\phn 0.1   & 0.17 && $<$0.1\ \phs\ \phn\phn\phd\phn & $>$500\ \phs\ \phn\phn\phn\phd\phn & 1.73 \\
NGC 3521          &  10.94& 0.52 & \nodata & \nodata & 2.50  $\pm$ 0.66  & 73.0    $\pm$ \phn 30.6   & 4.75 && 8.9 $\pm$ \phn 2.0 & 128.4  $\pm$ 16.4   & 5.55 \\
NGC 3621          &  10.14& 0.42 & \nodata & \nodata & 2.77   $\pm$ 0.10  & 48.9   $\pm$ \phn 2.8   & 1.09 && 7.8    $\pm$ \phn 0.2    & 120.2  $\pm$ \phn 1.4      & 0.55 \\
NGC 4736          &  10.12& 0.44 & 9.56\tablenotemark{a}& 0.31\tablenotemark{a}&$<$0.01   $\pm$ 0.65  & $>$1000\ \phs\ \phn\phd\phn\phn\phn  & 1.73 && 63.5   $\pm$ 24.2    & 42.4   $\pm$ \phn 1.7      & 1.41 \\
DDO 154           &  7.27 & 0.23 & \nodata & \nodata & 1.32  $\pm$ 0.06  & 28.5   $\pm$ \phn \phn1.7   & 0.44 &&4.5     $\pm$ \phn 0.4    & 58.0   $\pm$ \phn 4.1      & 0.83 \\
NGC 5055          &  10.94& 0.56 & 9.81\tablenotemark{b}& 0.34\tablenotemark{b}&11.73 $\pm$ 0.71  &4.8     $\pm$ \phn\phn 0.4   & 1.03 && 2.1 $\pm$ \phn 0.4  & 217.8   $\pm$  21.2      & 1.45\\
NGC 6946          &  10.62& 0.45 & 9.43    & 0.71    & 3.62 $\pm$ 0.16  & 45.7    $\pm$ \phn\phn 3.0   & 0.98 && 6.2 $\pm$ \phn 0.5 & 183.8 $\pm$ 11.1 & 1.03 \\
NGC 7331 (const)\tablenotemark{c}& 11.07& 0.50& 10.09   & 0.71    &5.40 $\pm$ 0.31   & 24.4   $\pm$  \phn\phn 2.1   &0.31  && 4.9 $\pm$ \phn 0.4 & 200.0 $\pm$ 10.7 & 0.24\\ 
NGC 7793          &  9.29 & 0.22 & \nodata & \nodata & 1.46  $\pm$ 0.10  & 126.0   $\pm$ \phn 12.2   & 3.56 && 9.1    $\pm$ \phn 1.1    & 114.1  $\pm$ 12.2     & 3.74 \\
NGC 7793 (rising) &  9.29 & 0.22 & \nodata & \nodata & 1.98  $\pm$ 0.22  & 95.7   $\pm$ \phn  10.7   & 4.08 && $<$0.1\ \phs\ \phn\phn\phd\phn & $>$500\ \phs\ \phn\phn\phn\phd\phn & 2.06 
\enddata 
\tablecomments{(2): Logarithm of the predicted stellar mass
  of the disk ($M_{\odot}$). (3): Average stellar mass-to-light ratio
  in the 3.6 $\mu$m band $(M_{\odot}/L_{\odot})$. (4): Logarithm of
  the predicted stellar mass of the bulge ($M_{\odot}$). (5): Average
  stellar mass-to-light ratio of the bulge component in the 3.6 $\mu$m band
  $(M_{\odot}/L_{\odot})$. (6) Core radius $R_C$ and associated
  uncertainty (kpc). (7) Central density $\rho_0$ and associated
  uncertainty ($10^{-3} M_{\odot}$ pc$^{-3}$). (8) Reduced
  $\chi^2_r$. (9) NFW $c$ parameter and associated
  uncertainty. (10) NFW parameter $V_{200}$ and associated
  uncertainty (km s$^{-1}$).  (11) Reduced $\chi^2_r$.}  \tablenotetext{a}{$M_{\rm
    bulge}$ is a free parameter. Predicted model values are $\Ups =
  0.6$ and $\log M_{\star}^B = 9.87$.}  \tablenotetext{b}{$M_{\rm bulge}$ is a
  free parameter. Predicted model values are $\Ups = 0.8$ and $\log
  M_{\star}^B = 10.25$.} \tablenotetext{c}{The model with the color gradient
  included severely over-predicts the data and is not listed here. Its
  parameters are $\Ups =0.61$ and $\log M_{\star}^D = 11.21$ for the
  main disk and $\Ups =0.69$ and $\log M_{\star}^B = 10.13$ for the
  inner disk.}
\label{table2kroupa}

\end{deluxetable*}

\begin{deluxetable*}{lrcrrrllrr}
\tabletypesize{\scriptsize}
\tablewidth{0pt}
\tablecaption{ISO Mass Models with free \Ups}
\tablehead{
\colhead{Name} & \colhead{$M_{\star}^D$} & \colhead{$\langle \Upsilon_{\star,D}^{3.6} \rangle$} & \colhead{$f^D$} & \colhead{$M_{\star}^B$}& \colhead{$\langle \Upsilon_{\star,B}^{3.6} \rangle$} &\colhead{$f^B$} & \colhead{$R_C$} & \colhead{$\rho_0$}& \colhead{$\chi^2_r$} \\
\colhead{(1)} &\colhead{(2)} & \colhead{(3)} &\colhead{(4)} &\colhead{(5)} &\colhead{(6)} &\colhead{(7)} &\colhead{(8)} &\colhead{(9)} &\colhead{(10)} 
}
\startdata
NGC 925           & 9.46 & 0.18  &0.28 $\pm$ 0.07 & \nodata & \nodata & \nodata & 5.65 $\pm$ 0.56    & 12.8  $\pm$ \phn\phn 1.8     & 0.68 \\
NGC 2366          &\nodata& 0.00\tablenotemark{a} &0.00\tablenotemark{a}& \nodata & \nodata & \nodata & 1.16 $\pm$ 0.06    & 50.8  $\pm$ \phn\phn 3.2     & 0.18 \\
NGC 2403 (1 comp) & 9.91 & 0.65  &1.58 $\pm$ 0.04 & \nodata & \nodata & \nodata & 3.76 $\pm$ 0.15    & 28.6  $\pm$ \phn\phn1.9     & 0.56 \\
NGC 2403 (2 comp) & 9.66 & 0.38  &0.97 $\pm$ 0.18 & 8.93    & 1.18    &1.98     $\pm$ 0.11     & 2.51 $\pm$ 0.32    & 59.1  $\pm$ \phn 14.3    & 0.49 \\
NGC 2841          & 10.96& 0.61  &0.83 $\pm$ 0.05 & 10.40   & 0.83    &0.99     $\pm$ 0.36     & 1.36 $\pm$ 0.75    & 674.8 $\pm$ 736.4   & 0.18 \\
NGC 2903 (outer)  & 10.50& 1.36  &2.23 $\pm$ 0.28 & \nodata & 0.00\tablenotemark{a}&0.00\tablenotemark{a}& 1.01 $\pm$ 0.45    & 541.1 $\pm$ 480.5   & 0.36 \\
NGC 2976          & 9.12 & 0.40  &0.73 $\pm$ 0.03 & \nodata & \nodata & \nodata &$\infty$& 30.7  $\pm$ \phn\phn 2.3     & 0.51 \\
NGC 3031          & 10.93 & 0.99 &1.24 $\pm$ 0.54 & 9.54    & 0.22    & 0.27    $\pm$ 0.56     & 4.13 $\pm$ 23.4    & 12.2  $\pm$ 124.6   & 3.26 \\
NGC 3198 (1 comp) & 10.34 & 0.70 &0.88 $\pm$ 0.07 & \nodata & \nodata &  \nodata & 2.71 $\pm$ 0.33    & 47.5  $\pm$ \phn 11.4    & 0.81 \\
NGC 3198 (2 comp) & 10.30 & 0.57 &0.71 $\pm$ 0.03 & 8.36    & 0.06    &0.08     $\pm$ 0.04     & 1.86 $\pm$ 0.12    & 97.1  $\pm$ \phn 12.2    & 0.36 \\
IC 2574           & 8.37  & 0.10 &0.23 $\pm$ 0.18 & \nodata & \nodata &\nodata & 4.99 $\pm$ 0.34    & 6.7   $\pm$ \phn\phn 0.7     & 0.16 \\
NGC 3521          & 10.77 & 0.34 &0.47 $\pm$ 0.13 & \nodata & \nodata & \nodata & 1.32 $\pm$ 0.76    & 370.2 $\pm$ 451.1   & 4.04 \\
NGC 3621          & 10.30 & 0.60 &1.02 $\pm$ 0.02 & \nodata & \nodata &  \nodata & 5.88 $\pm$ 0.32    & 13.0  $\pm$ \phn\phn 1.1     & 0.61 \\
NGC 4736          & 10.23 & 0.58 &0.92 $\pm$ 0.06 & 9.56    & 0.32    &0.35     $\pm$ 0.09     & 0.04 $\pm$ 1.23    & $>$500\ \phs\ \phn\phn\phn\phd\phn & 1.53 \\
DDO 154           & 8.45 & 3.46 &10.82 $\pm$ 1.05 & \nodata & \nodata &\nodata & 2.69 $\pm$ 0.24    & 9.0   $\pm$ \phn\phn 1.1     & 0.28 \\
NGC 5055          & 10.88 & 0.24 &0.62 $\pm$ 0.02 & 9.69    & 0.40    &0.31     $\pm$ 0.01     & 7.15 $\pm$ 1.12    & 11.1  $\pm$ \phn\phn 3.1     & 0.87 \\
NGC 6946          & 10.58 & 0.41 &0.64 $\pm$ 0.15 & 9.50    & 0.83    & 0.83    $\pm$ 0.06     & 3.32 $\pm$ 1.23    & 55.6  $\pm$ \phn 42.3    & 0.96 \\
NGC 7331 (grad)   & 10.88 & 0.39 &0.33 $\pm$ 0.04 & 10.22   & 0.95    &0.88     $\pm$ 0.08     & 3.88 $\pm$ 0.98    & 58.0  $\pm$ \phn 27.8    & 0.26 \\
NGC 7331 (const)  & 11.10 & 0.32 &0.76 $\pm$ 0.03 & 10.14   & 0.79    &0.79     $\pm$ 0.09     & 9.35 $\pm$ 1.43    & 10.1  $\pm$ \phn\phn 2.4     & 0.26 \\
NGC 7793          & 9.53  & 0.38 &1.22 $\pm$ 0.14 & \nodata & \nodata &  \nodata & 2.40 $\pm$ 0.48    & 53.3  $\pm$ \phn 15.7    & 2.83 \\
NGC 7793 (rising) & 9.57  & 0.42 &1.35 $\pm$ 0.04 & \nodata & \nodata & \nodata  &$\infty$ & 25.7  $\pm$ \phn\phn 1.0     & 1.01
\enddata
\tablecomments{(2): Logarithm of the predicted stellar mass of the
  disk ($M_{\odot}$). (3): Average stellar mass-to-light ratio in the
  3.6 $\mu$m band $(M_{\odot}/L_{\odot})$. (4): Scaling factor $f^D$
  with respect to the predicted diet-Salpeter \Ups\ value listed in
  Table 3. The factor with respect to the Kroupa \Ups\ value listed in
  Table 4 can be obtained by multiplying with a factor of 1.41. (5):
  Logarithm of the predicted stellar mass of the bulge
  ($M_{\odot}$). (6): Average stellar mass-to-light ratio in the 3.6
  $\mu$m band $(M_{\odot}/L_{\odot})$. (7): Scaling factor $f^B$ with
  respect to the predicted diet-Salpeter \Ups\ value listed in Table
  3.  The factor with respect to the Kroupa \Ups\ value listed in
  Table 4 can be obtained by multiplying with a factor of 1.41. (8)
  Core radius $R_C$ and associated uncertainty (kpc). (9) Central
  density $\rho_0$ and associated uncertainty ($10^{-3} M_{\odot}$
  pc$^{-3}$). (10) Reduced $\chi^2_r$.}  \tablenotetext{a}{The stellar
  mass-to-light ratio \Ups\ was fixed to $\Ups = 0$.}
\label{table3a}
\end{deluxetable*}

\begin{deluxetable*}{lrllrrlrrr}
\tabletypesize{\scriptsize}
\tablewidth{0pt}
\tablecaption{NFW Mass Models with free \Ups}
\tablehead{
\colhead{Name} &\colhead{$M_{\star}^D$} & \colhead{$\langle \Upsilon_{\star,D}^{3.6} \rangle$} & \colhead{$f^D$} &\colhead{$M_{\star}^B$} & \colhead{$\langle \Upsilon_{\star,B}^{3.6} \rangle$} & \colhead{$f^B$} &\colhead{$c$} & \colhead{$V_{200}$} & \colhead{$\chi^2_r$}\\
\colhead{(1)} &\colhead{(2)} & \colhead{(3)} &\colhead{(4)} &\colhead{(5)} &\colhead{(6)} &\colhead{(7)} &\colhead{(8)} &\colhead{(9)} &\colhead{(10)} 
}
\startdata
NGC 925           &\nodata& 0.00\tablenotemark{a}&0.00\tablenotemark{a}& \nodata & \nodata & \nodata & $<$0.1\ \phs\ \phn\phn\phd\phn & $>$500\ \phs\ \phn\phn\phn\phd\phn & 1.11 \\
NGC 2366          &\nodata& 0.00\tablenotemark{a}&0.00\tablenotemark{a}& \nodata & \nodata & \nodata & $<$0.1\ \phs\ \phn\phn\phd\phn & $>$500\ \phs\ \phn\phn\phn\phd\phn & 1.44 \\
NGC 2403 (1 comp) & 9.69  & 0.39  &0.96 $\pm$ 0.05  & \nodata & \nodata &\nodata & 10.2 $\pm$ \phn 0.5   & 108.3 $\pm$ \phn\phn 1.9   & 0.55 \\
NGC 2403 (2 comp) & 9.59  & 0.33  &0.83 $\pm$ 0.08  & 8.62   & 0.60    &0.99     $\pm$ 0.09   & 10.9 $\pm$ \phn 0.6   & 106.1 $\pm$ \phn\phn 1.9   & 0.55 \\
NGC 2841          & 11.02 & 0.89  &0.96 $\pm$ 0.04  & 10.58  & 1.26    &1.50     $\pm$ 0.05   & 13.7 $\pm$ \phn 0.8   & 190.9 $\pm$ \phn\phn 2.2   & 0.19 \\
NGC 2903 (outer)  & 10.45 & 1.23  &2.01 $\pm$ 0.23  & 9.46   & 1.78    &1.37     $\pm$ 0.85   & 18.7 $\pm$ \phn 1.4   & 119.3 $\pm$ \phn\phn 1.4   & 0.25 \\
NGC 2976          & 8.88  & 0.23  &0.42 $\pm$ 0.11  & \nodata & \nodata &\nodata  & $<$0.1\ \phs\ \phn\phn\phd\phn &$>$500\ \phs\ \phn\phn\phn\phd\phn  & 1.65 \\
NGC 3031          & 10.92 & 0.96  &1.20 $\pm$ 0.65  & 9.47   & 0.23    &0.23     $\pm$ 0.27   & 9.3  $\pm$ 60.4  & 77.1  $\pm$ \phn 96.5  & 3.31 \\
NGC 3198 (1 comp) & 10.27 & 0.60  &0.75 $\pm$ 0.14  & \nodata & \nodata &\nodata & 10.1 $\pm$ \phn 1.8   & 107.6 $\pm$ \phn\phn 2.8   & 1.30 \\
NGC 3198 (2 comp) & 10.29 & 0.55  &0.69 $\pm$ 0.03  & \nodata & 0.00\tablenotemark{a}&0.00\tablenotemark{a}& 11.2 $\pm$ 0.43  & 104.0 $\pm$ \phn\phn 0.7   & 0.37 \\
IC 2574           & \nodata&0.00\tablenotemark{a}&0.00\tablenotemark{a}& \nodata & \nodata &\nodata & $<$0.1\ \phs\ \phn\phn\phd\phn & $>$500\ \phs\ \phn\phn\phn\phd\phn& 1.66 \\
NGC 3521          & 10.87 & 0.43  &0.59 $\pm$ 0.21  & \nodata & \nodata &\nodata  & 14.0 $\pm$ 12.6  & 122.5 $\pm$ \phn 20.4  & 5.48 \\
NGC 3621          & 10.18 & 0.47  &0.80 $\pm$ 0.03  & \nodata & \nodata &\nodata  & 6.5  $\pm$ \phn 0.4   & 128.0 $\pm$ \phn\phn 3.3  & 0.52 \\
NGC 4736          & 10.09 & 0.42  &0.67 $\pm$ 0.18  & 9.51   & 0.15    &0.28     $\pm$ 0.08   & 72.3 $\pm$ 54.9  & 43.4  $\pm$ \phn\phn 4.6   & 1.44 \\
DDO 154           & 8.00  & 1.21  &3.79 $\pm$ 2.84   & \nodata & \nodata & \nodata & 3.4  $\pm$ \phn 1.3   & 68.0 $\pm$   \phn 16.8 & 0.81 \\
NGC 5055          &\nodata& 0.00\tablenotemark{a}&0.00\tablenotemark{a}& 9.60   & 0.21    &0.16     $\pm$ 0.01   & 33.3 $\pm$ \phn 0.6   & 121.6 $\pm$ \phn\phn 0.5   & 0.57 \\
NGC 6946          & 10.68 & 0.52  &0.82 $\pm$ 0.06  & 9.35   & 0.59    &0.59     $\pm$ 0.05   & 3.2  $\pm$ \phn 2.6   & 281.4 $\pm$ 183.2 & 1.01 \\
NGC 7331 (grad)   & 10.90 & 0.34  &0.35 $\pm$ 0.02  & 10.13  & 0.75    &0.69     $\pm$ 0.07   & 9.3  $\pm$ \phn 1.2   & 171.2 $\pm$ \phn\phn 8.1   & 0.27 \\
NGC 7331 (const)  & 11.08 & 0.64  &0.64 $\pm$ 0.03  & 10.10  & 0.90    &0.90     $\pm$ 0.09   & 3.3  $\pm$ \phn 1.1   & 257.3 $\pm$ \phn 56.9  & 0.24 \\
NGC 7793          & 8.22  & 0.02  &0.06 $\pm$ 0.96  & \nodata & \nodata & \nodata & 15.3 $\pm$ \phn 9.6  & 89.5  $\pm$ \phn 20.7  & 3.53 \\
NGC 7793 (rising) & 8.96  & 0.10  &0.33 $\pm$ 0.34  & \nodata & \nodata & \nodata & $<$0.1\ \phs\ \phn\phn\phd\phn & $>$500\ \phs\ \phn\phn\phn\phd\phn &  1.30 
\enddata
\tablecomments{(2): Logarithm of the predicted stellar mass of the
  disk ($M_{\odot}$). (3): Average stellar mass-to-light ratio in the
  3.6 $\mu$m band $(M_{\odot}/L_{\odot})$. (4): Scaling factor $f^D$
  with respect to the predicted \Ups\ value listed in Table 3.  The
  factor with respect to the Kroupa \Ups\ value listed in Table 4 can
  be obtained by multiplying with a factor of 1.41.  (5): Logarithm of
  the predicted stellar mass of the bulge ($M_{\odot}$). (6): Average
  stellar mass-to-light ratio in the 3.6 $\mu$m band
  $(M_{\odot}/L_{\odot})$. (7): Scaling factor $f^B$ with respect to
  the predicted \Ups\ value listed in Table 3.  The factor with
  respect to the Kroupa \Ups\ value listed in Table 4 can be obtained
  by multiplying with a factor of 1.41.  (8) NFW parameter $c$ and
  associated uncertainty (kpc). (9) NFW parameter $V_{200}$ and
  associated uncertainty (km s$^{-1}$). (10) Reduced $\chi^2_r$.}
\tablenotetext{a}{The stellar mass-to-light ratio \Ups\ was fixed to
  $\Ups = 0$.}
\label{table3}
\end{deluxetable*}

\section{Discussion}

We now put the results derived for individual galaxies in a broader
context by looking at the sample as a whole.  Sect.~7.1 deals with
some of the general properties of the rotation curves, and looks at
correlations with luminosity, as well as the mass fractions within the
galaxies. Sect.~7.2 deals with the values for \Ups\ and compares them
with other measures for the stellar mass-to-light ratio.  Sections 7.3
and 7.4 deal with the halo properties. In Sect.~7.3 we discuss the
halo rotation curves, and in Sect.~7.4 the parameters of the halo
models are compared with predictions based on cosmological models.

\subsection{General properties\label{sec:genprop}}

\begin{figure*}[t]
  \epsfxsize=0.45\hsize \epsfbox{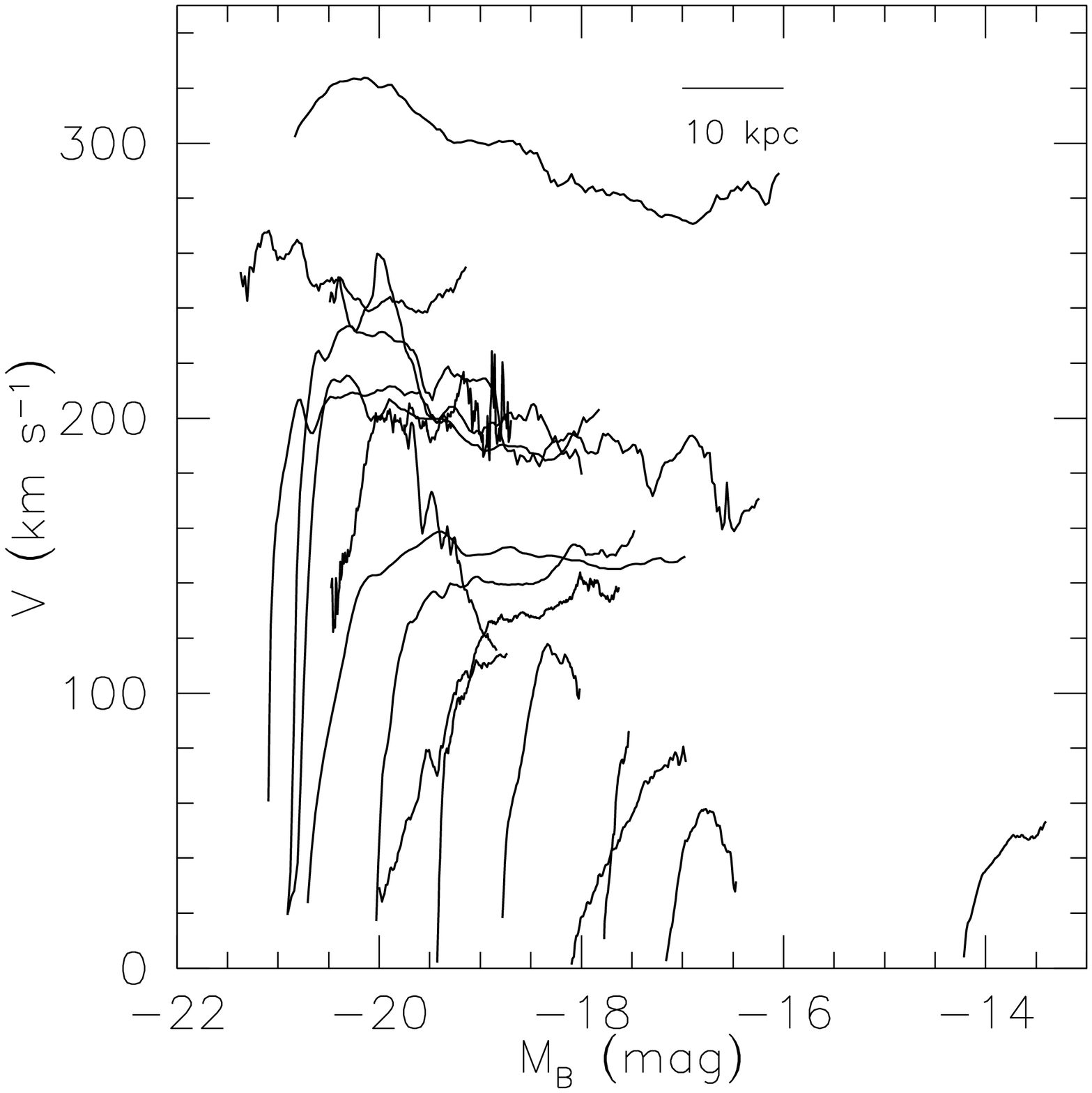} \hfil
  \epsfxsize=0.45\hsize \epsfbox{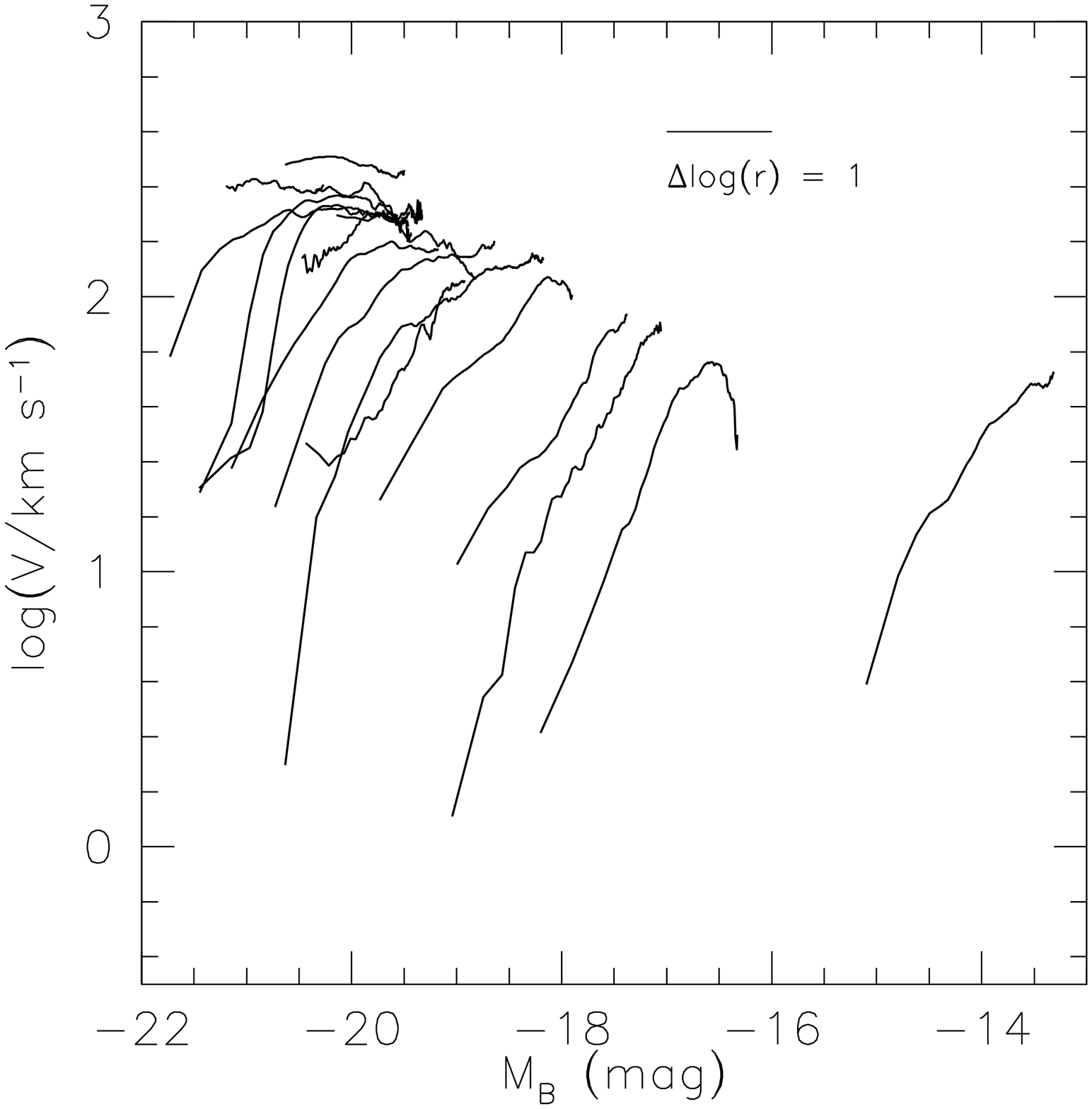} \hfil
  \figcaption{All THINGS rotation curves plotted in linear units in
    the left panel and in logarithmic units in the right panel. The
    origin of the rotation curves has been shifted according to their
    absolute luminosity as indicated on the horizontal axis. The bar
    in the respective panels indicates the radial scale.
\label{fig:allcurves}}
\end{figure*}

THINGS contains galaxies spanning a large range of luminosities and
Hubble types. Consequently, there is a large range in shape, amplitude
and extent of the rotation curves within our sample.  To illustrate 
this diversity, we plot in Fig.~\ref{fig:allcurves} all derived
rotation curves on the same physical scale, with the position of the
origin of the rotation curve offset using the absolute magnitude $M_B$
of the galaxy.  The well-known change in properties of the rotation
curves with luminosity (see, e.g.,\ \citealt{broeils_phd}) is readily
apparent. Luminous galaxies have rotation curves that rise steeply,
followed by a decline and an asymptotic approach to the flat outer
part of the curve; low-luminosity galaxies show a more gradual increase, never
quite reaching the flat part of the curve over the extent of their
\HI\ disks.

This is shown more clearly in the right hand panel in
Fig.~\ref{fig:allcurves} where we plot the logarithmic rotation
curves, $\log(V)$ \emph{vs} $\log(R)$.  Once again the starting point
of the rotation curve is determined by the absolute luminosity. The
change in slope as a function of luminosity can readily be made out.

Another well-known relation for disk galaxies, is that low-luminosity galaxies
(late-type disk and dwarf galaxies) are more dominated by dark matter
than luminous galaxies (early-type disk galaxies)
\citep[e.g.,][]{robertshaynes}. The flat rotation curves of spiral
galaxies are additionally an indication that dark matter should become
more and more dominant towards larger radii. We can explicitly evaluate
these relations by deriving the ratio of the mass of the baryons and
the mass of the dark matter halo at each radius as
\begin{equation}
  \left({{M_{\rm baryons}}\over{M_{\rm tot}}}\right) = {{V_{\rm gas}^2+\Upsilon_{\star}{V_{\star}^2}}\over{V_{\rm obs}^2}},
\end{equation}
where $M_{\rm baryons}/M_{\rm tot}$ is an indication for the degree
of dark matter dominance, $V_{\rm obs}$ is the observed rotation
curve, $V_{\rm gas}$ the rotation curve of the gas component as
described in Sect.\ 5.1, and $\sqrt{\Upsilon_{\star}}V_{\star}$ is the rotation velocity of
the stellar component (taken to be the quadratic sum of multiple
stellar components if present).

Note that for some of the galaxies discussed in Sect.~6, we derived
multiple mass models with, e.g., single or multiple stellar
components.  In the rest of the paper, we will, so as not to
needlessly confuse the discussion, restrict ourselves to one model per
galaxy.  For NGC 2403 and NGC 3198 we use the two-component model, for
NGC 7793 the model that uses the entire rotation curve, and for NGC
7331 the model without the color gradient. These particular choices do
not affect any of our conclusions.

\begin{figure*}[t]
  \epsfxsize=0.9\hsize \epsfbox{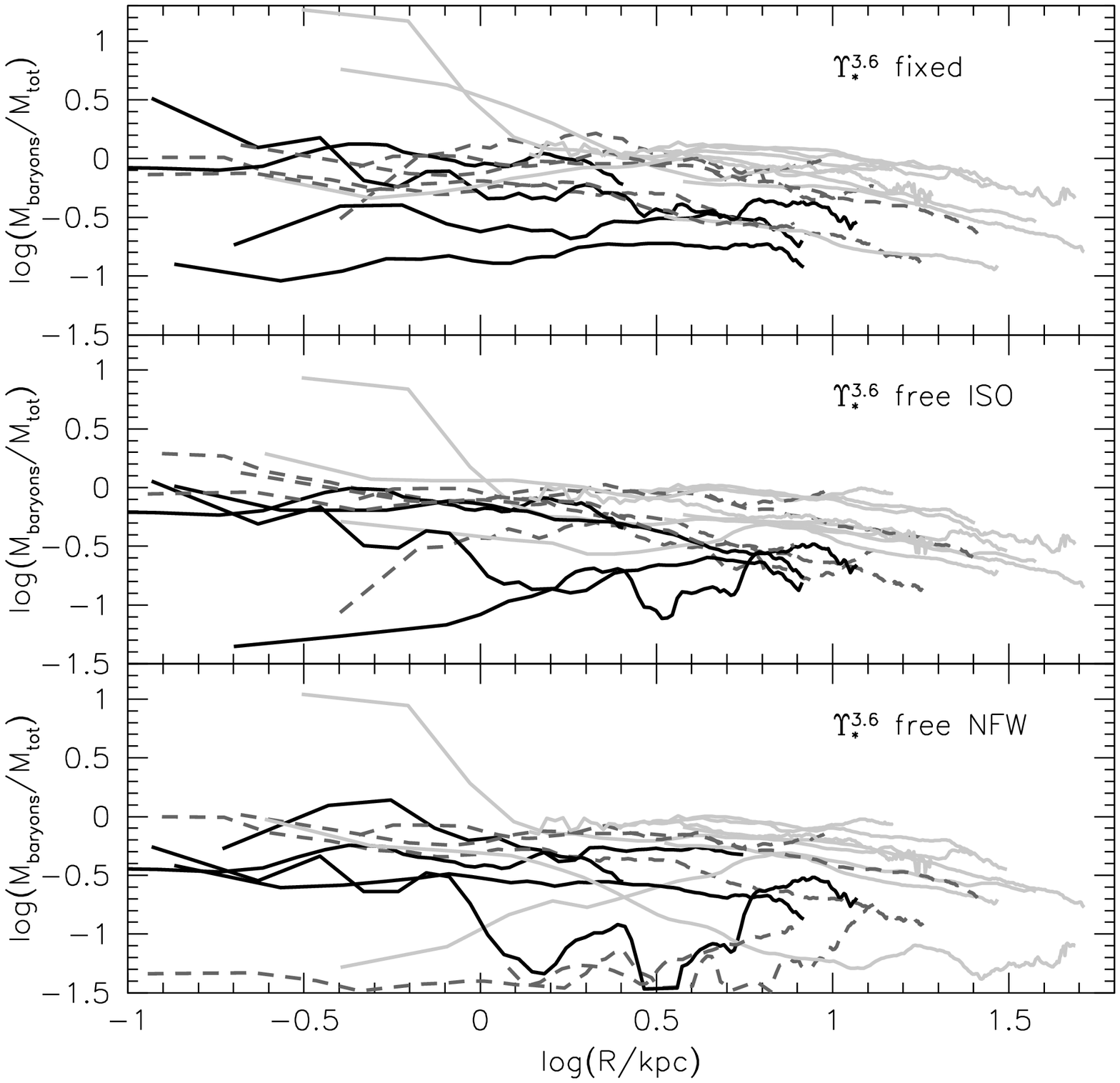} \hfil \figcaption{The
    radial distribution of $M_{\rm baryons}/M_{\rm tot}$ in the
    THINGS galaxies for several assumptions on \Ups. The top panel
    shows the distribution assuming the fixed, predicted diet-Salpeter
    \Ups\ values. The center and bottom panels show the distributions
    assuming the best-fitting \Ups\ values determined by using the ISO
    and NFW models. Light-gray full curves represent galaxies brighter
    than $M_B=-20.5$. Dashed dark-gray curves show galaxies with
    $<-20.5 \leq M_B < -18.5$. Black curves show galaxies fainter than
    $M_B=-18.5$.
\label{fig:dmdom}}
\end{figure*}

The top panel of Fig.~\ref{fig:dmdom} shows the radial dependence of
$M_{\rm baryons}/{M_{\rm tot}}$ for our sample galaxies, assuming
values for $\Ups$ as predicted from the photometry.  These
distributions are for all practical purposes independent of the choice
of IMF.  The largest variation in $M_{\rm baryons}/M_{\rm tot}$ is
about 0.1 dex (a factor 1.3) in the very inner parts of the luminous galaxies, and
quickly becomes totally negligible towards lower luminosities and
large radii.

There are a few things to note in Fig.~\ref{fig:dmdom}. Firstly,
galaxies with the highest absolute luminosity are in general the least
dominated by dark matter. This is not only true globally, but also
locally: even at the outermost radii, the baryons in these luminous
galaxies are more dominant than in the least luminous
galaxies. Secondly, while the most luminous galaxies show a clear
gradient in $M_{\rm baryons}/{M_{\rm tot}}$ with radius (with the
outer radii becoming progressively more dark-matter-dominated), this
trend is less pronounced in the low-luminosity galaxies. The latter
indeed are dominated by dark matter at all radii.  These results are
robust against any particular choice of dark matter model, but only
depend on the values of \Ups.

It is also interesting to explore the distribution of $M_{\rm
  baryons}/{M_{\rm tot}}$ for the case of \Ups\ as a free parameter
(i.e.\ as determined by the dynamics, rather than the photometry).  In
these cases the results are not entirely independent of the halo model
assumed, as the choice of model plays a role in determining \Ups\
during the fitting.  The middle and bottom panels of
Fig.~\ref{fig:dmdom} show $M_{\rm baryons}/{M_{\rm tot}}$ for the
best-fitting ISO and NFW \Ups\ values, respectively. The trends
discussed above are still present in the fixed \Ups\ ISO case, but they
become confused for the NFW model.  The latter is most likely a
reflection of the NFW model being an inappropriate fitting function
for at least some of the galaxies.

\subsection{Stellar mass-to-light ratio trends}

\begin{figure*}[t]
  \epsfxsize=0.9\hsize \epsfbox{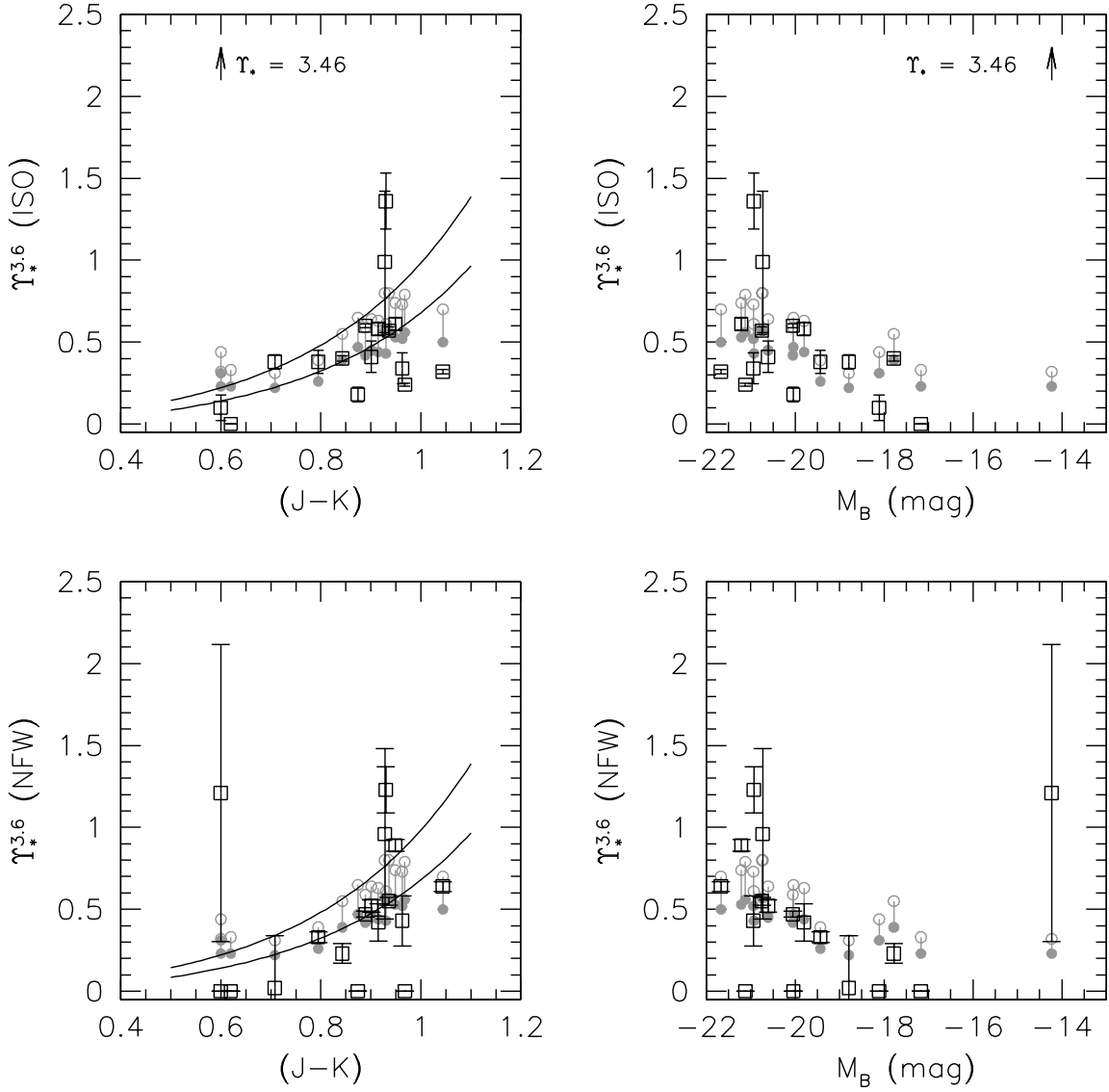} \hfil
  \figcaption{\Ups\ plotted against color and luminosity.  Results are
    shown as a function of colour $(J-K)$ in the lefthand column, as a
    function of absolute magnitude $M_B$ in the righthand column. Top row
    shows results derived for the ISO halo; bottom row those for the
    NFW halo. Open squares indicate the results derived for the free (dynamical)
    \Ups\ fits, the connected grey symbols show the fixed (photometric) \Ups\
    values. The upper symbols (open grey circles) indicate the results
    for a diet-Salpeter IMF, the lower symbols (filled grey circles)
    those for a Kroupa IMF.  In the lefthand panels the curves 
    shows the theoretical color-\Ups\ relations as derived from
    Eqs.~(4) and (5). The upper curve assumes a diet-Salpeter IMF, the
    lower curve a Kroupa IMF. In the top row the arrows indicate the
    high free \Ups\ value derived for DDO 154.
\label{fig:colourML}}
\end{figure*}

An overview of the derived \Ups\ values of the main disk components of
our sample galaxies is given in Fig.~\ref{fig:colourML}. Here we plot
the photometric (fixed) and dynamical (free) values for \Ups\ against
the colors and luminosities of the galaxies.  The left-hand panels
show the fixed and free \Ups\ values of the disks for both the ISO and
NFW model plotted against $(J-K)$ color. We also distinghuish between our
two choices for the IMF. Also shown is the predicted relation between
$\Ups$\ and color (Eqs.~4 and 5), again for both the diet-Salpeter and
the Kroupa IMFs. By definition, the photometric \Ups\ values follow
their respective relations.  The small scatter is caused by the color
gradients present in these galaxies which make the photometric \Ups\
differ slightly from those with a constant \Ups\ as a function of
radius.

In general, we see that the diet-Salpeter curve defines an approximate
upper limit to the distribution of the majority of the best fit values
(apart from a few obviously discrepant cases).  Accepting the best fit
(free) \Ups\ values at face value, this would suggest that a
diet-Salpeter \Ups\ analysis overestimates the disk masses slightly.
Indeed, the Kroupa fixed \Ups\ values seem to be a better match to the
free \Ups\ values.  Alternative explanations could be effects of star
formation history, or unexpectedly large contamination by PAHs, AGBs
or hot dust in the 3.6 $\mu$m maps, though it is likely that the IMF
plays the dominant role.  Future rigorous population synthesis
modeling should shed light on some of these issues. The NFW \Ups\
values tend to be lower than the ISO values, which is to be expected
due to the intrinsically steeper mass distribution of the NFW halo.

\begin{figure*}[t]
  \hfil \epsfxsize=0.9\hsize \epsfbox{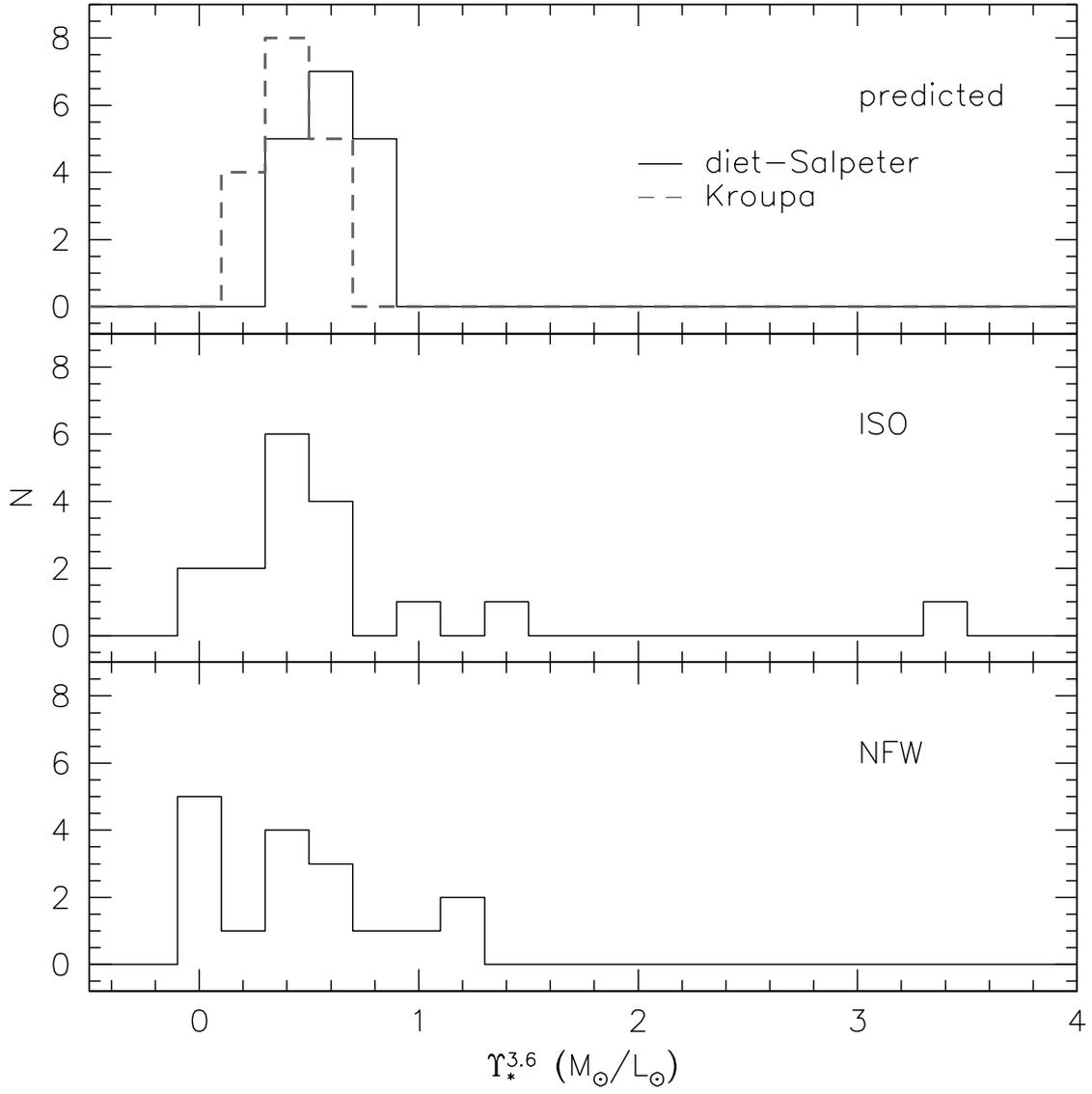} \hfil
  \figcaption{Histograms of the various \Ups\ values. Top: predicted
    values: full histogram assumes a diet-Salpeter IMF, dashed
    histogram assumes a Kroupa IMF; center: best-fit values assuming
    the ISO halo model; bottom: best-fit values assuming the NFW
    model.
\label{fig:histos}}
\end{figure*}

We compare the distributions of the \Ups\ values in
Fig.~\ref{fig:histos}.  The diet-Salpeter fixed \Ups\ values are
offset towards somewhat higher values compared to the ISO free \Ups\
distribution. The Kroupa fixed \Ups\ values are a much better match,
as already suggested by Fig.~\ref{fig:colourML}.  The distribution for
the NFW model is less well-defined, and has a larger number of
galaxies at very low \Ups\ values.  These are the cases where \Ups\
had to be set to zero manually, due to the preferred negative \Ups\
values.  The smaller spread of the \Ups\ values for the ISO case, and
the somewhat better agreement with the predicted values, especially
for the Kroupa IMF results, could be taken as tentative supporting
evidence that the ISO model is superior in describing the properties
of real galaxies.

In general, it is remarkable that despite our empirical and, from the
point of view of a population synthesis expert, undoubtedly
rudimentary modeling of the stellar mass-to-light ratios, the best-fit
dynamical \Ups\ values lie so close to the photometric \Ups\ values.
The photometric \Ups\ values ``don't know'' anything about dynamics,
and the best-fitting dynamical values, whereas the best-fitting \Ups\
ratios have no \emph{a priori} reason to stay close to the predicted
photometric stellar population values. It is therefore gratifying to
see that the calculated \Ups\ agree with the fitted \Ups\ to such a
good degree.

Plotting \Ups\ against luminosity (right-hand panels in
Fig.~\ref{fig:colourML}) we see that the brightest galaxies have the
highest \Ups\ ratios.  These large values are not found for the lower
luminosity galaxies.

The one galaxy that shows a different behavior is DDO 154.  Its
predicted population-synthesis \Ups\ value is low, but
the \Ups\ values derived from the mass-modeling do not agree and are
much higher.  This discrepancy can neither be due to the data quality
(THINGS observed galaxies at identical resolution and sensitivity),
nor due to large non-circular motions, as DDO 154 has a well-behaved,
``quiescent'' velocity field \citep{clemens2007}.

One can interpret this result at face value, and assume that DDO 154
does have a massive disk. This would, however, put it out of
odds with the rest of the sample, and imply a substantial change in
the nature of galaxy disks towards low luminosities.  An alternative
explanation could be that for galaxies like DDO 154 the disk is not
massive enough to have a noticeable effect on the total rotation
curve.  There would then be no small deviations induced by, e.g.,
spiral arms, and the dynamics are everywhere dominated by the dark
matter.  The fitting program is then unable to derive a sensible value
for \Ups.  Obviously, a larger sample of high-resolution observations
of galaxies fainter than $M_B \sim -17$ should then show the same
trend.  A similar, but slightly more prosaic, explanation, is related
to the fact that DDO 154 has the lowest surface brightness disk in 3.6
$\mu$m in our sample. The surface brightness profile is only measured
out to $\sim 100''$ ($\sim 2$ kpc), while the rotation curve extends
out to $\sim 400''$ ($\sim 8$ kpc). This means that most of the disk
rotation curve is based on a smooth exponential extrapolation of the
observed surface brightness profile. If the underlying (undetected)
disk does indeed induce small wiggles in the observed rotation curve,
then these will not be matched by the disk rotation curve (as derived
from the exponential extrapolation), and this will throw the fitting
program off track. Tests of replacing the observed disk rotation curve
of a few of the other THINGS galaxies with their smooth exponential
disk equivalents, indicate that this does indeed affect the derived
\Ups\ values, but the effects are not unambiguous enough to
conclusively show that this is the single cause of what is happening
in the DDO 154 mass modeling. Deeper observations of the stellar disk
should help pinpoint the exact cause of the \Ups\ discrepancy.

\begin{figure*}[t]
  \hfil \epsfxsize=0.9\hsize \epsfbox[18 410 592
  718]{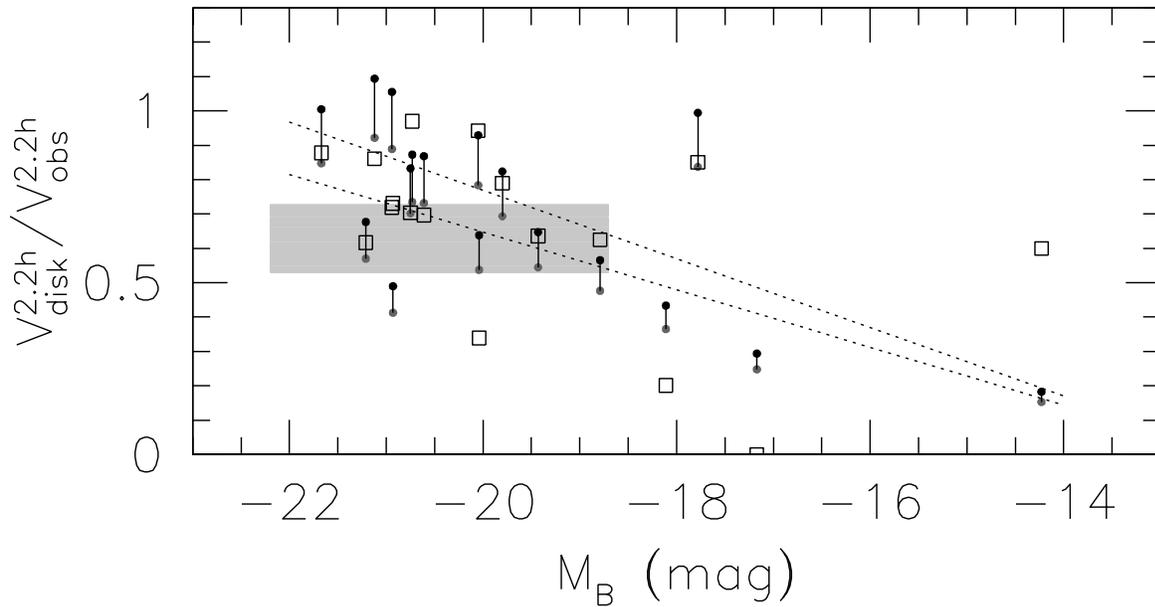} \hfil \figcaption{Ratio of the inferred
    maximum rotation velocity of the disk and the total rotation
    velocity occurring at that radius, plotted against absolute
    luminosity.  Open squares show the ratios assuming the free
    (dynamical) \Ups\ values. The connected symbols show those derived
    for the fixed (photometric) \Ups\ values. The upper symbols
    (filled black circles) show the results for a diet-Salpeter IMF,
    the lower (filled gray circles) show those for a Kroupa IMF.  The
    gray box indicates the range in luminosity and velocity ratios
    derived by \citet{botje97} on the basis of stellar velocity
    dispersion measurements.  The dotted line represents a
    least-squares fit assuming the predicted \Ups\ values. The upper
    line is derived for the diet-Salpeter IMF and decreases as $\sim
    0.25 \log L_B$.  The lower line assumes a Kroupa IMF and decreases
    as $\sim 0.20 \log L_B$.  These slopes are, however, to a large
    extent determined by the DDO 154 result at $M_B \sim  -14$.
\label{fig:ratio}}
\end{figure*}

Finally, in Fig.~\ref{fig:ratio} we show the dynamical importance of the
stellar disk as a function of luminosity.  We plot the ratio of the
maximum rotation velocity of the disk (assuming the photometric \Ups\
ratios and using both IMF assumptions) and the total rotation velocity at the radius where the
maximum disk velocity occurs ($2.2h$ for an exponential disk).  It is
clear that disks become dynamically less and less important with
decreasing luminosity. Note that a similar plot of the importance of
the disk versus luminosity for the case of free \Ups\ (also shown in
Fig.~\ref{fig:ratio}) does not show such a trend as clearly, but here
the result may be diluted by the effects just described.

In a previous analysis, \citet{botje97} used measurements of stellar
velocity dispersions in a number of early-type disk galaxies to
estimate the dynamical contribution of the disk.  He found for the
magnitude range he investigated ($-22 \la M_B \la -19$) that the ratio
of maximum rotation velocity of the stellar disk and total rotation
velocity at the corresponding radius (generally $2.2h$) is $V_{\rm
  disk}^{2.2h}/V_{\rm obs}^{2.2h} = 0.63 \pm 0.10$. These results are
also indicated in Fig.~\ref{fig:ratio}.  

Over the same range in magnitude we find that $V_{\rm
  disk}^{2.2h}/V_{\rm obs}^{2.2h} = 0.81 \pm 0.19$ for the fixed
diet-Salpeter \Ups\ case and $V_{\rm disk}^{2.2h}/V_{\rm obs}^{2.2h} =
0.68 \pm 0.16$ for the Kroupa IMF fits.  The latter is in reasonable
agreement with the \citet{botje97} values. The free \Ups\ results
yield $V_{\rm disk}^{2.2h}/V_{\rm obs}^{2.2h} = 0.73 \pm 0.17$.

\begin{figure*}[t]
  \hfil \epsfxsize=0.9\hsize \epsfbox{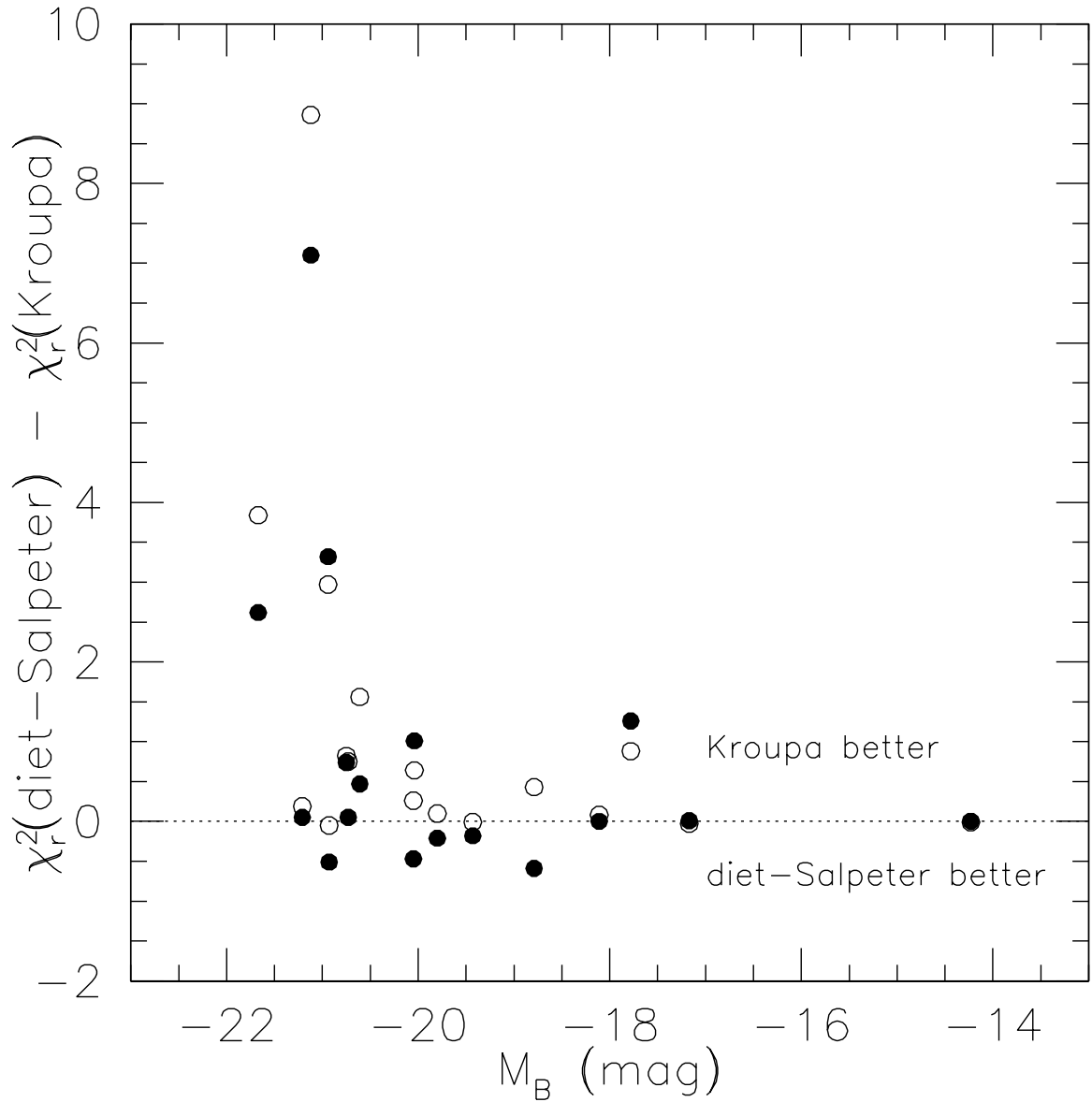} \hfil
  \figcaption{Comparison of the quality of the models
under various assumptions for IMF and dark matter halo model.
The open circles show the difference in goodness-of-fit 
between a diet-Salpeter \Ups\ model and a Kroupa \Ups\ model assuming
an NFW halo. The filled circles show the same for the ISO model.
\label{imf}}
\end{figure*}

In order to gauge whether our data show a preference for one
particular IMF, we compare in Fig.~\ref{imf} the goodness-of-fit
values for the fixed \Ups\ models (for both the diet-Salpeter and
Kroupa IMFs) and the ISO and NFW halos as a function of luminosity.
Several conclusions can be drawn from this comparison. Firstly, the
choice of IMF does not influence the goodness of fit for
low-luminosity galaxies. This is of course to be expected as the
dynamics of these galaxies are dominated by the dark matter
component. Secondly, for the intermediate luminosity galaxies the ISO
models prefer a diet-Salpeter IMF, while the NFW models prefer a
Kroupa IMF.  This is also easily understood, as the NFW halo has a
steeper mass-density profile (a steeper inner rotation curve) which
leaves less room for the luminous component and  results in
lower optimal \Ups\ values.  Thirdly, for the high-luminosity
galaxies, the Kroupa IMF is much preferred over the diet-Salpeter
fits. This reflects the fact that the diet-Salpeter fits result in
disks that are close to the maximum allowed by the dynamics. A small
change in the \Ups\ value can then result in a large difference in the
quality of the fit. The Kroupa IMF with its lower \Ups\ values is more
stable against these effects, as it still allows a dark matter
component in the inner parts which can compensate for any small \Ups\
uncertainties.

The photometrically determined, fixed \Ups\ values therefore do not
allow us to choose a particular IMF, without making additional
assumptions on the distribution of the dark matter.  It would thus be
interesting to independently confirm the \MLstar\ values of the disks
with measurements of the stellar velocity dispersions within the
THINGS disks, and thus calibrate our \MLstar\ measurements.

\subsection{Halo rotation curves}

In Sect.~\ref{sec:genprop} we used the derived \Ups\ values in
combination with our rotation curves to derive the ratio of baryonic
to dark matter as a function of radius. In a similar way we can derive
the rotation curves of the dark matter halos:
\begin{equation}
V_{\rm halo}^2 = V_{\rm obs}^2 - V_{\rm gas}^2 - \MLstar V_{\star}^2,
\end{equation}
where again we are, strictly speaking, dealing with the circular
rotation velocities of test particles in the plane of the galaxy.  The
velocities of the gas and stars have been calculated using the
three-dimensional mass distributions as described in Sect.~5.1 and
5.2. We again assume spherical halos.

\citet{stacy07} show that these empirical halo rotation curves, for a
realistic range of \MLstar\ values, yield halo dark matter densities
about $\sim 50$ percent lower than expected from $\Lambda$CDM
simulations. One interesting aspect of their conclusions is that they
are based on the \emph{outer} ($R>1$ kpc) parts of the rotation
curves, i.e., those regions where the usual concerns about 
resolution or slope of the curve are not relevant.

\begin{figure*}[t]
  \hfil \epsfxsize=0.9\hsize \epsfbox{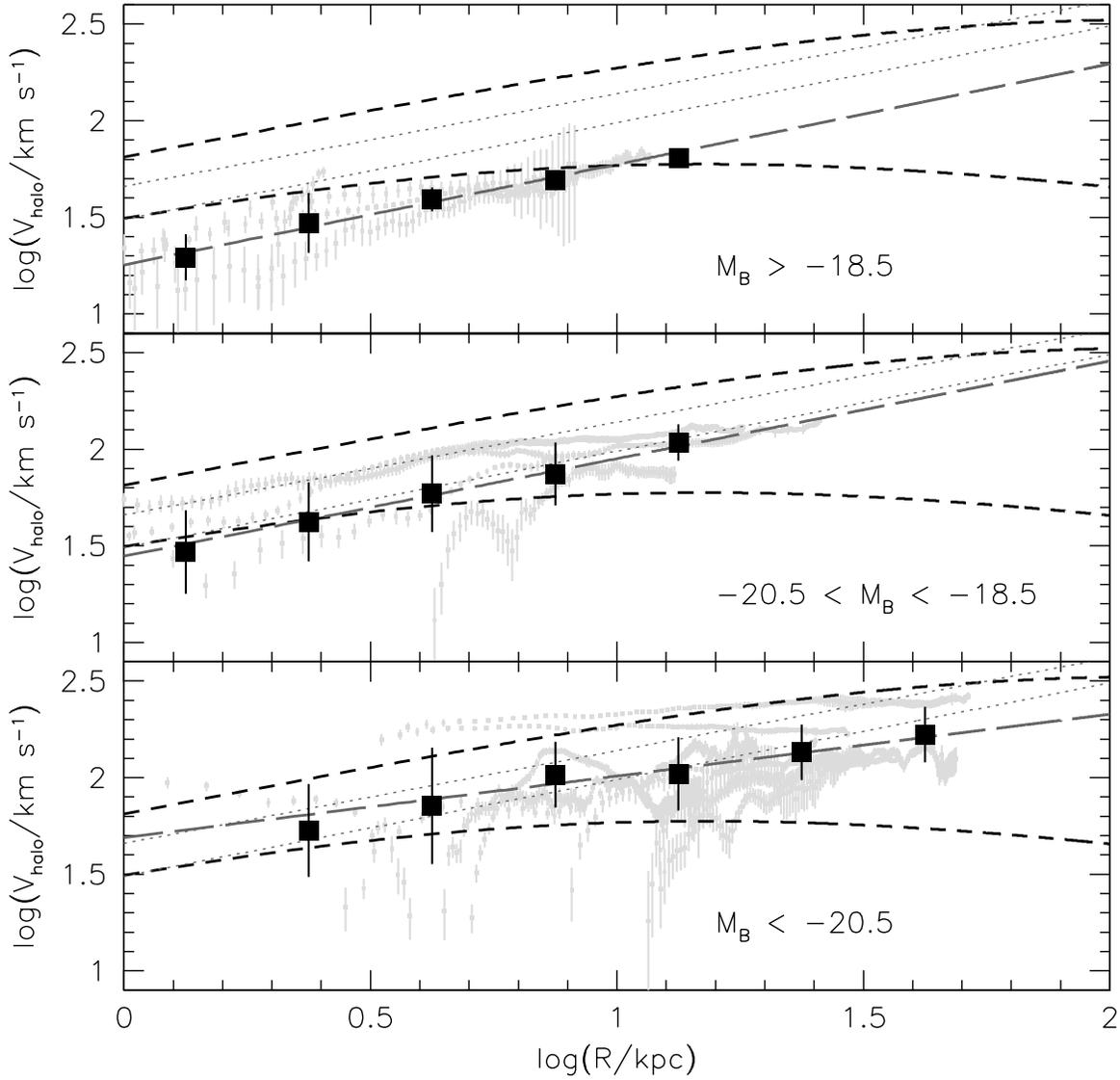} \hfil
  \figcaption{Halo rotation curves of the THINGS galaxies.  The
    galaxies are binned according to their luminosities, as indicated
    in the figure. Individual halo rotation curve points derived
    assuming fixed \Ups\ values are shown as the small lightgray
    points.  The average velocity values binned every 0.25 dex in
    radius are shown as the large black squares. The thick long-dashed
    dark grey line shows the best power-law fit to those points. The
    thick black dashed curves indicate the minimum and maximum
    rotation velocities one expects to find for NFW halos at each
    radius. The lower curve has parameters $(c,V_{200}) = (9.5,\, 50\,
    \kms)$. The upper curve has $(c, V_{200}) = (6.7,\,
    300\,\kms)$. All realistic and CDM-consistent NFW halos are
    expected to fall between these two curves.  The lower of the two
    thin dotted line indicates the best power-law fit to the empirical
    halo rotation curves derived in \citet{stacy07}. The upper one
    shows their best power-law fit to a comparable sample of NFW
    halos. These curves assume a diet-Salpeter IMF; results for the
    Kroupa IMF are very similar, see text for a more extensive
    description.
\label{fig:halocurve}}
\end{figure*}

Here we repeat the analysis of \citet{stacy07} for our current sample.
We derive the halo rotation curves for our galaxies assuming the
photometric (fixed) \Ups\ values for both IMF assumptions.  The
resulting diet-Salpeter IMF logarithmic halo rotation curves are shown
in Fig.~\ref{fig:halocurve}, where we have divided the galaxies into
three bins according to their luminosity.  The Kroupa IMF halo
rotation curves are very similar and we will not show them here,
though we do discuss their implications below.

For each luminosity bin we show the original data, as well as the
observed average halo rotation velocity, binned in steps of 0.25 dex
in radius.  Following \citet{stacy07}, we only consider radii $R>1$
kpc, so as not to confuse the issue with potential systematic effects
in the determination of the very inner slope.  Over the radial range
observed, the average velocities behave very nearly like a power-law.
Also shown in Fig.~\ref{fig:halocurve} are two NFW model rotation
curves with $V_{200}$ velocity values chosen so they encompass the
observed data set. The lowest velocity model has $V_{200} = 50$ \kms,
and is representative of the dwarf galaxies in our sample. On the
high-luminosity end of the range we show a model with $V_{200} = 300$
\kms.  Galaxies with intermediate luminosities or masses will have
curves lying between these two extremes. The corresponding
$c$-parameters of these curves have been chosen to be consistent with
current $\Lambda$CDM cosmological models (cf.~Sect.~5.3.1 and 7.4.2).

In Fig.~\ref{fig:halocurve} we also show the fit to the halo rotation
velocities derived for the $\sim 60$ galaxies in the \citet{stacy07}
sample, assuming their photometrically derived \MLstar\ ratios. They
find that over the radial range observed, the halo rotation velocity
can best be describe by $\log\,V = 0.50\, \log\,R + 1.49$. To compare
this with CDM predictions, \citet{stacy07} subjected a large number of
NFW halo models to observational selection effects and found that the
average halo rotation velocity in a $\Lambda$CDM universe is best
described as $\log\,V = 0.48\, \log\,R + 1.66$ (also shown in
Fig.~\ref{fig:halocurve}). From this they conclude that the
empirically determined densities of dark matter halos are $\sim 50$
percent lower than those predicted by CDM simulations.


We first consider the lowest luminosity bin with $M_B > -18.5$.  Over
the observed radial range the average diet-Salpeter halo velocities
can be described by a power-law $\log\,V = 0.52\, \log\,R + 1.25$
(Kroupa: $\log\,V = 0.41\, \log\,R + 1.35$).  The normalization is
thus 0.24 (0.15) dex (a factor 1.7 to 1.4) lower than the empirical
one derived by \citet{stacy07} and 0.41 (0.31) (a factor 2.5 to 2.0)
dex lower than the average velocity for CDM halos as derived in that
paper. The latter value was, however, derived for a much larger range
in mass than considered in this low-luminosity
sub-sample. Nevertheless, even if we compare with the value found at 1
kpc for a $V_{200} = 50$ \kms\ halo (which should be more typical for
the galaxies considered here; see Fig.~\ref{fig:halocurve}), we still
find a velocity that is $0.25$ (0.15) dex lower than predicted. Only
at $R\sim 10$ kpc do we start finding some agreement between the
empirical and simulated velocities.

The situation is similar for the intermediate luminosity galaxies with
$-20.5 < M_B < -18.5$. The average diet-Salpeter derived halo velocity
can be described as $\log\,V = 0.50\, \log\,R + 1.45$ (Kroupa:
$\log\,V = 0.47\, \log\,R + 1.50$), virtually indistinguishable from
the \citet{stacy07} fit. The same conclusions thus hold here: the
velocity normalization is about 0.21 (0.16) dex (a factor 1.6 to 1.4)
too low compared to average CDM halos, or about 0.25 (0.20) dex (a
factor 1.8 to 1.6) lower than that of a $V_{200} = 150$ \kms\ halo
(which has a value $\log V \sim 1.7$ at $R=1$ kpc). This difference in
normalization is illustrated by the fact that in
Fig.~\ref{fig:halocurve} the inner three observed average data points
overlap with those of the $V_{200} = 50$ \kms\ halo, even though the
average maximum rotation velocity of the galaxies in this sub-sample
is $\sim 130$ \kms.

At the high-luminosity end of the sample, $M_B < -20.5$, all observed
average data points are now within the range described by NFW halos,
but there is still a normalization discrepancy for the diet-Salpeter
case. The data extend further out in radius and are best described by
a power-law fit $\log\,V = 0.32\, \log\,R + 1.69$ (Kroupa: $\log\,V =
0.27\, \log\,R + 1.83$). The slope is thus more shallow than in the
other luminosity bins, perhaps indicating we are starting to see the
velocity turn-over or flattening expected in the ISO and NFW models.
If we look at halos with $V_{200}$ values between 200 and 300 \kms, we
find intercept values at $R=1$ kpc between $\log V = 1.75$ and
$1.80$. This discrepancy is therefore less pronounced than for the
lower luminosity bins, but still amounts to $\sim 0.1$ dex, or a
factor 1.3 difference, for the diet-Salpeter case.  For the Kroupa
case, the intercepts agree, but note that the slope of the average
halo rotation curve is much shallower than the NFW model curves. The
real rotation curves flatten much faster than NFW models with
comparable mass.

For completeness we note that the halo velocities of the entire THINGS
rotation curve sample (assuming a diet-Salpeter IMF) are best
described as $\log\,V = 0.46\, \log\,R + 1.49$, in good agreement with
the \citet{stacy07} result.  For the Kroupa case we find $\log\,V =
0.45\, \log\,R + 1.58$.  The best-fit dynamical, free \Ups\ values (as
derived for the ISO model) give a fit $\log\,V = 0.50\, \log\,R +
1.50$, indistinguishable from the fixed \Ups\ diet-Salpeter IMF
result.

In summary, the empirically derived rotation velocities associated
with dark matter halos of galaxies fainter than $M_B \sim -19$ are
about 0.2 dex or $\sim 1.6$ times lower than predicted by CDM
simulations, independent of IMF, with the largest discrepancies found
in the lowest luminosity galaxies. The agreement with simulated CDM
halos improves towards higher luminosities, but the degree of
improvement strongly depends on the assumed IMF.

If we translate the 0.2 dex in difference in velocity to density
differences, we find an average density that is $\sim 40$ to $\sim 60$
percent of the predicted values. These numbers do depend on the
precise density distribution, but they suggest that dark matter halos
of late-type galaxies have densities that are about half of those
predicted by CDM.  This may also hold true of the luminous galaxies in
our sample, but here the results depend on the assumed IMF.

\subsection{Mass models and model parameters}

\subsubsection{Treatment of uncertainties}

In the preceding discussion we have analyzed the importance of dark
matter in the THINGS galaxies without any strong assumptions on its
actual distribution. However, rotation curves are most often used to
test particular dark matter models, and, as described in Sect.~1, the
debate in recent years has focused on the apparent inability of
cosmological CDM simulations to describe the distribution of dark
matter on the scale of galaxies.  With the high resolution and
excellent sensitivity of the THINGS velocity fields and rotation
curves, we are now in a position to revisit some of the
outstanding questions, and compare the applicability of both the
observationally motivated ISO model and the cosmologically motivated
NFW model (cf.\ Sect.~5.3).

Comparison of the fits derived here can tell us which of the dark
matter models, if any, is preferred. Note that with the sampling of
our velocity fields of two points per beam, the individual rotation
curve data points are not entirely independent. This obviously affects
the absolute $\chi^2$ values, but with the non-Gaussian distribution
of the uncertainties in the rotation curve data points, we will not
dwell on an interpretation of the absolute $\chi^2$ values.  More
importantly, our conservative estimates of the uncertainties, taking
into account the dispersion of velocity values found along tilted
rings, as well as the differences between approaching and receding
sides (Sect.~\ref{deriverotcur}) will also affect the absolute
$\chi^2$ values.  More relevant for our discussion are therefore the
\emph{differences} between the \emph{reduced} $\chi^2$ values, which
tell us the relative merits of any particular model.  In this respect,
the $\chi^2$-\emph{statistic} should  not be confused with the
$\chi^2$-\emph{test}.  The statistic simply quantifies the sum of the
absolute differences between the observed velocities and the
best-fitting model velocities, expressed in terms of the measurement
uncertainties and normalized using the effective number of data
points.  This reduced $\chi^2$ is here only used as a proxy for the
average excess velocity per data point.

In determining the halo parameters and the corresponding uncertainties
the {\sc rotmas} software uses a simple least-squares procedure.  The
uncertainties in the rotation velocity should therefore be reflected
in the uncertainties in the halo parameters, and our definition of the
rotation velocity uncertainties could thus potentially have an impact
on the values we derive. We evaluated the effect of adopting a
different definition for the uncertainties (as discussed in
Sect.~\ref{deriverotcur}) on the fit parameters, and found the
resulting differences to be small, and certainly negligible when
compared to the differences introduced by small variations in \Ups.
We illustrate this for two typical galaxies in Table \ref{modelpars},
where we list the fitted parameters for the ISO and NFW halos,
assuming fixed \Ups\ values, and adopting respectively, \emph{(i)} our
original definition for the uncertainties, \emph{(ii)} the dispersion
of velocities along the rings, and \emph{(iii)} the formal fit error
in the rotation velocity.

The extreme values for $\chi^2_r$ as found when adopting the formal
fit error from {\sc rotcur}, immediately show that these severely
underestimate the true uncertainties, as already discussed in
Sect.~\ref{deriverotcur}. It is clear that for the other two
definitions of the errors small differences in the resulting halo
parameters exist, but it is also clear that different (reasonable)
definitions of the uncertainties do not dominate the error budget: the
largest uncertainty is introduced by the stellar mass-to-light-ratio
$\MLstar$ (cf.\ the results in Table \ref{modelpars} with those in
Tables \ref{table2}-\ref{table3}).  In summary, our definition and
treatment of the uncertainties in the rotation velocities does not
dominate the final uncertainties in the halo parameters.

\begin{deluxetable*}{llrrrrrrrr}
\tablewidth{0pt} 
\tablecaption{Uncertainties in model parameters}
\tablehead{
& & & \multicolumn{3}{c}{ISO halo}&&\multicolumn{3}{c}{NFW halo}\\
\cline{4-6} \cline{8-10}\\
\colhead{Name} & \colhead{type} & \colhead{$\Delta$} & \colhead{$R_C$} &
\colhead{$\rho_0$} & \colhead{$\chi^2_r$} && \colhead{$c$} &
\colhead{$V_{200}$} & \colhead{$\chi^2_r$}\\ 
\colhead{(1)}&\colhead{(2)}&\colhead{(3)}&\colhead{(4)}&\colhead{(5)}
&\colhead{(6)}&&\colhead{(7)}  & \colhead{(8)} & \colhead{(9)} } 
\startdata 
NGC 2403 & total& $4.64 \pm 1.09$  & $2.14 \pm 0.05$ & $77.9 \pm 3.3$ & 0.79 && $9.8 \pm 0.2$ & $110.2 \pm 1.0$ & 0.56 \\ 
& disp & $4.22 \pm 1.21$ & $2.16 \pm 0.05$ & $ 77.1 \pm 3.3$ & 0.94 && $9.9 \pm 0.2$ & $109.5 \pm 0.9$ & 0.70 \\ 
& fit & $0.21 \pm 0.15$ & $2.07 \pm 0.05$ & $ 80.2 \pm 3.0$ & 399.24 && $10.5 \pm 0.2$ & $105.2 \pm 1.2$ & 403.93 \\ 
DDO 154 & total & $2.40 \pm 0.91$ & $1.34 \pm 0.06$ & $ 27.6 \pm 1.6$ & 0.44 && $4.4 \pm 0.4$ & $58.7 \pm 4.3$ & 0.82 \\
 & disp& $2.25 \pm 0.90$  & $1.31 \pm 0.05$ & $ 28.4 \pm 1.7$ & 0.51 && $4.4 \pm 0.4$ & $59.0 \pm 4.2$ & 0.87 \\ 
& fit & $ 0.13 \pm 0.07$ & $1.44 \pm 0.06$ & $24.7 \pm 1.5$ & 141.15 && $5.6 \pm 0.3$ & $49.5 \pm 2.0$ & 168.37
\enddata 
\tablecomments{(1) Name of galaxy; (2) definition of uncertainties;
  ``total'': definition adopted in this paper; ``disp'': dispersion in
  velocities along tilted rings; ``fit'': formal fit error in rotation
  velocity; (3) Average value and dispersion of uncertainties; (4)-(9) halo fit parameters, definitions as in Tables
  \ref{table2}-\ref{table3}.}
\label{modelpars}
\end{deluxetable*}

\subsubsection{Halo parameters}

\begin{figure*}[t]
  \epsfxsize=0.9\hsize \epsfbox[18 410 592 718]{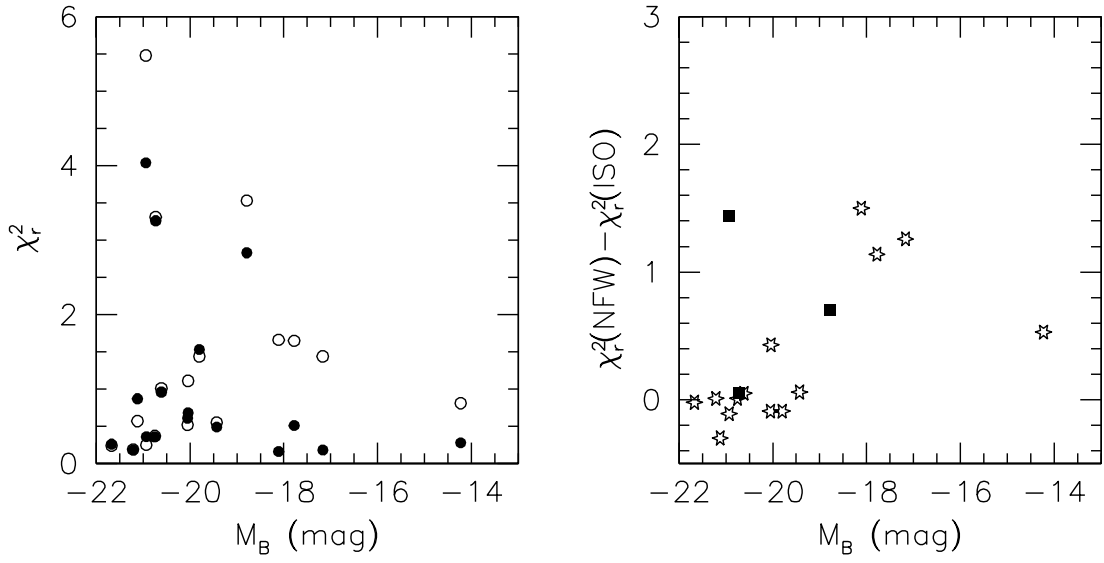} \hfil
  \figcaption{Left: Distribution of $\chi^2_r$ as a function of
    luminosity with \Ups\ as a free parameter. Filled circles: ISO
    fits. Open circles NFW fits.  Right: Difference in ISO and NFW
    $\chi^2$ values. Stars: good fits with $\chi^2_R$(ISO)$ < 2$ and
    $\chi^2_R$(NFW)$ < 2$. Filled squares: fits with $\chi^2_R$(ISO)$ > 2$ or
    $\chi^2_R$(NFW)$ > 2$. 
\label{fig:chi2diff}}
\end{figure*}

We now consider the decompositions with $\Ups$ as a free parameter
(Tables \ref{table3a} and \ref{table3}).  The corresponding $\chi^2_r$
values are compared in Fig.~\ref{fig:chi2diff}.  The left panel
compares the values of $\chi^2_r$ for both models, plotted against
absolute luminosity. Three galaxies ($\chi^2_r > 2$) are obviously not well described
with either model.
Concentrating on the other galaxies, we see a systematic difference
between high- and low-luminosity galaxies.  The luminous galaxies show
comparable $\chi^2_r$ values for either model; the low-luminosity
galaxies with $M_B \ga -19$ are clearly better described using the ISO
model.  This is shown more clearly in the right hand panel of
Fig.~\ref{fig:chi2diff}, where we show the difference between the
$\chi^2_r$ values for both models as a function of luminosity.

NFW and ISO models thus fit equally well for high luminosity galaxies,
with reasonable \Ups\ values. The extent to which this is determined
by subtle trade-offs between, e.g., dynamical contributions of bulge,
disk and halo still needs to be determined, but such an analysis is
likely limited by the accuracy with which stellar mass-to-light ratios
can be determined from photometric data and models.  Alternative
methods (via, e.g., dispersion measurements) will be needed to
constrain the importance of the disk further.

Regarding the low-luminosity galaxies, \citet{clemens2007} show, from
harmonic decompositions of the velocity fields, that non-circular
motions are lowest in these galaxies.  The increasing incompatibility
of the NFW model model towards lower luminosities, is therefore 
difficult to explain purely in terms of these non-circular
motions. Note also that the current sample has been observed at
identical and homogeneous sensitivity and resolution and has been
analyzed in a consistent manner.  The differences found here can
therefore not be attributed to different data qualities or analysis
methods.

\begin{figure*}[t]
  \epsfxsize=0.9\hsize \epsfbox{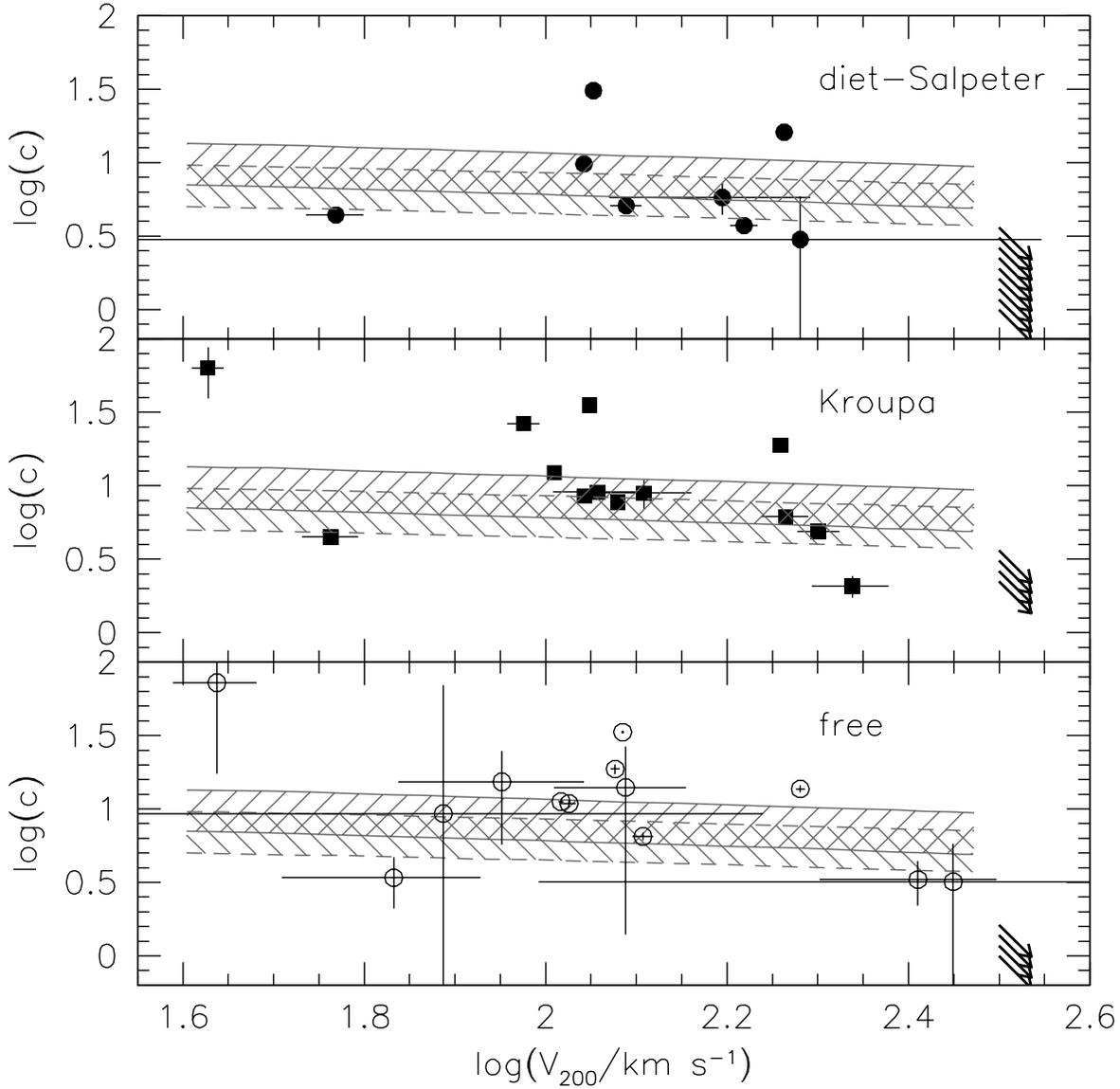} \hfil
  \figcaption{Left: Distribution of $c$ as a function of $V_{200}$ for
    the various assumptions on \Ups.  Upper panel shows the results
    assuming the fixed (photometric) \Ups\ values with a diet-Salpeter
    IMF. The center panel show the same but for a Kroupa IMF. The
    bottom panels uses the free \Ups\ values.  Arrows in the
    bottom-right corners indicate the fits with $c<1$.  The hatched
    area bordered by the full curves indicates the predicted
    $c-V_{200}$ relation based on the ``vanilla'' $\Lambda$CDM
    cosmology presented in \citet{tegmark}.  The counter-hatched area
    bordered by the dashed curves shows the relation that can be
    derived from the 3-year WMAP results \citep{spergel2006}. The
    widths of the distributions correspond to the $\pm 1 \sigma$
    scatter in $c$ as derived from CDM simulations \citep{bullock01}.
\label{fig:cv200}}
\end{figure*}

Lastly, we investigate the model parameters. To start with the NFW
model, as described before, the fit parameters $c$ and $V_{200}$ are
related and their values are determined by the assumed cosmology
(Sect~5.3.1).  \citet{stacy07} show the relation between $c$ and
$V_{200}$ for the ``vanilla $\Lambda$CDM'' model presented in
\citet{tegmark}, and the 3-year WMAP results \citet{spergel2006}. Both
show a fairly flat distribution of $c$ as a function of $V_{200}$, but
with a different offset (due to a different power spectrum
normalization).  Both distributions are shown in Fig.~\ref{fig:cv200}
along with the $1\sigma$ spread expected from simulations
\citep{bullock01}. It is clear that for rotation curves which are
consistent with CDM, we expect to find concentration values between $c
\sim 4$ and $c\sim 10$, fairly independent of $V_{200}$.  In
Fig.~\ref{fig:cv200} we also plot the values for $c$ and $V_{200}$
derived from our rotation curve fits, both for photometrical
(diet-Salpeter and Kroupa IMF) and dynamical \Ups\ values.  Also
indicated are the galaxies for which no sensible fitting parameters
could be derived. It is immediately clear that the distribution of $c$
values we find shows much more scatter than the CDM relations. At
first glance one could conclude that we are finding $c$-values that
are on average higher than predicted, but it is not clear to what
extent our distribution is affected by the absence of galaxies for
which sensible NFW parameters could not be derived.

\begin{figure*}[t]
  \epsfxsize=0.9\hsize \epsfbox{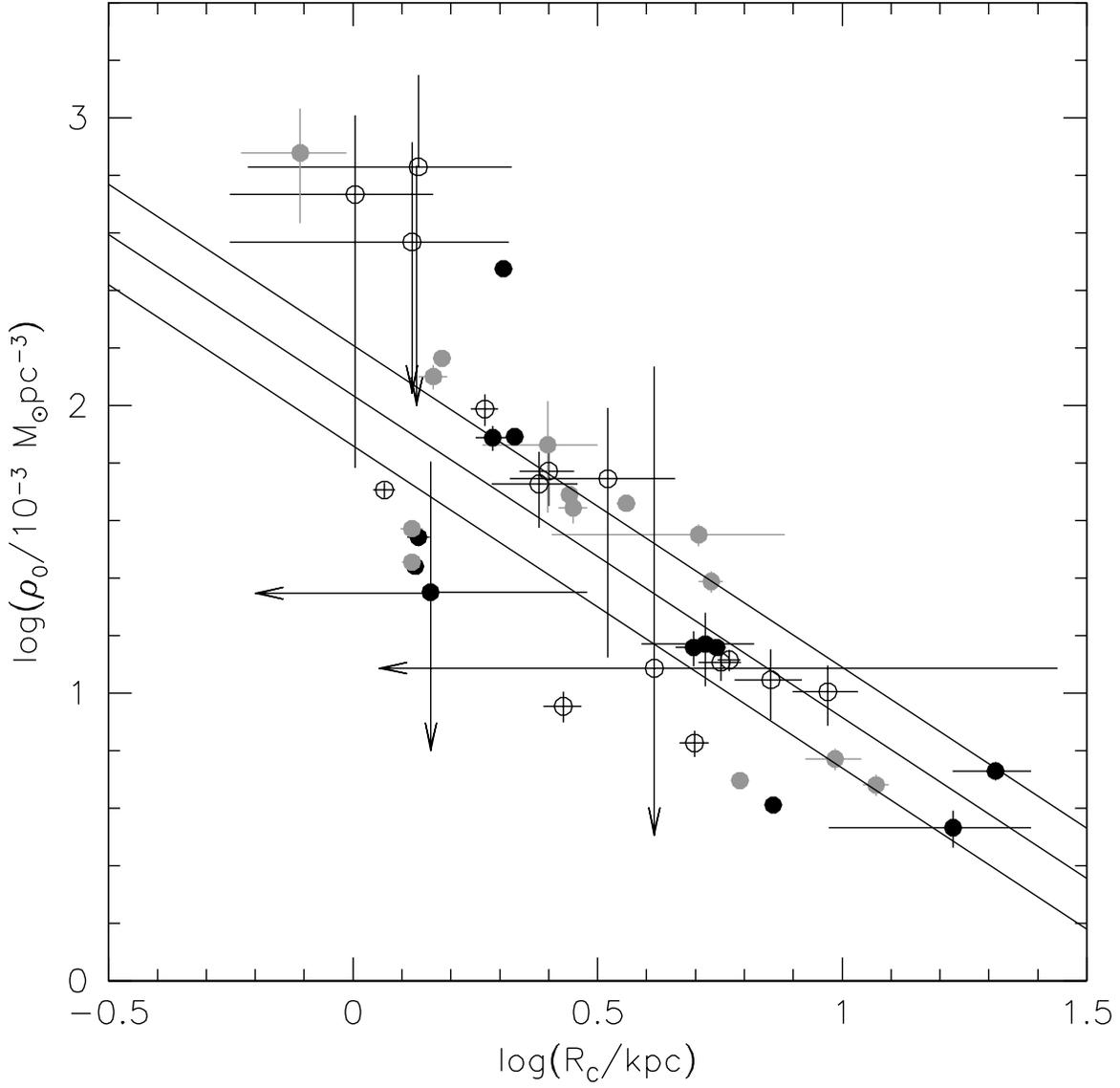} \figcaption{Left:
    Distribution of core density $\rho$ of the ISO halo against core
    radius $R_C$. Filled circles show the results derived assuming
    fixed \Ups\ values, with black symbols assuming a diet-Salpeter
    IMF and grey circles assuming a Kroupa IMF.  Open circles show the
    results derived using free \Ups\ values. The full lines show the
    relation derived in \citet{kormendyfreeman} along with the
    $1\sigma$ scatter.
\label{fig:rcrho}}
\end{figure*}

As the ISO halo has no basis in current cosmology, there are no
\emph{a priori} expectations for its model parameters. One can however
still derive them and check for possible trends with other fundamental
galaxy parameters, as \citet{kormendyfreeman} have done.
Fig.~\ref{fig:rcrho} shows the distribution of the central density
$\rho_0$ against the core radius $R_C$.  The distribution is
consistent with the \citet{kormendyfreeman} relation (their Eq.~20),
but with slightly larger scatter.

\begin{figure*}[t]
  \epsfxsize=0.4\hsize \epsfbox{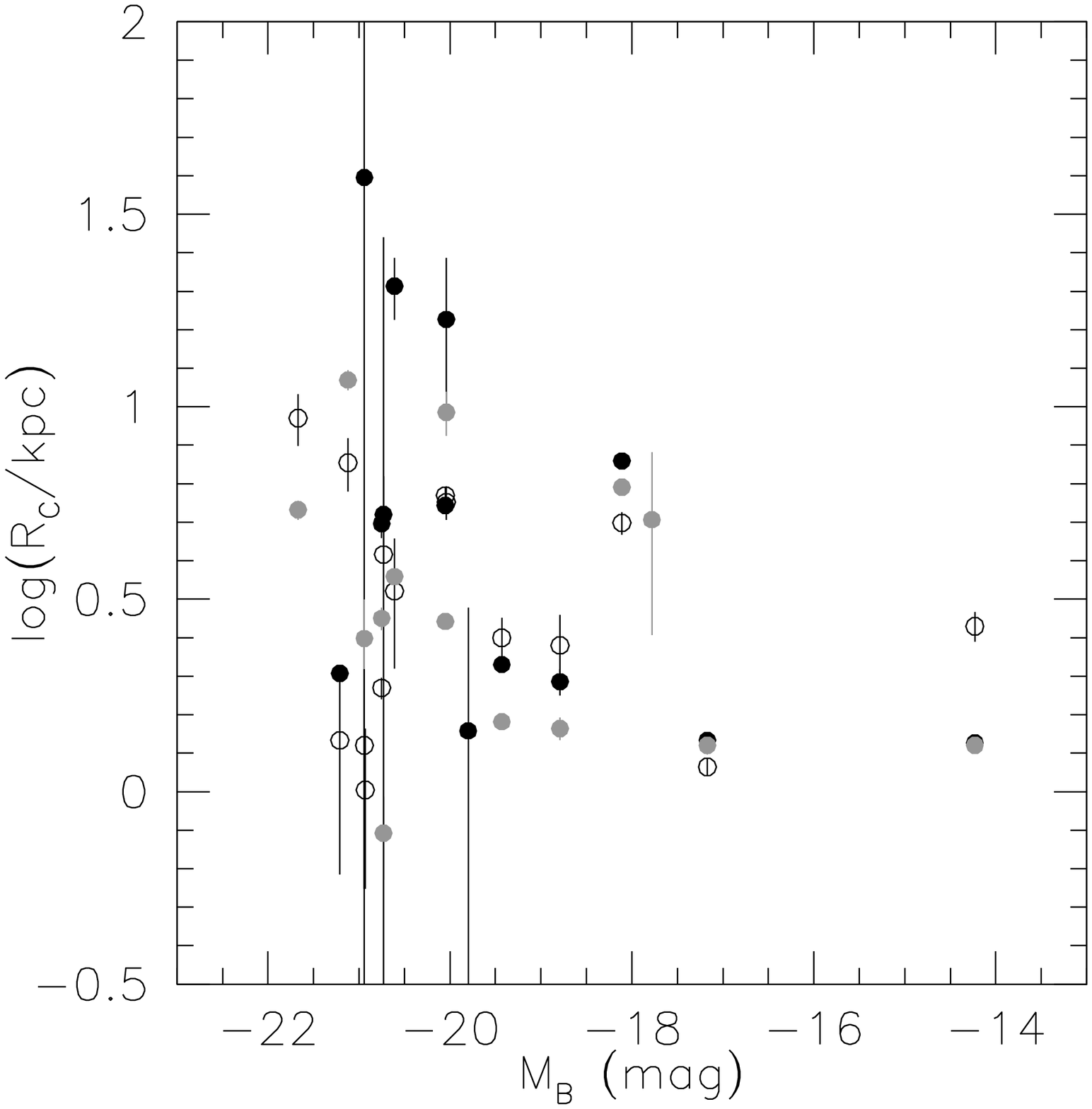}
  \epsfxsize=0.4\hsize \epsfbox{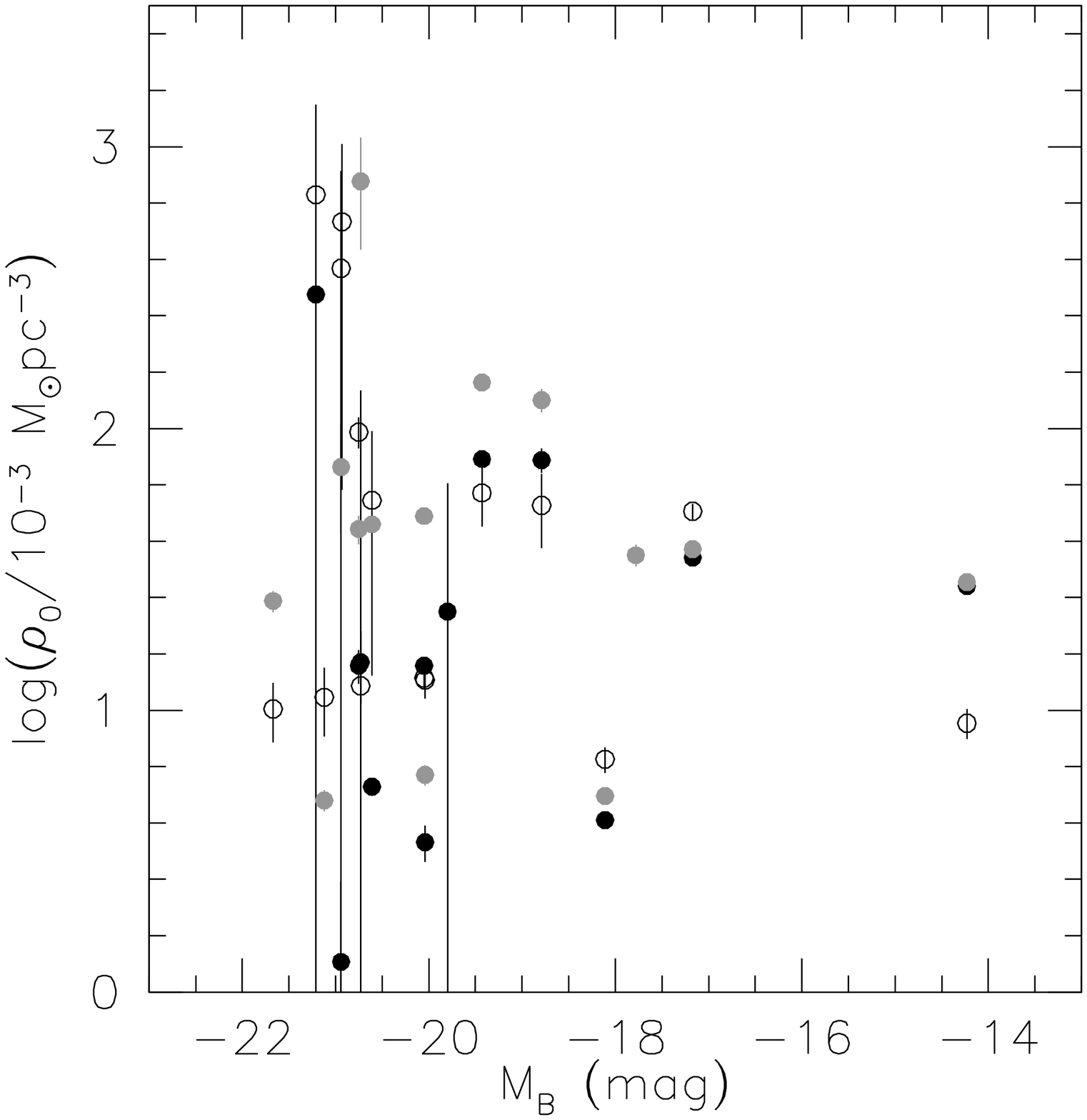}
  \figcaption{Left: Distribution of ISO halo core radius $R_C$ (left panel) and
core density $\rho_0$ (right panel), plotted against absolute luminosity $M_B$.
Filled symbols indicate the fixed \Ups\ results, with black circles representing
the diet-Salpeter fits and  grey circles the Kroupa fits. The open circles
show the free \Ups\ outcomes. 
\label{fig:lumrcrho}}
\end{figure*}

Similar relations with the absolute luminosity (shown in
Fig.~\ref{fig:lumrcrho}) are less well-defined.  The most noteworthy
aspect is that most galaxies have a core radius $R_C>1$ kpc,
independent of the adopted value for \Ups, showing that the effects of
the ``soft'' core are already clearly noticeable well outside the
central regions.  In other words, while scales smaller than $\sim 1$
kpc are needed for a clear distinction between cusp and core,
``symptoms'' of the core should already be visible at scales larger
than that.

\section{Summary}

We present a uniquely high-resolution analysis of the kinematics of 19
galaxies. The galaxies form part of THINGS, and have therefore been
observed and analyzed in a homogeneous and consistent manner.  The high
spatial and velocity resolution, as well as the exquisite sensitivity
enable us to derive the tilted-ring parameters based solely on the
\HI\ data.  There is therefore no dependence on, e.g., optical
inclinations or axis ratios, which generally carry their own
systematic effects with them.  We summarize our results as follows:

\begin{itemize}

\item We determine the 3.6 $\mu$m stellar mass-to-light ratio \Ups\ in
  two fully independent ways: one is based on population synthesis
  modeling \citep{oh2007}, the other is a purely `dynamical' \Ups,
  determined solely by the best fit \Ups\ value in the rotation curve
  mass model.  For the population synthesis \Ups\ values we consider
  two assumptions for the IMF: the ``diet''-Salpeter IMF
  \citep{bell_dejong}, designed to fit maximum disk models, and the
  \citet{kroupa} IMF, which yields disks that are $\sim 40$ percent
  lighter.  We find good agreement between the photometric and
  dynamical \Ups\ values, especially for the Kroupa IMF.  The close
  match between the dynamical and population synthesis results
  indicates the stellar disks of the galaxies in the THINGS sample do
  not contain large amounts of dark matter associated with the
  disks themselves. The dark matter content of the THINGS galaxies is
  associated with the halo, not the disk. Our analysis also obviates
  the need for so-called minimum or maximum disk approaches.

\item We do not find any steeply declining rotation curves out to the
  last measured point. Specifically, previous observations seemed to
  indicate declining rotation curves for NGC 3521, NGC 7793, NGC 2366
  and DDO 154. With the increased resolution of the data we can now
  better trace the inclination and PA trends.  Any gentle decreases in
  the velocity that we still find can now be attributed to
  uncertainties in the rotation velocity or inclination.  There is
  therefore no evidence from THINGS that we have probed the ``edge''
  of a dark matter halo.

\item We fit NFW and ISO models to our rotation curves using our best
  estimates for \Ups. We find that for galaxies $M_B < -19$ both
  models statistically fit equally well. Note, however, that for the
  11 galaxies in this high-luminosity bin, the fixed \Ups\ diet-Salpeter
  NFW fits result in 4 galaxies with unphysical (severely inconsistent
  with CDM) halo parameters. These unphysical cases disappear when a
  Kroupa IMF is assumed. However, the resulting halo parameters still
  show a scatter much larger than the simulation predict.

\item For galaxies with $M_B > -19$ the core-dominated ISO model fits
  significantly better than the NFW model (and the fixed \Ups\ NFW
  fits yield 4 out of 5 galaxies with unphysical halo parameters
  independent of IMF assumption). The success of the ISO halo for
  these galaxies is remarkable, given the conclusion by
  \citet{clemens2007} that the THINGS low-luminosity galaxies have the
  smallest non-circular motions: the observed dynamics most likely
  reflect that of the dark matter halo.
	
\item We find that the low-luminosity galaxies in our sample are dark
  matter-dominated throughout, based on both the population synthesis
  \Ups\ values, as well as the `dynamical' \Ups\ results. As noted
  above, the agreement between these two sets of values indicates the
  dark matter is not associated with the stellar disk.  The dark
  matter rotation curves of these galaxies are about 0.2 dex (a factor
  1.6) lower than predicted by CDM over their entire disks. This
  implies that the density of dark matter in real late-type galaxies
  is $\sim 50$ percent of what is predicted by CDM simulations.

\end{itemize}

\acknowledgements

We thank the anonymous referee for carefully reviewing this paper.  It
is a pleasure to thank Adam Leroy for the many IMF discussions,
Nicolas Bonne for his initial work on the NGC 2841 rotation curve and
Karen Lewis for her work on NGC 3621.  The work of WJGdB is based upon
research supported by the South African Research Chairs Initiative of
the Department of Science and Technology and National Research
Foundation.  EB gratefully acknowledges financial support through an
EU Marie Curie International Reintegration Grant (Contract
No. MIRG-CT-6-2005-013556) The work of CT is supported by the German
Ministry for Education and Science (BMBF) through grant 05 AV5PDA/3.
This research has made use of the NASA/IPAC Extragalactic Database
(NED) which is operated by the Jet Propulsion Laboratory, California
Institute of Technology, under contract with the National Aeronautics
and Space Administration.  This publication makes use of data products
from the Two Micron All Sky Survey, which is a joint project of the
University of Massachusetts and the Infrared Processing and Analysis
Center/California Institute of Technology, funded by the National
Aeronautics and Space Administration and the National Science
Foundation.  We acknowledge the usage of the HyperLeda database
(http://leda.univ-lyon1.fr).



\clearpage

\appendix

\section{Data and rotation curve derivation}

We present detailed descriptions of the derivations of the tilted ring
models and rotation curves of the individual galaxies.  For each
galaxy we show the Hermite velocity field, the resulting tilted-ring
model velocity field, and the difference (residual) velocity
field. Additionally we show overlays of the derived rotation curves on
minor- and major-axis position-velocity ($pV$) diagrams, the rotation
curves themselves, the radial distributions of PA and $i$, and the
positions of the dynamical center in comparison with the integrated
\HI\ maps and the SINGS 3.6 $\mu$m images\footnote{Note that NGC 2366,
  NGC 2903 and DDO 154 are not part of the SINGS sample. For these
  galaxies we have retrieved data from the \emph{Spitzer}
  archive.}. We also briefly discuss the derivation of the various
tilted ring parameters for each galaxy.

\begin{figure*}[t]
\epsfxsize=0.95\hsize \epsfbox{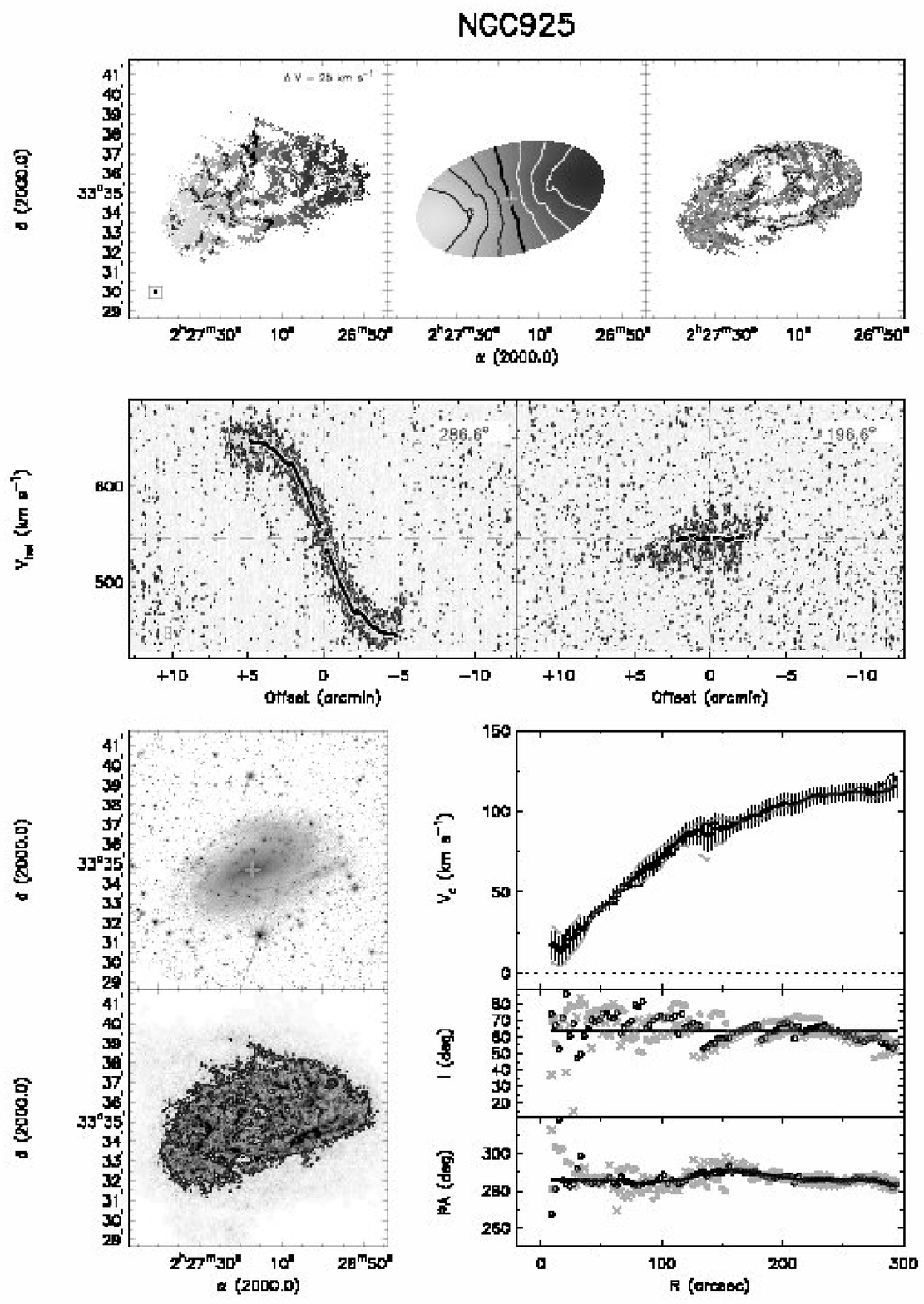} \figcaption{
Summary panel for NGC 925. See the Appendix for more information.
\label{fig:n925}}
\end{figure*}

\begin{figure*}[t]
\epsfxsize=0.95\hsize \epsfbox{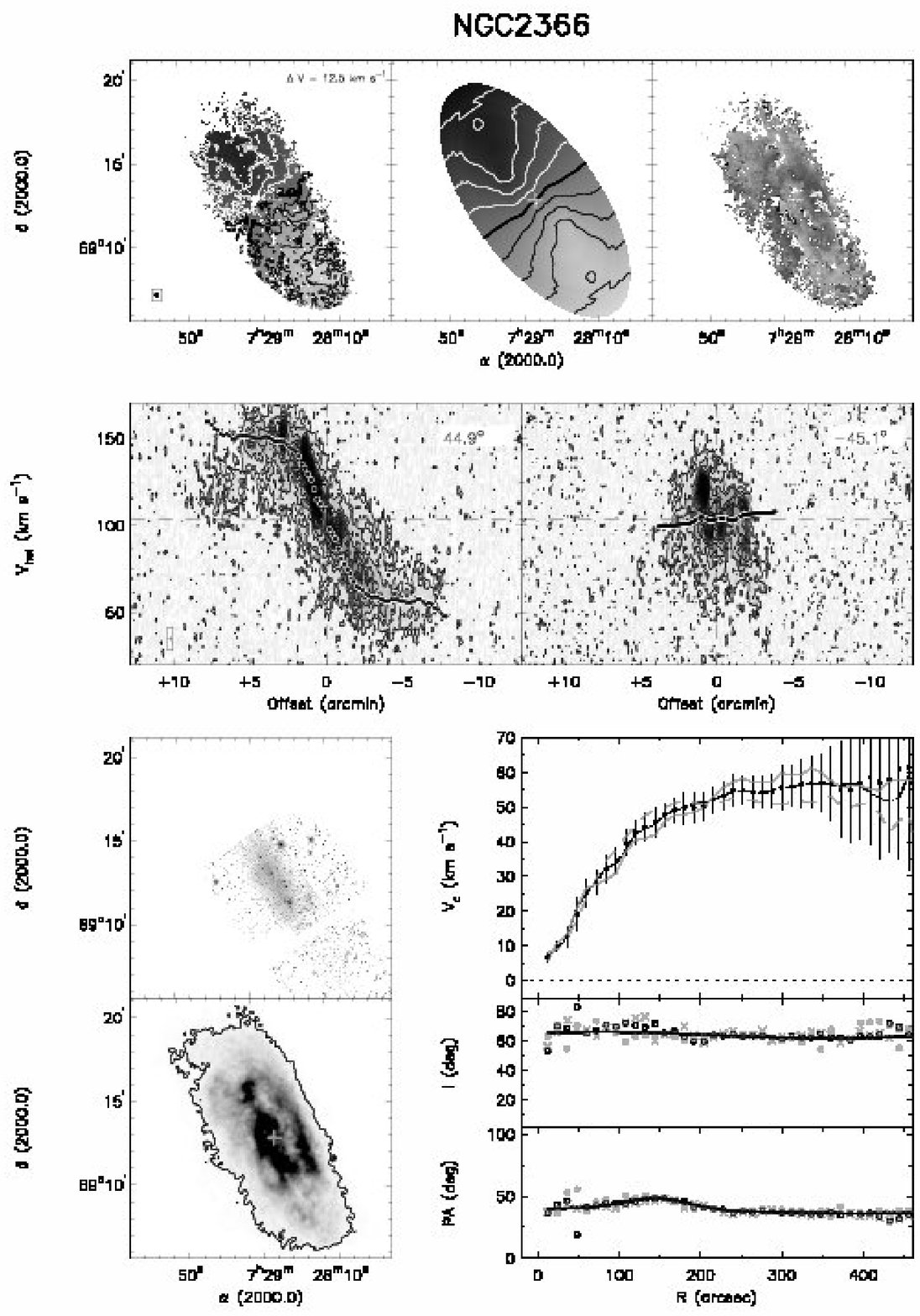} \figcaption{
Summary panel for NGC 2366. See the Appendix for more information.
\label{fig:n2366}}
\end{figure*}

\begin{figure*}[t]
\epsfxsize=0.95\hsize \epsfbox{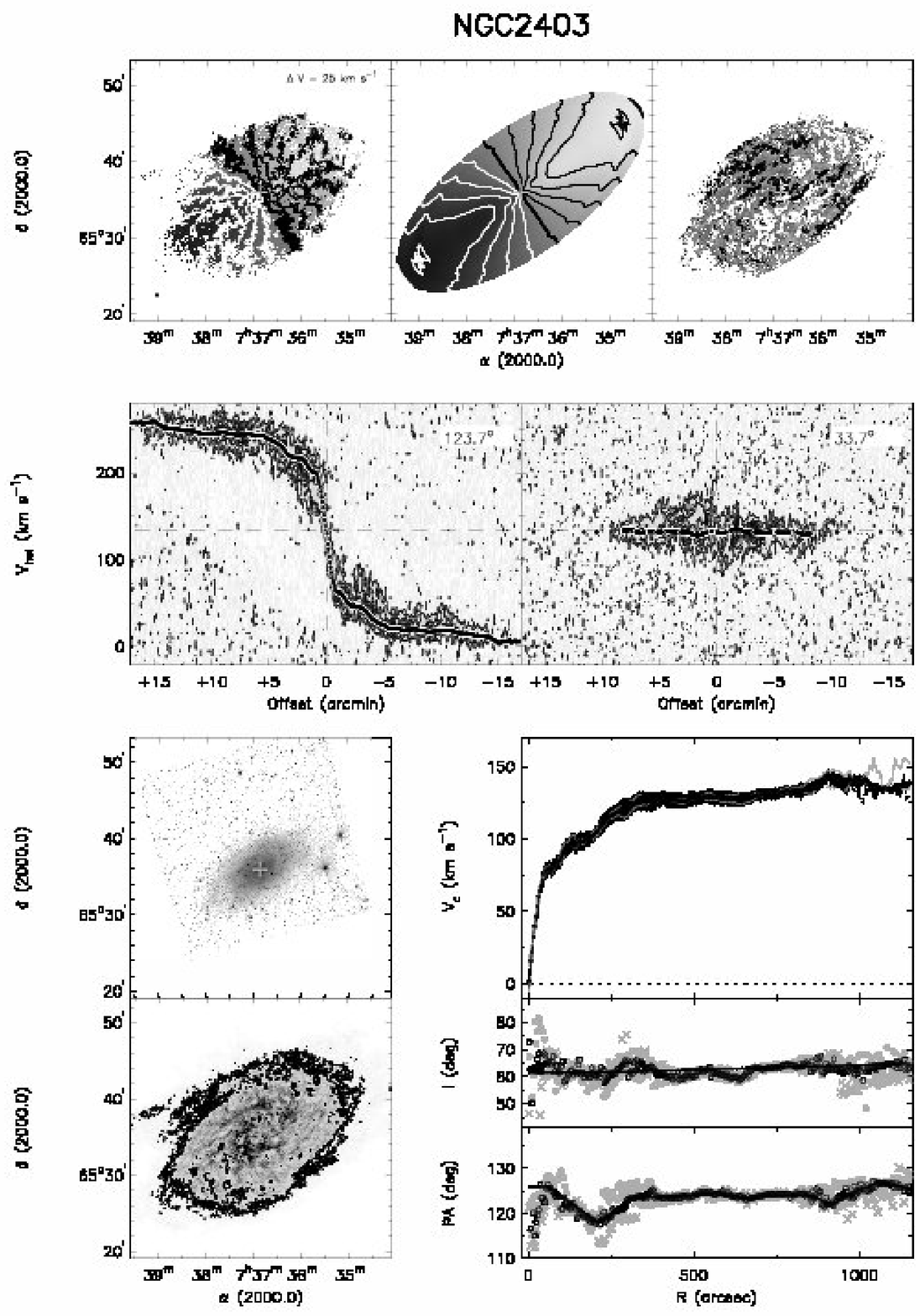} \figcaption{
Summary panel for NGC 2403. See the Appendix for more information.
\label{fig:n2403}}
\end{figure*}

\begin{figure*}[t]
\epsfxsize=0.95\hsize \epsfbox{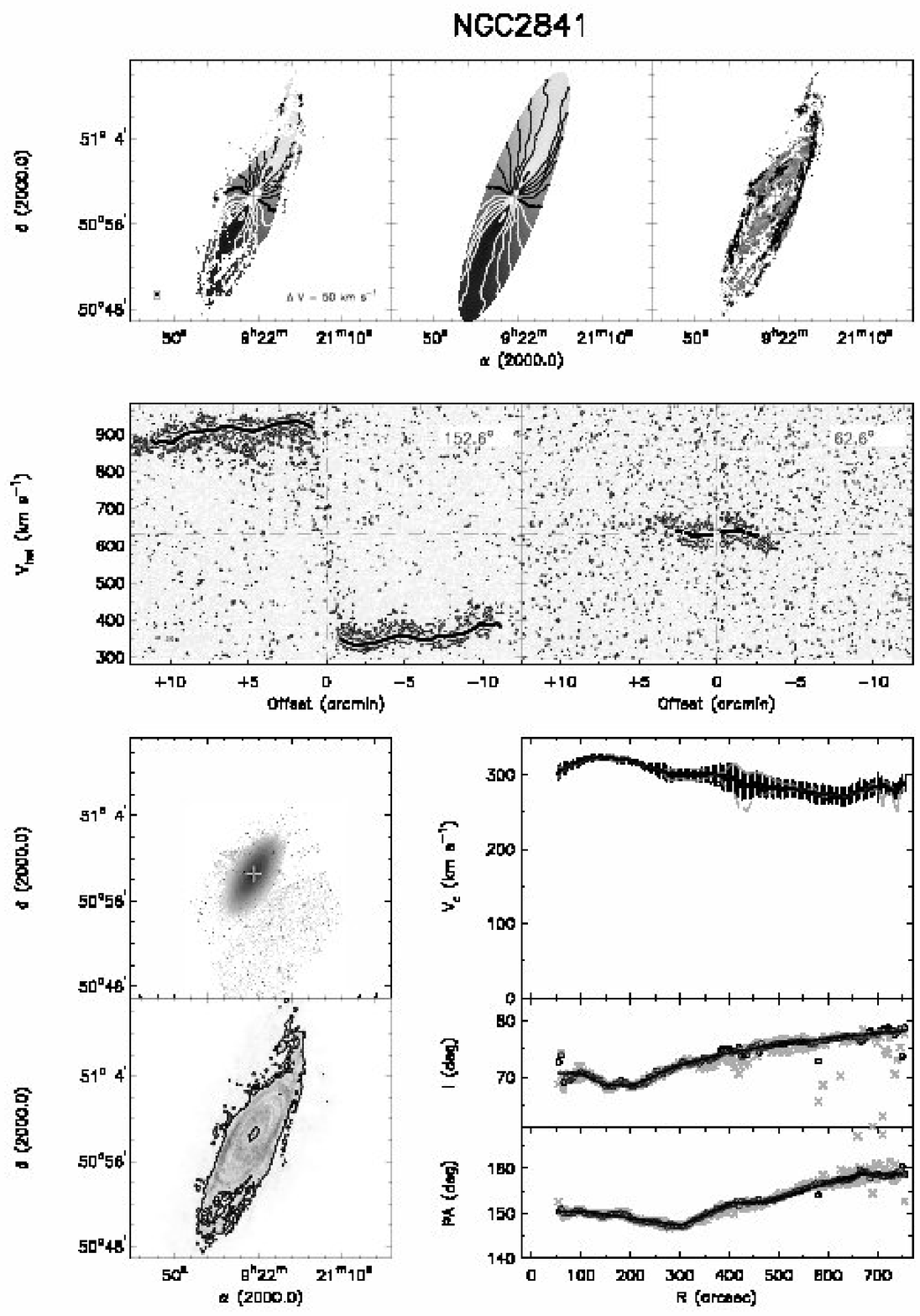} \figcaption{
Summary panel for NGC 2841. See the Appendix for more information.
\label{fig:n2841}}
\end{figure*}

\begin{figure*}[t]
\epsfxsize=0.95\hsize \epsfbox{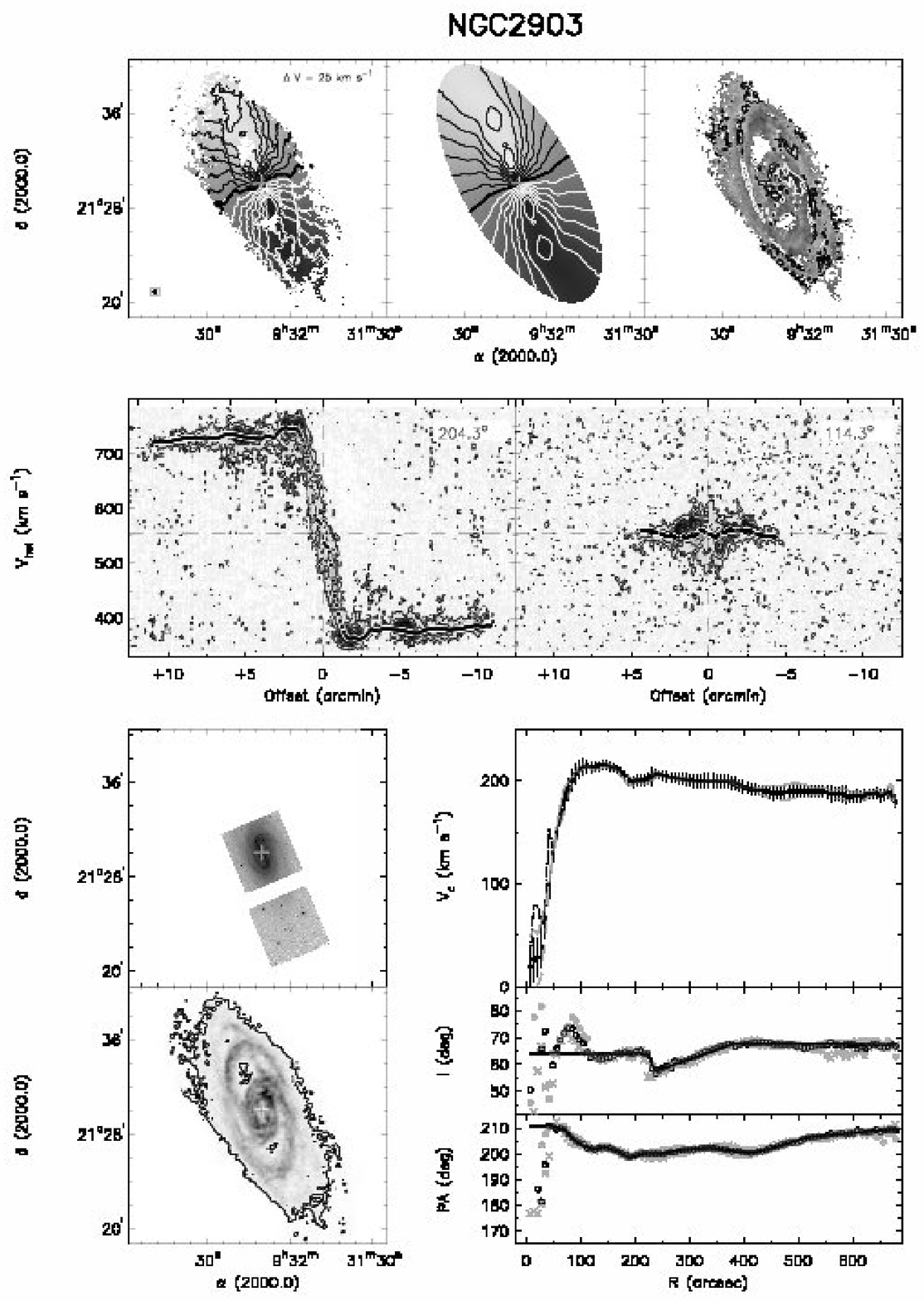} \figcaption{
Summary panel for NGC 2903. See the Appendix for more information.
\label{fig:n2903}}
\end{figure*}

\begin{figure*}[t]
\epsfxsize=0.95\hsize \epsfbox{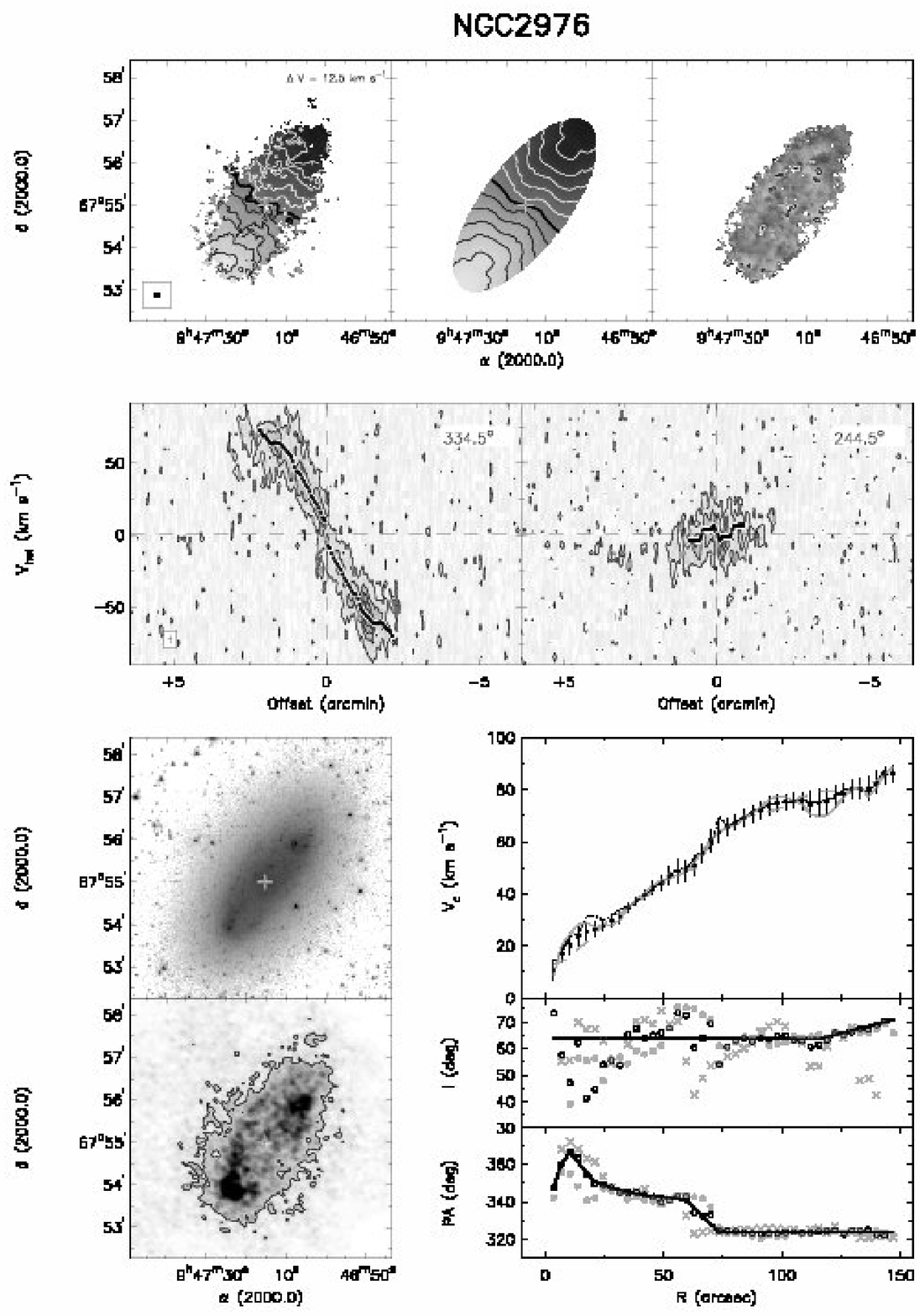} \figcaption{
Summary panel for NGC 2976. See the Appendix for more information.
\label{fig:n2976}}
\end{figure*}

\begin{figure*}[t]
\epsfxsize=0.95\hsize \epsfbox{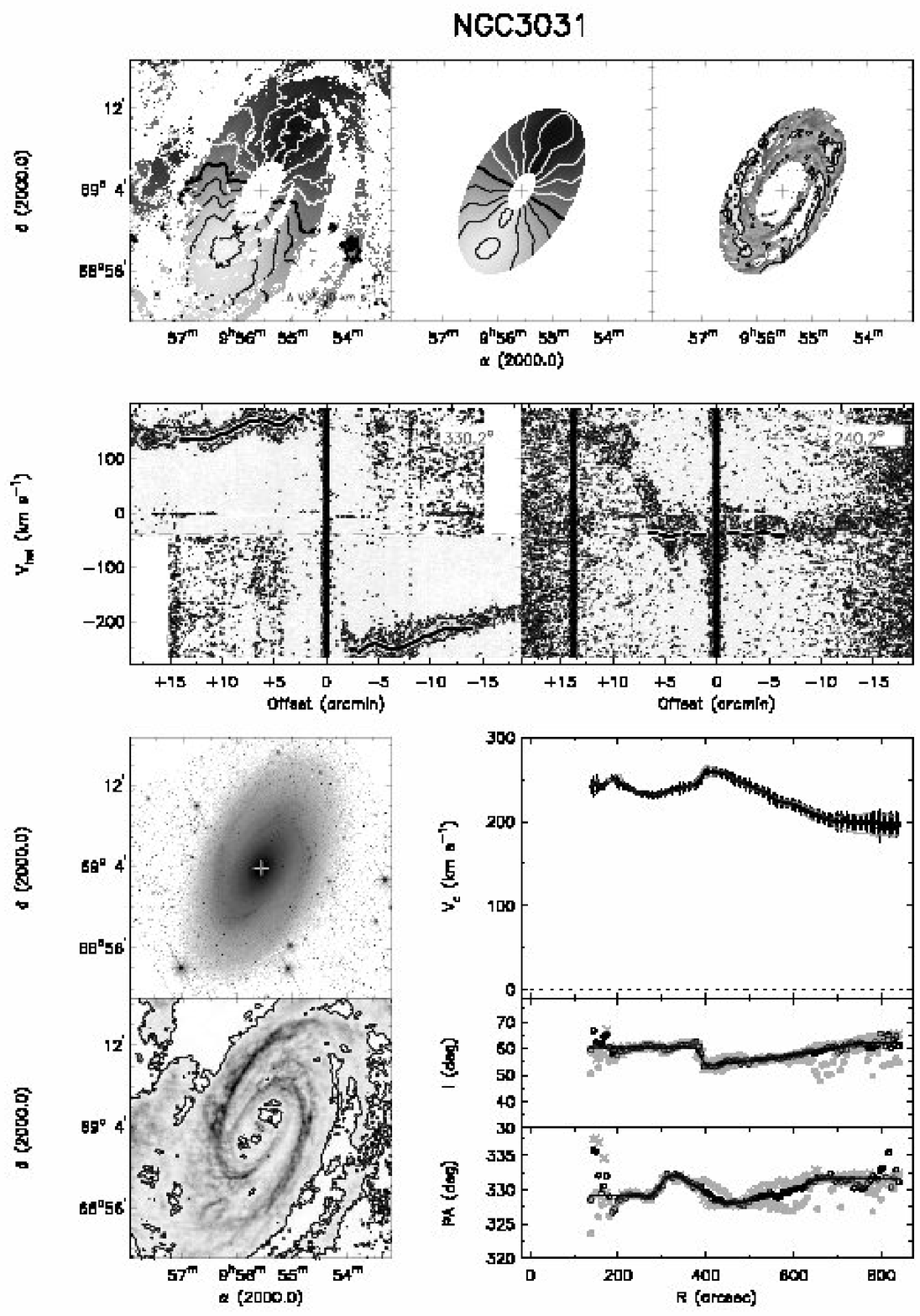} \figcaption{
Summary panel for NGC 3031. See the Appendix for more information.
\label{fig:n3031}}
\end{figure*}

\begin{figure*}[t]
\epsfxsize=0.95\hsize \epsfbox{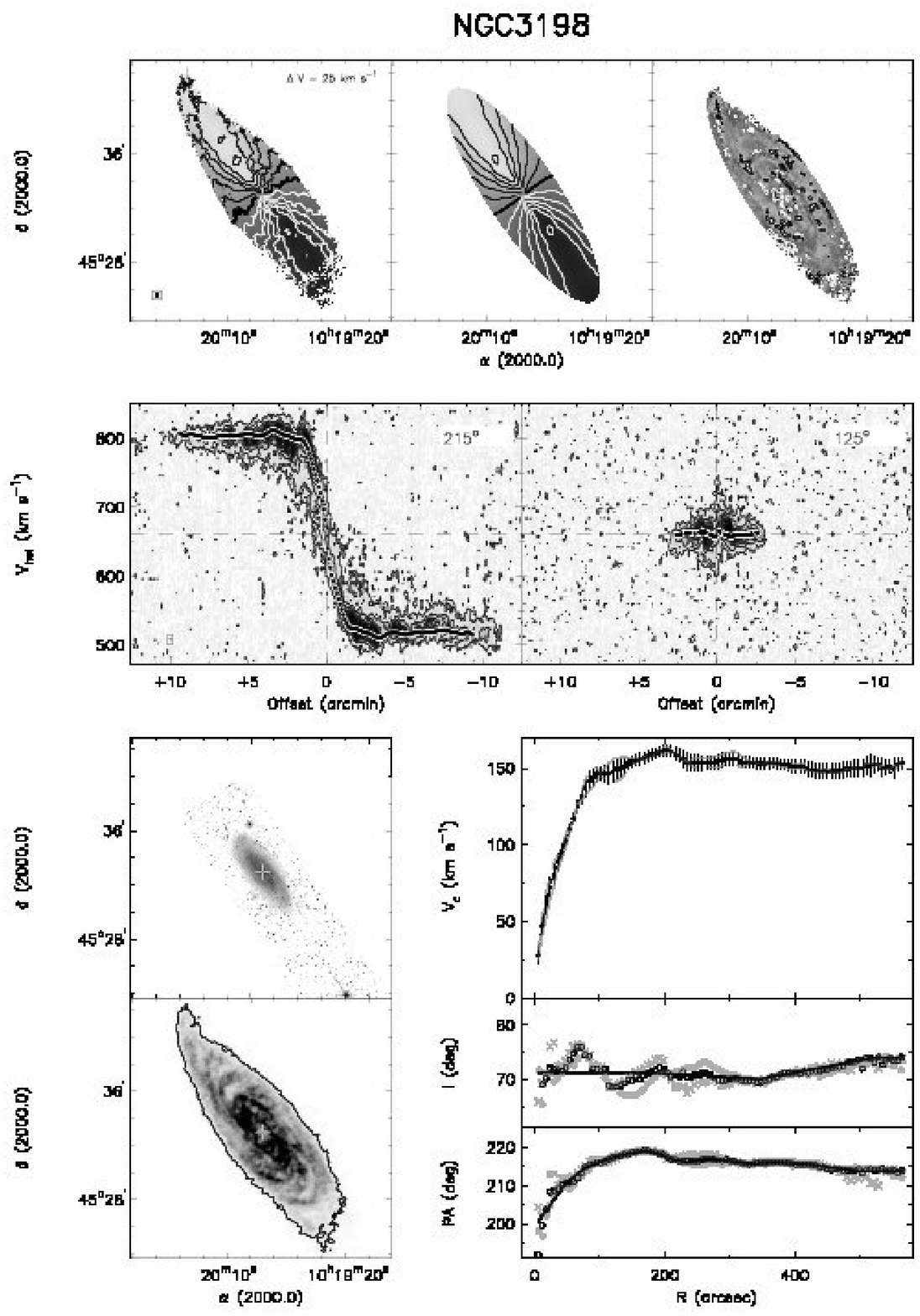} \figcaption{
Summary panel for NGC 3198. See the Appendix for more information.
\label{fig:n3198}}
\end{figure*}

\begin{figure*}[t]
\epsfxsize=0.95\hsize \epsfbox{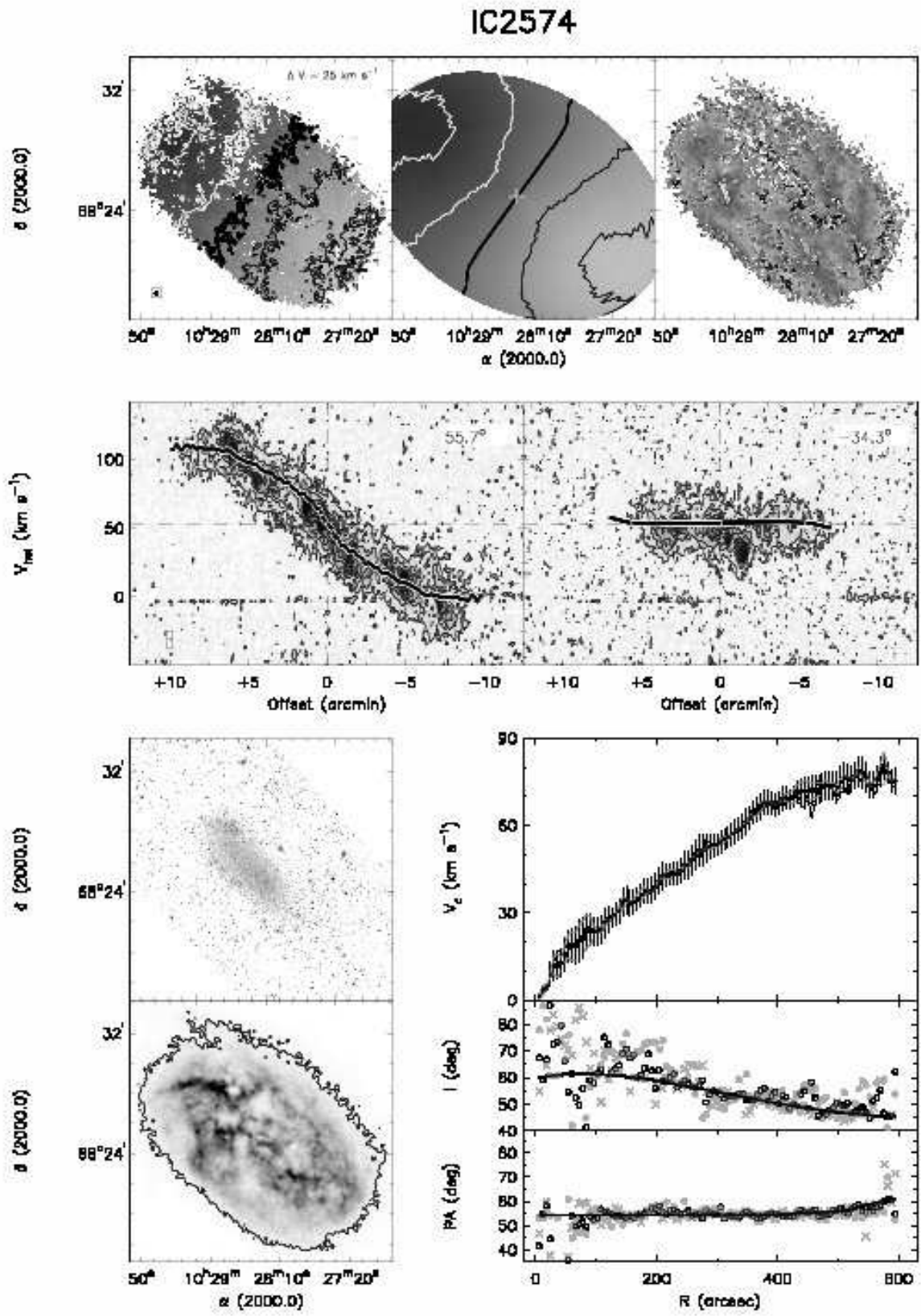} \figcaption{
Summary panel for IC2574. See the Appendix for more information.
\label{fig:ic2574}}
\end{figure*}

\begin{figure*}[t]
\epsfxsize=0.95\hsize \epsfbox{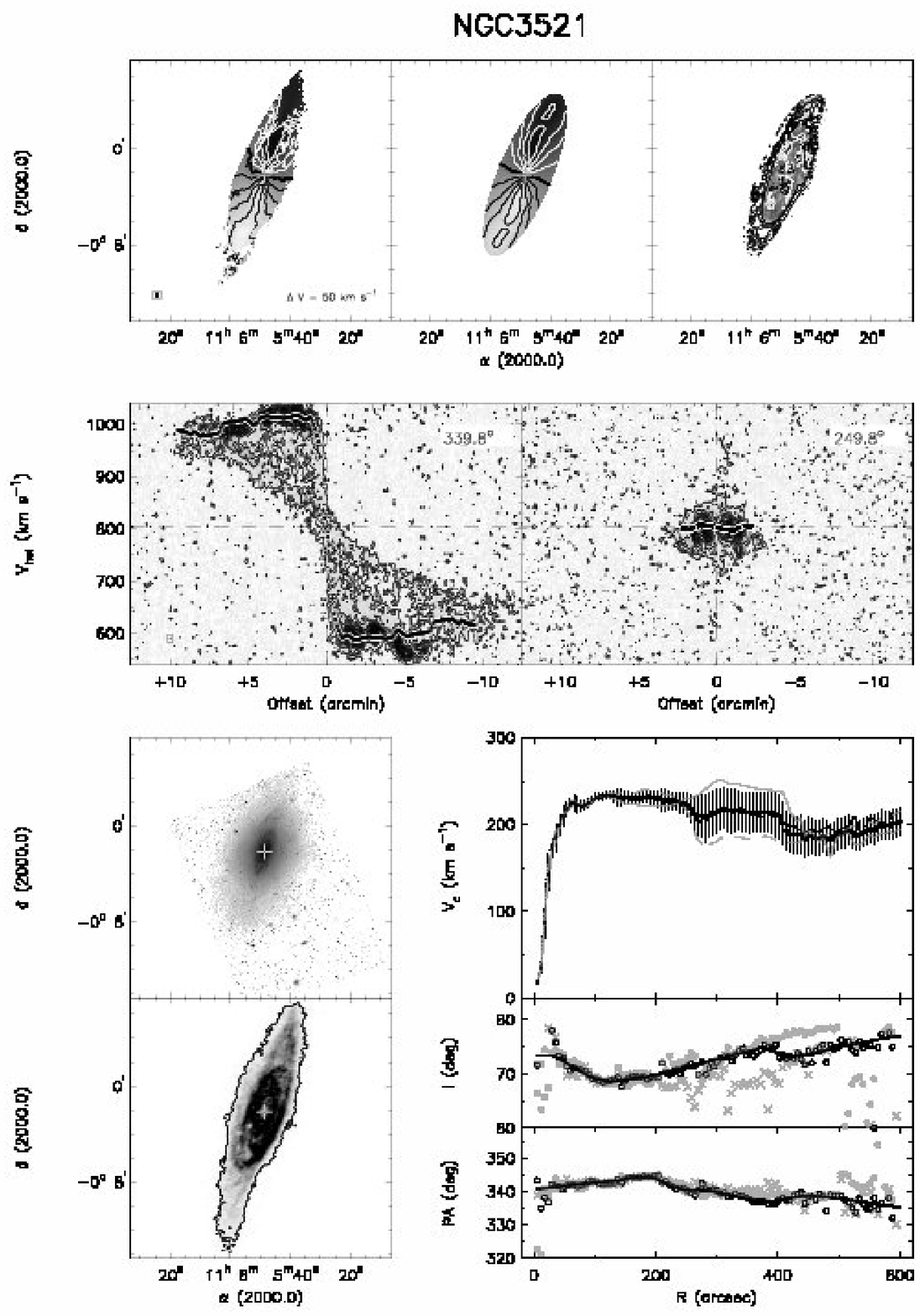} \figcaption{
Summary panel for NGC 3521. See the Appendix for more information.
\label{fig:n3521}}
\end{figure*}

\begin{figure*}[t]
\epsfxsize=0.95\hsize \epsfbox{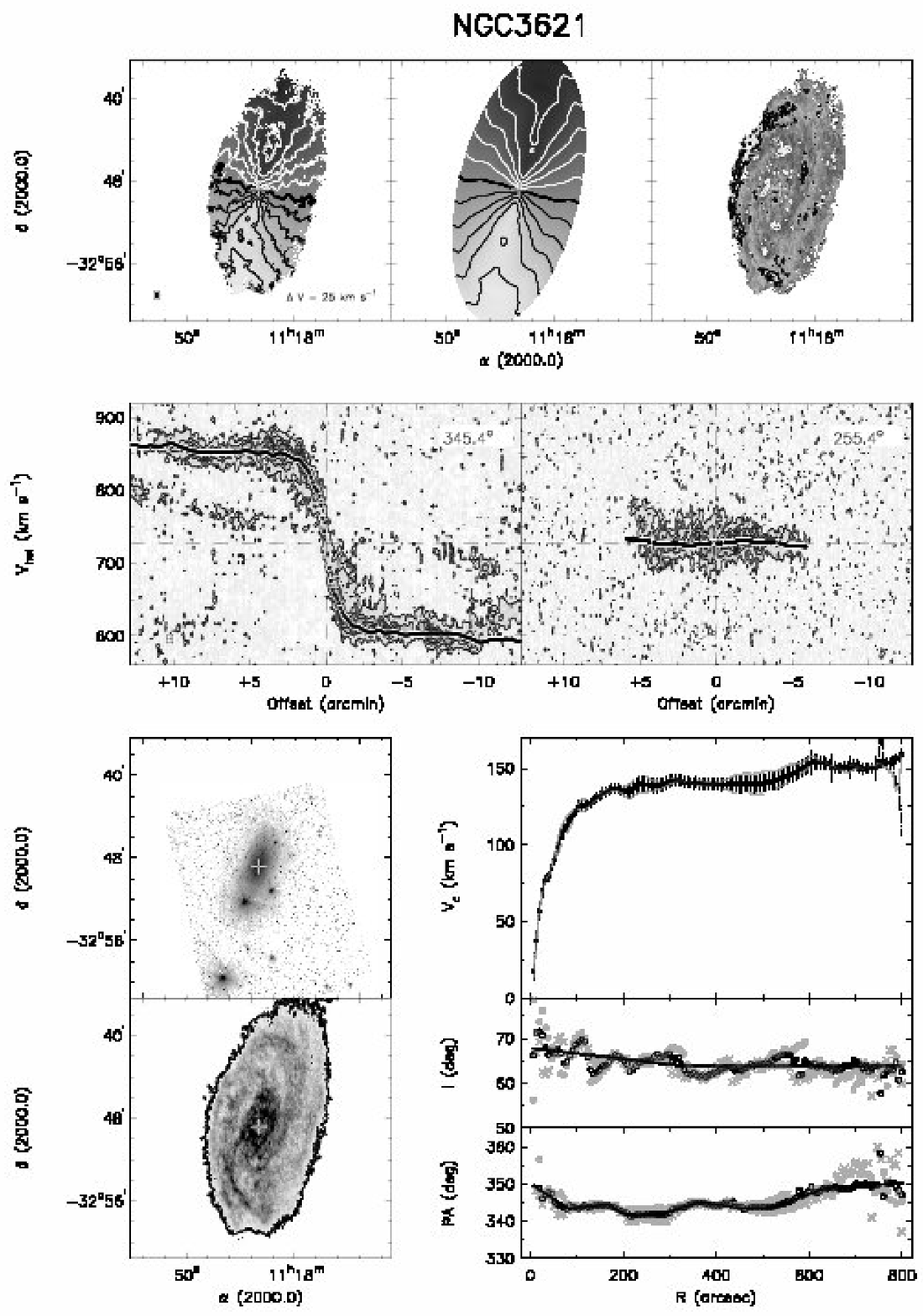} \figcaption{
Summary panel for NGC 3621. See the Appendix for more information.
\label{fig:n3621}}
\end{figure*}

\begin{figure*}[t]
\epsfxsize=0.95\hsize \epsfbox{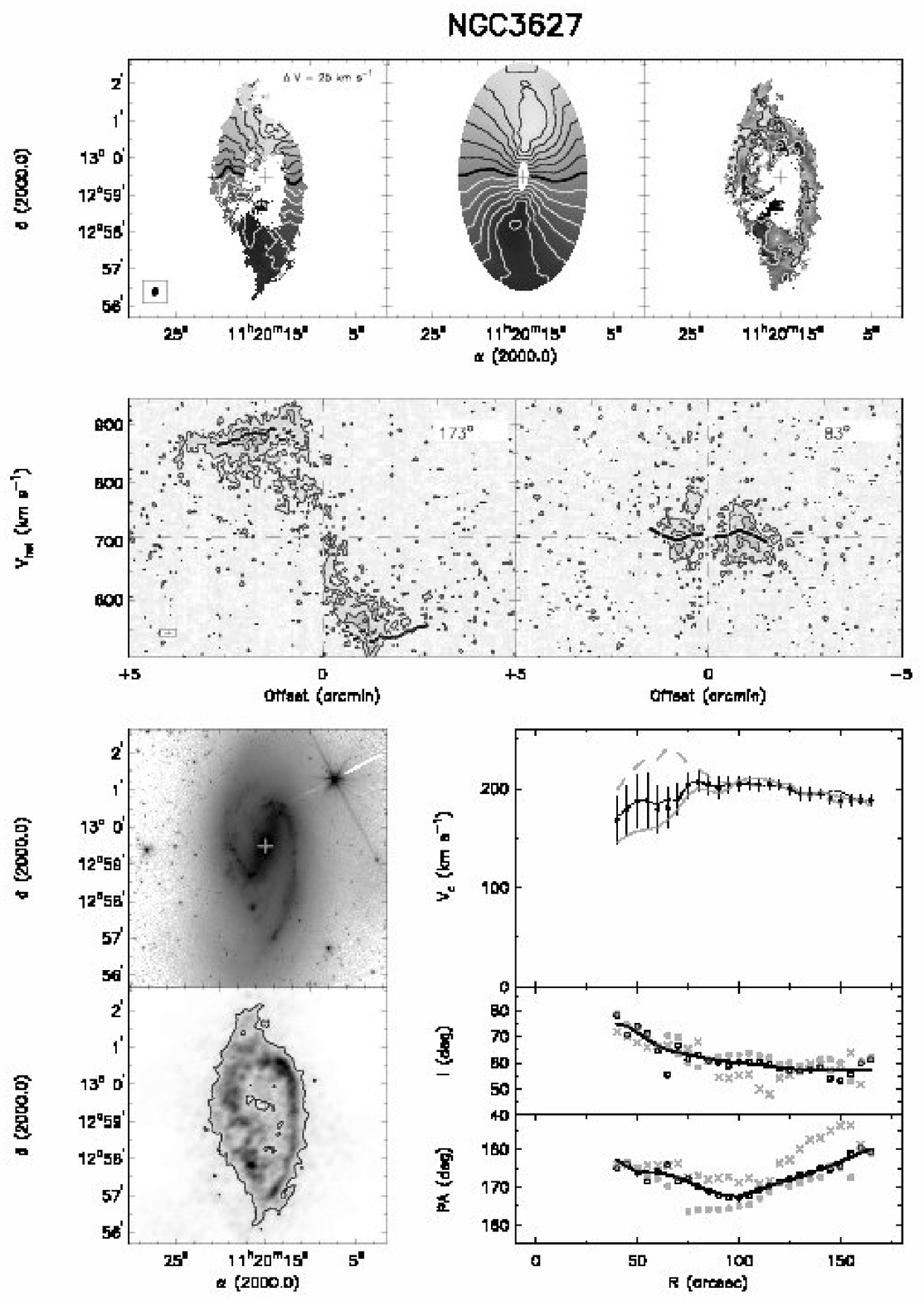} \figcaption{
Summary panel for NGC 3627. See the Appendix for more information.
\label{fig:n3627}}
\end{figure*}

\begin{figure*}[t]
\epsfxsize=0.95\hsize \epsfbox{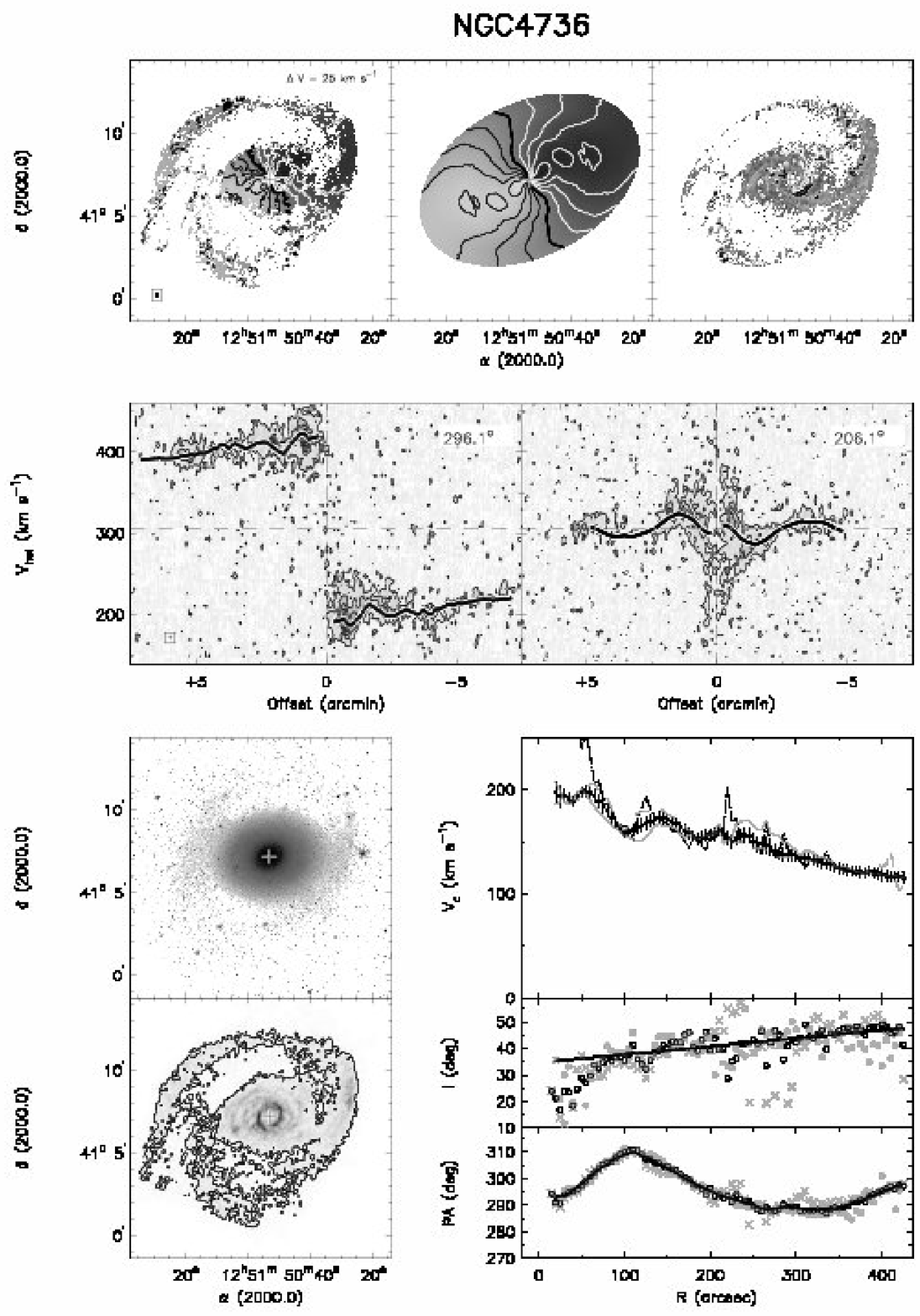} \figcaption{
Summary panel for NGC 4736. See the Appendix for more information.
\label{fig:n4736}}
\end{figure*}

\begin{figure*}[t]
\epsfxsize=0.95\hsize \epsfbox{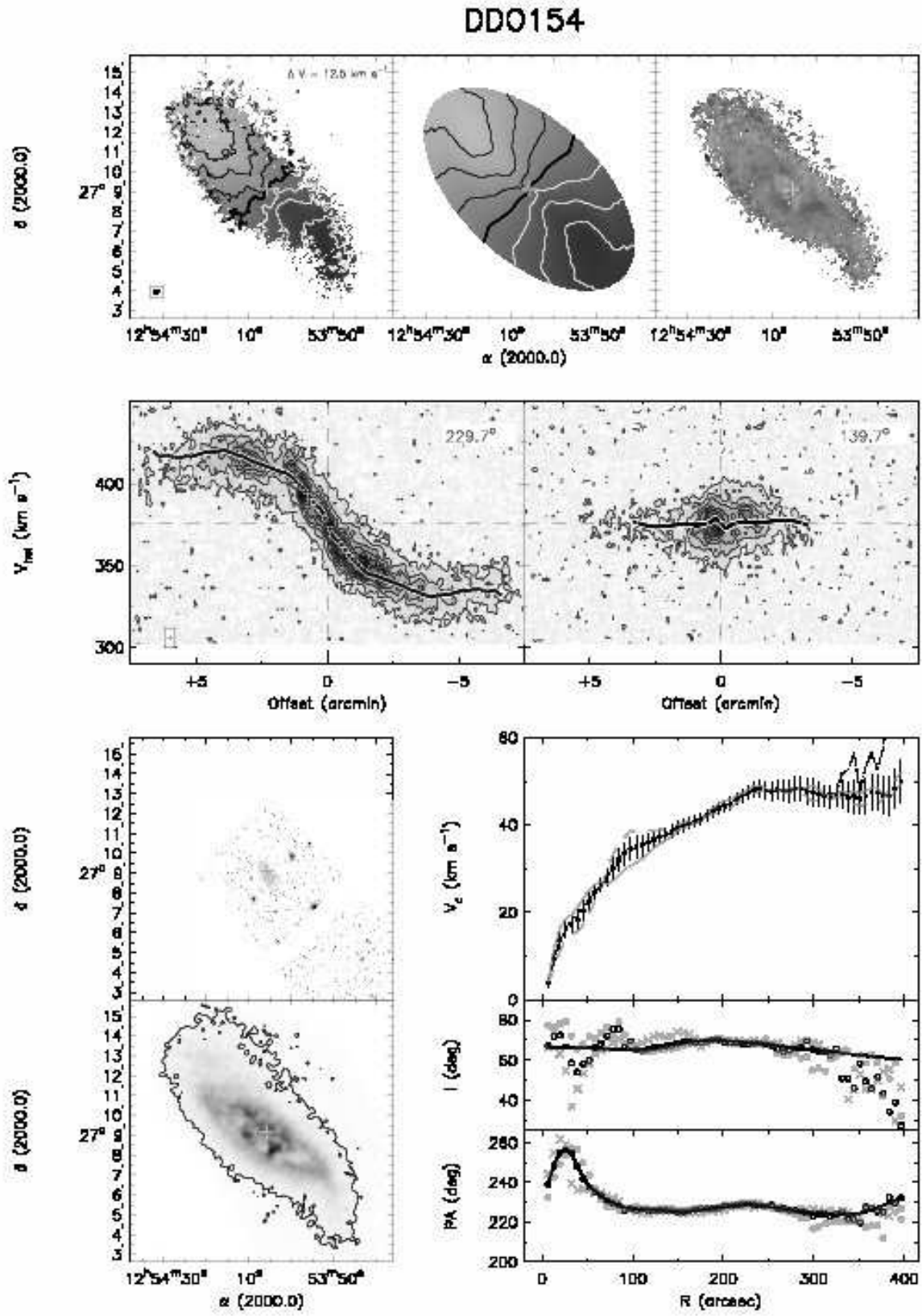} \figcaption{
Summary panel for DDO 154. See the Appendix for more information.
\label{fig:ddo154}}
\end{figure*}

\begin{figure*}[t]
\epsfxsize=0.95\hsize \epsfbox{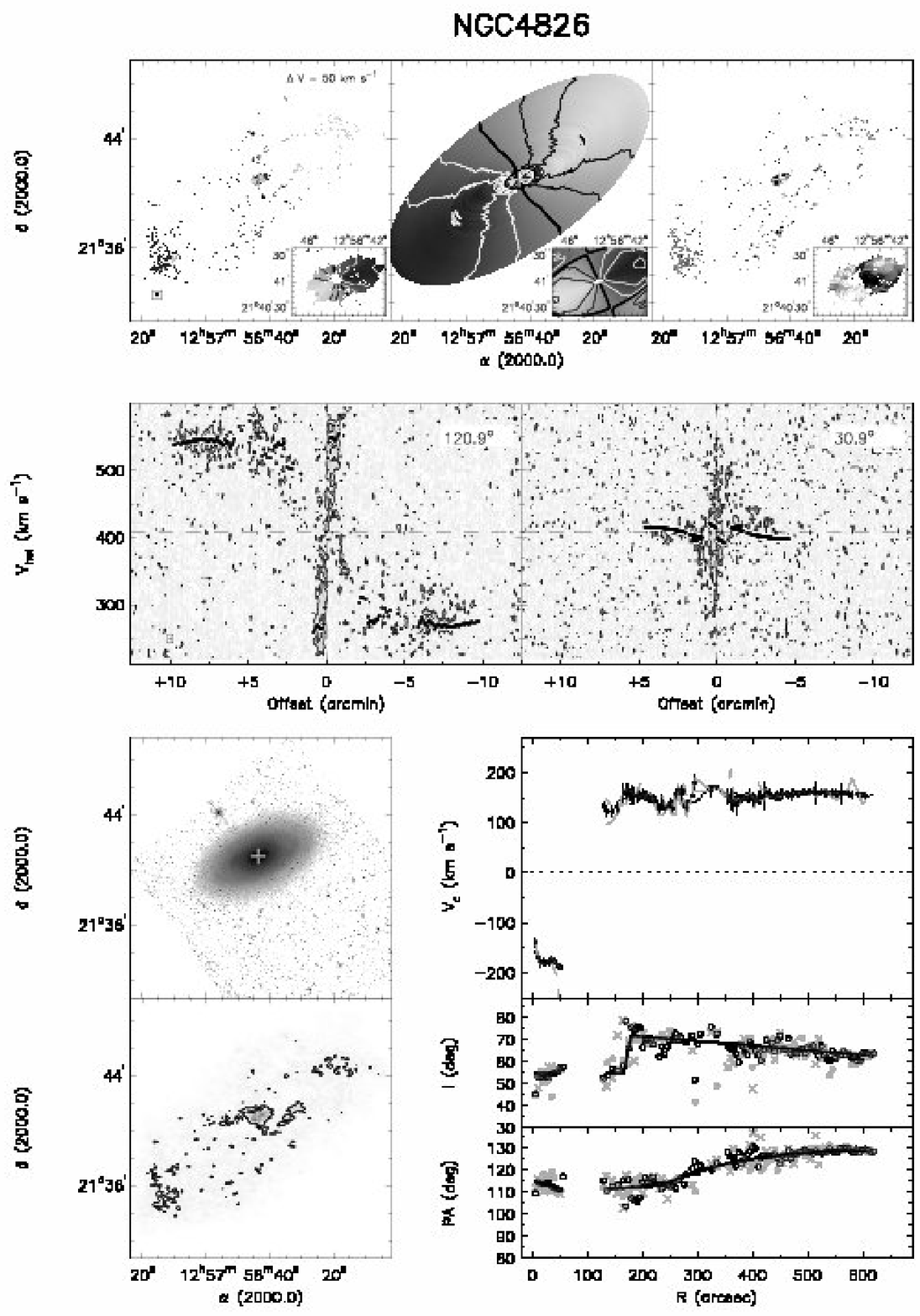} \figcaption{
Summary panel for NGC 4826. See the Appendix for more information.
\label{fig:n4826}}
\end{figure*}

\begin{figure*}[t]
\epsfxsize=0.95\hsize \epsfbox{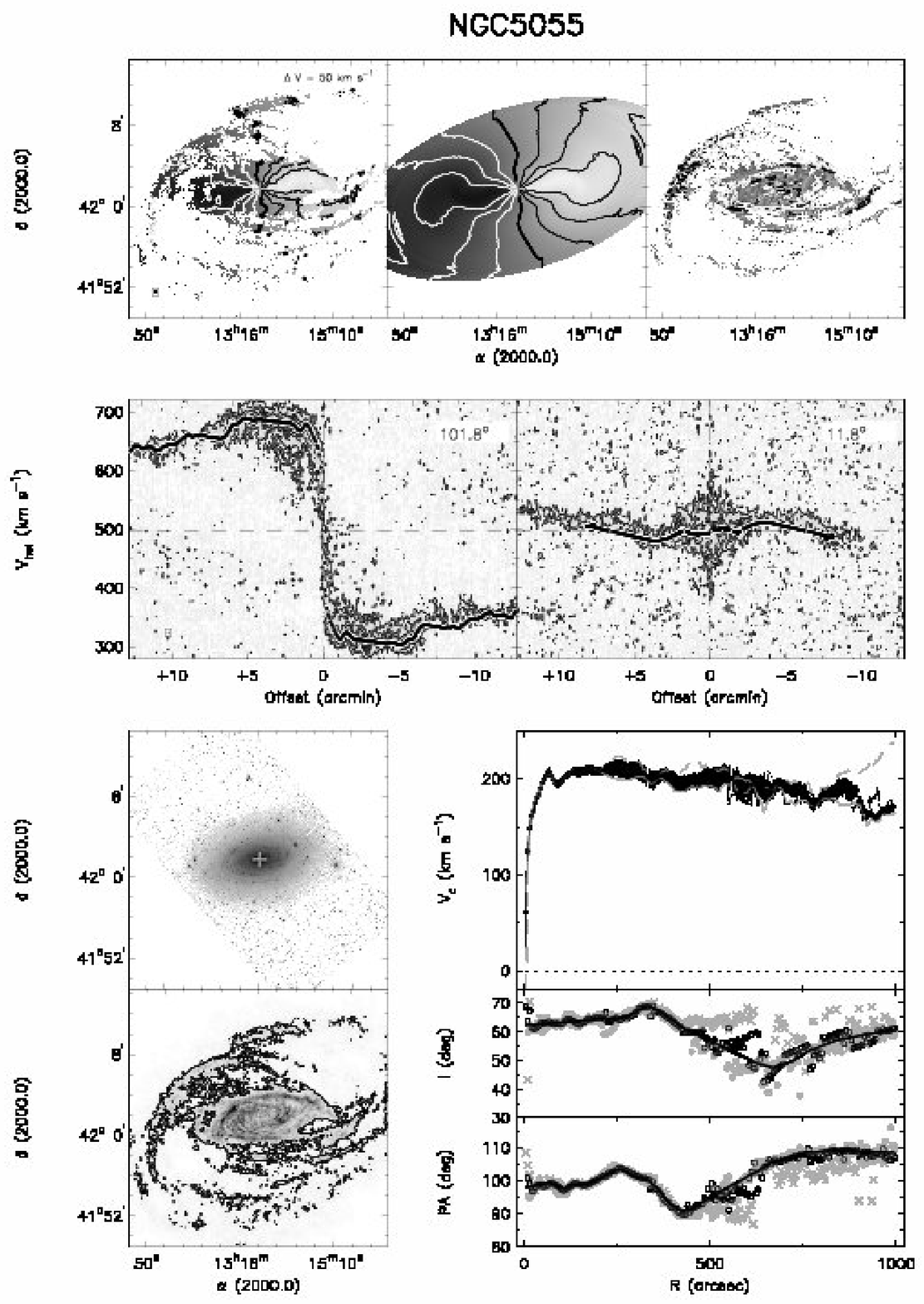} \figcaption{
Summary panel for NGC 5055. See the Appendix for more information.
\label{fig:n5055}}
\end{figure*}

\begin{figure*}[t]
\epsfxsize=0.95\hsize \epsfbox{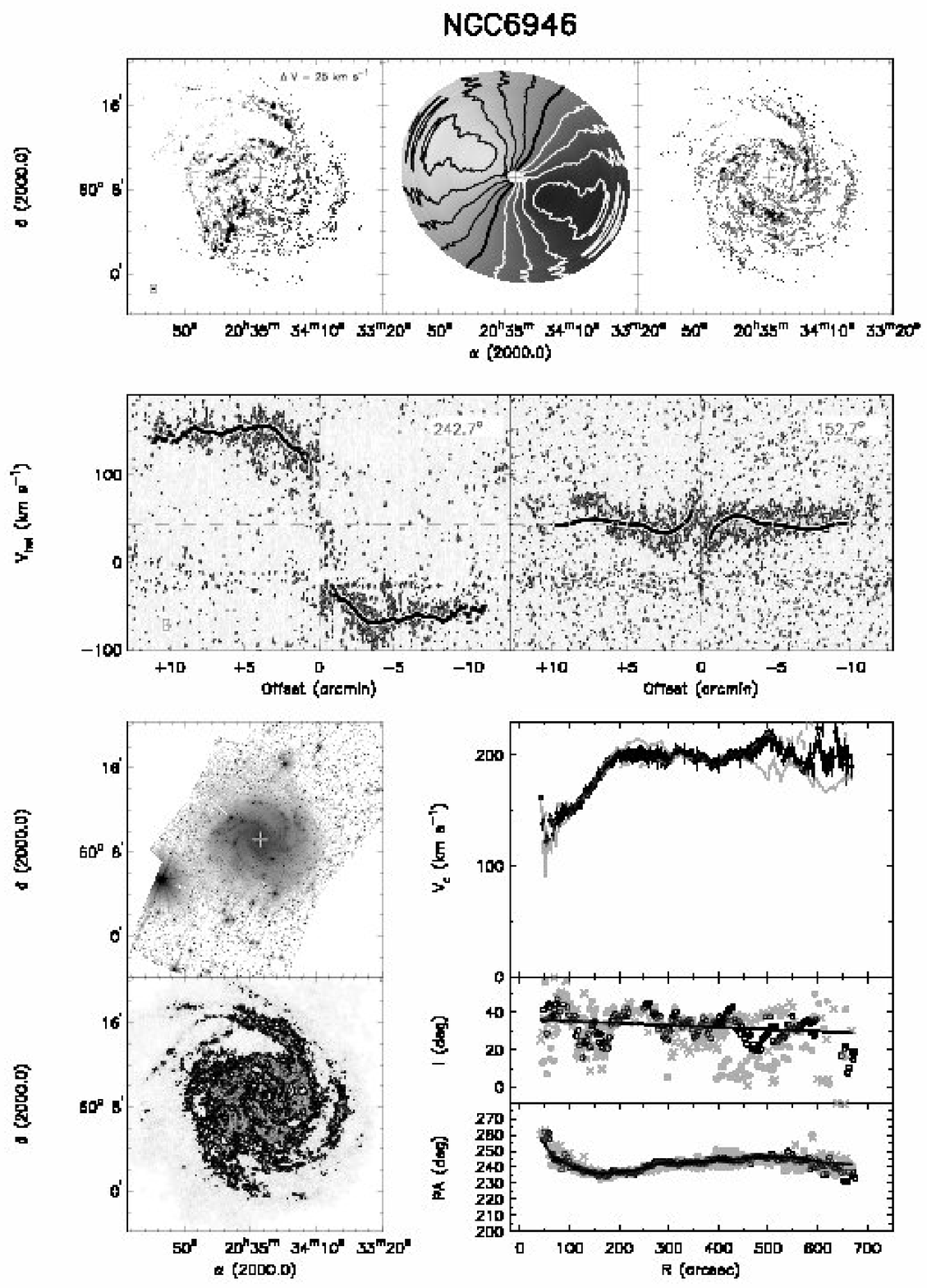} \figcaption{
Summary panel for NGC 6946. See the Appendix for more information.
\label{fig:n6946}}
\end{figure*}

\begin{figure*}[t]
\epsfxsize=0.95\hsize \epsfbox{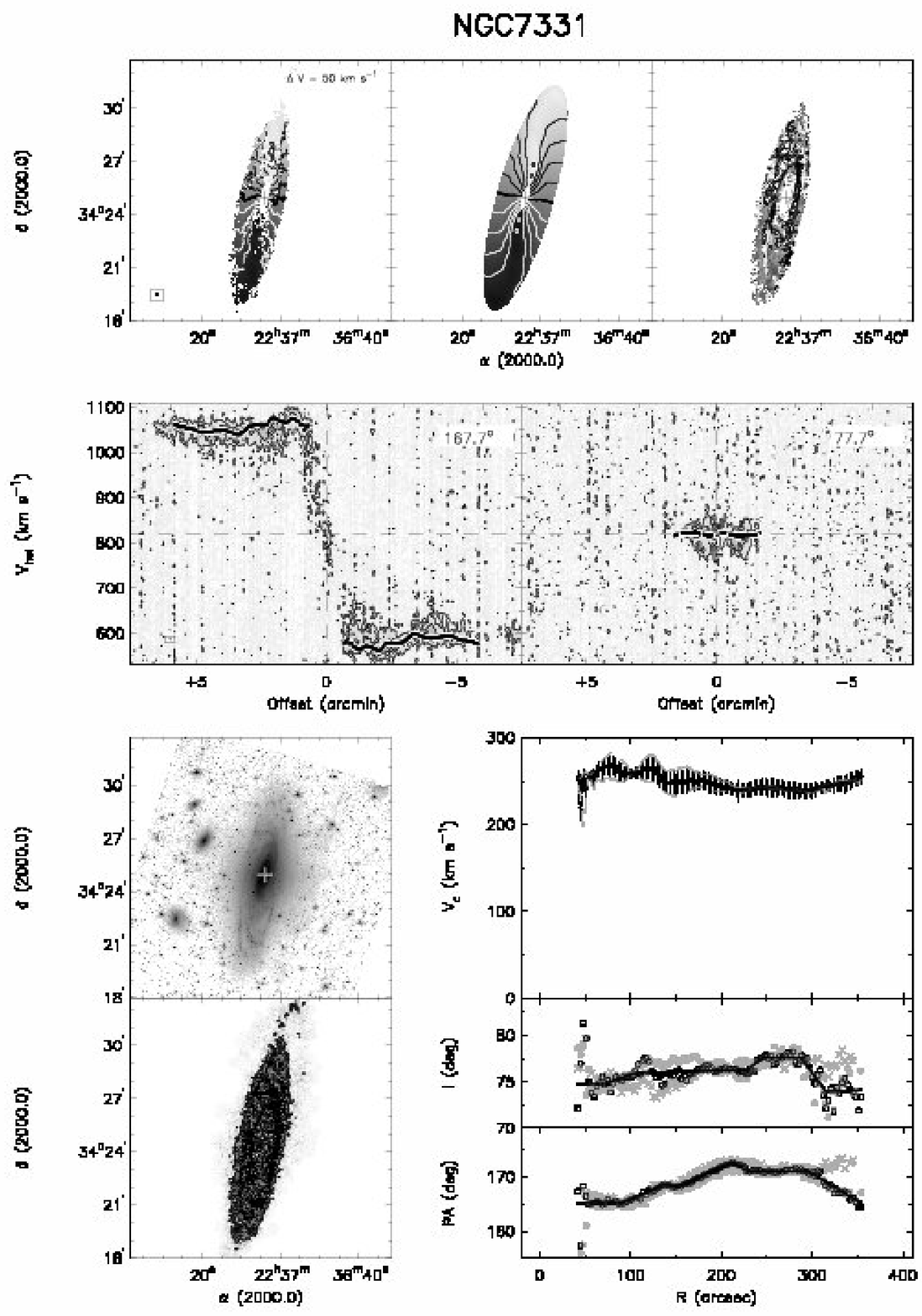} \figcaption{
Summary panel for NGC 7331. See the Appendix for more information.
\label{fig:n7331}}
\end{figure*}

\begin{figure*}[t]
\epsfxsize=0.95\hsize \epsfbox{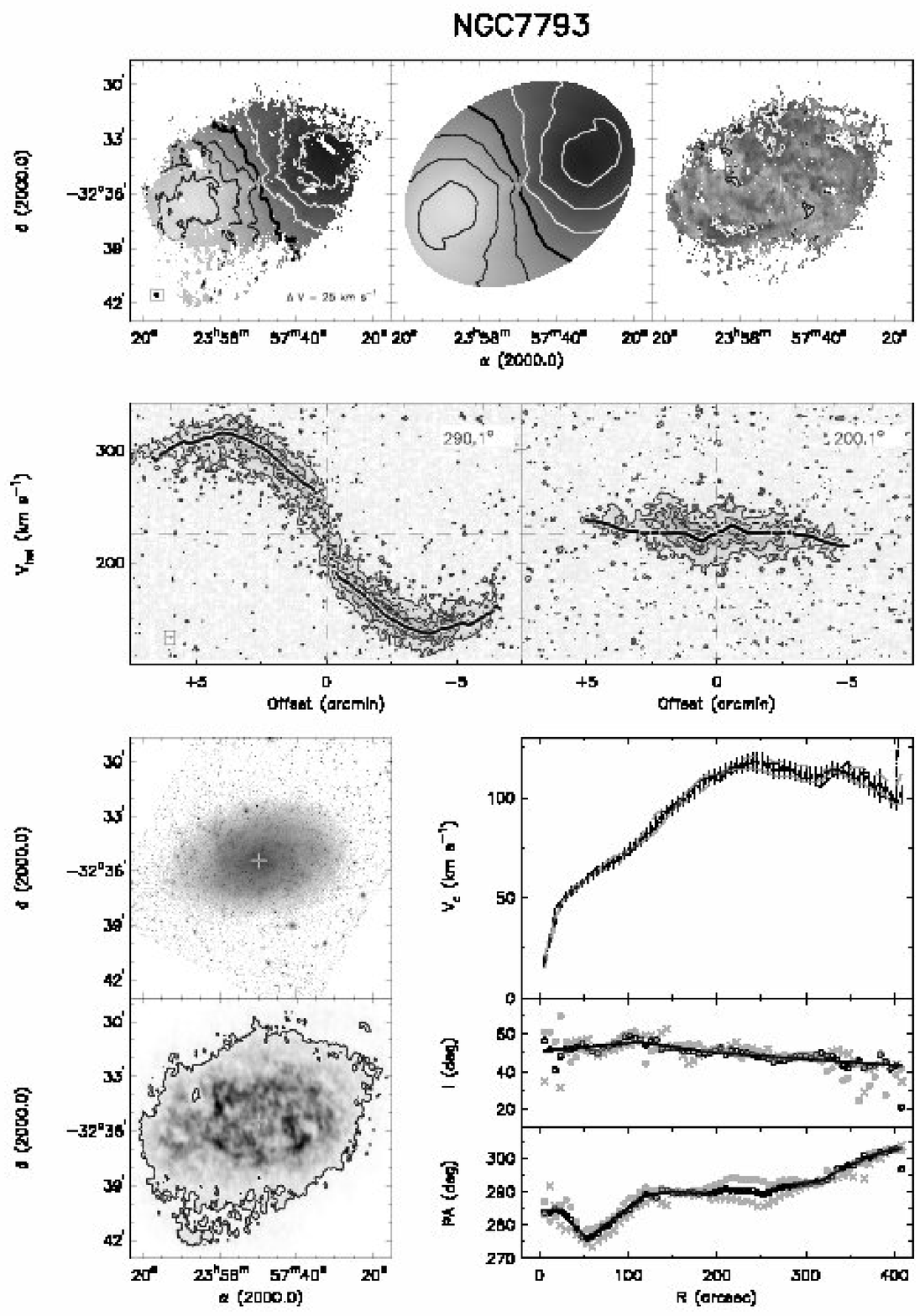} \figcaption{
Summary panel for NGC 7793. See the Appendix for more information.
\label{fig:n7793}}
\end{figure*}

The rotation curves, velocity fields and other supplementary
information are shown in Figs.~\ref{fig:n925}-\ref{fig:n7793}.
The following is a general description of the contents of these figures.

{\bf Top row:} \emph{Left:} Velocity field derived from fitting Hermite polynomials
to the natural-weighted data cube. See Sect.\
\ref{sect:velfi} for a description.  The systemic velocity is
indicated by the thick contour. The spacing $\Delta V$ between
velocity contours is indicated in the figure. The approaching side can
be identified by the light gray-scales and the dark contours. The
receding side can be identified by the dark gray-scales and the white
contours.  The adopted dynamical center is indicated with a cross. The
beam size is indicated by the ellipse enclosed by the rectangle in the
bottom-left corner. \emph{Center:} Model velocity field derived from
the tilted-ring model. Gray-scales and contour levels are
identical to those in the observed velocity field in the left
panel. \emph{Right:} Residual velocity field defined as the observed
velocity field minus the model velocity field. The gray-scale range
runs from $-20$ \kms\ (white) to $+20$ \kms\ (black). Contour levels
are $-10, -20, -30, ...$ \kms\ (black) and $+10, +20, +30, ...$ \kms\
(white).

{\bf Centre row:} \emph{Left:} Position-velocity diagram taken along
the average PA of the major axis as listed in Table~\ref{bigtable}.
This PA is also indicated in the top-left of the panel.  The thickness
of the slice equals one pixel (typically 1.5$''$, cf.\
\citealt{THINGS1}) in the corresponding cube. Contours start at
$+2\sigma$ in steps of $4\sigma$ (full contours), and $-2\sigma$ in
steps of $-4\sigma$ (dotted contours).  The systemic velocity and
position of the center are indicated by dashed lines. Over-plotted is
the rotation curve projected onto the average major axis using the
derived radial variations of PA and $i$.  The spatial and velocity
resolutions are indicated by the cross enclosed by the rectangle in
the bottom-left corner. \emph{Right:} Position-velocity diagram taken
along the average minor axis. Contours and symbols are as in the major
axis diagram. Over-plotted is again the rotation curve but projected
onto the average minor axis using the derived radial variations of PA
and $i$.

{\bf Bottom-left panel:} \emph{Top:} The 3.6 $\mu$m IRAC Spitzer
image.  For all galaxies the same logarithmic intensity scale was
used, running from $\log (I/({\rm{MJy\, ster}}^{-1})) = -2$ (white) to
$\log (I/({\rm{MJy\, ster}}^{-1})) = +1$ (black). The dynamical center
is indicated by a cross. \emph{Bottom:} Integrated \HI\ map. The
dynamical center is indicated by a cross. The maximum column density
level displayed (black) is $2 \cdot 10^{21}$ cm$^{-2}$. The contour
indicates the  $3\sigma$ level. This level was computed as
follows: for the standard THINGS cubes $\sigma_{\rm tot} =
\sqrt{N}\sigma_{\rm chan}$, where $\sigma_{\rm tot}$ is the noise in the
integrated \HI\ map, $\sigma_{\rm chan}$ the noise in one channel, and
$N$ the number of channels contributing to a pixel. 

{\bf Bottom-right panel:} \emph{Top:} The rotation curve corresponding
to the tilted-ring model is represented by the black filled
circles. The error-bars correspond to the dispersion of the velocity
values found along the corresponding ring. This rotation curve was
derived with all parameters fixed to their final values (as indicated
by the thick black curves in the PA and inclination panels below). The
full drawn gray line shows the corresponding rotation curve of the
approaching side derived using these adopted distributions of
inclination and PA. The dashed gray line shows the equivalent rotation
curve of the receding side. The dotted black curve indicates the
rotation curve of the entire disk derived with PA and inclination as
free parameters. The uncertainties in the rotation velocity are
defined as the quadratic addition of the dispersion in velocities
found along each ring and one quarter of the difference between the
approaching and receding sides velocity values. See Sect.~3.5 for a
full discussion.
 \emph{Center and bottom:} Inclination and PA values
used in the tilted-ring models. The open circles show the distribution
of inclination and PA when left as free parameters.  These values
result in the dotted black rotation curve described above.  The
crosses show same for the approaching side, the gray filled circles
for the receding side.  The thick black lines indicate the
distributions for PA and inclination that were ultimately adopted to
derive the rotation curves in the panel above.

We now discuss details of the derivation of tilted-ring models of
individual galaxies.

{\bf NGC 925:} The results for NGC 925 are shown in
Fig.~\ref{fig:n925}.  NGC 925 is classified as a late-type barred
spiral.  A small weak central bar-shaped component is indeed visible
in the IRAC 3.6 $\mu$m image. The outer parts are dominated by a
two-armed spiral, with the southern arm also prominently visible in
\HI.
With the dynamical center fixed, we find that $V_{\rm sys}$ varies
between $\sim 555$ \kms\ in the center to $\sim 545$ \kms\ at $R\sim
120''$. At larger radii $V_{\rm sys}$ settles at a constant value of
$546.6 \pm 0.6$ \kms. This value is very close to the average value
determined over the entire radial range ($546.3 \pm 3.9$ \kms), and we
adopt the latter value as the systemic velocity. This value is slightly
lower than the central value of the global profile of $552.5$
$(551.5)$ \kms\ as measured at the 20 (50) percent level.
The PA distribution is continuous and well-defined. The inclination
shows some scatter, but a global trend from $i \sim 75^{\circ}$ in the
inner parts to $i \sim 60^{\circ}$ in the outer parts is clearly
visible, indicating a slight warp in NGC 925.
The rotation curve derived using the PA and inclination models is
almost identical to the one with PA and $i$ as free parameters.
Despite the asymmetric appearance of the \HI\ disk, the approaching
and receding sides are fairly symmetric in terms of their rotation
velocity.

{\bf NGC 2366:} The data and analysis of NGC 2366 are given in
\citet{oh2007}. For completeness, we summarize and the data and
analysis in Fig.~\ref{fig:n2366}, but refer refer to their paper for a
complete discussion.

{\bf NGC 2403:} NGC 2403 is a late-type Sc spiral and member of the
M81 group.  We present the data in Fig.~\ref{fig:n2403}.  The IRAC 3.6
$\mu$m image shows multiple spiral arms, some of which are also
evident in the integrated \HI\ map.  The velocity field of NGC 2403 is
very regular as shown by the perpendicular major and minor axes and
the close agreement between approaching and receding sides. The small
wiggles in the inclination distribution are caused by streaming
motions induced by the spiral arms, also visible as residuals in the
residual velocity field.
After fixing the position of the dynamical center, we find the
systemic velocity to be well-behaved, showing no large-scale trend
with radius. The average value is $V_{\rm sys}=132.8 \pm 1.6$ \kms,
which is the value we adopt here. This value is close to the central
velocity of 133.1 (134.8) \kms\ derived from the global profile at the
20 (50) percent level.  The inclination and position angle are
well-behaved. A comparison of the curve with PA and $i$ as free
parameters with the curve which has these parameters fixed using the
model distributions shows no appreciable differences.  Note the
remarkable symmetry between approaching and receding sides in rotation
velocity, inclination and position angle between $450'' \la R \la
850''$.

{\bf NGC 2841:} NGC 2841 is an early-type (Sb) spiral which is
dominated by a prominent central bulge, as well as a central hole in
the \HI\ distribution, as shown in Fig.~\ref{fig:n2841}.  The
tightly-wound spiral structure is also visible in the inner part of
the \HI\ distribution; in the outer parts the disk is dominated by two
arms which seem to bend away out of the plane of the inner disk.  The
velocity field is regular in the inner parts but in the outer regions
shows clear evidence for the presence of a warp. The major-axis slice
shows that at the outer edge of the central \HI\ hole the rotation has
already almost reached its maximum value.
After fixing the central position, we find a systemic velocity that is
constant within $R<350''$, but starts to deviate at larger radii. Due
to the obvious presence of the warp at these larger radii we adopt the
average value within $R= 350''$ as our best value: $V_{\rm sys} =
633.7 \pm 1.8$ \kms. This agrees well with the central velocity
derived from the global profile at the 20 (50) percent level of 635.2
(636.2) \kms.
With the systemic velocity and dynamical center fixed, we find
well-defined distributions of PA and $i$ with radius. Despite the
warp, NGC 2841 is very symmetrical with no obvious differences in PA
and $i$ for the approaching and receding sides. Our best model for the
inclination shows an approximately constant value of $i \sim
70^{\circ}$ for $R \la 200''$, with a gradual but systematic increase
to $i \sim 78^{\circ}$ in the outer parts.  The rotation curve with PA
and $i$ as free parameter is identical (within the uncertainties) to
the one derived with PA and $i$ fixed.

{\bf NGC 2903:} As can be seen in Fig.~\ref{fig:n2903}, the appearance
of NGC 2903 is dominated by tightly-wound spiral arms and a bar as
visible in the $3.6\ \mu$m image.  The velocity field and integrated
\HI\ map can be divided into two distinct regions: in the inner parts
we find an \HI\ ring, corresponding to the bar and inner spiral arms
as seen in the IRAC image. At the outer radii we see a lower surface
density disk which is dominated by two broad \HI\ spiral arms. These
two regions are clearly separated around $R\sim 240''$ by minima in
the \HI\ column density.  The kinks in the velocity contours near the
outer spiral arms clearly indicate the presence of non-circular
motions there. The very inner parts of the disk also show a strong
twist in the PA of the kinematical major axis, very likely due to the
effect of the bar and associated streaming motions. Their presence is
also implied by the motions that can be seen in the minor axis profile
and the increased width of the profiles in the inner parts (visible in
the major axis position-velocity diagram).  With the position of the
dynamical center fixed, the systemic velocity still shows large
changes in the inner parts, most likely due to the effects of the bar.
We fixed $V_{\rm sys}$ to the average value between $200'' < R <
400''$ and find $V_{\rm sys} = 555.6 \pm 1.3$ \kms.  This agrees well
with the central velocity from the global profile derived at the 20
(50) percent level of 556.6 (555.7) \kms.  The inclination and PA for
$R \ga 100''$ are straightforward to model, but at smaller radii both
show large variations. The sudden change in inclination at $R\sim
110''$ corresponds to the outer radius of the bright, compact spiral
arms associated with the central bar, whereas the sudden change in PA
at $R \sim 60''$ occurs where the bright tips of the straight central
bar are found.  As both of these changes are associated with
non-rotational motions, we have for our tilted-ring models simply
extrapolated the innermost unaffected values inwards.

{\bf NGC 2976:} NGC 2976 is classified as an Sc galaxy. Its analysis
is illustrated in Fig.~\ref{fig:n2976}.  The IRAC 3.6 $\mu$m image
shows no sign of spiral arms or bar component.  The stellar component
of the galaxy is characterized by two regions of enhanced star
formation at either end of the disk, which correspond to the two
density enhancements seen in \HI.
Having fixed the dynamical center, we do not detect systematic
trends of $V_{\rm sys}$ with radius except for a slightly larger
scatter in the innermost few points. We therefore fix it to the
average value for $R>20''$ and find $V_{\rm sys} = 1.1 \pm 1.3$ \kms,
which compares well with the central value of 2.6 (4.2) \kms\ derived
from the global profile at the 20 (50) percent level.
The trend of PA with radius is well defined, and can can be
approximated by a constant PA for $R>70''$, with a modest increase
within that radius.  The inclination shows increased scatter within
$R<70''$, and we fix it to the average value derived for
$80''<R<110''$.  The final rotation curve agrees well with the curve
derived with PA and $i$ free.  The residual field also shows no large
residuals, indicating that circular motions describe the dynamics of
this galaxy well.

{\bf NGC 3031:} The analysis of NGC 3031 in illustrated in Fig.~\ref{fig:n3031}.
The IRAC 3.6 $\mu$m image
as well as the \HI\ total intensity map show two well-defined spiral
arms. The \HI\ observations used here consist of a mosaic of two
separate pointings. Because of this, it was necessary to use a
primary-beam corrected data cube.  These corrections cause an apparent
increase of the noise towards the edges of the field, apparent in the
major- and minor-axis slices. Careful blanking of these regions
ensured that they did not affect the analysis presented here.  Also
note that due to the very large velocity width of the gas in the M81
system it was not possible to define line-free channels in order to
subtract the continuum (see \citealt{THINGS1}).  The total
intensity \HI\ map, as well as the position velocity slices therefore
still contain the radio continuum emission. The central radio
continuum source can be seen as the vertical high-intensity feature in the
position-velocity slices.
An inspection of the velocity field shows that M81 has clear
non-circular motions associated with the prominent spiral arms.  The
residual velocity field clearly shows the effects of these streaming
motions.  It is also clear that beyond $R \ga 800''$ the \HI\ gas is
no longer in regular rotation around M81. At these large radii the
motion of the gas is starting to become dominated by the tidal
interaction processes within the group. For this reason we do not
attempt to derive the rotation curve beyond that radius. More evidence
for these tidal effects is evident in the minor axis slice where we
see a sudden and large deviation of the gas from the systemic
velocity. This corresponds to the strong twist in the systemic
velocity contour seen in the eastern part of the outer velocity field.
After fixing the position of the dynamical center, we determine the
systemic velocity by first fixing $i$ and PA to indicative values of
$60^{\circ}$ and ${330^{\circ}}$, respectively. We find $V_{\rm sys}$
to be constant within $R<350''$ and use the average value of $-39.4
\pm 2.8$ \kms. This agrees with the central value of the global
\HI\ profile of $-39.4\ (-41.3)$ \kms\ at the 20 (50) percent level.
After fixing $V_{\rm sys}$, and re-running {\sc rotcur} we find
well-defined distributions of $i$ and PA. The radius of $\sim 400''$
at which the sudden change in inclination occurs corresponds to the
outer edge of the innermost \HI\ ``ring'' visible in the integrated
\HI\ map. From the PA distribution, it is clear that the approaching
and receding sides are not entirely symmetric. The rotation curve
itself, by contrast, \emph{is}.  The curve shows a very pronounced
drop in the outer parts; however given the position of M81 within an
interacting system, this very likely carries no implication for the
distribution of the dark matter.

{\bf NGC 3198:} The analysis of NGC 3031 in illustrated in
Fig.~\ref{fig:n3198}.  The IRAC 3.6 $\mu$m image shows two
well-defined strong spiral arms, with several less prominent arms
branching off from the two main ones.  The velocity field is regular
with indications of a very modest warp. Small streaming motions due to
the two dominant spiral arms show up as small kinks in the velocity
contours. These are also visible in the residual velocity field.  With
the position of the center fixed, we find a well-behaved systemic
velocity with an average value $V_{\rm sys} = 660.7 \pm 2.6$ \kms,
which compares favorably with the central velocity of $661.2$
$(661.7)$ \kms\ as derived from the global profile at the 20 (50)
percent level.
The PA is well-defined for all radii; for $R>300''$, the distribution
of $i$ is also well-defined, and shows a small but systematic
increase. For $R<300''$ the inclination shows some variations due to
motions along the spiral arms (cf.\ the low-amplitude pattern visible
in the residual velocity field). We therefore adopt an almost constant
inclination for $R<300''$. Note that the rotation curve does not
depend on these assumptions: the curve with $i$ and PA as free
parameters equals the final curve almost point by point. The
approaching and receding sides are very symmetrical.

{\bf IC 2574:} The data and analysis of IC 2574 are given in
\citet{oh2007}. For completeness, we show a summary of their results
in Fig.~\ref{fig:ic2574}, but refer to their paper for a complete
discussion.

{\bf NGC 3521: } NGC 3521 is a disk galaxy showing a flocculent spiral
structure. The inner \HI\ disk seems to be in regular rotation, but
the outer disk shows a much different behavior. This is most clearly
visible in the channel maps \citep[see][]{THINGS1}, but can also be
seen in the major-axis position-velocity slice, as shown in
Fig.~\ref{fig:n3521}. The central velocity derived from the
symmetrical global profile is $798.2$ (798.6) \kms\ at the 20 (50)
percent level.  After fixing the position of the dynamical center, the
systemic velocity shows a systematic trend with radius and the largest
radial variation observed within the THINGS sample.  It is constant at
a value of $\sim 800$ \kms\ for $R \la 250''$ (8 kpc) and $R\ga 450''$
(15 kpc), but drops to a value of $\sim 777$ \kms\ at intermediate
radii.  Due to the good agreement between $V_{\rm sys}$ values found
in the inner and outer parts, we fix $V_{\rm sys}$ to the average
value for $R<250''$ and $R>450''$ and get $V_{\rm sys} = 803.5 \pm
4.5$ \kms.  This choice for $V_{\rm sys}$ leads to a rotation curve
which shows a large difference between approaching and receding sides
at the intermediate radii.

{\bf NGC 3621:} NGC 3621 is a late-type spiral with an
\HI\ distribution that is characterized by a regularly rotating disk
and a warp in the line of sight (see the channel maps presented in
\citealt{THINGS1}). This feature is also visible in the major-axis
position-velocity diagram, as shown on Fig.~\ref{fig:n3621}. To derive
the velocity field of the disk component only, we blanked all emission
associated with the warp prior to deriving the velocity field.  This
velocity field and the associated integrated \HI\ map presented here,
therefore only show the disk component. For the full velocity field
and \HI\ map see \citet{THINGS1}.  After fixing the position of the
center, we find $V_{\rm sys}$ to be on average constant with
radius. Small variations with a maximum amplitude of $\sim 5$
\kms\ are visible, but show no systematic trends with radius. We adopt
the average value over the entire radial range as our best value:
$V_{\rm sys} = 728.5 \pm 2.7$ \kms. This is consistent with the
central value of $730.1$ $(728.8)$ \kms\ as derived from the global
\HI\ profile at the 20 (50) percent level.  The PA is well-defined and
only varies slightly with radius. The inclination shows significant
variations, but in general decreases with radius.

{\bf NGC 3627:} NGC 3627, or M66, is a barred, interacting spiral
galaxy that is part of the Leo Triplet. Our analysis of this galaxy is
illustrated in Fig.~\ref{fig:n3627}.  The IRAC 3.6 $\mu$m image shows
a pronounced bar, with a two-armed, asymmetric spiral structure. The
western arm is also clearly visible in the integrated \HI\ map. The
eastern arm is confused with other high-column density features.  The
pronounced spiral arms and asymmetry of NGC 3627 do not make this a
prime candidate for detailed studies that assume azimuthal symmetry.
In the central parts the \HI\ column density lies below the 3$\sigma$
level that we applied during the profile fitting, hence the absence of
velocity information there.
After fixing the position of the dynamical center, we fix PA and $i$
to indicative values of $175^{\circ}$ and $60^{\circ}$, respectively, and
find  $V_{\rm sys}$ values that vary slightly within $R\sim 100''$, but
are constant outside that radius.  We use the mean value for $R>100''$
and find $V_{\rm sys} = 708.2 \pm 1.1$ \kms. This differs somewhat
from the 20 (50) percent level central value derived from the global
\HI\ profile of $717.3$ (720.3) \kms, but given the asymmetry of this
galaxy this should not come as a surprise. Trends of $i$ and PA with
radius are fairly straightforward to determine with both parameters
showing modest variations. The approaching and receding sides show
different radial trends, as might have been expected from the obvious
asymmetry visible in the IRAC image.  Nevertheless, the respective
rotation velocities converge in the outer parts, leading to a
well-determined flat outer rotation curve. 

{\bf NGC 4736:} NGC 4736 is dominated by an inner high surface
brightness disk in the IRAC 3.6 $\mu$m image as shown in
Fig.~\ref{fig:n4736}. Prominent arms are visible in the integrated
\HI\ map.  After fixing the position of the dynamical center, the
systemic velocity shows no systematic trends with radius and we adopt
the average value $V_{\rm sys} = 306.7 \pm 3.7$ \kms. This corresponds
well with the values derived from the global profile at the 20 (50)
percent level of $307.6$ $(306.6)$ \kms.
The PA is very well defined and constructing a model is
straight-forward.  The inclination shows a well-defined trend from
$\sim 45^{\circ}$ in the outer parts to $\sim 35^{\circ}$ at $R\sim
80''$. At this radius the slope of the inclination steepens, and the
value appears to decrease to $\sim 20^{\circ}$ in the center.  We
consider this break in the inclination slope not physical; the inner
region of NGC 4736 is known to be a region of non-circular motions and
the radius where the break occurs coincides with the radius of the
inner star formation ring (the ``expanding H{\sc ii} regions ring''
described in \citealt{munoz}). The minor-axis position-velocity
diagram also shows clear evidence for strong non-circular motions in
the inner arc-minute.  For our inclination model we simply extrapolate
inward the trend found for $R>80''$.
Despite the disturbed distribution of the \HI, the approaching and
receding sides of the final rotation curve are remarkably similar. The
velocity systematically decreases with radius: it peaks at $V
= 211$ \kms, and declines to $V=115$ \kms\ in the outermost parts.  As
with NGC 3627, it is clear from the morphology that the assumption of
azimuthal symmetry is not appropriate. \citet{clemens2007} show
that NGC 4736 has the second highest non-circular motions in the
current sample (after NGC 3627) and any interpretation of its rotation
curve should therefore be treated with care.

{\bf DDO 154:} Our analysis of DDO 154 is illustrated in
Fig.~\ref{fig:ddo154}.  DDO 154 has no central radio continuum source
or central light concentration to help pinpoint the dynamical center.
The position of the dynamical center was therefore derived purely
using the velocity field (see also \citealt{clemens2007}).  The
systemic velocity shows a stable behavior. Using the average value
within $R=260''$ we find $V_{\rm sys} = 375.9 \pm 1.4$ \kms.  This
corresponds very well with the central values derived from the global
\HI\ profile at the 20 (50) percent level of 375.5 (375.4) \kms.
The PA is straightforward to model. The inclination shows some
variation in the inner parts, mainly due to the small number of
independent pixels per ring; here we simply extrapolated the model
inwards from the smallest radius where $i$ was well-behaved. The outer
parts show an apparent steep drop in inclination. Most of the gas
contributing to these points is however found in the outermost
flocculent parts of the warp, and given the small filling factor of
the rings there, it is not clear how realistic this steep drop in
inclination is.  We therefore extrapolated our model outwards from the
largest radii with well-determined inclinations.

{\bf NGC 4826:} As can be seen in Fig.~\ref{fig:n4826}, the outer \HI\
disk of NGC 4826 has a very low column density -- one of the lowest in
the entire THINGS sample. Only a small fraction of the disk has
profiles with a peak value $>3\sigma$. Nevertheless, this small
filling factor, combined with the very regular kinematics in the outer
disk is still sufficient to derive a well-defined outer rotation
curve.  Small insets in Fig.~\ref{fig:n4826} show the region of the
counter-rotating disk, as well as the corresponding model and residual
velocity fields in more detail.
After fixing the position of the dynamical center, we find no
systematic trends of the systemic velocity with radius, and adopt the
average value: $V_{\rm sys} = 407.4 \pm 7.0$ \kms. This value
corresponds very well with the central value of $407.9$ \kms\ as
derived from the global \HI\ profile for both the 20 and 50 percent
level.  The inclination and PA show clear trends with radius. A good
description for the inclination is a constant value of $\sim
55^{\circ}$ in the inner parts, followed by break and a sudden
increase to $\sim 70^{\circ}$ at $R\sim 180''$, in turn followed by a
gradual decline in the outer parts to a value of $\sim 63^{\circ}$.

{\bf NGC 5055: } As illustrated in Fig.~\ref{fig:n5055}, the
major-axis position-velocity diagram shows evidence for the presence
of gas at velocities lower than the local rotation velocity. The
structure of NGC 5055 as seen in the 3.6 $\mu$m image is regular, with
signs of well-defined flocculent spiral arms.  After fixing the
position of the dynamical center we find that the systemic velocity
shows a systematic variation, from a constant value of $\sim 496$
\kms\ within $R\la 200''$ to a value of $\sim 510$ \kms\ in the outer
parts. We make the assumption that the gas in the inner parts is
likely to be more tightly bound to the system than the tenuous outer
arms, and adopt the average value between $50'' < R < 180''$. We find
$V_{\rm sys} = 496.8 \pm 0.7$ \kms. The central velocity as derived
from the global profile at the 20 (50) percent level is 499.3 (497.4)
\kms.  The PA and $i$ values are well-defined for the inner \HI\
disk. The distributions for the approaching and receding sides of the
outer warped part differ significantly, leading to different
approaching and receding rotation curves for $R \ga 250''$.

{\bf NGC 6946: } Our analysis of NGC 6946 is illustrated in
Fig.~\ref{fig:n6946}.  For the position of the dynamical center we
adopt the position of the central radio continuum source.  The
systemic velocity is fairly constant with radius. We adopt the average
value within $500''$ of $V_{\rm sys} = 43.7 \pm 3.3$ \kms.  This
agrees well with the central value of 42.0 (45.1) \kms\ derived from
the 20 (50) percent level of the global \HI\ profile.  The PA is
well-behaved and straight-forward to describe.  The inclination, when
left free, shows a large scatter, showing the increased difficulty of
determining the inclination using only dynamical information at low
inclination values. Nevertheless, the large number of independent data
points allow us to accurately model it as a linear radial dependence.
The assumption we make is that the scatter around this average value
is caused by small non-circular motions that have a disproportionally
large influence due to the low inclination.  Due to this inclination
scatter we do not consider here the rotation curve derived using PA
and $i$ as free parameters.

{\bf NGC 7331:} We illustrate our analysis of NGC 7331 in
Fig.~\ref{fig:n7331}.  The velocity field, though regular on large
scales, shows many wiggles in the velocity contours, often coinciding
with the positions of the spiral arms.
After fixing the position of the dynamical center, we find that the
systemic velocity is well-defined over the range
$80''<R<280''$. Between $80''$ and $200''$ it varies systematically
from $\sim 830$ \kms\ to $\sim 805$ \kms, settling on a constant
value of $818.3 \pm 0.9$ \kms\ between $200'' < R < 280''$. We have
chosen the latter value as our $V_{\rm sys}$ value, comparing
favorably with the central velocity from the global profile of $815.7$
$(816.5)$ \kms\ as measured at the 20 (50) percent level.  The
varying $V_{\rm sys}$ value could indicate a kinematic lopsidedness
(e.g., \citealt{remco97,remco}). A harmonic analysis of the velocity field to
quantify this is presented in \citet{clemens2007}.
In the outer parts the PA values for approaching and receding sides
differ distinctly.  At $R\sim 150''$ the inclinations are different as
well, most likely due to streaming motions along the spiral
arms. These show up clearly in the residual velocity field. Also note
the large difference between the rotation curves of approaching and
receding sides within $R \la 200''$. The typical difference is $\sim
25$ \kms, and shows why $V_{\rm sys}$ is difficult to constrain over
this radial range.  The major axis position-velocity diagram shows
some \HI\ outlining the rising part of the rotation curve, but the
emission is too faint to fit a rotation curve.
NGC 7331 was previously observed in \HI\ by \citet{begeman87}, who
already noted many of the features we have described above.  The
dynamical center as determined by \citet{begeman87}, coincides with
ours to within the errors.  \citet{begeman87} also notes the same
varying $V_{\rm sys}$ in the inner parts. He determines $V_{\rm sys}$
using only points at $R>210''$ and finds a value $820.1 \pm 2.7$ \kms.
For the PA a trend very similar to ours was found. The trend we find
in the inclination distribution is also present in the
\citet{begeman87} data, but with the much smaller number of resolution
elements in that data set, a constant inclination model was chosen
there.

{\bf NGC 7793: } As shown in Fig.~\ref{fig:n7793}, the velocity field
of NGC 7793 is well-defined with indications of a small-amplitude
warp.  The major-axis slice shows little evidence for extensive
non-circular motions. The residuals are small, and prominent in only a
few small areas in the outermost part of the disk.  After fixing the
position of the dynamical center we find the systemic velocity to be
well-defined: we adopt the average value between $40'' < R < 250''$
and find $V_{\rm sys} = 226.2 \pm 1.2$ \kms, agreeing well with a
central velocity of $227.2$ $(226.6)$ \kms\ as derived from the global
profile at the 20 (50) percent level.  The PA and $i$ values are
straight-forward to model: the PA is well defined; the inclination
shows a modest increase in the inner parts, followed by a gradual
decline in the outer parts.  The curve as derived with $i$ and PA as
free parameters, is within the uncertainties equal to the one with $i$
and PA fixed to their model values.

\end{document}